\documentclass[11pt]{article}

\usepackage[a4paper, total={6in, 10in}]{geometry}

\usepackage{preamble/packages}
\usepackage{preamble/commands}

\begin{document}

\begin{titlepage}
   \begin{center}
       \vspace*{1cm}
        \Huge
       \textbf{Alakazam: Symbolic Tensor Algebra for Field Theory in Julia}
       
       \vspace{0.3cm}
         \LARGE
       \vspace{0.2cm}
     
\begin{center}

{\bf
Jesse Woods
} \\
\vspace{2mm}

{\small{\it 
	Albert Einstein Center for Fundamental Physics,\\
	Institute for Theoretical Physics, University of Bern,\\
	Sidlerstrasse 5, CH-3012 Bern, Switzerland}
}
\vspace{2mm}
~\\
\texttt{jesse.woods@unibe.ch}\\
\vspace{1mm}
\end{center}
       \vfill
             \Large
\includegraphics[width=0.6\textwidth]{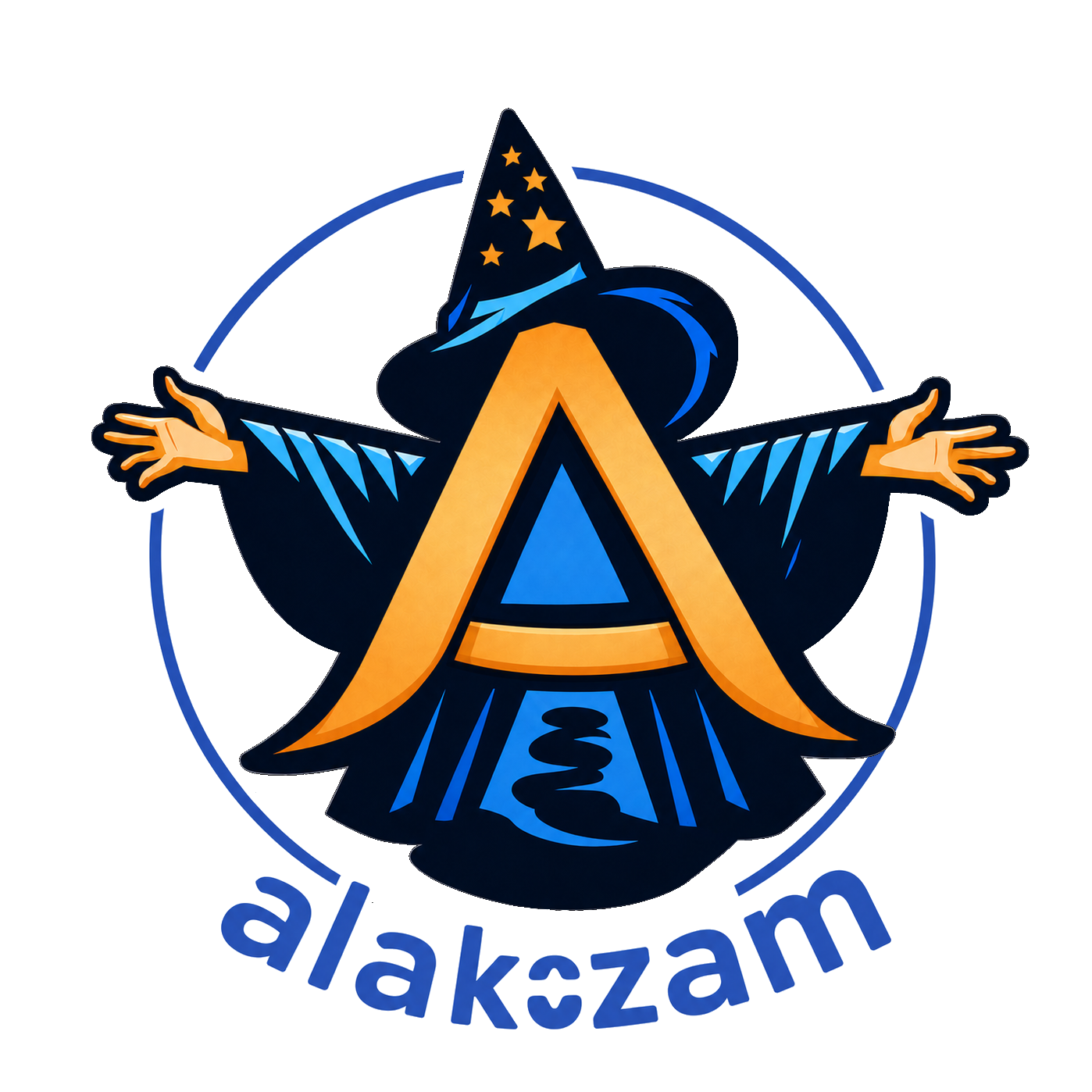}
       \vspace{0.2cm}
  
   \end{center}
   \begin{abstract}
\baselineskip=14pt
We present ``Alakazam'', a Julia package for performing symbolic manipulations of tensor expressions arising in the study of field theories and quantum mechanics. This will serve as a powerful tool for studying spacetime and supersymmetric structures, including support for calculations in superspace. The package implements many key features such as raising and lowering indices, simplification of terms with dummy indices, replacement patterns, differential and variational calculus, bosonic and fermionic grading, gamma matrix identities, and Lie brackets and commutators. We present notable speed improvements over established systems. Some possible applications include the study of descendants and invariants.

\end{abstract}
\newpage

\vspace{5mm}
\end{titlepage}

\tableofcontents
\newpage
\section{Introduction}
A computer algebra system (CAS) allows for the manipulation of symbolic mathematical expressions, and are indispensable tools for many computations in mathematics and physics. Calculations in field theory require specialised tools not generically available in general purpose symbolic packages, primarily due to the index summation notation on tensors. Furthermore, calculations with supersymmetric objects also require attentiveness to the anticommutativity of symbols.

Alakazam is a CAS designed for the manipulation of symbolic ``tensorial expressions'' that occur in field theory and quantum mechanics. It is furthermore explicitly designed from the ground up to correctly support grading on fermionic objects, including operators. The package consists of tensors, which carry symbolic abstract indices, as the basic building blocks from which larger expressions can be built. These basic tensors serve as ``supertypes'' for more specialised structures, for example operators which additionally carry an internal tensor expression (like in the case of derivatives or Lie brackets) or matrices which are non-commutative - just to name a few. Among its features, it has built-in understanding of Einstein dummy indices, versatile simplification algorithms, custom macros for writing easier code, import/export of expressions and other data via JSON, and conversion to LaTeX markup.

While the package is general (nothing in the core algebra assumes supersymmetry), its design was driven by the demands of superspace calculations, where grading and expression length are important considerations. In the superspace formalism, the number of Grassmann coordinates required grows with the number of supersymmetry generators, and since a superfield expansion contains every product of these coordinates, the number of terms grows rapidly. After many applications of derivatives, the number of terms in such an expression can quickly become impractical to manipulate by hand.
Grading also compounds this difficulty, as every object carries indices that may be bosonic or fermionic. In such large expressions, it is easy to introduce sign errors from line to line. These are difficult to detect and can propagate throughout a calculation.

This motivates the use of computer algebra software to perform such computations. There are many instances in the literature where CAS packages have been an indispensable tool for calculations, especially for supersymmetry. Historically, CAS tools have been exceptionally useful to tackle problems related to higher dimensional supergravity \cite{Peeters:2000qj,Peeters:2003pv}, and more recently problems in conformal supergravity \cite{Butter:2017jqu,Butter:2019edc,Casarin:2024qdn,Gold:2024gsj,Kennedy:2025nzm} and the calculation of invariants \cite{Butter:2018wss,Gold:2023ykx,Gold:2024nbw}.

Although there are several pieces of software that can do similar manipulations already (as well as some standalone packages for specialised features), see for example \cite{Peeters:2006kp,Peeters:2007wn,Peeters2018,xAct,Maple10,Gomez-Lobo:2011kaw,Bolotin:2013qgr,kuusela2019gammamap,Feger:2019tvk,Gates:2020qtw,Fonseca:2020vke,Frob:2020gdh,Gaudel:2025vce,Chen:2025zcm}, each having their own strengths, they are not without limitations; sometimes they are unintuitive to use or use idiosyncratic syntax, require licensed software, no longer receive updates or support, are difficult to adapt or extend, don't handle graded objects natively (or act inconsistently when performing operations like the graded Leibniz rule), don't allow for the use of multi-index derivatives, have incomplete support for some common useful operations, or are too slow for large expressions. For these reasons, and inspired by these systems, we opted to create our own software package that tries to address some of these issues while supporting the operations needed to perform algebraic manipulations, explicitly with the supersymmetric setting in mind. In particular, Alakazam distinguishes itself from adjacent systems via being open-source, its compatibility with unicode, customisability of index properties, robust grading, multi-index derivatives, treatment of commutators, and the use of graph isomorphism over the Butler-Portugal algorithm. It is also open-source and without licensing. Via our implementation, we achieve consistent, tested behaviour, as well as up to orders-of-magnitude speed increases compared to some popular existing systems on the operations we have benchmarked. 

We note that Alakazam is not without its own limitations, which we discuss in the implementation section. Notably, we do not currently have component computations implemented, and expressions are stored as lists rather than as a tree. It is also a new package, and may lack some features and optimisations of more mature systems. As it currently stands, it seems that no single CAS is entirely satisfactory according to the criteria of \cite{Korolkova:2013diz}. They suggest that a CAS should be able to handle non-index computations, where symmetry problems of the objects are direct properties of the tensor; abstract index computations, which express richer algebraic information and are the primary sort of operations we consider here; and component calculations, in specific coordinate system with a given metric. We aim to further address such limitations in future releases.

This package is written in Julia. Julia is a high-level computer language that can achieve speeds comparable to low-level languages \cite{bezanson2012julia,bezanson2015julia} for well-written, type-stable code. It aims to address the two-language problem: what is convenient for a developer to write is not necessarily convenient for the machine run. It was designed with scientific computing in mind, and thus includes many convenient features for those in the sciences. This language was thus chosen as it possesses the advantages of being fast, modular (and exists in an environment with many useful packages), easy to program in, has the ability to use Unicode characters when writing code, and has strong meta-programming capabilities. It is also syntactically simple for basic code - comparable to Python, which aids accessibility. Speed is important when analysing long and complex expressions like those that arise in our use cases. For example, this matters when manipulating the long expressions that arise in extended superspace calculations, where a single integration can generate hundreds if not thousands of terms. The use of Unicode characters is also convenient in a mathematics/physics context as the user can name variables with Greek letters and other mathematical characters such as $\nabla$. One main challenge in the design of tensor algebra software is the soft requirement that user inputs should try to resemble ``what one would do with pen and paper". Many systems for this reason use some form of LaTeX-style inputs. However, this has the downside of not being particularly readable. As the Julia REPL, and many editors like VSCode, allow for the user to type things like \jil{\alpha} which can be tab-completed to the unicode symbol \jil{α}, this allows the user to not only transcribe mathematics in a familiar way, but also for the resulting code to be much more readable than explicit TeX markup. This is further extended by the ability for Julia string macros to automatically parse text to structures, which can be used to convert LaTeX expressions directly to usable code. Additionally, Julia's type system is a natural fit for mathematical structures since mathematical objects are already organised hierarchically. For example, a partial derivative is a type of derivative that commutes with other partial derivatives; a derivative is an operator that obeys a derivation (Leibniz) rule; a Leibniz operator is an operator. Julia's abstract types let hierarchies of this sort be stated directly, and methods defined on one abstract type descend to all subtypes unless explicitly overwritten. A Lie bracket, for instance, is not a derivative but \emph{is} a derivation, so it shares the Leibniz rule - in Alakazam, the product rule is implemented once, against the abstract type of Leibniz operators, and is inherited by partial derivatives, covariant derivatives, variational derivatives and Lie brackets alike, despite their behaviour differing in every other respect. This also makes Alakazam easily extensible: any user can create their own flavour of tensor as a subtype of the existing hierarchy, and it will work with the whole library - all existing functions dispatch on supertypes, so no existing code needs modifying. The only genuinely new code that would need to be written is that encoding new behaviour specific to the user-defined type. 

In this paper, we aim to outline the basic features of Alakazam with enough examples to serve as a tutorial of sorts. We will also present an overview of the details of implementation for the larger, more sophisticated algorithms. At the end of the document, we present a more technical breakdown of the internals of Alakazam and the design choices that have been made. We include here also benchmarks against the popular computer algebra packages Cadabra\cite{Peeters2018}, xAct\cite{xAct} (in Mathematica, based on xPerm\cite{MartinGarcia:2008xPerm}), and Redberry\cite{Bolotin:2013qgr}, for some basic operations to serve as indicative examples. We find Alakazam tends to outperform the other packages in terms of speed for these operations. 

In Section \ref{sec:basic-usage} we overview installation of Alakazam and introduce the basic objects of the package, acting as a quick-start guide. In Section \ref{sec:special-tensors} we introduce tensors that have specialised functionality such as symmetries, non-commutativity, or contractions. We also explain the macro system for easier creation of tensor objects. In Section \ref{sec:tensor-operations} we introduce some key utility functions for the manipulation and simplification of tensor expressions. Section \ref{sec:operators} explains the special ``Operator'' class of tensors which encode behaviour such as derivatives, Lie brackets, and traces. In Section \ref{sec:pattern-matching-and-replacements} we explain the syntax for pattern matching to perform replacements in tensor expressions. Section \ref{sec:variational-calculus} introduced variational calculus and Euler-Lagrange equation functionality. In Section \ref{sec:differential-geometry} we catalogue the basic differential geometry operations included with Alakazam. In Section \ref{sec:serialisation-and-saving} we overview Alakazam's ability to export and import expressions to and from JSON. In Section \ref{sec:supersymmetry-and-superspace} we introduce a set of features specialised for performing operations in superspace, as well as the \jil{Alakazam.Lightcone} submodule. In Section \ref{sec:extended-examples} we provide explicit examples of problems that can be solved using a combination of features in the package. In Section \ref{sec:implementation} we discuss the more technical side of the implementation and comment on design choices made to improve efficiency. We provide some indicative benchmarks in Section \ref{sec:benchmarks} alongside other systems. We summarise in Section \ref{sec:conclusion} with some future directions. In the Appendix \ref{sec:export-index}, we include an index of all the functions and structures exported by the package for user-level access, including useful short-hand aliases, and back-links to relevant sections of the text.
\section{Basic Usage}\label{sec:basic-usage}
In this section we will introduce how to install and use the software package from the first run, highlighting the various features. We assume some small familiarity with \jil{Julia} syntax.

\subsection{Dependencies}\label{sec:dependencies}
Note that the package has dependencies 
\jil{Combinatorics}, \jil{DataStructures}, \jil{Graphs}, \jil{InteractiveUtils}, \jil{JSON}, \jil{Latexify}, \jil{OrderedCollections}, 
\jil{PrecompileTools}, \jil{SymbolicUtils}, and \jil{Symbolics}. These are installed automatically with the installation of Alakazam.
\subsection{Setup}\label{sec:setup}
The package can be installed directly from the Julia package manager within the terminal or script. Alternatively, the package can be found on GitLab \href{https://gitlab.com/B0bGary/alakazam.jl}{here}. Inside the Julia REPL (Read-Eval-Print Loop), press \jil{]} to enter package mode and run
\begin{Verbatim}[commandchars=\\\{\},xleftmargin=\parindent,numbers=left,bgcolor=bg]
\PYG{+w}{    }\PYG{n}{add}\PYG{+w}{ }\PYG{n}{Alakazam}
\end{Verbatim}
To install from inside a script, one may instead include at the beginning of the file
\begin{Verbatim}[commandchars=\\\{\},xleftmargin=\parindent,numbers=left,bgcolor=bg]
\PYG{+w}{    }\PYG{k}{using}\PYG{+w}{ }\PYG{n}{Pkg}
\PYG{+w}{    }\PYG{n}{Pkg}\PYG{o}{.}\PYG{n}{add}\PYG{p}{(}\PYG{l+s}{\PYGZdq{}}\PYG{l+s}{Alakazam}\PYG{l+s}{\PYGZdq{}}\PYG{p}{)}
\end{Verbatim}
Once installed, the package is loaded with
\begin{Verbatim}[commandchars=\\\{\},xleftmargin=\parindent,numbers=left,bgcolor=bg]
\PYG{+w}{    }\PYG{k}{using}\PYG{+w}{ }\PYG{n}{Alakazam}
\end{Verbatim}
This will automatically import all functions and internal variables. For example, the package automatically creates shorthand variables \jil{upper} and \jil{lower} (or \jil{↑} and \jil{↓} for even shorter-hand), which are used to denote index positions when creating a tensor, and an instance of the empty expression accessed with \jil{zero()}.

\subsection{Fundamentals}\label{sec:fundamentals}
At the core level, this software is designed to manipulate tensor expressions containing indices. Thus, unless your theory contains only scalars, one must define indices and index sets. 

\subsubsection{\jil{IndexSet}}\label{sec:indexset}
Firstly, one must create an \jil{IndexSet} object. Index sets are necessary, as they tell the package which indices can be drawn upon as dummy indices in Einstein summation notation, or whether a replacement pattern (discussed later) is equivalent to an expression modulo index names. One is permitted to use multiple kinds of index sets in the creation of tensors, for example, spacetime indices, spinor indices, flavour indices, etc. An \jil{IndexSet} is also permitted to include a mix of character types (e.g. Latin, Greek) if the user so desires. By separating into different sets, these indices can be assigned extra properties such as grading and allow for greater flexibility.
The basic constructor is given by
\begin{Verbatim}[commandchars=\\\{\},xleftmargin=\parindent,numbers=left,bgcolor=bg]
\PYG{+w}{    }\PYG{n}{IndexSet}\PYG{p}{(}\PYG{n}{name}\PYG{o}{::}\PYG{k+kt}{String}\PYG{p}{,}\PYG{+w}{ }\PYG{n}{range}\PYG{o}{::}\PYG{k+kt}{Int}\PYG{o}{=}\PYG{l+m+mi}{4}\PYG{p}{)}
\end{Verbatim}
Here, \jil{name} is the name of the index set, \jil{range} is the number of values the index ranges over in summation (i.e. the dimension of the space the indices represent). There are, in addition, many optional arguments that can be included in the constructor for more customisable index behaviour. Arguments are dispatched by type and are thus order independent, allowing for flexibility. We collect these in Table~\ref{tab:indexset} and thus establish our conventions. Combinations of these fields also imply complex conjugation behaviour of tensors (see~\ref{sec:dagger-and-complex-conjugation}).

\begin{table}[htbp]
  \centering
  \small
  \begin{tabular}{@{}p{2.1cm} p{3.5cm} p{3.4cm} p{4.6cm}@{}}
    \toprule
    \textbf{Property} & \textbf{Type} & \textbf{Example} & \textbf{Default} \\
    \midrule
    \textbf{Range}, or dimension &
      \jil{Int} &
      \jil{4} &
      \jil{4} \\\midrule
    \textbf{Signature} $(p,q)$, for $p$ minuses and $q$ pluses &
      \jil{Tuple{Int,Int}} or \jil{Vector{Int}} of length $2$ &
      \jil{(1,3)}, \jil{[0,4]} &
      \jil{(1,range-1)} for a symmetric metric (i.e., Lorentzian); \jil{(0,range)} otherwise \\\midrule
    \textbf{Metric} &
      \jil{Type{<:MetricSuperType}} &
      \jil{Metric}, \jil{EpsilonTensor}, \jil{SymplecticForm} &
      \jil{Metric} \\\midrule
    Canonical \textbf{Position} &
      \jil{IndexPosition} &
      \jil{upper}, \jil{lower} (or \jil{↑}, \jil{↓}) &
      \jil{upper} \\\midrule
    \textbf{Grading} &
      \jil{IndexGrading} &
      \jil{bosonic}, \jil{fermionic} &
      \jil{bosonic} \\\midrule
    \textbf{Lie algebra}, for invariant tensors and complex conjugation behaviour &
      \jil{LieAlgebraSuperType} &
      \jil{SUAlgebra(2)}, \jil{SOAlgebra(5)}, \jil{SpAlgebra(4)}, \jil{Spin(4)} &
      \jil{none}; when given, it also fixes the metric, grading, and the range if not supplied\\\midrule
      Complex \textbf{Conjugate} index set &
      \jil{IndexSet} &
      \jil{IndexSet("spinordot")} &
      \jil{none} \\ \midrule
    \textbf{Lightcone} &
      \jil{Bool} &
      \jil{true} &
      \jil{true} exactly when \jil{name} is \jil{"lightcone"}, \jil{false} otherwise \\
    \bottomrule
  \end{tabular}
    \caption{Optional arguments to the \jil{IndexSet} constructor.}
    \label{tab:indexset}
\end{table}

To instantiate an object of this kind, one could use the below code, for example
\begin{Verbatim}[commandchars=\\\{\},xleftmargin=\parindent,numbers=left,bgcolor=bg]
\PYG{+w}{    }\PYG{n}{spacetime\PYGZus{}indices}\PYG{+w}{ }\PYG{o}{=}\PYG{+w}{ }\PYG{n}{IndexSet}\PYG{p}{(}\PYG{l+s}{\PYGZdq{}}\PYG{l+s}{spacetime}\PYG{l+s}{\PYGZdq{}}\PYG{p}{,}\PYG{+w}{ }\PYG{l+m+mi}{4}\PYG{p}{)}
\PYG{+w}{    }\PYG{n}{euclidean\PYGZus{}indices}\PYG{+w}{ }\PYG{o}{=}\PYG{+w}{ }\PYG{n}{IndexSet}\PYG{p}{(}\PYG{l+s}{\PYGZdq{}}\PYG{l+s}{euclidean\PYGZus{}spacetime}\PYG{l+s}{\PYGZdq{}}\PYG{p}{,}\PYG{+w}{ }\PYG{l+m+mi}{4}\PYG{p}{,}\PYG{+w}{ }\PYG{p}{(}\PYG{l+m+mi}{0}\PYG{p}{,}\PYG{l+m+mi}{4}\PYG{p}{)}\PYG{p}{)}
\PYG{+w}{    }\PYG{n}{su2flavour\PYGZus{}indices}\PYG{+w}{ }\PYG{o}{=}\PYG{+w}{ }\PYG{n}{IndexSet}\PYG{p}{(}\PYG{l+s}{\PYGZdq{}}\PYG{l+s}{flavour}\PYG{l+s}{\PYGZdq{}}\PYG{p}{,}\PYG{n}{lower}\PYG{p}{,}\PYG{n}{SUAlgebra}\PYG{p}{(}\PYG{l+m+mi}{2}\PYG{p}{)}\PYG{p}{)}
\PYG{+w}{    }\PYG{n}{spinor\PYGZus{}indices}\PYG{+w}{ }\PYG{o}{=}\PYG{+w}{ }\PYG{n}{IndexSet}\PYG{p}{(}\PYG{l+s}{\PYGZdq{}}\PYG{l+s}{spinor}\PYG{l+s}{\PYGZdq{}}\PYG{p}{,}\PYG{+w}{ }\PYG{l+m+mi}{2}\PYG{p}{,}\PYG{n}{upper}\PYG{p}{,}\PYG{n}{EpsilonTensor}\PYG{p}{,}\PYG{n}{fermionic}\PYG{p}{)}
\PYG{+w}{    }\PYG{n}{majorana\PYGZus{}indices}\PYG{+w}{ }\PYG{o}{=}\PYG{+w}{ }\PYG{n}{IndexSet}\PYG{p}{(}\PYG{l+s}{\PYGZdq{}}\PYG{l+s}{(2+1)\PYGZhy{}majorana}\PYG{l+s}{\PYGZdq{}}\PYG{p}{,}\PYG{n}{Spin}\PYG{p}{(}\PYG{l+m+mi}{3}\PYG{p}{)}\PYG{p}{)}
\end{Verbatim}

Importantly, in this software, information on the grading, i.e. whether tensors commute or anti-commute, is encoded in the indices\footnote{There also exist explicit \jil{Boson} and \jil{Fermion} types \jil{<:TensorSuperType} that override this grading if needed, see Section~\ref{sec:bosonfermion}.}. 

\subsubsection{\jil{Index}}\label{sec:index}

Now, one must create the indices that can be attached to tensors. The user should ensure that they create enough indices in order to perform all algorithms (for example, many extra indices to be used as dummy indices may be needed depending on the algorithm) required for their problem. If there are not enough indices defined for internal use, Alakazam will automatically try  when possible to generate new indices, using the same set of characters (e.g. Greek lowercase, dotted Latin uppercase, etc) as the other indices in the \jil{IndexSet}.
The constructor is given by
\begin{Verbatim}[commandchars=\\\{\},xleftmargin=\parindent,numbers=left,bgcolor=bg]
\PYG{+w}{    }\PYG{n}{Index}\PYG{p}{(}\PYG{n}{name}\PYG{o}{::}\PYG{k+kt}{String}\PYG{p}{,}\PYG{n}{index\PYGZus{}set}\PYG{o}{::}\PYG{k+kt}{IndexSet}\PYG{p}{)}
\end{Verbatim}

Here, \jil{name} will be used to identify the index and should be a single unique character. Attempting to create a new index with the same name will return the original.
There are no restrictions on the name of this index, except that it should not use special characters \jil{* ? _ {  } [ ] ^ ( ) ∘} which are all reserved for internal pattern matching logic. This means that unicode characters are also valid, including dotted, hatted, barred, and tilded characters.
Upon construction, the new index will be automatically assigned to the specified \jil{IndexSet}.
For example,
\begin{Verbatim}[commandchars=\\\{\},xleftmargin=\parindent,numbers=left,bgcolor=bg]
\PYG{+w}{    }\PYG{n}{i}\PYG{+w}{ }\PYG{o}{=}\PYG{+w}{ }\PYG{n}{Index}\PYG{p}{(}\PYG{l+s}{\PYGZdq{}}\PYG{l+s}{i}\PYG{l+s}{\PYGZdq{}}\PYG{p}{,}\PYG{+w}{ }\PYG{n}{flavour\PYGZus{}indices}\PYG{p}{)}
\PYG{+w}{    }\PYG{n}{j}\PYG{+w}{ }\PYG{o}{=}\PYG{+w}{ }\PYG{n}{Index}\PYG{p}{(}\PYG{l+s}{\PYGZdq{}}\PYG{l+s}{j}\PYG{l+s}{\PYGZdq{}}\PYG{p}{,}\PYG{+w}{ }\PYG{n}{flavour\PYGZus{}indices}\PYG{p}{)}
\PYG{+w}{    }\PYG{n}{α}\PYG{+w}{ }\PYG{o}{=}\PYG{+w}{ }\PYG{n}{Index}\PYG{p}{(}\PYG{l+s}{\PYGZdq{}}\PYG{l+s}{α}\PYG{l+s}{\PYGZdq{}}\PYG{p}{,}\PYG{+w}{ }\PYG{n}{spinor\PYGZus{}indices}\PYG{p}{)}
\PYG{+w}{    }\PYG{n}{αdot}\PYG{+w}{ }\PYG{o}{=}\PYG{+w}{ }\PYG{n}{Index}\PYG{p}{(}\PYG{l+s}{\PYGZdq{}}\PYG{l+s}{̇α}\PYG{l+s}{\PYGZdq{}}\PYG{p}{,}\PYG{+w}{ }\PYG{n}{spinor\PYGZus{}conj\PYGZus{}indices}\PYG{p}{)}
\end{Verbatim}

In order to create many indices at once, we can also do something like
\begin{Verbatim}[commandchars=\\\{\},xleftmargin=\parindent,numbers=left,bgcolor=bg]
\PYG{+w}{    }\PYG{n}{α}\PYG{p}{,}\PYG{n}{β}\PYG{p}{,}\PYG{n}{γ}\PYG{p}{,}\PYG{n}{χ}\PYG{+w}{ }\PYG{o}{=}\PYG{+w}{ }\PYG{n}{Index}\PYG{p}{(}\PYG{p}{[}\PYG{l+s}{\PYGZdq{}}\PYG{l+s}{α}\PYG{l+s}{\PYGZdq{}}\PYG{p}{,}\PYG{l+s}{\PYGZdq{}}\PYG{l+s}{β}\PYG{l+s}{\PYGZdq{}}\PYG{p}{,}\PYG{l+s}{\PYGZdq{}}\PYG{l+s}{γ}\PYG{l+s}{\PYGZdq{}}\PYG{p}{,}\PYG{l+s}{\PYGZdq{}}\PYG{l+s}{χ}\PYG{l+s}{\PYGZdq{}}\PYG{p}{]}\PYG{p}{,}\PYG{+w}{ }\PYG{n}{spinor\PYGZus{}indices}\PYG{p}{)}
\end{Verbatim}
To create a large number of indices, there is also a helpful macro
\begin{Verbatim}[commandchars=\\\{\},xleftmargin=\parindent,numbers=left,bgcolor=bg]
\PYG{+w}{    }\PYG{n+nd}{@indices}\PYG{+w}{ }\PYG{n}{a}\PYG{o}{:}\PYG{n}{f∈flavour\PYGZus{}indices}
\PYG{+w}{    }\PYG{n+nd}{@indices}\PYG{+w}{ }\PYG{n}{α}\PYG{o}{:}\PYG{n}{γ∈spinor\PYGZus{}indices}
\end{Verbatim}
which will automatically declare indices with names between the specified range of characters, and are immediately assigned as variables in the user scope. Due to limitations of prefixing, one cannot name a variable literally \jil{̇α}, however the macro still supports names of the form
\begin{Verbatim}[commandchars=\\\{\},xleftmargin=\parindent,numbers=left,bgcolor=bg]
\PYG{+w}{    }\PYG{n+nd}{@indices}\PYG{+w}{ }\PYG{n}{αdot}\PYG{o}{:}\PYG{n}{γdot∈some\PYGZus{}indexset}
\PYG{+w}{    }\PYG{n+nd}{@indices}\PYG{+w}{ }\PYG{n}{abar}\PYG{o}{:}\PYG{n}{cbar∈some\PYGZus{}indexset}
\PYG{+w}{    }\PYG{n+nd}{@indices}\PYG{+w}{ }\PYG{n}{αhat}\PYG{o}{:}\PYG{n}{γhat∈some\PYGZus{}indexset}
\PYG{+w}{    }\PYG{n+nd}{@indices}\PYG{+w}{ }\PYG{n}{itilde}\PYG{o}{:}\PYG{n}{ktilde∈some\PYGZus{}indexset}
\end{Verbatim}
which will name the indices with the appropriately prefixed character. Such naming is also supported in macros (see~\ref{sec:macros}). 

\subsubsection{\jil{Coordinate}}\label{sec:coordinate}
If the user wishes to study tensor fields, rather than constant tensors, coordinates can be defined. Tensors can then be specified to be a function of these coordinates. This information is used later in computing derivatives, for example. The coordinate constructor is given by

\begin{Verbatim}[commandchars=\\\{\},xleftmargin=\parindent,numbers=left,bgcolor=bg]
\PYG{+w}{    }\PYG{n}{Coordinate}\PYG{p}{(}\PYG{n}{name}\PYG{o}{::}\PYG{k+kt}{String}\PYG{p}{,}\PYG{+w}{ }\PYG{n}{index\PYGZus{}set}\PYG{o}{::}\PYG{k+kt}{IndexSet}\PYG{p}{)}
\PYG{+w}{    }\PYG{n}{Coordinate}\PYG{p}{(}\PYG{n}{name}\PYG{o}{::}\PYG{k+kt}{String}\PYG{p}{,}\PYG{+w}{ }\PYG{n}{index\PYGZus{}set}\PYG{o}{::}\PYG{k+kt}{Vector}\PYG{p}{\PYGZob{}}\PYG{k+kt}{IndexSet}\PYG{p}{\PYGZcb{}}\PYG{p}{)}
\end{Verbatim}

Here, \jil{name} is the name of the coordinate, which will be used when printing an expression. \jil{index_set} can be either a single \jil{IndexSet} or a vector of \jil{IndexSet}s, which allow for coordinates to have multiple associated indices. For example, one may wish to represent a vector with one spacetime index in terms of two spinor indices (bispinor convention), or to represent Grassmann coordinates on superspace. As an example in practice,
\begin{Verbatim}[commandchars=\\\{\},xleftmargin=\parindent,numbers=left,bgcolor=bg]
\PYG{+w}{    }\PYG{n}{x}\PYG{+w}{ }\PYG{o}{=}\PYG{+w}{ }\PYG{n}{Coordinate}\PYG{p}{(}\PYG{l+s}{\PYGZdq{}}\PYG{l+s}{x}\PYG{l+s}{\PYGZdq{}}\PYG{p}{,}\PYG{+w}{ }\PYG{n}{spacetime\PYGZus{}indices}\PYG{p}{)}
\PYG{+w}{    }\PYG{n}{θ}\PYG{+w}{ }\PYG{o}{=}\PYG{+w}{ }\PYG{n}{Coordinate}\PYG{p}{(}\PYG{l+s}{\PYGZdq{}}\PYG{l+s}{θ}\PYG{l+s}{\PYGZdq{}}\PYG{p}{,}\PYG{+w}{ }\PYG{p}{[}\PYG{n}{flavour\PYGZus{}indices}\PYG{p}{,}\PYG{+w}{ }\PYG{n}{spinor\PYGZus{}indices}\PYG{p}{]}\PYG{p}{)}
\end{Verbatim}

\subsubsection{\jil{Tensor}}\label{sec:tensor}
The package contains a hierarchy of \jil{Tensorial} objects to build expressions on which to perform tensor algebra. At the lowest level, there are \jil{Tensor} objects, representing a single tensor. These single tensors can then be multiplied or contracted together to create a \jil{TensorTerm}. These terms can then be added together with numerical or symbolic coefficients to get a \jil{TensorExpression}. Each of these is a subtype of \jil{Tensorial}. Internally, the package automatically will parse a \jil{Tensor} or \jil{TensorTerm} as a \jil{TensorExpression} as needed.

There are many ways to create a \jil{Tensor} in \jil{Alakazam}, including many optional parameters designed to accommodate different use cases. For a simple case of a tensor field with indices, one can use the constructor
\begin{Verbatim}[commandchars=\\\{\},xleftmargin=\parindent,numbers=left,bgcolor=bg]
\PYG{n}{Tensor}\PYG{p}{(}\PYG{n}{name}\PYG{o}{::}\PYG{k+kt}{String}\PYG{p}{,}\PYG{+w}{ }\PYG{n}{indices}\PYG{o}{::}\PYG{k+kt}{Vector}\PYG{p}{\PYGZob{}}\PYG{k+kt}{Pair}\PYG{p}{\PYGZob{}}\PYG{k+kt}{IndexSuperType}\PYG{p}{,}\PYG{+w}{ }\PYG{k+kt}{IndexPosition}\PYG{p}{\PYGZcb{}}\PYG{p}{\PYGZcb{}}\PYG{p}{,}\PYG{+w}{ }
\PYG{+w}{    }\PYG{n}{function\PYGZus{}of}\PYG{o}{::}\PYG{k+kt}{Coordinate}\PYG{p}{)}
\end{Verbatim}
Here, \jil{name} is the string that represents the tensor when doing pattern matching or printing to display. \jil{indices} is a vector containing pairs of indices and their positions. \jil{function_of} is used to specify that the \jil{Tensor} is a tensor field, dependent on the specified coordinates.
For a basic usage example, consider
\begin{Verbatim}[commandchars=\\\{\},xleftmargin=\parindent,numbers=left,bgcolor=bg]
\PYG{+w}{     }\PYG{n}{A}\PYG{+w}{ }\PYG{o}{=}\PYG{+w}{ }\PYG{n}{Tensor}\PYG{p}{(}\PYG{l+s}{\PYGZdq{}}\PYG{l+s}{A}\PYG{l+s}{\PYGZdq{}}\PYG{p}{,}\PYG{+w}{ }\PYG{p}{[}\PYG{n}{i}\PYG{+w}{ }\PYG{o}{=\PYGZgt{}}\PYG{+w}{ }\PYG{n}{upper}\PYG{p}{,}\PYG{+w}{ }\PYG{n}{j}\PYG{+w}{ }\PYG{o}{=\PYGZgt{}}\PYG{+w}{ }\PYG{n}{lower}\PYG{p}{]}\PYG{p}{,}\PYG{+w}{ }\PYG{n}{x}\PYG{p}{)}
\end{Verbatim}
This encodes the tensor field 
\begin{equation}
    A^{i}{ }_{j}(x).
\end{equation}
It is important to stress that index order and placement is important. There are no implicit assumptions about symmetry in the indices, which means
\begin{equation}
    A^{i}{ }_{j}(x)\neq A_{j}{ }^{i}(x)\neq  A_{i}{ }^{j}(x)\neq  A^{j}{}_{i}(x)\neq A_{ij}(x)\neq A_{ji}(x)\neq A^{ij}(x)\neq A^{ji}(x).
\end{equation}
One should keep this in mind when using the package, as it will generically treat a tensor with a lower then upper index, differently from a tensor with an upper then lower index (for example, when doing algebra or pattern matching).

In addition to the above, one may also use the \jil{@Tensor} macro to allow for shorter inputs. For example, we can encode the tensor
\begin{equation}
    A_{ij}{}^{k}{}_l(x,\theta)
\end{equation}
using macros as
\begin{Verbatim}[commandchars=\\\{\},xleftmargin=\parindent,numbers=left,bgcolor=bg]
\PYG{+w}{   }\PYG{n+nd}{@Tensor}\PYG{+w}{ }\PYG{n}{A\PYGZus{}ij}\PYG{o}{\PYGZca{}}\PYG{k+kt}{k\PYGZus{}l}\PYG{p}{\PYGZob{}}\PYG{k+kt}{x}\PYG{p}{,}\PYG{k+kt}{θ}\PYG{p}{\PYGZcb{}}
\end{Verbatim}
which is interpreted exactly the same as the less compact expression
\begin{Verbatim}[commandchars=\\\{\},xleftmargin=\parindent,numbers=left,bgcolor=bg]
\PYG{+w}{    }\PYG{n}{Tensor}\PYG{p}{(}\PYG{l+s}{\PYGZdq{}}\PYG{l+s}{A}\PYG{l+s}{\PYGZdq{}}\PYG{p}{,}\PYG{+w}{ }\PYG{p}{[}\PYG{n}{i}\PYG{+w}{ }\PYG{o}{=\PYGZgt{}}\PYG{+w}{ }\PYG{n}{lower}\PYG{p}{,}\PYG{+w}{ }\PYG{n}{j}\PYG{+w}{ }\PYG{o}{=\PYGZgt{}}\PYG{+w}{ }\PYG{n}{lower}\PYG{p}{,}\PYG{n}{k}\PYG{o}{=\PYGZgt{}}\PYG{n}{upper}\PYG{p}{,}\PYG{n}{l}\PYG{o}{=\PYGZgt{}}\PYG{n}{lower}\PYG{p}{]}\PYG{p}{,}\PYG{p}{[}\PYG{n}{x}\PYG{p}{,}\PYG{n}{θ}\PYG{p}{]}\PYG{p}{)}
\end{Verbatim}
For a more extensive explanation of the macro syntax, see~\ref{sec:macros}. 

Similar to the \jil{IndexSet}, the \jil{Tensor} constructor can accept many arguments, which are dispatched by type. See Table~\ref{tab:tensor} for a list of the possible parameters.

\begin{table}[H]
  \centering
  \small
  \begin{tabular}{@{}p{2.7cm} p{3.9cm} p{3.3cm} p{3.6cm}@{}}
    \toprule
    \textbf{Property} & \textbf{Type} & \textbf{Example} & \textbf{Default} \\
    \midrule
    \textbf{Index}, in a given position &
      \jil{Pair{Index,IndexPosition}} &
      \jil{μ=>upper}, \jil{α=>↓} &
      none \\
    \midrule
    Several \textbf{Indices} at once &
      \jil{Vector} of the above &
      \jil{[μ=>upper, ν=>lower]} &
      no indices, i.e. a scalar \\
    \midrule
    \textbf{Coordinate} dependence &
      \jil{Coordinate}, or a \jil{Vector} of them &
      \jil{x}, \jil{[x,θ]} &
      none, constant \\
    \midrule
    Symmetry encoded by \textbf{Young tableaux} &
      \jil{YoungTableau}, or \jil{Vector{YoungTableau}} &
      See~\ref{sec:symmetries-and-youngtableau} &
      no symmetry \\
    \midrule
    \textbf{Label}, distinguishing objects of the same name &
      \jil{String}, or \jil{Pair{String,IndexPosition}}, or a \jil{Vector} of those &
      \jil{"1"}, \jil{"bare"=>upper} &
      unlabelled \\
    \midrule
    \textbf{Weights} encoding extra properties &
      \jil{Dict{String,Number}} &
      \jil{Dict("x"=>2)} &
      $1$ for each coordinate the tensor is a function of \\
    \midrule
    \textbf{Number field} over which the tensor is defined &
      \jil{NumberSet}, or the bare alias &
      \jil{ℂ}, \jil{iℝ}, \jil{ℝ()} &
      \jil{ℝ} for tensors, \jil{ℂ} for matrices \\
    \bottomrule
  \end{tabular}
  \caption{Optional arguments to the \jil{Tensor} constructor, shared by every tensor-like type in  the package. Their order is immaterial.}
  \label{tab:tensor}
\end{table}

\subsubsection{Einstein Summation and Dummy Indices}\label{sec:einstein-summation-and-dummy-indices}

When an index is repeated in a tensorial expression, Einstein summation notation is implied. By convention, this summation is on a pair of upper and lower (or covariant and contravariant) indices. Alakazam respects these conventions. Additionally, repeated indices may only appear twice in an expression, otherwise it is ambiguous as to between which indices the summation is over. Alakazam detects these repetitions, and an \jil{OverusedIndices} exception will be thrown. Furthermore, one of the indices must be in the \jil{upper} position and the other \jil{lower}, otherwise a \jil{SameLevelDummyError} exception will be thrown. This is true even for Euclidean indices.
\begin{gather}
    A_{ij}\implies\textcolor{mygreen}{\text{valid}}\\
    A_{i}{}^{i}\implies\textcolor{mygreen}{\text{valid}}\\
    A_{i}{}_{i}\implies\textcolor{red}{\text{invalid - \jil{SameLevelDummyError}}}\\
    A_{i}{}_{j}B^i{}^jC^{i}{}_kD_i{}^{k}\implies\textcolor{red}{\text{invalid - \jil{OverusedIndices}}}
\end{gather}
An exception is made for the specialised lightcone indices, which we will discuss later in~\ref{sec:lightcone}. Dummy indices in expressions can be renamed with \jil{rename_dummies()} to help simplify expressions (see~\ref{sec:rename-dummies}).

\subsubsection{\jil{TensorTerm}}\label{sec:tensorterm}

Products of \jil{Tensor}s are stored as a \jil{TensorTerm}. This can be done simply through multiplication of two or more \jil{Tensor}s, for example
\begin{Verbatim}[commandchars=\\\{\},xleftmargin=\parindent,numbers=left,bgcolor=bg]
\PYG{+w}{      }\PYG{n}{A}\PYG{+w}{ }\PYG{o}{=}\PYG{+w}{ }\PYG{n}{Tensor}\PYG{p}{(}\PYG{l+s}{\PYGZdq{}}\PYG{l+s}{A}\PYG{l+s}{\PYGZdq{}}\PYG{p}{,}\PYG{+w}{ }\PYG{p}{[}\PYG{n}{i}\PYG{+w}{ }\PYG{o}{=\PYGZgt{}}\PYG{+w}{ }\PYG{n}{upper}\PYG{p}{,}\PYG{+w}{ }\PYG{n}{j}\PYG{+w}{ }\PYG{o}{=\PYGZgt{}}\PYG{+w}{ }\PYG{n}{lower}\PYG{p}{]}\PYG{p}{,}\PYG{+w}{ }\PYG{n}{x}\PYG{p}{)}
\PYG{+w}{      }\PYG{n}{B}\PYG{+w}{ }\PYG{o}{=}\PYG{+w}{ }\PYG{n}{Tensor}\PYG{p}{(}\PYG{l+s}{\PYGZdq{}}\PYG{l+s}{B}\PYG{l+s}{\PYGZdq{}}\PYG{p}{,}\PYG{+w}{ }\PYG{p}{[}\PYG{n}{j}\PYG{+w}{ }\PYG{o}{=\PYGZgt{}}\PYG{+w}{ }\PYG{n}{upper}\PYG{p}{,}\PYG{+w}{ }\PYG{n}{k}\PYG{+w}{ }\PYG{o}{=\PYGZgt{}}\PYG{+w}{ }\PYG{n}{lower}\PYG{p}{]}\PYG{p}{,}\PYG{+w}{ }\PYG{n}{x}\PYG{p}{)}
\PYG{+w}{      }\PYG{n}{AB}\PYG{+w}{ }\PYG{o}{=}\PYG{+w}{ }\PYG{n}{A}\PYG{o}{*}\PYG{n}{B}
\end{Verbatim}

Here, \jil{AB} will be a \jil{TensorTerm}, representing
\begin{equation}
A^{i}{ }_{j}(x)B^{j}{ }_{k}(x).
\end{equation}
 A \jil{TensorTerm} can also be constructed directly using either of the constructors
\begin{Verbatim}[commandchars=\\\{\},xleftmargin=\parindent,numbers=left,bgcolor=bg]
\PYG{+w}{    }\PYG{n}{TensorTerm}\PYG{p}{(}\PYG{n}{terms}\PYG{o}{::}\PYG{k+kt}{Vector}\PYG{p}{\PYGZob{}}\PYG{k+kt}{T}\PYG{p}{\PYGZcb{}}\PYG{+w}{ }\PYG{k}{where}\PYG{+w}{ }\PYG{p}{\PYGZob{}}\PYG{k+kt}{T}\PYG{o}{\PYGZlt{}:}\PYG{k+kt}{TensorSuperType}\PYG{p}{\PYGZcb{}}\PYG{p}{)}
\PYG{+w}{    }\PYG{n}{TensorTerm}\PYG{p}{(}\PYG{n}{term}\PYG{o}{::}\PYG{k+kt}{TensorSuperType}\PYG{p}{)}
\end{Verbatim}
Note that upon construction, the field \jil{indices} in \jil{TensorTerm} will be initialised to contain the full set of indices of each of the \jil{Tensor}s it contains in \jil{terms}.

\subsubsection{\jil{TensorExpression}}\label{sec:tensorexpression}

In order to represent a full equation involving tensors, one must also be able to add tensors and products of tensors. Such equation objects are stored as a \jil{TensorExpression}. The \jil{TensorExpression} object has many possible constructors for all possible ways of creating such an equation. However, it is more intuitive to create a \jil{TensorExpression} using arithmetic operations on \jil{Tensor}s.

\begin{Verbatim}[commandchars=\\\{\},xleftmargin=\parindent,numbers=left,bgcolor=bg]
\PYG{+w}{    }\PYG{n}{A}\PYG{+w}{ }\PYG{o}{=}\PYG{+w}{ }\PYG{n}{Tensor}\PYG{p}{(}\PYG{l+s}{\PYGZdq{}}\PYG{l+s}{A}\PYG{l+s}{\PYGZdq{}}\PYG{p}{,}\PYG{+w}{ }\PYG{p}{[}\PYG{n}{i}\PYG{+w}{ }\PYG{o}{=\PYGZgt{}}\PYG{+w}{ }\PYG{n}{upper}\PYG{p}{,}\PYG{+w}{ }\PYG{n}{j}\PYG{+w}{ }\PYG{o}{=\PYGZgt{}}\PYG{+w}{ }\PYG{n}{lower}\PYG{p}{]}\PYG{p}{,}\PYG{+w}{ }\PYG{n}{x}\PYG{p}{)}
\PYG{+w}{    }\PYG{n}{B}\PYG{+w}{ }\PYG{o}{=}\PYG{+w}{ }\PYG{n}{Tensor}\PYG{p}{(}\PYG{l+s}{\PYGZdq{}}\PYG{l+s}{B}\PYG{l+s}{\PYGZdq{}}\PYG{p}{,}\PYG{+w}{ }\PYG{p}{[}\PYG{n}{j}\PYG{+w}{ }\PYG{o}{=\PYGZgt{}}\PYG{+w}{ }\PYG{n}{upper}\PYG{p}{,}\PYG{+w}{ }\PYG{n}{k}\PYG{+w}{ }\PYG{o}{=\PYGZgt{}}\PYG{+w}{ }\PYG{n}{lower}\PYG{p}{]}\PYG{p}{,}\PYG{+w}{ }\PYG{n}{x}\PYG{p}{)}
\PYG{+w}{    }\PYG{n}{AB}\PYG{+w}{ }\PYG{o}{=}\PYG{+w}{ }\PYG{n}{A}\PYG{o}{*}\PYG{n}{B}\PYG{+w}{ }\PYG{c}{\PYGZsh{}This will be a TensorTerm}
\PYG{+w}{    }\PYG{n}{C}\PYG{+w}{ }\PYG{o}{=}\PYG{+w}{ }\PYG{n}{Tensor}\PYG{p}{(}\PYG{l+s}{\PYGZdq{}}\PYG{l+s}{C}\PYG{l+s}{\PYGZdq{}}\PYG{p}{,}\PYG{+w}{ }\PYG{p}{[}\PYG{n}{i}\PYG{+w}{ }\PYG{o}{=\PYGZgt{}}\PYG{+w}{ }\PYG{n}{upper}\PYG{p}{,}\PYG{+w}{ }\PYG{n}{k}\PYG{+w}{ }\PYG{o}{=\PYGZgt{}}\PYG{+w}{ }\PYG{n}{lower}\PYG{p}{]}\PYG{p}{,}\PYG{+w}{ }\PYG{n}{x}\PYG{p}{)}
\PYG{+w}{    }
\PYG{+w}{    }\PYG{n}{exp}\PYG{o}{=}\PYG{n}{AB}\PYG{o}{+}\PYG{l+m+mi}{2}\PYG{o}{*}\PYG{n}{C}\PYG{+w}{ }\PYG{c}{\PYGZsh{}This will be a TensorExpression}
\end{Verbatim}

The above example encodes the expression 
\begin{equation}
    A^{i}{ }_{j}(x)B^{j}{ }_{k}(x)+2C^{i}{ }_{k}(x).
\end{equation}
Note that any type of number, or symbol from the \jil{Symbolics} package~\cite{gowda2022highperformancesymbolicnumericsmultipledispatch}, may be used as coefficients for terms in an expression.
When creating a \jil{TensorExpression} in this way, the package will throw a \jil{FreeIndexMismatch} exception when trying to add together terms that have different free indices:
\begin{Verbatim}[commandchars=\\\{\},xleftmargin=\parindent,numbers=left,bgcolor=bg]
\PYG{+w}{    }\PYG{n}{AB}\PYG{+w}{ }\PYG{o}{=}\PYG{+w}{ }\PYG{p}{(}\PYG{n}{Tensor}\PYG{p}{(}\PYG{l+s}{\PYGZdq{}}\PYG{l+s}{A}\PYG{l+s}{\PYGZdq{}}\PYG{p}{,}\PYG{+w}{ }\PYG{p}{[}\PYG{n}{i}\PYG{+w}{ }\PYG{o}{=\PYGZgt{}}\PYG{+w}{ }\PYG{n}{upper}\PYG{p}{,}\PYG{+w}{ }\PYG{n}{j}\PYG{+w}{ }\PYG{o}{=\PYGZgt{}}\PYG{+w}{ }\PYG{n}{lower}\PYG{p}{]}\PYG{p}{,}\PYG{+w}{ }\PYG{n}{x}\PYG{p}{)}
\PYG{+w}{        }\PYG{o}{*}\PYG{n}{Tensor}\PYG{p}{(}\PYG{l+s}{\PYGZdq{}}\PYG{l+s}{B}\PYG{l+s}{\PYGZdq{}}\PYG{p}{,}\PYG{+w}{ }\PYG{p}{[}\PYG{n}{j}\PYG{+w}{ }\PYG{o}{=\PYGZgt{}}\PYG{+w}{ }\PYG{n}{upper}\PYG{p}{,}\PYG{+w}{ }\PYG{n}{k}\PYG{+w}{ }\PYG{o}{=\PYGZgt{}}\PYG{+w}{ }\PYG{n}{lower}\PYG{p}{]}\PYG{p}{,}\PYG{+w}{ }\PYG{n}{x}\PYG{p}{)}\PYG{p}{)}

\PYG{+w}{    }\PYG{n}{CD}\PYG{+w}{ }\PYG{o}{=}\PYG{+w}{  }\PYG{p}{(}\PYG{n}{Tensor}\PYG{p}{(}\PYG{l+s}{\PYGZdq{}}\PYG{l+s}{C}\PYG{l+s}{\PYGZdq{}}\PYG{p}{,}\PYG{+w}{ }\PYG{p}{[}\PYG{n}{i}\PYG{+w}{ }\PYG{o}{=\PYGZgt{}}\PYG{+w}{ }\PYG{n}{upper}\PYG{p}{,}\PYG{+w}{ }\PYG{n}{l}\PYG{+w}{ }\PYG{o}{=\PYGZgt{}}\PYG{+w}{ }\PYG{n}{lower}\PYG{p}{]}\PYG{p}{,}\PYG{+w}{ }\PYG{n}{x}\PYG{p}{)}
\PYG{+w}{        }\PYG{o}{*}\PYG{n}{Tensor}\PYG{p}{(}\PYG{l+s}{\PYGZdq{}}\PYG{l+s}{D}\PYG{l+s}{\PYGZdq{}}\PYG{p}{,}\PYG{+w}{ }\PYG{p}{[}\PYG{n}{l}\PYG{+w}{ }\PYG{o}{=\PYGZgt{}}\PYG{+w}{ }\PYG{n}{upper}\PYG{p}{,}\PYG{+w}{ }\PYG{n}{k}\PYG{+w}{ }\PYG{o}{=\PYGZgt{}}\PYG{+w}{ }\PYG{n}{lower}\PYG{p}{]}\PYG{p}{,}\PYG{+w}{ }\PYG{n}{x}\PYG{p}{)}\PYG{p}{)}
\PYG{+w}{    }\PYG{n}{exp1}\PYG{+w}{ }\PYG{o}{=}\PYG{+w}{ }\PYG{n}{AB}\PYG{o}{+}\PYG{n}{CD}\PYG{+w}{ }\PYG{c}{\PYGZsh{}This is ok, terms have the same free indices,}
\PYG{+w}{                }\PYG{c}{\PYGZsh{} even though they use different dummy indices}
\PYG{+w}{    }
\PYG{+w}{    }\PYG{n}{EF}\PYG{+w}{ }\PYG{o}{=}\PYG{+w}{  }\PYG{p}{(}\PYG{n}{Tensor}\PYG{p}{(}\PYG{l+s}{\PYGZdq{}}\PYG{l+s}{E}\PYG{l+s}{\PYGZdq{}}\PYG{p}{,}\PYG{+w}{ }\PYG{p}{[}\PYG{n}{i}\PYG{+w}{ }\PYG{o}{=\PYGZgt{}}\PYG{+w}{ }\PYG{n}{upper}\PYG{p}{,}\PYG{+w}{ }\PYG{n}{j}\PYG{+w}{ }\PYG{o}{=\PYGZgt{}}\PYG{+w}{ }\PYG{n}{lower}\PYG{p}{]}\PYG{p}{,}\PYG{+w}{ }\PYG{n}{x}\PYG{p}{)}
\PYG{+w}{        }\PYG{o}{*}\PYG{n}{Tensor}\PYG{p}{(}\PYG{l+s}{\PYGZdq{}}\PYG{l+s}{F}\PYG{l+s}{\PYGZdq{}}\PYG{p}{,}\PYG{+w}{ }\PYG{p}{[}\PYG{n}{j}\PYG{+w}{ }\PYG{o}{=\PYGZgt{}}\PYG{+w}{ }\PYG{n}{upper}\PYG{p}{,}\PYG{+w}{ }\PYG{n}{l}\PYG{+w}{ }\PYG{o}{=\PYGZgt{}}\PYG{+w}{ }\PYG{n}{lower}\PYG{p}{]}\PYG{p}{,}\PYG{+w}{ }\PYG{n}{x}\PYG{p}{)}\PYG{p}{)}
\PYG{+w}{    }\PYG{n}{exp2}\PYG{+w}{ }\PYG{o}{=}\PYG{+w}{ }\PYG{n}{AB}\PYG{o}{+}\PYG{n}{EF}\PYG{+w}{ }\PYG{c}{\PYGZsh{}This will error, FreeIndexMismatch}

\PYG{+w}{    }\PYG{n}{GH}\PYG{+w}{ }\PYG{o}{=}\PYG{+w}{  }\PYG{p}{(}\PYG{n}{Tensor}\PYG{p}{(}\PYG{l+s}{\PYGZdq{}}\PYG{l+s}{G}\PYG{l+s}{\PYGZdq{}}\PYG{p}{,}\PYG{+w}{ }\PYG{p}{[}\PYG{n}{i}\PYG{+w}{ }\PYG{o}{=\PYGZgt{}}\PYG{+w}{ }\PYG{n}{lower}\PYG{p}{,}\PYG{+w}{ }\PYG{n}{j}\PYG{+w}{ }\PYG{o}{=\PYGZgt{}}\PYG{+w}{ }\PYG{n}{lower}\PYG{p}{]}\PYG{p}{,}\PYG{+w}{ }\PYG{n}{x}\PYG{p}{)}
\PYG{+w}{        }\PYG{o}{*}\PYG{n}{Tensor}\PYG{p}{(}\PYG{l+s}{\PYGZdq{}}\PYG{l+s}{H}\PYG{l+s}{\PYGZdq{}}\PYG{p}{,}\PYG{+w}{ }\PYG{p}{[}\PYG{n}{j}\PYG{+w}{ }\PYG{o}{=\PYGZgt{}}\PYG{+w}{ }\PYG{n}{upper}\PYG{p}{,}\PYG{+w}{ }\PYG{n}{k}\PYG{+w}{ }\PYG{o}{=\PYGZgt{}}\PYG{+w}{ }\PYG{n}{upper}\PYG{p}{]}\PYG{p}{,}\PYG{+w}{ }\PYG{n}{x}\PYG{p}{)}\PYG{p}{)}
\PYG{+w}{    }\PYG{n}{exp3}\PYG{+w}{ }\PYG{o}{=}\PYG{+w}{ }\PYG{n}{AB}\PYG{o}{+}\PYG{n}{GH}\PYG{+w}{ }\PYG{c}{\PYGZsh{}This will error, FreeIndexMismatch}
\PYG{+w}{    }
\end{Verbatim}

Note also that the tensors need not be a \jil{function_of} the same coordinates in order to be added together.

\subsubsection{Weights}\label{sec:weights}

All \jil{Tensor}s allow for arbitrary numerical \jil{weight} properties to be defined upon construction. For example, by default, a \jil{Tensor} will be assigned a weight of \jil{1} labelled by each \jil{Coordinate} that the \jil{Tensor} is a \jil{function_of}. Weights are additive across products; thus the weight of a \jil{TensorTerm} will be the sum of the weights of its component \jil{Tensors}. The aforementioned example will determine the polynomial weight of a term for each coordinate, allowing for example for truncation at some fixed order.

The \jil{get_weighted_terms()} function allows for a particular weight to be extracted from an expression. This property can be useful for example in getting the $n$th degree of a polynomial. For example,

\begin{Verbatim}[commandchars=\\\{\},xleftmargin=\parindent,numbers=left,bgcolor=bg]
\PYG{+w}{    }\PYG{n}{x}\PYG{+w}{ }\PYG{o}{=}\PYG{+w}{ }\PYG{n}{Coordinate}\PYG{p}{(}\PYG{l+s}{\PYGZdq{}}\PYG{l+s}{x}\PYG{l+s}{\PYGZdq{}}\PYG{p}{,}\PYG{+w}{ }\PYG{p}{[}\PYG{n}{spacetime\PYGZus{}indices}\PYG{p}{]}\PYG{p}{)}
\PYG{+w}{    }\PYG{n}{X}\PYG{o}{=}\PYG{n}{Tensor}\PYG{p}{(}\PYG{l+s}{\PYGZdq{}}\PYG{l+s}{X}\PYG{l+s}{\PYGZdq{}}\PYG{p}{,}\PYG{n}{x}\PYG{p}{)}
\PYG{+w}{    }\PYG{n}{exp}\PYG{o}{=}\PYG{+w}{ }\PYG{l+m+mi}{2}\PYG{o}{*}\PYG{n}{X}\PYG{o}{*}\PYG{n}{X}\PYG{o}{+}\PYG{l+m+mi}{8}\PYG{o}{*}\PYG{n}{X}\PYG{o}{+}\PYG{l+m+mi}{9}\PYG{o}{*}\PYG{n}{X}\PYG{o}{*}\PYG{n}{X}\PYG{o}{*}\PYG{n}{X}
\PYG{+w}{    }\PYG{n}{get\PYGZus{}weighted\PYGZus{}terms}\PYG{p}{(}\PYG{n}{exp}\PYG{p}{,}\PYG{l+s}{\PYGZdq{}}\PYG{l+s}{x}\PYG{l+s}{\PYGZdq{}}\PYG{p}{,}\PYG{l+m+mi}{3}\PYG{p}{)}
\end{Verbatim}
This will have output \jil{9*XXX}. This is a simple example, however the weights allow for the encoding of any sort of extra numerical property associated with the tensor, for example, the engineering scaling dimension.

Additionally, there is also the \jil{drop_weight_above()} function, which could be useful for example to extract terms up to a certain order $\leq n$ in a Taylor series, as well as \jil{drop_weight_below()} to drop terms below a given weight.

\subsubsection{Labels}\label{sec:labels}
In addition to indices and weights, it is often useful to distinguish tensors from each other by additional labels, for example $k_1^\mu$ and $k_2^\mu$ or $A_\text{bare}$ and $A_\text{ren}$. For this purpose, Alakazam is able to associate label strings to tensors during construction. Labels are attached, similarly to indices, as \jil{String => position} pairs; a
bare \jil{String} is taken as a lower label. They are ordinary constructor arguments, so a tensor
may carry any number of them, in either position.

\begin{Verbatim}[commandchars=\\\{\},xleftmargin=\parindent,numbers=left,bgcolor=bg]
\PYG{+w}{    }\PYG{n}{k1}\PYG{+w}{ }\PYG{o}{=}\PYG{+w}{ }\PYG{n}{Tensor}\PYG{p}{(}\PYG{l+s}{\PYGZdq{}}\PYG{l+s}{k}\PYG{l+s}{\PYGZdq{}}\PYG{p}{,}\PYG{+w}{ }\PYG{p}{[}\PYG{n}{μ}\PYG{+w}{ }\PYG{o}{=\PYGZgt{}}\PYG{+w}{ }\PYG{n}{upper}\PYG{p}{]}\PYG{p}{,}\PYG{+w}{ }\PYG{l+s}{\PYGZdq{}}\PYG{l+s}{1}\PYG{l+s}{\PYGZdq{}}\PYG{+w}{ }\PYG{o}{=\PYGZgt{}}\PYG{+w}{ }\PYG{n}{lower}\PYG{p}{)}\PYG{+w}{     }\PYG{c}{\PYGZsh{}  k\PYGZca{}\PYGZob{}μ\PYGZcb{}\PYGZus{}\PYGZob{}1\PYGZcb{}}
\PYG{+w}{    }\PYG{n}{k2}\PYG{+w}{ }\PYG{o}{=}\PYG{+w}{ }\PYG{n}{Tensor}\PYG{p}{(}\PYG{l+s}{\PYGZdq{}}\PYG{l+s}{k}\PYG{l+s}{\PYGZdq{}}\PYG{p}{,}\PYG{+w}{ }\PYG{p}{[}\PYG{n}{μ}\PYG{+w}{ }\PYG{o}{=\PYGZgt{}}\PYG{+w}{ }\PYG{n}{upper}\PYG{p}{]}\PYG{p}{,}\PYG{+w}{ }\PYG{l+s}{\PYGZdq{}}\PYG{l+s}{2}\PYG{l+s}{\PYGZdq{}}\PYG{+w}{ }\PYG{o}{=\PYGZgt{}}\PYG{+w}{ }\PYG{n}{lower}\PYG{p}{)}\PYG{+w}{     }\PYG{c}{\PYGZsh{}  k\PYGZca{}\PYGZob{}μ\PYGZcb{}\PYGZus{}\PYGZob{}2\PYGZcb{}}
\PYG{+w}{    }\PYG{n}{A}\PYG{+w}{  }\PYG{o}{=}\PYG{+w}{ }\PYG{n}{Tensor}\PYG{p}{(}\PYG{l+s}{\PYGZdq{}}\PYG{l+s}{A}\PYG{l+s}{\PYGZdq{}}\PYG{p}{,}\PYG{+w}{ }\PYG{p}{[}\PYG{n}{μ}\PYG{+w}{ }\PYG{o}{=\PYGZgt{}}\PYG{+w}{ }\PYG{n}{upper}\PYG{p}{]}\PYG{p}{,}\PYG{+w}{ }\PYG{l+s}{\PYGZdq{}}\PYG{l+s}{bare}\PYG{l+s}{\PYGZdq{}}\PYG{p}{)}\PYG{+w}{           }\PYG{c}{\PYGZsh{}  A\PYGZca{}\PYGZob{}μ\PYGZcb{}\PYGZus{}\PYGZob{}bare\PYGZcb{}}
\PYG{+w}{    }\PYG{n+nd}{@show}\PYG{+w}{ }\PYG{n}{labels}\PYG{p}{(}\PYG{n}{k1}\PYG{p}{)}
\PYG{+w}{    }\PYG{c}{\PYGZsh{}   labels(k1) = Pair\PYGZob{}String, IndexPosition\PYGZcb{}[\PYGZdq{}1\PYGZdq{} =\PYGZgt{} lower]}
\end{Verbatim}

A label is part of a tensor's identity rather than decoration: \jil{==} compares labels, so
\jil{k1} and \jil{k2} are distinct objects and \jil{k1 + k2} is a genuine two-term sum, where
\jil{k1 + k1} collapses to \jil{2k1}. They are otherwise inert. Labels can also be
written directly in the tensor string macros introduced in~\ref{sec:macros}.

\begin{landscape}
\begin{figure}[t]
    \centering
      \vspace*{-.3cm}
    \includegraphics[width=1.8\textwidth]{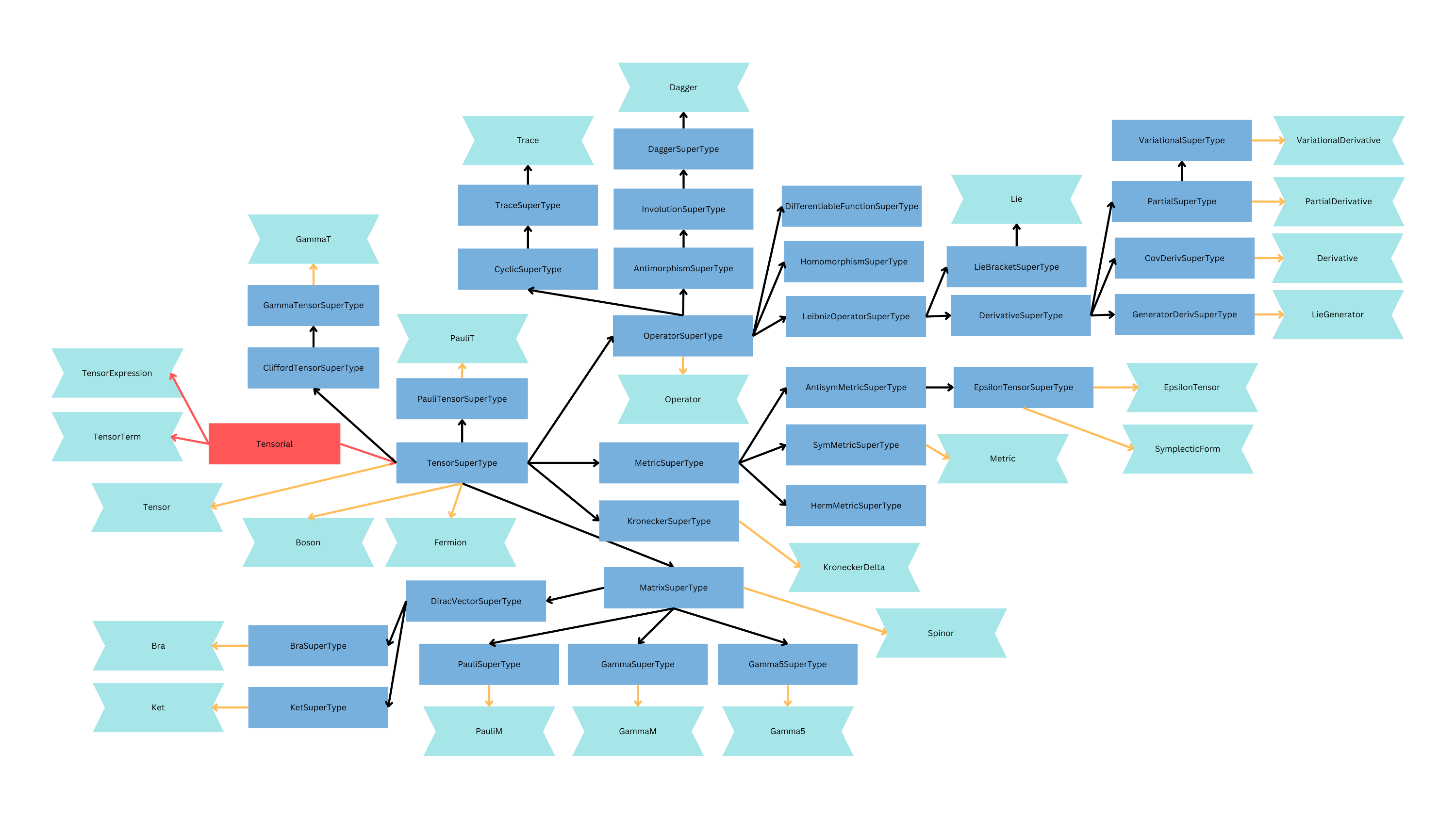}
    \caption{Typing Hierarchy. Arrows denote extends. Blue types are abstract types and can be extended.  Cyan types are concrete and cannot be extended. All types extend the red \jil{Tensorial} supertype.}
    \label{fig:types}
      \vspace*{-.35cm}
\end{figure}
\end{landscape}

\subsection{Typing Hierarchy}\label{sec:typing-hierarchy}

In Julia, structures can only extend abstract types. Thus, at each level, abstract ``super''-types are used that possess the necessary methods that all subtypes can extend and use. This means that if the user would like to define a new type of tensor with custom behaviour, they do not extend the concrete \jil{Tensor} type, but must extend \jil{TensorSuperType} in order to use any default methods defined on tensors. The current typing hierarchy in the package can be seen in Figure~\ref{fig:types}.

For instance, the method \jil{product_rule()} is defined for the \jil{LeibnizOperatorSuperType}. Since both \jil{DerivativeSuperType} and \jil{LieBracketSuperType} are subtypes of this type, all instances of derivatives and Lie brackets automatically inherit a graded derivation rule.  This makes the creation of new types with more specific behaviour very easy.

 \subsection{Arithmetic Operations}\label{sec:arithmetic-operations}

 \subsubsection{Basic Arithmetic}\label{sec:basic-arithmetic}
 Basic arithmetic functions \jil{+},\jil{-},\jil{*} are implemented as expected. 
 
 Addition and subtraction are defined between \jil{Tensorial} objects as long as they have the same free indices. This includes constants, which may be numerical or symbolic, which will automatically be interpreted as a constant tensor.

 Multiplication between \jil{Tensor} objects is allowed when there are no repeated indices between the terms, or when repeated indices are on different levels and occur only twice such that Einstein summation is implied.

 Additionally, division \jil{/} and \jil{//} of \jil{Tensorial} objects by symbols or numbers is allowed, where the latter denotes explicit fractional division (which is recommended over the use of floats for accuracy).

\subsubsection{Tensor Appension (\jil{×})}\label{sec:tensor-appension}

Suppose one has tensors,
\begin{Verbatim}[commandchars=\\\{\},xleftmargin=\parindent,numbers=left,bgcolor=bg]
\PYG{+w}{    }\PYG{n}{A}\PYG{+w}{ }\PYG{o}{=}\PYG{+w}{ }\PYG{n}{Tensor}\PYG{p}{(}\PYG{l+s}{\PYGZdq{}}\PYG{l+s}{A}\PYG{l+s}{\PYGZdq{}}\PYG{p}{,}\PYG{+w}{ }\PYG{p}{[}\PYG{n}{i}\PYG{+w}{ }\PYG{o}{=\PYGZgt{}}\PYG{+w}{ }\PYG{n}{upper}\PYG{p}{,}\PYG{+w}{ }\PYG{n}{j}\PYG{+w}{ }\PYG{o}{=\PYGZgt{}}\PYG{+w}{ }\PYG{n}{lower}\PYG{p}{]}\PYG{p}{,}\PYG{+w}{ }\PYG{n}{x}\PYG{p}{)}
\PYG{+w}{    }\PYG{n}{B}\PYG{+w}{ }\PYG{o}{=}\PYG{+w}{ }\PYG{n}{Tensor}\PYG{p}{(}\PYG{l+s}{\PYGZdq{}}\PYG{l+s}{B}\PYG{l+s}{\PYGZdq{}}\PYG{p}{,}\PYG{+w}{ }\PYG{p}{[}\PYG{n}{i}\PYG{+w}{ }\PYG{o}{=\PYGZgt{}}\PYG{+w}{ }\PYG{n}{upper}\PYG{p}{]}\PYG{p}{,}\PYG{+w}{ }\PYG{n}{x}\PYG{p}{)}
\PYG{+w}{    }\PYG{n}{C}\PYG{+w}{ }\PYG{o}{=}\PYG{+w}{ }\PYG{n}{Tensor}\PYG{p}{(}\PYG{l+s}{\PYGZdq{}}\PYG{l+s}{C}\PYG{l+s}{\PYGZdq{}}\PYG{p}{,}\PYG{+w}{ }\PYG{p}{[}\PYG{n}{i}\PYG{+w}{ }\PYG{o}{=\PYGZgt{}}\PYG{+w}{ }\PYG{n}{lower}\PYG{p}{,}\PYG{+w}{ }\PYG{n}{j}\PYG{+w}{ }\PYG{o}{=\PYGZgt{}}\PYG{+w}{ }\PYG{n}{lower}\PYG{p}{]}\PYG{p}{,}\PYG{+w}{ }\PYG{n}{x}\PYG{p}{)}
\end{Verbatim}
Then, if the user would like to multiply these tensors, this would produce an error as the resulting expression would have an invalid index structure. If the user would like to multiply such tensors, without performing any contractions over the indices, instead of manually creating new tensors such that the indices differ, the tensor appension binary operator \jil{×} (\jil{\times}) can be used:
\begin{Verbatim}[commandchars=\\\{\},xleftmargin=\parindent,numbers=left,bgcolor=bg]
\PYG{+w}{    }\PYG{n}{exp}\PYG{o}{=}\PYG{n}{A×B×C}
\end{Verbatim}
which produces output \jil{A^{i∘}_{∘j}B^{k}C_{lm}}. Indices are automatically labelled by indices from the correct \jil{IndexSet} and dummy indices within any single tensor are preserved. For example,

\begin{Verbatim}[commandchars=\\\{\},xleftmargin=\parindent,numbers=left,bgcolor=bg]
\PYG{+w}{    }\PYG{n}{E}\PYG{+w}{ }\PYG{o}{=}\PYG{+w}{ }\PYG{n}{Tensor}\PYG{p}{(}\PYG{l+s}{\PYGZdq{}}\PYG{l+s}{E}\PYG{l+s}{\PYGZdq{}}\PYG{p}{,}\PYG{+w}{ }\PYG{p}{[}\PYG{n}{i}\PYG{+w}{ }\PYG{o}{=\PYGZgt{}}\PYG{+w}{ }\PYG{n}{upper}\PYG{p}{,}\PYG{+w}{ }\PYG{n}{j}\PYG{+w}{ }\PYG{o}{=\PYGZgt{}}\PYG{+w}{ }\PYG{n}{lower}\PYG{p}{,}\PYG{+w}{ }\PYG{n}{α}\PYG{+w}{ }\PYG{o}{=\PYGZgt{}}\PYG{+w}{ }\PYG{n}{upper}\PYG{p}{]}\PYG{p}{,}\PYG{+w}{ }\PYG{n}{x}\PYG{p}{)}
\PYG{+w}{    }\PYG{n}{F}\PYG{+w}{ }\PYG{o}{=}\PYG{+w}{ }\PYG{n}{Tensor}\PYG{p}{(}\PYG{l+s}{\PYGZdq{}}\PYG{l+s}{F}\PYG{l+s}{\PYGZdq{}}\PYG{p}{,}\PYG{+w}{ }\PYG{p}{[}\PYG{n}{i}\PYG{+w}{ }\PYG{o}{=\PYGZgt{}}\PYG{+w}{ }\PYG{n}{upper}\PYG{p}{,}\PYG{+w}{ }\PYG{n}{α}\PYG{+w}{ }\PYG{o}{=\PYGZgt{}}\PYG{+w}{ }\PYG{n}{lower}\PYG{p}{,}\PYG{+w}{ }\PYG{n}{β}\PYG{+w}{ }\PYG{o}{=\PYGZgt{}}\PYG{+w}{ }\PYG{n}{upper}\PYG{p}{]}\PYG{p}{,}\PYG{+w}{ }\PYG{n}{x}\PYG{p}{)}
\PYG{+w}{    }\PYG{n}{G}\PYG{+w}{ }\PYG{o}{=}\PYG{+w}{ }\PYG{n}{Tensor}\PYG{p}{(}\PYG{l+s}{\PYGZdq{}}\PYG{l+s}{G}\PYG{l+s}{\PYGZdq{}}\PYG{p}{,}\PYG{+w}{ }\PYG{p}{[}\PYG{n}{i}\PYG{+w}{ }\PYG{o}{=\PYGZgt{}}\PYG{+w}{ }\PYG{n}{lower}\PYG{p}{,}\PYG{+w}{ }\PYG{n}{i}\PYG{+w}{ }\PYG{o}{=\PYGZgt{}}\PYG{+w}{ }\PYG{n}{upper}\PYG{p}{,}\PYG{+w}{ }\PYG{n}{j}\PYG{+w}{ }\PYG{o}{=\PYGZgt{}}\PYG{+w}{ }\PYG{n}{lower}\PYG{p}{]}\PYG{p}{,}\PYG{+w}{ }\PYG{n}{x}\PYG{p}{)}

\PYG{+w}{    }\PYG{n}{exp}\PYG{o}{=}\PYG{n}{E×F×G}
\end{Verbatim}

produces output \jil{E^{i∘α}_{∘j∘}F^{k∘β}_{∘γ∘}G^{∘l∘}_{l∘m}}. Appension can also be performed via \jil{append(A,B)}.

\subsubsection{Tensor Powers (\jil{^})}\label{sec:tensor-powers}
Taking the powers of a tensor also has a naive implementation. Raising to an even power $p=2k$ will return $k$ pairs of fully contracted tensors, where the contraction is between indices in the same slot on either tensor. For odd powers $p=2k+1$, the operation will return the original uncontracted tensor appended by $k$ pairs of contracted tensors. For example,
\begin{Verbatim}[commandchars=\\\{\},xleftmargin=\parindent,numbers=left,bgcolor=bg]
\PYG{+w}{    }\PYG{n}{F}\PYG{o}{=}\PYG{n}{Tensor}\PYG{p}{(}\PYG{l+s}{\PYGZdq{}}\PYG{l+s}{F}\PYG{l+s}{\PYGZdq{}}\PYG{p}{,}\PYG{p}{[}\PYG{n}{a}\PYG{o}{=\PYGZgt{}}\PYG{n}{lower}\PYG{p}{,}\PYG{n}{b}\PYG{o}{=\PYGZgt{}}\PYG{n}{lower}\PYG{p}{]}\PYG{p}{)}
\PYG{+w}{    }\PYG{n+nd}{@show}\PYG{+w}{ }\PYG{n}{F}\PYG{o}{\PYGZca{}}\PYG{l+m+mi}{2}
\PYG{+w}{    }\PYG{c}{\PYGZsh{}  F\PYGZus{}\PYGZob{}ab\PYGZcb{}F\PYGZca{}\PYGZob{}ab\PYGZcb{}}
\PYG{+w}{    }\PYG{n+nd}{@show}\PYG{+w}{ }\PYG{n}{F}\PYG{o}{\PYGZca{}}\PYG{l+m+mi}{3}
\PYG{+w}{    }\PYG{c}{\PYGZsh{}  F\PYGZus{}\PYGZob{}ab\PYGZcb{}F\PYGZus{}\PYGZob{}cd\PYGZcb{}F\PYGZca{}\PYGZob{}cd\PYGZcb{}}
\end{Verbatim}

\subsection{Solving Equations}\label{sec:solving-equations}
Alakazam has the ability to solve basic linear equations for a single tensor. This may seem like a trivial operation, but is powerful when combined with the integration by parts and pattern-matching features seen in Sections~\ref{sec:ibp} and~\ref{sec:pattern-matching-and-replacements}.

If we understand a \jil{TensorExpression} as the left-hand side of \(\text{eqn}=0\), \jil{solve_linear} returns the factor solved for, together with what it is equal to,
\begin{Verbatim}[commandchars=\\\{\},xleftmargin=\parindent,numbers=left,bgcolor=bg]
\PYG{+w}{    }\PYG{n}{solve\PYGZus{}linear}\PYG{p}{(}\PYG{l+m+mi}{2}\PYG{n}{A}\PYG{+w}{ }\PYG{o}{+}\PYG{+w}{ }\PYG{n}{B}\PYG{p}{,}\PYG{+w}{ }\PYG{n}{A}\PYG{p}{)}\PYG{+w}{         }\PYG{c}{\PYGZsh{} (A\PYGZus{}\PYGZob{}ν\PYGZcb{}, \PYGZhy{}1//2B\PYGZus{}\PYGZob{}ν\PYGZcb{})}
\PYG{+w}{    }\PYG{n}{solve\PYGZus{}linear}\PYG{p}{(}\PYG{n}{∂}\PYG{p}{(}\PYG{n}{A}\PYG{p}{)}\PYG{+w}{ }\PYG{o}{\PYGZhy{}}\PYG{+w}{ }\PYG{n}{C}\PYG{p}{,}\PYG{+w}{ }\PYG{n}{∂}\PYG{p}{(}\PYG{n}{A}\PYG{p}{)}\PYG{p}{)}\PYG{+w}{    }\PYG{c}{\PYGZsh{} (∂\PYGZus{}\PYGZob{}μ\PYGZcb{}⟦A\PYGZus{}\PYGZob{}ν\PYGZcb{}⟧, C\PYGZus{}\PYGZob{}μν\PYGZcb{})}
\end{Verbatim}
The target may be any single factor, an operator included, so a derivative
\(\partial_\mu A_\nu\) is solved for as readily as a bare field.  The equation
must be linear in the target and the target must actually appear; when either
fails the expression is returned unchanged alongside \jil{nothing},
\begin{Verbatim}[commandchars=\\\{\},xleftmargin=\parindent,numbers=left,bgcolor=bg]
\PYG{+w}{    }\PYG{n}{solve\PYGZus{}linear}\PYG{p}{(}\PYG{n}{B}\PYG{p}{,}\PYG{+w}{ }\PYG{n}{A}\PYG{p}{)}\PYG{+w}{              }\PYG{c}{\PYGZsh{} (nothing, B\PYGZus{}\PYGZob{}ν\PYGZcb{})  \PYGZsh{} A does not appear}
\PYG{+w}{    }\PYG{n}{solve\PYGZus{}linear}\PYG{p}{(}\PYG{n}{A}\PYG{+w}{ }\PYG{o}{×}\PYG{+w}{ }\PYG{n}{C}\PYG{+w}{ }\PYG{o}{+}\PYG{+w}{ }\PYG{n}{B}\PYG{+w}{ }\PYG{o}{×}\PYG{+w}{ }\PYG{n}{C}\PYG{p}{,}\PYG{+w}{ }\PYG{n}{A}\PYG{p}{)}\PYG{+w}{  }\PYG{c}{\PYGZsh{} (nothing, ...)    \PYGZsh{} not a single factor A}
\end{Verbatim}
so that a caller can always test what happened.
\newpage
\section{Special Tensors}\label{sec:special-tensors}
There are several useful special types of tensors in field theory,
for instance tensors with symmetries, metrics, epsilon tensors, Pauli $\sigma$ matrices, and $\gamma$ matrices. These obey certain algebraic relations or have special properties. Alakazam can recognise and simplify terms involving these special tensors using known identities.

\subsection{\jil{SymmetricTensor} and \jil{AntisymmetricTensor}}\label{sec:symmetrictensor-and-antisymmetrictensor}

For totally symmetric and totally antisymmetric tensors, there exist functions to create tensors that have these symmetries encoded. These symmetries will be treated appropriately when performing calculations and simplifications. These can be constructed the same way as a regular \jil{Tensor} type, for example for symmetric tensors
\begin{Verbatim}[commandchars=\\\{\},xleftmargin=\parindent,numbers=left,bgcolor=bg]
\PYG{n}{S}\PYG{o}{=}\PYG{n}{SymmetricTensor}\PYG{p}{(}\PYG{l+s}{\PYGZdq{}}\PYG{l+s}{S}\PYG{l+s}{\PYGZdq{}}\PYG{p}{,}\PYG{+w}{ }\PYG{p}{[}\PYG{n}{a}\PYG{+w}{ }\PYG{o}{=\PYGZgt{}}\PYG{+w}{ }\PYG{n}{lower}\PYG{p}{,}\PYG{+w}{ }\PYG{n}{b}\PYG{+w}{ }\PYG{o}{=\PYGZgt{}}\PYG{+w}{ }\PYG{n}{lower}\PYG{p}{,}\PYG{+w}{ }\PYG{n}{c}\PYG{+w}{ }\PYG{o}{=\PYGZgt{}}\PYG{+w}{ }\PYG{n}{lower}\PYG{p}{,}\PYG{+w}{ }\PYG{n}{d}\PYG{+w}{ }\PYG{o}{=\PYGZgt{}}\PYG{+w}{ }\PYG{n}{lower}\PYG{p}{]}\PYG{p}{,}\PYG{+w}{ }\PYG{p}{[}\PYG{n}{x}\PYG{p}{]}\PYG{p}{)}
\end{Verbatim}
and totally antisymmetric
\begin{Verbatim}[commandchars=\\\{\},xleftmargin=\parindent,numbers=left,bgcolor=bg]
\PYG{n}{A}\PYG{o}{=}\PYG{n}{AntisymmetricTensor}\PYG{p}{(}\PYG{l+s}{\PYGZdq{}}\PYG{l+s}{A}\PYG{l+s}{\PYGZdq{}}\PYG{p}{,}\PYG{+w}{ }\PYG{p}{[}\PYG{n}{a}\PYG{+w}{ }\PYG{o}{=\PYGZgt{}}\PYG{+w}{ }\PYG{n}{lower}\PYG{p}{,}\PYG{+w}{ }\PYG{n}{b}\PYG{+w}{ }\PYG{o}{=\PYGZgt{}}\PYG{+w}{ }\PYG{n}{lower}\PYG{p}{,}\PYG{+w}{ }\PYG{n}{c}\PYG{+w}{ }\PYG{o}{=\PYGZgt{}}\PYG{+w}{ }\PYG{n}{lower}\PYG{p}{,}\PYG{+w}{ }\PYG{n}{d}\PYG{+w}{ }\PYG{o}{=\PYGZgt{}}\PYG{+w}{ }\PYG{n}{lower}\PYG{p}{]}\PYG{p}{,}\PYG{+w}{ }\PYG{p}{[}\PYG{n}{x}\PYG{p}{]}\PYG{p}{)}
\end{Verbatim}
tensors. In the simplification algorithms in the software, terms involving \jil{SymmetricTensor} or \jil{AntisymmetricTensor}s will be merged appropriately if the terms are equivalent up to permutation of indices.

\subsection{Symmetries and \jil{YoungTableau}}\label{sec:symmetries-and-youngtableau}
Often, one may like to consider tensors that have some other symmetry in the indices other than totally symmetric or antisymmetric. These can also be encoded and will be respected in simplification algorithms. These are encoded in a (or vector of) \jil{YoungTableau} object(s).
To illustrate, suppose we have a tensor that has 4 indices, and is symmetric in the first and last index, with no other symmetries on the remaining indices:
\begin{equation}
    A_{\alpha\beta\gamma\delta}(x)=   A_{(\alpha|\beta\gamma|\delta)}(x) .
\end{equation}
To encode these relations, one may create a simple ''Young Tableau". In these diagrams, rows represent symmetric relations, while columns represent antisymmetric relations. These are created by the constructor
\begin{Verbatim}[commandchars=\\\{\},xleftmargin=\parindent,numbers=left,bgcolor=bg]
\PYG{n}{YoungTableau}\PYG{p}{(}\PYG{n}{row\PYGZus{}lengths}\PYG{o}{::}\PYG{k+kt}{Vector}\PYG{p}{\PYGZob{}}\PYG{k+kt}{Int64}\PYG{p}{\PYGZcb{}}\PYG{p}{,}\PYG{+w}{ }\PYG{n}{index\PYGZus{}assignment}\PYG{o}{::}\PYG{k+kt}{Vector}\PYG{p}{\PYGZob{}}\PYG{k+kt}{Int64}\PYG{p}{\PYGZcb{}}\PYG{p}{)}
\end{Verbatim}
The first argument will tell us the length of each row, and thus implicitly the number of rows. For our example $A_{\alpha\beta\gamma\delta}(x)$, this will just be the vector \jil{[2]}, creating the tableau
\begin{center}
\ydiagram{2}.
\end{center}
The second argument \jil{index_assignment} will tell us how to fill each box of the diagram. The boxes are filled left to right, top to bottom. In this simple case, we have an assignment \jil{[1,4]}, thus yielding the final tableau
\begin{center}
    \begin{ytableau}
1 & 4\\
\end{ytableau}.
\end{center}
This example is very simple. We could instead consider a more complicated pattern of symmetries. Suppose now that we have a 4-index tensor, which lives in some irreducible representation of $S_4$, such that the tensor components obey non-trivial constraints. For example, this could be the tableau
\begin{center}
    \begin{ytableau}
1 & 2&4\\ 3
\end{ytableau}.
\end{center}
In this case, the symmetry properties can be encoded via
\begin{Verbatim}[commandchars=\\\{\},xleftmargin=\parindent,numbers=left,bgcolor=bg]
\PYG{+w}{    }\PYG{n}{tab}\PYG{o}{=}\PYG{n}{YoungTableau}\PYG{p}{(}\PYG{p}{[}\PYG{l+m+mi}{3}\PYG{p}{,}\PYG{l+m+mi}{1}\PYG{p}{]}\PYG{p}{,}\PYG{+w}{ }\PYG{p}{[}\PYG{l+m+mi}{1}\PYG{p}{,}\PYG{l+m+mi}{2}\PYG{p}{,}\PYG{l+m+mi}{4}\PYG{p}{,}\PYG{l+m+mi}{3}\PYG{p}{]}\PYG{p}{)}
\end{Verbatim}
The tableau can be supplied to the tensor constructor upon instantiation
\begin{Verbatim}[commandchars=\\\{\},xleftmargin=\parindent,numbers=left,bgcolor=bg]
\PYG{+w}{        }\PYG{n}{A}\PYG{+w}{ }\PYG{o}{=}\PYG{+w}{ }\PYG{n}{Tensor}\PYG{p}{(}\PYG{l+s}{\PYGZdq{}}\PYG{l+s}{A}\PYG{l+s}{\PYGZdq{}}\PYG{p}{,}\PYG{+w}{ }\PYG{p}{[}\PYG{n}{i}\PYG{+w}{ }\PYG{o}{=\PYGZgt{}}\PYG{+w}{ }\PYG{n}{lower}\PYG{p}{,}\PYG{+w}{ }\PYG{n}{j}\PYG{+w}{ }\PYG{o}{=\PYGZgt{}}\PYG{+w}{ }\PYG{n}{lower}\PYG{p}{,}\PYG{+w}{ }\PYG{n}{k}\PYG{o}{=\PYGZgt{}}\PYG{n}{lower}\PYG{p}{,}\PYG{+w}{ }\PYG{n}{l}\PYG{o}{=\PYGZgt{}}\PYG{n}{lower}\PYG{p}{]}\PYG{p}{,}\PYG{+w}{ }\PYG{n}{x}\PYG{p}{,}\PYG{+w}{ }\PYG{n}{tab}\PYG{p}{)}
\end{Verbatim}

Any such tensor can be put in a form where the symmetries are manifest via the \jil{young_projection()} function,
\begin{Verbatim}[commandchars=\\\{\},xleftmargin=\parindent,numbers=left,bgcolor=bg]
\PYG{+w}{      }\PYG{n+nd}{@show}\PYG{+w}{ }\PYG{n}{young\PYGZus{}projection}\PYG{p}{(}\PYG{n}{A}\PYG{p}{)}
\PYG{+w}{      }\PYG{c}{\PYGZsh{}\PYGZhy{}1//8A\PYGZus{}\PYGZob{}kjil\PYGZcb{}+1//8A\PYGZus{}\PYGZob{}ljki\PYGZcb{}\PYGZhy{}1//8A\PYGZus{}\PYGZob{}klij\PYGZcb{}}
\PYG{+w}{      }\PYG{c}{\PYGZsh{}+1//8A\PYGZus{}\PYGZob{}likj\PYGZcb{}\PYGZhy{}1//8A\PYGZus{}\PYGZob{}jkil\PYGZcb{}\PYGZhy{}1//8A\PYGZus{}\PYGZob{}lkij\PYGZcb{}}
\PYG{+w}{      }\PYG{c}{\PYGZsh{}+1//8A\PYGZus{}\PYGZob{}jlki\PYGZcb{}+1//8A\PYGZus{}\PYGZob{}ijkl\PYGZcb{}+1//8A\PYGZus{}\PYGZob{}ilkj\PYGZcb{}}
\PYG{+w}{      }\PYG{c}{\PYGZsh{}+1//8A\PYGZus{}\PYGZob{}jikl\PYGZcb{}\PYGZhy{}1//8A\PYGZus{}\PYGZob{}ljik\PYGZcb{}\PYGZhy{}1//8A\PYGZus{}\PYGZob{}jlik\PYGZcb{}}
\end{Verbatim}
This can be used to simplify some expressions and prove known identities.

\subsection{Riemann Tensor}\label{sec:riemann-tensor}
The preceding example was just one irreducible representation of $S_4$. Perhaps one that is more useful is the tableau that encodes the symmetries of the Riemann tensor:
\begin{Verbatim}[commandchars=\\\{\},xleftmargin=\parindent,numbers=left,bgcolor=bg]
\PYG{+w}{    }\PYG{n}{labels}\PYG{o}{=}\PYG{p}{[}\PYG{l+m+mi}{1}\PYG{p}{,}\PYG{l+m+mi}{3}\PYG{p}{,}\PYG{l+m+mi}{2}\PYG{p}{,}\PYG{l+m+mi}{4}\PYG{p}{]}
\PYG{+w}{    }\PYG{n}{row\PYGZus{}lengths}\PYG{o}{=}\PYG{p}{[}\PYG{l+m+mi}{2}\PYG{p}{,}\PYG{l+m+mi}{2}\PYG{p}{]}
\PYG{+w}{    }\PYG{n}{riemann\PYGZus{}tableau}\PYG{o}{=}\PYG{n}{YoungTableau}\PYG{p}{(}\PYG{n}{row\PYGZus{}lengths}\PYG{p}{,}\PYG{n}{labels}\PYG{p}{)}
\end{Verbatim}
which corresponds to the tableau
\begin{center}
    \begin{ytableau}
1 & 3\\ 2 & 4
\end{ytableau}.
\end{center}
If one creates a tensor with this tableau,
\begin{Verbatim}[commandchars=\\\{\},xleftmargin=\parindent,numbers=left,bgcolor=bg]
\PYG{n}{R}\PYG{o}{=}\PYG{n}{Tensor}\PYG{p}{(}\PYG{l+s}{\PYGZdq{}}\PYG{l+s}{R}\PYG{l+s}{\PYGZdq{}}\PYG{p}{,}\PYG{+w}{ }\PYG{p}{[}\PYG{n}{a}\PYG{+w}{ }\PYG{o}{=\PYGZgt{}}\PYG{+w}{ }\PYG{n}{lower}\PYG{p}{,}\PYG{+w}{ }\PYG{n}{b}\PYG{+w}{ }\PYG{o}{=\PYGZgt{}}\PYG{+w}{ }\PYG{n}{lower}\PYG{p}{,}\PYG{+w}{ }\PYG{n}{c}\PYG{+w}{ }\PYG{o}{=\PYGZgt{}}\PYG{+w}{ }\PYG{n}{lower}\PYG{p}{,}\PYG{+w}{ }\PYG{n}{d}\PYG{+w}{ }\PYG{o}{=\PYGZgt{}}\PYG{+w}{ }\PYG{n}{lower}\PYG{p}{]}\PYG{p}{,}\PYG{+w}{ }\PYG{p}{[}\PYG{n}{riemann\PYGZus{}tableau}\PYG{p}{]}\PYG{p}{)}
\end{Verbatim}
then manifestly, it will obey independent antisymmetry in slots $1\leftrightarrow 2$ and $3\leftrightarrow 4$ and pairwise symmetry $(1,2)\leftrightarrow (3,4)$,
\begin{equation}
    R_{abcd}=-R_{bacd}=-R_{abdc}=R_{badc}=R_{dcba}=-R_{cdba}=-R_{dcab}=R_{cdab},
\end{equation}
as well as the Bianchi identity (after Young projection)
\begin{equation}
    R_{abcd}+R_{acdb}+R_{adbc}=0.
\end{equation}

For convenience, one may also generate a Riemann tensor automatically with \jil{RiemannTensor()}
\begin{Verbatim}[commandchars=\\\{\},xleftmargin=\parindent,numbers=left,bgcolor=bg]
\PYG{+w}{    }\PYG{n}{R}\PYG{o}{=}\PYG{n}{RiemannTensor}\PYG{p}{(}\PYG{l+s}{\PYGZdq{}}\PYG{l+s}{R}\PYG{l+s}{\PYGZdq{}}\PYG{p}{,}\PYG{p}{[}\PYG{n}{a}\PYG{+w}{ }\PYG{o}{=\PYGZgt{}}\PYG{+w}{ }\PYG{n}{lower}\PYG{p}{,}\PYG{+w}{ }\PYG{n}{b}\PYG{+w}{ }\PYG{o}{=\PYGZgt{}}\PYG{+w}{ }\PYG{n}{lower}\PYG{p}{,}\PYG{+w}{ }\PYG{n}{c}\PYG{+w}{ }\PYG{o}{=\PYGZgt{}}\PYG{+w}{ }\PYG{n}{lower}\PYG{p}{,}\PYG{+w}{ }\PYG{n}{d}\PYG{+w}{ }\PYG{o}{=\PYGZgt{}}\PYG{+w}{ }\PYG{n}{lower}\PYG{p}{]}\PYG{p}{)}
\end{Verbatim}

Other types of tensors relating to General Relativity will be added in future releases.

\subsection{Tensor Creation Macros}\label{sec:macros}
As mentioned earlier above, tensors can be created using macros, for example
\begin{Verbatim}[commandchars=\\\{\},xleftmargin=\parindent,numbers=left,bgcolor=bg]
\PYG{+w}{    }\PYG{n+nd}{@Tensor}\PYG{+w}{ }\PYG{n}{A\PYGZus{}ij}\PYG{o}{\PYGZca{}}\PYG{k+kt}{k\PYGZus{}l}\PYG{p}{\PYGZob{}}\PYG{k+kt}{x}\PYG{p}{,}\PYG{k+kt}{θ}\PYG{p}{\PYGZcb{}}
\end{Verbatim}
to encode the tensor
\begin{equation}
    A_{ij}{}^{k}{}_l(x,\theta)
\end{equation}
in the user scope.
This syntax is unable to encode more advanced information, like the tableaux, due to unavoidable limitations of the compiler. We can instead however parse tensors directly from strings, via the macro
\begin{Verbatim}[commandchars=\\\{\},xleftmargin=\parindent,numbers=left,bgcolor=bg]
\PYG{+w}{    }\PYG{l+s+sa}{T}\PYG{l+s}{\PYGZdq{}}\PYG{l+s}{A\PYGZus{}ij\PYGZca{}k\PYGZus{}l\PYGZob{}x,θ\PYGZcb{}}\PYG{l+s}{\PYGZdq{}}
\end{Verbatim}
which creates the same tensor. Here, the braces encode the \jil{function_of} coordinates. Note that indices should be declared before using these macros. Additionally, via this macro we may also specify basic symmetries, for instance of our previous example
\begin{equation}
    A_{\alpha\beta\gamma\delta}(x)=   A_{(\alpha|\beta\gamma|\delta)}(x)
\end{equation}
via
\begin{Verbatim}[commandchars=\\\{\},xleftmargin=\parindent,numbers=left,bgcolor=bg]
\PYG{+w}{    }\PYG{l+s+sa}{T}\PYG{l+s}{\PYGZdq{}}\PYG{l+s}{A\PYGZus{}(α|βγ|δ)\PYGZob{}x\PYGZcb{}}\PYG{l+s}{\PYGZdq{}}
\end{Verbatim}
for symmetrisation of indices in the 1st and 4th slots. This works analogously for antisymmetrisation, including across levels, for example
\begin{Verbatim}[commandchars=\\\{\},xleftmargin=\parindent,numbers=left,bgcolor=bg]
\PYG{+w}{    }\PYG{l+s+sa}{T}\PYG{l+s}{\PYGZdq{}}\PYG{l+s}{A\PYGZus{}(a\PYGZca{}b)\PYGZus{}[c\PYGZca{}de]f\PYGZob{}x,θ\PYGZcb{}}\PYG{l+s}{\PYGZdq{}}
\end{Verbatim}
which has two associated tableaux
\begin{center}
    \begin{ytableau}
1 & 2\\
\end{ytableau}, \quad    \begin{ytableau}
3 \\ 4 \\ 5
\end{ytableau}.
\end{center}

We can also include simple labels in the construction here. When parsing the macro, if the character is not a valid index, it will be parsed as a label.

\begin{Verbatim}[commandchars=\\\{\},xleftmargin=\parindent,numbers=left,bgcolor=bg]
\PYG{+w}{     }\PYG{n}{k}\PYG{+w}{ }\PYG{o}{=}\PYG{+w}{ }\PYG{l+s+sa}{T}\PYG{l+s}{\PYGZdq{}}\PYG{l+s}{k\PYGZus{}1\PYGZca{}a}\PYG{l+s}{\PYGZdq{}}
\PYG{+w}{     }\PYG{n+nd}{@show}\PYG{+w}{ }\PYG{n}{labels}\PYG{p}{(}\PYG{n}{k}\PYG{p}{)}
\PYG{+w}{        }\PYG{c}{\PYGZsh{}  Pair\PYGZob{}String, IndexPosition\PYGZcb{}[\PYGZdq{}1\PYGZdq{} =\PYGZgt{} lower]}
\PYG{+w}{     }\PYG{n+nd}{@show}\PYG{+w}{ }\PYG{n}{indices}\PYG{p}{(}\PYG{n}{k}\PYG{p}{)}\PYG{+w}{ }
\PYG{+w}{        }\PYG{c}{\PYGZsh{} Pair\PYGZob{}Union\PYGZob{}DummyPatternIndex, Index\PYGZcb{}, IndexPosition\PYGZcb{}[a =\PYGZgt{} upper]}
\end{Verbatim}
For parsing accented indices, there are a few available syntaxes. Index names longer than one character are read by placing a comma after each, which switches the parser out of character-at-a-time mode for that group. Decorated indices may be written with the suffixes \jil{dot}, \jil{hat}, \jil{bar}, and \jil{tilde}, which will be interpreted automatically as the appropriate accented index in the user's scope, when defined.
\begin{Verbatim}[commandchars=\\\{\},xleftmargin=\parindent,numbers=left,bgcolor=bg]
\PYG{+w}{    }\PYG{n+nd}{@show}\PYG{+w}{ }\PYG{n}{indices}\PYG{p}{(}\PYG{l+s+sa}{T}\PYG{l+s}{\PYGZdq{}}\PYG{l+s}{A\PYGZus{}adot,bhat}\PYG{l+s}{\PYGZdq{}}\PYG{p}{)}
\PYG{+w}{    }\PYG{c}{\PYGZsh{}   [̇a =\PYGZgt{} lower, ̂b =\PYGZgt{} lower]}
\end{Verbatim}

For users more familiar with a LaTeX-style input, Alakazam also has a custom macro permitting this syntax. Indeed, we can use the ``LaTeX-Tensor'' \jil{LT_str} macro to parse TeX markup as a \jil{Tensor} or \jil{TensorTerm}. This behaves in the expected way.
\begin{Verbatim}[commandchars=\\\{\},xleftmargin=\parindent,numbers=left,bgcolor=bg]
\PYG{+w}{    }\PYG{n+nd}{@show}\PYG{+w}{ }\PYG{l+s+sa}{LT}\PYG{l+s}{\PYGZdq{}}\PYG{l+s}{F\PYGZus{}\PYGZob{}}\PYG{l+s}{\PYGZbs{}}\PYG{l+s}{mu}\PYG{l+s+se}{\PYGZbs{}n}\PYG{l+s}{u\PYGZcb{}}\PYG{l+s}{\PYGZdq{}}
\PYG{+w}{        }\PYG{c}{\PYGZsh{}   F\PYGZus{}\PYGZob{}μν\PYGZcb{}}
\PYG{+w}{    }\PYG{n+nd}{@show}\PYG{+w}{ }\PYG{l+s+sa}{LT}\PYG{l+s}{\PYGZdq{}}\PYG{l+s}{A\PYGZca{}}\PYG{l+s}{\PYGZbs{}}\PYG{l+s}{mu B\PYGZus{}}\PYG{l+s}{\PYGZbs{}}\PYG{l+s}{mu}\PYG{l+s}{\PYGZdq{}}
\PYG{+w}{        }\PYG{c}{\PYGZsh{}   A\PYGZca{}\PYGZob{}μ\PYGZcb{}B\PYGZus{}\PYGZob{}μ\PYGZcb{}}
\PYG{+w}{    }\PYG{n+nd}{@show}\PYG{+w}{ }\PYG{l+s+sa}{LT}\PYG{l+s}{\PYGZdq{}}\PYG{l+s}{\PYGZbs{}}\PYG{l+s}{psi\PYGZus{}\PYGZob{}}\PYG{l+s}{\PYGZbs{}}\PYG{l+s}{dot}\PYG{l+s+se}{\PYGZbs{}a}\PYG{l+s}{lpha\PYGZcb{}}\PYG{l+s}{\PYGZdq{}}
\PYG{+w}{        }\PYG{c}{\PYGZsh{}   ψ\PYGZus{}\PYGZob{}̇α\PYGZcb{}}
\PYG{+w}{    }\PYG{n+nd}{@show}\PYG{+w}{ }\PYG{l+s+sa}{LT}\PYG{l+s}{\PYGZdq{}}\PYG{l+s}{A\PYGZus{}}\PYG{l+s}{\PYGZbs{}}\PYG{l+s}{mu(x)}\PYG{l+s}{\PYGZdq{}}
\PYG{+w}{        }\PYG{c}{\PYGZsh{}   A\PYGZus{}\PYGZob{}μ\PYGZcb{}(x)}
\PYG{+w}{    }\PYG{n+nd}{@show}\PYG{+w}{ }\PYG{l+s+sa}{LT}\PYG{l+s}{\PYGZdq{}}\PYG{l+s+se}{\PYGZbs{}f}\PYG{l+s}{rac\PYGZob{}1\PYGZcb{}\PYGZob{}2\PYGZcb{} A\PYGZus{}}\PYG{l+s}{\PYGZbs{}}\PYG{l+s}{mu}\PYG{l+s}{\PYGZdq{}}
\PYG{+w}{        }\PYG{c}{\PYGZsh{}   1//2A\PYGZus{}\PYGZob{}μ\PYGZcb{}}
\end{Verbatim}
The main limitation to this macro is that any symbolic coefficient is interpreted as a tensor when parsing. This is unavoidable, as there is no way to determine if the symbol should be interpreted as a symbolic coefficient or an index-less tensor. Note also that these macros parse all tensors as the plain \jil{Tensor} type. Specialised types (like matrices and operators) should be constructed explicitly.

\subsection{\jil{Metric}}\label{sec:metric}
The ability to raise and lower tensor indices is a requirement for any useful sort of manipulation of tensor expressions for field theory. To create a \jil{Metric}, one can supply two indices (which must be on the same level), and optionally a name
\begin{Verbatim}[commandchars=\\\{\},xleftmargin=\parindent,numbers=left,bgcolor=bg]
\PYG{+w}{    }\PYG{n+nd}{@show}\PYG{+w}{ }\PYG{n}{Metric}\PYG{p}{(}\PYG{l+s}{\PYGZdq{}}\PYG{l+s}{g}\PYG{l+s}{\PYGZdq{}}\PYG{p}{,}\PYG{n}{μ}\PYG{o}{=\PYGZgt{}}\PYG{n}{lower}\PYG{p}{,}\PYG{n}{ν}\PYG{o}{=\PYGZgt{}}\PYG{n}{lower}\PYG{p}{)}
\PYG{+w}{        }\PYG{c}{\PYGZsh{}   g\PYGZus{}\PYGZob{}μν\PYGZcb{}}
\PYG{+w}{    }\PYG{n+nd}{@show}\PYG{+w}{ }\PYG{n}{Metric}\PYG{p}{(}\PYG{n}{a}\PYG{o}{=\PYGZgt{}}\PYG{n}{lower}\PYG{p}{,}\PYG{n}{b}\PYG{o}{=\PYGZgt{}}\PYG{n}{lower}\PYG{p}{)}
\PYG{+w}{        }\PYG{c}{\PYGZsh{}   η\PYGZus{}\PYGZob{}ab\PYGZcb{}}
\end{Verbatim}
If no name is given, the name will be either \jil{η} or \jil{δ} depending on the signature of the \jil{IndexSet}. Note that this \jil{Metric} is a subtype of \jil{SymMetricSuperType}. As a side note, if the user needs an ``antisymmetric metric'' (for example, for $SL(2)$ indices used for spinors), they can use an \jil{EpsilonTensor} (see~\ref{sec:epsilontensor}), or extend \jil{AntisymMetricSuperType} to create a metric with unique functionality. 

The \jil{Metric} tensor can be multiplied with expressions. To perform a contraction, i.e. explicitly raise and lower indices, one can use the \jil{contract_metrics()} function,

\begin{Verbatim}[commandchars=\\\{\},xleftmargin=\parindent,numbers=left,bgcolor=bg]
\PYG{+w}{    }\PYG{n}{contract\PYGZus{}metrics}\PYG{p}{(}\PYG{n}{T}\PYG{o}{::}\PYG{k+kt}{TensorExpression}\PYG{p}{,}\PYG{+w}{ }\PYG{n}{typ}\PYG{o}{::}\PYG{k+kt}{Type}\PYG{o}{=}\PYG{n}{Metric}\PYG{p}{)}
\end{Verbatim}
Note that the parameter \jil{typ} can be used to selectively contract only certain kinds of metrics depending on the desired application. To contract all sorts of metrics, one can use \jil{SymMetricSuperType} and \jil{AntisymMetricSuperType} as \jil{typ}, as all metrics are a subtype of either one of these. 

As an example of the usage, consider

\begin{Verbatim}[commandchars=\\\{\},xleftmargin=\parindent,numbers=left,bgcolor=bg]
\PYG{+w}{    }\PYG{n}{A}\PYG{+w}{ }\PYG{o}{=}\PYG{+w}{ }\PYG{n}{Tensor}\PYG{p}{(}\PYG{l+s}{\PYGZdq{}}\PYG{l+s}{A}\PYG{l+s}{\PYGZdq{}}\PYG{p}{,}\PYG{+w}{ }\PYG{n}{i}\PYG{+w}{ }\PYG{o}{=\PYGZgt{}}\PYG{+w}{ }\PYG{n}{upper}\PYG{p}{,}\PYG{+w}{ }\PYG{n}{j}\PYG{+w}{ }\PYG{o}{=\PYGZgt{}}\PYG{+w}{ }\PYG{n}{lower}\PYG{p}{,}\PYG{+w}{ }\PYG{n}{x}\PYG{p}{)}
\PYG{+w}{    }\PYG{n}{ηjk}\PYG{+w}{ }\PYG{o}{=}\PYG{+w}{ }\PYG{n}{Metric}\PYG{p}{(}\PYG{n}{j}\PYG{+w}{ }\PYG{o}{=\PYGZgt{}}\PYG{+w}{ }\PYG{n}{upper}\PYG{p}{,}\PYG{+w}{ }\PYG{n}{k}\PYG{+w}{ }\PYG{o}{=\PYGZgt{}}\PYG{+w}{ }\PYG{n}{upper}\PYG{p}{)}
\PYG{+w}{    }\PYG{n}{exp}\PYG{+w}{ }\PYG{o}{=}\PYG{+w}{ }\PYG{n}{ηjk}\PYG{o}{*}\PYG{n}{A}
\PYG{+w}{    }\PYG{n}{out}\PYG{+w}{ }\PYG{o}{=}\PYG{+w}{ }\PYG{n}{contract\PYGZus{}metrics}\PYG{p}{(}\PYG{n}{exp}\PYG{p}{)}
\end{Verbatim}

This will have the output \jil{A^{ik}}. Note that a contraction of a metric with another metric will yield the \jil{KroneckerDelta}.

\subsection{\jil{KroneckerDelta}}\label{sec:kroneckerdelta}
The \jil{KroneckerDelta} type in this package is a tensor that has two symmetric indices, one upper and one lower. To create a \jil{KroneckerDelta}, one can use the constructor
\begin{Verbatim}[commandchars=\\\{\},xleftmargin=\parindent,numbers=left,bgcolor=bg]
\PYG{+w}{    }\PYG{n}{KroneckerDelta}\PYG{p}{(}\PYG{n}{indices}\PYG{o}{::}\PYG{k+kt}{Vector}\PYG{p}{\PYGZob{}}\PYG{k+kt}{Pair}\PYG{p}{\PYGZob{}}\PYG{k+kt}{Index}\PYG{p}{,}\PYG{+w}{ }\PYG{k+kt}{IndexPosition}\PYG{p}{\PYGZcb{}}\PYG{p}{\PYGZcb{}}\PYG{p}{)}
\end{Verbatim}
Note that an error will be thrown if the two indices contained in \jil{indices} are not on opposite levels, or if they are from different \jil{IndexSet}s. Expressions containing contractions between \jil{KroneckerDelta}s and \jil{Tensorial} objects can be simplified using the \jil{eliminate_kronecker()} function, a single-argument function taking a \jil{TensorExpression}. Note that when \jil{eliminate_kronecker()} is executed, \jil{KroneckerDelta}s with repeated indices will be replaced by a constant determined by the \jil{range} parameter of the \jil{IndexSet} associated with the indices of the \jil{KroneckerDelta} (i.e., a summation, or trace, will be performed over the indices).

\begin{Verbatim}[commandchars=\\\{\},xleftmargin=\parindent,numbers=left,bgcolor=bg]
\PYG{+w}{    }\PYG{n}{δ}\PYG{o}{=}\PYG{n}{KroneckerDelta}\PYG{p}{(}\PYG{n}{μ}\PYG{o}{=\PYGZgt{}}\PYG{n}{upper}\PYG{p}{,}\PYG{n}{ν}\PYG{o}{=\PYGZgt{}}\PYG{n}{lower}\PYG{p}{)}
\PYG{+w}{    }\PYG{n}{A}\PYG{o}{=}\PYG{n}{Tensor}\PYG{p}{(}\PYG{l+s}{\PYGZdq{}}\PYG{l+s}{A}\PYG{l+s}{\PYGZdq{}}\PYG{p}{,}\PYG{n}{ν}\PYG{o}{=\PYGZgt{}}\PYG{n}{upper}\PYG{p}{,}\PYG{n}{ρ}\PYG{o}{=\PYGZgt{}}\PYG{n}{upper}\PYG{p}{)}
\PYG{+w}{    }\PYG{n+nd}{@show}\PYG{+w}{ }\PYG{n}{eliminate\PYGZus{}kronecker}\PYG{p}{(}\PYG{n}{δ}\PYG{o}{*}\PYG{n}{A}\PYG{p}{)}
\PYG{+w}{        }\PYG{c}{\PYGZsh{} A\PYGZca{}\PYGZob{}μρ\PYGZcb{}}
\end{Verbatim}

\subsection{\jil{EpsilonTensor}}\label{sec:epsilontensor}
In this package, the \jil{EpsilonTensor} is a totally antisymmetric tensor with other special properties. Generically, an epsilon tensor will have indices equal in number to the range of the \jil{IndexSet}, acting as a volume form on that space. They can be created via the constructor
\begin{Verbatim}[commandchars=\\\{\},xleftmargin=\parindent,numbers=left,bgcolor=bg]
\PYG{+w}{    }\PYG{n}{EpsilonTensor}\PYG{p}{(}\PYG{n}{indices}\PYG{o}{::}\PYG{k+kt}{Vector}\PYG{p}{\PYGZob{}}\PYG{k+kt}{Pair}\PYG{p}{\PYGZob{}}\PYG{k+kt}{Index}\PYG{p}{,}\PYG{+w}{ }\PYG{k+kt}{IndexPosition}\PYG{p}{\PYGZcb{}}\PYG{p}{\PYGZcb{}}\PYG{p}{)}
\end{Verbatim}
Expressions containing products of \jil{EpsilonTensor}s can be reduced by the \jil{eliminate_epsilon()} function, which replaces them with either a constant or a sum of \jil{KroneckerDelta}s where appropriate according to the identity
\begin{equation}
    \epsilon^{i_1 i_2 \cdots i_n}
\epsilon_{j_1 j_2 \cdots j_n}
=
\operatorname{sgn}(\eta)\sum_{\sigma\in S_n}
\operatorname{sgn}(\sigma)
\prod_{k=1}^{n}
\delta^{i_k}_{j_{\sigma(k)}}
\end{equation}
Indeed, for example in $3+1D$,
\begin{Verbatim}[commandchars=\\\{\},xleftmargin=\parindent,numbers=left,bgcolor=bg]
\PYG{+w}{    }\PYG{n}{e1}\PYG{o}{=}\PYG{n}{EpsilonTensor}\PYG{p}{(}\PYG{p}{[}\PYG{n}{ν}\PYG{o}{=\PYGZgt{}}\PYG{n}{lower}\PYG{p}{,}\PYG{n}{μ}\PYG{o}{=\PYGZgt{}}\PYG{n}{lower}\PYG{p}{,}\PYG{n}{ρ}\PYG{o}{=\PYGZgt{}}\PYG{n}{lower}\PYG{p}{,}\PYG{n}{σ}\PYG{o}{=\PYGZgt{}}\PYG{n}{lower}\PYG{p}{]}\PYG{p}{)}
\PYG{+w}{    }\PYG{n}{e2}\PYG{o}{=}\PYG{n}{EpsilonTensor}\PYG{p}{(}\PYG{p}{[}\PYG{n}{ν}\PYG{o}{=\PYGZgt{}}\PYG{n}{upper}\PYG{p}{,}\PYG{n}{μ}\PYG{o}{=\PYGZgt{}}\PYG{n}{upper}\PYG{p}{,}\PYG{n}{ρ}\PYG{o}{=\PYGZgt{}}\PYG{n}{upper}\PYG{p}{,}\PYG{n}{τ}\PYG{o}{=\PYGZgt{}}\PYG{n}{upper}\PYG{p}{]}\PYG{p}{)}
\PYG{+w}{    }\PYG{n+nd}{@show}\PYG{+w}{ }\PYG{n}{simplify}\PYG{p}{(}\PYG{n}{eliminate\PYGZus{}epsilon}\PYG{p}{(}\PYG{n}{e1}\PYG{o}{*}\PYG{n}{e2}\PYG{p}{)}\PYG{p}{)}
\end{Verbatim}
will produce \jil{-6δ^{τ}_{σ}}. The overall sign is fixed by the metric signature specified by the \jil{IndexSet}.

In this package, objects of the type \jil{EpsilonTensor} are a subtype of the \jil{AntisymMetricSuperType}, and thus can be used as a $2D$ antisymmetric metric if needed for two-component spinors. Expressions containing contractions between \jil{EpsilonTensor}s and an arbitrary tensor will also be simplified by the \jil{eliminate_epsilon()} function. If possible, this will raise and lower indices, with antisymmetry in mind.
In this case, index raising and lowering use the convention
\begin{equation}
    \psi^\alpha=\epsilon^{\alpha\beta}\psi_\beta,\quad \psi_{\alpha}=\epsilon_{\alpha\beta}\psi^\beta, \implies \psi^\alpha\chi_\alpha=-\psi_\alpha\chi^\alpha.
\end{equation}
To use a different convention, for example that of $C_{\alpha\beta}$ of~\cite{gates2001superspace}, one can create a subtype of \jil{AntisymMetricSuperType} and define custom contraction behaviour by implementing \jil{antisym_contraction()} dispatched on the custom type, for example to admit north-west-south-east behaviour or otherwise. See also~\ref{sec:dagger-and-complex-conjugation} for overriding complex conjugation convention.

\subsection{Pauli Matrices}\label{sec:pauli-matrices}
Beyond graded tensors, it is also useful to consider objects that generically neither commute nor anti-commute. We refer to these as matrices, and are subtypes of \jil{MatrixSuperType}. A prototypical example is the Pauli-$\sigma$ matrices, obeying the commutation relations
\begin{equation}
    [\sigma^i,\sigma^j]=2i\epsilon^{ijk}\sigma_k.
\end{equation}
Furthermore, products of these matrices are given by
\begin{equation}
    \sigma^i\sigma^j=\delta^{ij}+i\epsilon^{ijk}\sigma_k.
\end{equation}
\jil{Alakazam} will automatically reduce the products of such matrices via the \jil{contract_sigma()} method.
For example, one can create these matrices via
\begin{Verbatim}[commandchars=\\\{\},xleftmargin=\parindent,numbers=left,bgcolor=bg]
\PYG{+w}{    }\PYG{n}{σ1}\PYG{o}{=}\PYG{n}{Pauli}\PYG{p}{(}\PYG{n}{μ}\PYG{o}{=\PYGZgt{}}\PYG{n}{upper}\PYG{p}{)}
\PYG{+w}{    }\PYG{n}{σ2}\PYG{o}{=}\PYG{n}{Pauli}\PYG{p}{(}\PYG{n}{ν}\PYG{o}{=\PYGZgt{}}\PYG{n}{upper}\PYG{p}{)}
\PYG{+w}{    }\PYG{n}{σ3}\PYG{o}{=}\PYG{n}{Pauli}\PYG{p}{(}\PYG{n}{ρ}\PYG{o}{=\PYGZgt{}}\PYG{n}{upper}\PYG{p}{)}
\end{Verbatim}
which will generate tensors of the type \jil{PauliM<:MatrixSuperType}. Then, calling
\begin{Verbatim}[commandchars=\\\{\},xleftmargin=\parindent,numbers=left,bgcolor=bg]
\PYG{+w}{    }\PYG{n+nd}{@show}\PYG{+w}{ }\PYG{n}{contract\PYGZus{}sigma}\PYG{p}{(}\PYG{n}{σ1}\PYG{o}{*}\PYG{n}{σ2}\PYG{o}{*}\PYG{n}{σ3}\PYG{p}{)}
\PYG{+w}{    }\PYG{c}{\PYGZsh{}   δ\PYGZca{}\PYGZob{}μν\PYGZcb{}σ\PYGZca{}\PYGZob{}ρ\PYGZcb{}+iϵ\PYGZca{}\PYGZob{}μνρ\PYGZcb{}+δ\PYGZca{}\PYGZob{}ρν\PYGZcb{}σ\PYGZca{}\PYGZob{}μ\PYGZcb{}\PYGZhy{}δ\PYGZca{}\PYGZob{}ρμ\PYGZcb{}σ\PYGZca{}\PYGZob{}ν\PYGZcb{}}
\end{Verbatim}
fully contracts the product of Pauli matrices.

Note that calling \jil{Pauli} with additional indices from a different set will produce a ``component form'' tensor \jil{PauliT} which acts as a regular commuting tensor.

\subsection{Gamma Matrices}\label{sec:gamma-matrices}
Gamma matrices are also pivotal in field theory manipulations. One may also generate gamma matrices, including the $\gamma^5$ matrix, with
\begin{Verbatim}[commandchars=\\\{\},xleftmargin=\parindent,numbers=left,bgcolor=bg]
\PYG{+w}{    }\PYG{n}{gμ}\PYG{o}{=}\PYG{n}{Gamma}\PYG{p}{(}\PYG{n}{μ}\PYG{o}{=\PYGZgt{}}\PYG{n}{upper}\PYG{p}{)}
\PYG{+w}{    }\PYG{n}{gν}\PYG{o}{=}\PYG{n}{Gamma}\PYG{p}{(}\PYG{n}{ν}\PYG{o}{=\PYGZgt{}}\PYG{n}{upper}\PYG{p}{)}
\PYG{+w}{    }\PYG{n}{g5}\PYG{o}{=}\PYG{n}{Gamma5}\PYG{p}{(}\PYG{p}{)}
\end{Verbatim}
This generates \jil{GammaM<:MatrixSuperType} objects. These obey the expected anticommutation relations,
\begin{equation}
    \{\gamma^\mu,\gamma^\nu\}=2\eta^{\mu\nu},\quad \{\gamma^\mu,\gamma^5\}=0,
\end{equation}
indeed, by fixing a canonical order and sorting the expression, we see
\begin{Verbatim}[commandchars=\\\{\},xleftmargin=\parindent,numbers=left,bgcolor=bg]
\PYG{+w}{    }\PYG{n}{custom\PYGZus{}sort\PYGZus{}order!}\PYG{p}{(}\PYG{p}{[}\PYG{l+s}{\PYGZdq{}}\PYG{l+s}{γ5}\PYG{l+s}{\PYGZdq{}}\PYG{p}{,}\PYG{l+s}{\PYGZdq{}}\PYG{l+s}{γ}\PYG{l+s}{\PYGZdq{}}\PYG{p}{]}\PYG{p}{)}
\PYG{+w}{    }\PYG{n}{exp}\PYG{o}{=}\PYG{n}{gν}\PYG{o}{*}\PYG{n}{g5}
\PYG{+w}{    }\PYG{n+nd}{@show}\PYG{+w}{ }\PYG{n}{sort}\PYG{p}{(}\PYG{n}{exp}\PYG{p}{)}
\end{Verbatim}
yields \jil{-γ^{5}γ^{ν}}, while
\begin{Verbatim}[commandchars=\\\{\},xleftmargin=\parindent,numbers=left,bgcolor=bg]
\PYG{+w}{    }\PYG{n+nd}{@show}\PYG{+w}{ }\PYG{n}{sort}\PYG{p}{(}\PYG{n}{gν}\PYG{o}{*}\PYG{n}{gμ}\PYG{p}{)}
\end{Verbatim}
gives \jil{-γ^{μ}γ^{ν}+2η^{μν}}.
Gamma matrices also obey various product identities. These can be applied by the \jil{contract_adjacent()} function. For example, for
\begin{Verbatim}[commandchars=\\\{\},xleftmargin=\parindent,numbers=left,bgcolor=bg]
\PYG{+w}{    }\PYG{n}{gμ}\PYG{o}{=}\PYG{n}{Gamma}\PYG{p}{(}\PYG{n}{μ}\PYG{o}{=\PYGZgt{}}\PYG{n}{upper}\PYG{p}{)}
\PYG{+w}{    }\PYG{n}{gν}\PYG{o}{=}\PYG{n}{Gamma}\PYG{p}{(}\PYG{n}{ν}\PYG{o}{=\PYGZgt{}}\PYG{n}{upper}\PYG{p}{)}
\PYG{+w}{    }\PYG{n}{g\PYGZus{}μ}\PYG{o}{=}\PYG{n}{Gamma}\PYG{p}{(}\PYG{n}{μ}\PYG{o}{=\PYGZgt{}}\PYG{n}{lower}\PYG{p}{)}

\PYG{+w}{    }\PYG{n+nd}{@show}\PYG{+w}{ }\PYG{n}{contract\PYGZus{}adjacent}\PYG{p}{(}\PYG{n}{gμ}\PYG{o}{*}\PYG{n}{gν}\PYG{o}{*}\PYG{n}{g\PYGZus{}μ}\PYG{p}{,}\PYG{n}{Gamma}\PYG{p}{)}
\PYG{+w}{    }\PYG{c}{\PYGZsh{}   \PYGZhy{}2γ\PYGZca{}\PYGZob{}ν\PYGZcb{}}
\end{Verbatim}
we can reduce the expression. Generically, \jil{contract_adjacent()} will try to apply the identity that involves the largest number of matrices first, moving to smaller subsets when this fails, by calling a function \jil{reduce_matrix()} which can take a varying number of arguments. Note that this reduction works even when the matrices are not adjacent - they will first be dragged (respecting commutation relations with other tensors in between where needed) to adjacency before reducing, maintaining consistency.

Similarly to the Pauli matrices, calling \jil{Gamma} with additional indices from a different set will produce a ``component form'' tensor \jil{GammaT} which acts as a regular commuting tensor. Using indices from the same set will also generate higher rank elements of the Clifford algebra (either in matrix or component form), which in turn can have their own commutation relations and reduction rules.
\subsection{General Matrix Relations}\label{sec:general-matrix-relations}
Beyond $\sigma$ and $\gamma$ matrices, one may also like to encode commutation relations, as well as product identities, between other generic matrices. This can be done straightforwardly in Alakazam using Julia macros. 

For example, let us consider commutation relations. Let us suppose we're interested in studying quantum mechanics, where
\begin{equation}
    [x^\mu,p^\nu]=i\hbar\eta^{\mu\nu}.
\end{equation}
This can be implemented quite simply in tandem with \jil{Symbolics} by defining a commutation relation between generic \jil{Mat} objects using the \jil{@commutator} macro
\begin{Verbatim}[commandchars=\\\{\},xleftmargin=\parindent,numbers=left,bgcolor=bg]
\PYG{+w}{    }\PYG{n}{X}\PYG{o}{=}\PYG{n}{Mat}\PYG{p}{(}\PYG{l+s}{\PYGZdq{}}\PYG{l+s}{x}\PYG{l+s}{\PYGZdq{}}\PYG{p}{,}\PYG{n}{μ}\PYG{o}{=\PYGZgt{}}\PYG{n}{upper}\PYG{p}{)}
\PYG{+w}{    }\PYG{n}{P}\PYG{o}{=}\PYG{n}{Mat}\PYG{p}{(}\PYG{l+s}{\PYGZdq{}}\PYG{l+s}{p}\PYG{l+s}{\PYGZdq{}}\PYG{p}{,}\PYG{n}{ν}\PYG{o}{=\PYGZgt{}}\PYG{n}{upper}\PYG{p}{)}
\PYG{+w}{    }\PYG{n+nd}{@syms}\PYG{+w}{ }\PYG{n}{ħ}
\PYG{+w}{    }\PYG{n}{result}\PYG{o}{=}\PYG{n+nb}{im}\PYG{o}{*}\PYG{n}{ħ}\PYG{o}{*}\PYG{n}{Metric}\PYG{p}{(}\PYG{n}{μ}\PYG{o}{=\PYGZgt{}}\PYG{n}{upper}\PYG{p}{,}\PYG{n}{ν}\PYG{o}{=\PYGZgt{}}\PYG{n}{upper}\PYG{p}{)}
\PYG{+w}{    }\PYG{n+nd}{@commutator}\PYG{+w}{ }\PYG{p}{[}\PYG{n}{X}\PYG{p}{,}\PYG{n}{P}\PYG{p}{]}\PYG{o}{=}\PYG{n}{result}
\end{Verbatim}
This internally stores a symbolic template involving \jil{x^{μ}} and \jil{p^{ν}}.
One could then, for example, consider matrices with different explicit indices
\begin{Verbatim}[commandchars=\\\{\},xleftmargin=\parindent,numbers=left,bgcolor=bg]
\PYG{+w}{    }\PYG{n}{X2}\PYG{o}{=}\PYG{n}{Mat}\PYG{p}{(}\PYG{l+s}{\PYGZdq{}}\PYG{l+s}{x}\PYG{l+s}{\PYGZdq{}}\PYG{p}{,}\PYG{+w}{ }\PYG{n}{ρ}\PYG{o}{=\PYGZgt{}}\PYG{n}{upper}\PYG{p}{)}
\PYG{+w}{    }\PYG{n}{P2}\PYG{o}{=}\PYG{n}{Mat}\PYG{p}{(}\PYG{l+s}{\PYGZdq{}}\PYG{l+s}{p}\PYG{l+s}{\PYGZdq{}}\PYG{p}{,}\PYG{+w}{ }\PYG{n}{σ}\PYG{o}{=\PYGZgt{}}\PYG{n}{upper}\PYG{p}{)}
\PYG{+w}{    }\PYG{n}{custom\PYGZus{}sort\PYGZus{}order!}\PYG{p}{(}\PYG{p}{[}\PYG{l+s}{\PYGZdq{}}\PYG{l+s}{x}\PYG{l+s}{\PYGZdq{}}\PYG{p}{,}\PYG{l+s}{\PYGZdq{}}\PYG{l+s}{p}\PYG{l+s}{\PYGZdq{}}\PYG{p}{]}\PYG{p}{)}
\PYG{+w}{    }\PYG{n+nd}{@show}\PYG{+w}{ }\PYG{n}{sort}\PYG{p}{(}\PYG{n}{P2}\PYG{o}{*}\PYG{n}{X2}\PYG{p}{)}
\end{Verbatim}
This yields \jil{x^{ρ}p^{σ}-iħη^{ρσ}}, where the indices have been replaced respecting the registered commutation relation pattern.

We may also be interested in reducing products of matrices. Let us show how this can be implemented by a user for the Gell-Mann matrices $\lambda_a$ in QCD, satisfying
\begin{equation}
    \lambda_a\lambda_b=\frac{2}{3}\delta_{ab}+(d_{(abc)}+if_{[abc]})\lambda_c.
\end{equation}
To register a reduction rule, we must implement a rudimentary new type of matrix corresponding to these operators. This is because our macro will automatically generate new functions \jil{reduce_matrix()} that dispatch on specific types, for a variable number of arguments, and thus can be called generically internally. This step is essential but is minimal: it can be done directly with the \jil{@matrix} macro
\begin{Verbatim}[commandchars=\\\{\},xleftmargin=\parindent,numbers=left,bgcolor=bg]
\PYG{+w}{    }\PYG{n+nd}{@matrix}\PYG{+w}{ }\PYG{n}{GMMatrix}
\end{Verbatim}
which will behind the scenes generate the minimal code for a new struct
\begin{Verbatim}[commandchars=\\\{\},xleftmargin=\parindent,numbers=left,bgcolor=bg]
\PYG{+w}{    }\PYG{k}{mutable}\PYG{+w}{ }\PYG{k}{struct} \PYG{k+kt}{GMMatrix}\PYG{+w}{ }\PYG{o}{\PYGZlt{}:}\PYG{+w}{ }\PYG{k+kt}{MatrixSuperType}
\PYG{+w}{        }\PYG{n}{data}\PYG{o}{::}\PYG{k+kt}{TensorData}
\PYG{+w}{        }\PYG{n}{indices}\PYG{o}{::}\PYG{k+kt}{Vector}\PYG{p}{\PYGZob{}}\PYG{k+kt}{IndexPair}\PYG{p}{\PYGZcb{}}
\PYG{+w}{        }\PYG{n}{GMMatrix}\PYG{p}{(}\PYG{n}{args}\PYG{o}{...}\PYG{p}{)}\PYG{+w}{ }\PYG{o}{=}\PYG{+w}{ }\PYG{n}{new}\PYG{p}{(}\PYG{n}{build\PYGZus{}tensor}\PYG{p}{(}\PYG{n}{args}\PYG{o}{...}\PYG{p}{)}\PYG{o}{...}\PYG{p}{)}\PYG{+w}{ }
\PYG{+w}{        }\PYG{n}{GMMatrix}\PYG{p}{(}\PYG{n}{d}\PYG{o}{::}\PYG{k+kt}{TensorData}\PYG{p}{,}\PYG{+w}{ }\PYG{n}{i}\PYG{o}{::}\PYG{k+kt}{Vector}\PYG{p}{\PYGZob{}}\PYG{k+kt}{IndexPair}\PYG{p}{\PYGZcb{}}\PYG{p}{)}\PYG{+w}{ }\PYG{o}{=}\PYG{+w}{ }\PYG{n}{new}\PYG{p}{(}\PYG{n}{d}\PYG{p}{,}\PYG{+w}{ }\PYG{n}{i}\PYG{p}{)}
\PYG{+w}{    }\PYG{k}{end}
\end{Verbatim}
We can then create a template, 
\begin{Verbatim}[commandchars=\\\{\},xleftmargin=\parindent,numbers=left,bgcolor=bg]
\PYG{+w}{    }\PYG{n}{su3\PYGZus{}indices}\PYG{+w}{ }\PYG{o}{=}\PYG{+w}{ }\PYG{n}{IndexSet}\PYG{p}{(}\PYG{l+s}{\PYGZdq{}}\PYG{l+s}{su3\PYGZus{}flavour}\PYG{l+s}{\PYGZdq{}}\PYG{p}{,}\PYG{p}{(}\PYG{l+m+mi}{0}\PYG{p}{,}\PYG{l+m+mi}{8}\PYG{p}{)}\PYG{p}{,}\PYG{n}{lower}\PYG{p}{)}
\PYG{+w}{    }\PYG{n+nd}{@indices}\PYG{+w}{ }\PYG{n}{a}\PYG{o}{:}\PYG{n}{f∈su3\PYGZus{}indices}
\PYG{+w}{    }\PYG{n}{λa}\PYG{o}{=}\PYG{n}{GMMatrix}\PYG{p}{(}\PYG{l+s}{\PYGZdq{}}\PYG{l+s}{λ}\PYG{l+s}{\PYGZdq{}}\PYG{p}{,}\PYG{+w}{ }\PYG{n}{a}\PYG{o}{=\PYGZgt{}}\PYG{n}{lower}\PYG{p}{)}
\PYG{+w}{    }\PYG{n}{λb}\PYG{o}{=}\PYG{n}{GMMatrix}\PYG{p}{(}\PYG{l+s}{\PYGZdq{}}\PYG{l+s}{λ}\PYG{l+s}{\PYGZdq{}}\PYG{p}{,}\PYG{+w}{ }\PYG{n}{b}\PYG{o}{=\PYGZgt{}}\PYG{n}{lower}\PYG{p}{)}
\PYG{+w}{    }\PYG{n}{rhs}\PYG{o}{=}\PYG{p}{(}\PYG{l+m+mi}{2}\PYG{o}{//}\PYG{l+m+mi}{3}\PYG{o}{*}\PYG{n}{Metric}\PYG{p}{(}\PYG{n}{a}\PYG{o}{=\PYGZgt{}}\PYG{n}{lower}\PYG{p}{,}\PYG{+w}{ }\PYG{n}{b}\PYG{o}{=\PYGZgt{}}\PYG{n}{lower}\PYG{p}{)}
\PYG{+w}{         }\PYG{o}{+}\PYG{p}{(}\PYG{l+s+sa}{T}\PYG{l+s}{\PYGZdq{}}\PYG{l+s}{D\PYGZus{}(abc)}\PYG{l+s}{\PYGZdq{}}\PYG{o}{+}\PYG{n+nb}{im}\PYG{o}{*}\PYG{l+s+sa}{T}\PYG{l+s}{\PYGZdq{}}\PYG{l+s}{F\PYGZus{}[abc]}\PYG{l+s}{\PYGZdq{}}\PYG{p}{)}\PYG{o}{*}\PYG{n}{GMMatrix}\PYG{p}{(}\PYG{l+s}{\PYGZdq{}}\PYG{l+s}{λ}\PYG{l+s}{\PYGZdq{}}\PYG{p}{,}\PYG{+w}{ }\PYG{n}{c}\PYG{o}{=\PYGZgt{}}\PYG{n}{upper}\PYG{p}{)}\PYG{p}{)}
\end{Verbatim}
and register a reduction rule with the \jil{@mrule} macro
\begin{Verbatim}[commandchars=\\\{\},xleftmargin=\parindent,numbers=left,bgcolor=bg]
\PYG{+w}{    }\PYG{n+nd}{@mrule}\PYG{+w}{ }\PYG{n}{λa}\PYG{o}{*}\PYG{n}{λb}\PYG{+w}{ }\PYG{o}{=}\PYG{+w}{ }\PYG{n}{rhs}
\end{Verbatim}
This \jil{@mrule} macro will  automatically generate a \jil{reduce_matrix()} method that takes, in this case, two \jil{GMMatrix} as arguments and will generate code to reduce the product. We can then apply this to an expression,
\begin{Verbatim}[commandchars=\\\{\},xleftmargin=\parindent,numbers=left,bgcolor=bg]
\PYG{+w}{    }\PYG{n}{λc}\PYG{o}{=}\PYG{n}{GMMatrix}\PYG{p}{(}\PYG{l+s}{\PYGZdq{}}\PYG{l+s}{λ}\PYG{l+s}{\PYGZdq{}}\PYG{p}{,}\PYG{+w}{ }\PYG{n}{c}\PYG{o}{=\PYGZgt{}}\PYG{n}{lower}\PYG{p}{)}
\PYG{+w}{    }\PYG{n}{λd}\PYG{o}{=}\PYG{n}{GMMatrix}\PYG{p}{(}\PYG{l+s}{\PYGZdq{}}\PYG{l+s}{λ}\PYG{l+s}{\PYGZdq{}}\PYG{p}{,}\PYG{+w}{ }\PYG{n}{d}\PYG{o}{=\PYGZgt{}}\PYG{n}{lower}\PYG{p}{)}
\PYG{+w}{    }\PYG{n}{λe}\PYG{o}{=}\PYG{n}{GMMatrix}\PYG{p}{(}\PYG{l+s}{\PYGZdq{}}\PYG{l+s}{λ}\PYG{l+s}{\PYGZdq{}}\PYG{p}{,}\PYG{+w}{ }\PYG{n}{e}\PYG{o}{=\PYGZgt{}}\PYG{n}{lower}\PYG{p}{)}
\PYG{+w}{    }\PYG{n+nd}{@show}\PYG{+w}{ }\PYG{n}{contract\PYGZus{}adjacent}\PYG{p}{(}\PYG{n}{λc}\PYG{o}{*}\PYG{n}{λd}\PYG{o}{*}\PYG{n}{λe}\PYG{p}{,}\PYG{n}{GMMatrix}\PYG{p}{)}
\end{Verbatim}
which will completely reduce the expression to
\begin{Verbatim}[commandchars=\\\{\},xleftmargin=\parindent,numbers=left,bgcolor=bg]
\PYG{+w}{     }\PYG{c}{\PYGZsh{}2//3η\PYGZus{}\PYGZob{}cd\PYGZcb{}λ\PYGZus{}\PYGZob{}e\PYGZcb{}+2//3d\PYGZus{}\PYGZob{}cde\PYGZcb{}+D\PYGZca{}\PYGZob{}∘∘b\PYGZcb{}\PYGZus{}\PYGZob{}cd∘\PYGZcb{}D\PYGZus{}\PYGZob{}bea\PYGZcb{}λ\PYGZca{}\PYGZob{}a\PYGZcb{}}
\PYG{+w}{     }\PYG{c}{\PYGZsh{}+iD\PYGZca{}\PYGZob{}∘∘b\PYGZcb{}\PYGZus{}\PYGZob{}cd∘\PYGZcb{}F\PYGZus{}\PYGZob{}bea\PYGZcb{}λ\PYGZca{}\PYGZob{}a\PYGZcb{}+2//3iF\PYGZus{}\PYGZob{}cde\PYGZcb{}+iF\PYGZca{}\PYGZob{}∘∘b\PYGZcb{}\PYGZus{}\PYGZob{}cd∘\PYGZcb{}D\PYGZus{}\PYGZob{}bea\PYGZcb{}λ\PYGZca{}\PYGZob{}a\PYGZcb{}}
\PYG{+w}{     }\PYG{c}{\PYGZsh{}\PYGZhy{}F\PYGZca{}\PYGZob{}∘∘b\PYGZcb{}\PYGZus{}\PYGZob{}cd∘\PYGZcb{}F\PYGZus{}\PYGZob{}bea\PYGZcb{}λ\PYGZca{}\PYGZob{}a\PYGZcb{}}
\end{Verbatim}
We are not limited to reduction rules for only a product of two matrices - one may define different rules for each product of $n$ matrices, analogously to Gamma matrix identities. This will generate a different \jil{reduce_matrix()} method that takes $n$ arguments instead. When reducing an expression, \jil{Alakazam} will first aim to apply the rule to reduce to the largest number of matrices before attempting smaller rules.

\subsection{Bosons and Fermions}\label{sec:bosonfermion}
As mentioned in~\ref{sec:indexset}, grading information in Alakazam is stored on indices. We recognise, however, this is not always satisfactory - as a simple example, ghost fields obey opposite statistics than to what indices may suggest. For this reason, we include explicit \jil{Boson} and \jil{Fermion} types, which behave exactly as a plain \jil{Tensor}, but with fixed grading, regardless of indices. In fact, the function that determines the grading dispatches on types, so users are free to override such behaviour on custom tensors with any sort of logic to return the $\mathbb{Z}_2$ grading
\begin{Verbatim}[commandchars=\\\{\},xleftmargin=\parindent,numbers=left,bgcolor=bg]
\PYG{+w}{    }\PYG{n}{Alakazam}\PYG{o}{.}\PYG{n}{grading}\PYG{p}{(}\PYG{n}{T}\PYG{o}{::}\PYG{k+kt}{MyTensor}\PYG{p}{)}\PYG{o}{=}\PYG{+w}{ }\PYG{c}{\PYGZsh{}...}
\end{Verbatim}

\subsection{\jil{Spinor}}\label{sec:spinors}
For utility purposes, we include a \jil{Spinor<:MatrixSuperType} which has explicit fermionic grading. This allows one to encode spinors without indices while maintaining grading, which will not commute with gamma matrices, nor other matrices, by default. This is useful for example expressing barred spinors which include a gamma matrix implicitly. 

\subsection{Bras and Kets}\label{sec:braket}
We have included a base implementation of bras and kets via \jil{Bra<:BraSuperType} and \jil{Ket<:KetSuperType}, which are a \jil{<:DiracVectorSuperType<:MatrixSuperType}, which have special rendering properties to use Dirac notation and to display adjacent bra-ket pairs as an inner product. These currently have no specialised functionality, other than what users can already encode using custom matrix reduction rules. These will be expanded upon at a future date.
\begin{Verbatim}[commandchars=\\\{\},xleftmargin=\parindent,numbers=left,bgcolor=bg]
\PYG{+w}{    }\PYG{n+nd}{@show}\PYG{+w}{ }\PYG{n}{Bra}\PYG{p}{(}\PYG{l+s}{\PYGZdq{}}\PYG{l+s}{ψ}\PYG{l+s}{\PYGZdq{}}\PYG{p}{)}
\PYG{+w}{    }\PYG{c}{\PYGZsh{}   ⟨ψ|}
\PYG{+w}{    }\PYG{n+nd}{@show}\PYG{+w}{ }\PYG{n}{Ket}\PYG{p}{(}\PYG{l+s}{\PYGZdq{}}\PYG{l+s}{ϕ}\PYG{l+s}{\PYGZdq{}}\PYG{p}{)}
\PYG{+w}{    }\PYG{c}{\PYGZsh{}   |ϕ⟩}
\PYG{+w}{    }\PYG{n+nd}{@show}\PYG{+w}{ }\PYG{n}{Bra}\PYG{p}{(}\PYG{l+s}{\PYGZdq{}}\PYG{l+s}{ψ}\PYG{l+s}{\PYGZdq{}}\PYG{p}{)}\PYG{o}{*}\PYG{n}{Ket}\PYG{p}{(}\PYG{l+s}{\PYGZdq{}}\PYG{l+s}{ϕ}\PYG{l+s}{\PYGZdq{}}\PYG{p}{)}
\PYG{+w}{    }\PYG{c}{\PYGZsh{}   ⟨ψ|ϕ⟩}
\PYG{+w}{    }\PYG{n+nd}{@show}\PYG{+w}{ }\PYG{n}{Bra}\PYG{p}{(}\PYG{l+s}{\PYGZdq{}}\PYG{l+s}{ψ}\PYG{l+s}{\PYGZdq{}}\PYG{p}{)}\PYG{o}{\PYGZsq{}}
\PYG{+w}{    }\PYG{c}{\PYGZsh{}   |ψ⟩}
\end{Verbatim}

\newpage

\section{Tensor Operations}\label{sec:tensor-operations}

In this section, we will describe various common functions to manipulate tensors and tensor expressions. We list here just a few routine examples - a full list of all function handles exposed to the user in the current version at the time of writing can be found in the appendix.

\subsection{Accessors and Properties}\label{sec:accessors-and-properties}
Properties of tensorial objects can be accessed by appropriate accessor functions. This is the safe way to access a tensors internal fields, and is stable against any changes to the internal implementation. The basic accessors to be aware of are
\begin{Verbatim}[commandchars=\\\{\},xleftmargin=\parindent,numbers=left,bgcolor=bg]
\PYG{+w}{    }\PYG{n}{name}\PYG{p}{(}\PYG{n}{T}\PYG{p}{)}\PYG{+w}{         }\PYG{c}{\PYGZsh{} the tensor\PYGZsq{}s name, as a String}
\PYG{+w}{    }\PYG{n}{indices}\PYG{p}{(}\PYG{n}{T}\PYG{p}{)}\PYG{+w}{      }\PYG{c}{\PYGZsh{} Vector of index =\PYGZgt{} position pairs, in slot order}
\PYG{+w}{    }\PYG{n}{function\PYGZus{}of}\PYG{p}{(}\PYG{n}{T}\PYG{p}{)}\PYG{+w}{  }\PYG{c}{\PYGZsh{} Vector of Coordinates the tensor depends on}
\PYG{+w}{    }\PYG{n}{weights}\PYG{p}{(}\PYG{n}{T}\PYG{p}{)}\PYG{+w}{      }\PYG{c}{\PYGZsh{} Dict of weight name =\PYGZgt{} value, one entry per coordinate}
\PYG{+w}{                    }\PYG{c}{\PYGZsh{}   by default}
\PYG{+w}{    }\PYG{n}{tableaux}\PYG{p}{(}\PYG{n}{T}\PYG{p}{)}\PYG{+w}{     }\PYG{c}{\PYGZsh{} Set of YoungTableaux encoding the index symmetries}
\PYG{+w}{    }\PYG{n}{labels}\PYG{p}{(}\PYG{n}{T}\PYG{p}{)}\PYG{+w}{       }\PYG{c}{\PYGZsh{} Vector of label =\PYGZgt{} position pairs}
\PYG{+w}{    }\PYG{n}{free\PYGZus{}indices}\PYG{p}{(}\PYG{n}{T}\PYG{p}{)}\PYG{+w}{ }\PYG{c}{\PYGZsh{} multiset of the uncontracted index =\PYGZgt{} position pairs,}
\PYG{+w}{                    }\PYG{c}{\PYGZsh{}   with multiplicities}
\PYG{+w}{    }\PYG{n}{dummies}\PYG{p}{(}\PYG{n}{T}\PYG{p}{)}\PYG{+w}{      }\PYG{c}{\PYGZsh{} Vector of the Index objects that are summed over}
\PYG{+w}{    }\PYG{n}{grading}\PYG{p}{(}\PYG{n}{T}\PYG{p}{)}\PYG{+w}{      }\PYG{c}{\PYGZsh{} 0 if bosonic, 1 if fermionic}
\PYG{+w}{    }\PYG{n}{isAntisymmetric}\PYG{p}{(}\PYG{n}{T}\PYG{p}{)}\PYG{+w}{  }\PYG{c}{\PYGZsh{} true if T is totally antisymmetric}
\PYG{+w}{    }\PYG{n}{isSymmetric}\PYG{p}{(}\PYG{n}{T}\PYG{p}{)}\PYG{+w}{      }\PYG{c}{\PYGZsh{} true if T is totally symmetric}
\PYG{+w}{    }\PYG{n}{hasIndices}\PYG{p}{(}\PYG{n}{T}\PYG{p}{)}\PYG{+w}{       }\PYG{c}{\PYGZsh{} true if T carries any indices}
\PYG{+w}{    }\PYG{n}{data}\PYG{p}{(}\PYG{n}{T}\PYG{p}{)}\PYG{+w}{             }\PYG{c}{\PYGZsh{} the underlying TensorData}
\end{Verbatim}
For \jil{TensorTerm} and \jil{TensorExpression} there are also
\begin{Verbatim}[commandchars=\\\{\},xleftmargin=\parindent,numbers=left,bgcolor=bg]
\PYG{+w}{    }\PYG{n}{isTensor}\PYG{p}{(}\PYG{n}{T}\PYG{p}{)}\PYG{+w}{     }\PYG{c}{\PYGZsh{} the single Tensor if T is one, otherwise nothing}
\PYG{+w}{    }\PYG{n}{terms}\PYG{p}{(}\PYG{n}{T}\PYG{p}{)}\PYG{+w}{        }\PYG{c}{\PYGZsh{} Vector of the TensorTerms (or, for a TensorTerm,}
\PYG{+w}{                    }\PYG{c}{\PYGZsh{}   its TensorSuperType factors)}
\PYG{+w}{    }\PYG{n}{summands}\PYG{p}{(}\PYG{n}{T}\PYG{p}{)}\PYG{+w}{     }\PYG{c}{\PYGZsh{} Vector of coefficient =\PYGZgt{} TensorTerm pairs}
\PYG{+w}{    }\PYG{n}{coefficients}\PYG{p}{(}\PYG{n}{T}\PYG{p}{)}\PYG{+w}{ }\PYG{c}{\PYGZsh{} Vector of just the coefficients of a TensorExpression}
\PYG{+w}{    }\PYG{n}{is\PYGZus{}const}\PYG{p}{(}\PYG{n}{T}\PYG{p}{)}\PYG{+w}{     }\PYG{c}{\PYGZsh{} true if T is a bare coefficient with no tensor content}
\PYG{+w}{    }\PYG{n}{first\PYGZus{}summand}\PYG{p}{(}\PYG{n}{T}\PYG{p}{)}\PYG{c}{\PYGZsh{} the first TensorTerm, or nothing if empty}
\PYG{+w}{    }\PYG{n}{first\PYGZus{}coef}\PYG{p}{(}\PYG{n}{T}\PYG{p}{)}\PYG{+w}{   }\PYG{c}{\PYGZsh{} the coefficient of the first summand}
\end{Verbatim}

\subsubsection{\jil{flip_all()}, \jil{raise_all()}, \jil{lower_all()}}\label{sec:flip-all-raise-all-lower-all}
These utilities change the positions of the indices of a \jil{Tensorial} expression.
\jil{flip_all()} reverses every position, upper becoming lower and lower upper, while
\jil{raise_all()} and \jil{lower_all()} move them all to a single level. They act on a
\jil{Tensor}, \jil{TensorTerm} or \jil{TensorExpression} alike, so a whole expression can be
converted at once, and \jil{flip_all()} is its own inverse. Note that all three act on the free indices only, leaving dummy pairs invariant. This avoids sign ambiguities for spinors in \jil{flip_all}, or a \jil{SameLevelDummyError} for \jil{raise_all()} and \jil{lower_all()}.
These operations relabel index slots; they insert no metric factors, and are not a substitute for
\jil{pop_metric()} or \jil{contract_metrics()} where explicit metrics are wanted. Their typical use
is to prepare one copy of an expression for contraction against another:

\begin{Verbatim}[commandchars=\\\{\},xleftmargin=\parindent,numbers=left,bgcolor=bg]
\PYG{+w}{    }\PYG{n}{A}\PYG{+w}{ }\PYG{o}{=}\PYG{+w}{ }\PYG{n}{Tensor}\PYG{p}{(}\PYG{l+s}{\PYGZdq{}}\PYG{l+s}{A}\PYG{l+s}{\PYGZdq{}}\PYG{p}{,}\PYG{+w}{ }\PYG{p}{[}\PYG{n}{μ}\PYG{+w}{ }\PYG{o}{=\PYGZgt{}}\PYG{+w}{ }\PYG{n}{upper}\PYG{p}{,}\PYG{+w}{ }\PYG{n}{ν}\PYG{+w}{ }\PYG{o}{=\PYGZgt{}}\PYG{+w}{ }\PYG{n}{lower}\PYG{p}{]}\PYG{p}{)}
\PYG{+w}{    }\PYG{n+nd}{@show}\PYG{+w}{ }\PYG{n}{flip\PYGZus{}all}\PYG{p}{(}\PYG{n}{A}\PYG{p}{)}\PYG{+w}{                     }\PYG{c}{\PYGZsh{}   A\PYGZca{}\PYGZob{}∘ν\PYGZcb{}\PYGZus{}\PYGZob{}μ∘\PYGZcb{}}

\PYG{+w}{    }\PYG{n}{M}\PYG{+w}{ }\PYG{o}{=}\PYG{+w}{ }\PYG{l+m+mi}{2}\PYG{o}{*}\PYG{n}{Tensor}\PYG{p}{(}\PYG{l+s}{\PYGZdq{}}\PYG{l+s}{M}\PYG{l+s}{\PYGZdq{}}\PYG{p}{,}\PYG{+w}{ }\PYG{p}{[}\PYG{n}{μ}\PYG{+w}{ }\PYG{o}{=\PYGZgt{}}\PYG{+w}{ }\PYG{n}{upper}\PYG{p}{,}\PYG{+w}{ }\PYG{n}{ν}\PYG{+w}{ }\PYG{o}{=\PYGZgt{}}\PYG{+w}{ }\PYG{n}{upper}\PYG{p}{]}\PYG{p}{)}
\PYG{+w}{    }\PYG{n+nd}{@show}\PYG{+w}{ }\PYG{n}{simplify}\PYG{p}{(}\PYG{n}{flip\PYGZus{}all}\PYG{p}{(}\PYG{n}{M}\PYG{p}{)}\PYG{+w}{ }\PYG{o}{*}\PYG{+w}{ }\PYG{n}{M}\PYG{p}{)}\PYG{+w}{       }\PYG{c}{\PYGZsh{}   4M\PYGZca{}\PYGZob{}μν\PYGZcb{}M\PYGZus{}\PYGZob{}μν\PYGZcb{}}

\PYG{+w}{    }\PYG{n}{T}\PYG{+w}{ }\PYG{o}{=}\PYG{+w}{ }\PYG{n}{Tensor}\PYG{p}{(}\PYG{l+s}{\PYGZdq{}}\PYG{l+s}{S}\PYG{l+s}{\PYGZdq{}}\PYG{p}{,}\PYG{+w}{ }\PYG{p}{[}\PYG{n}{μ}\PYG{+w}{ }\PYG{o}{=\PYGZgt{}}\PYG{+w}{ }\PYG{n}{upper}\PYG{p}{]}\PYG{p}{)}\PYG{+w}{ }\PYG{o}{*}\PYG{+w}{ }\PYG{n}{Tensor}\PYG{p}{(}\PYG{l+s}{\PYGZdq{}}\PYG{l+s}{T}\PYG{l+s}{\PYGZdq{}}\PYG{p}{,}\PYG{+w}{ }\PYG{p}{[}\PYG{n}{μ}\PYG{+w}{ }\PYG{o}{=\PYGZgt{}}\PYG{+w}{ }\PYG{n}{lower}\PYG{p}{,}\PYG{+w}{ }\PYG{n}{ν}\PYG{+w}{ }\PYG{o}{=\PYGZgt{}}\PYG{+w}{ }\PYG{n}{lower}\PYG{p}{]}\PYG{p}{)}
\PYG{+w}{    }\PYG{n+nd}{@show}\PYG{+w}{ }\PYG{n}{raise\PYGZus{}all}\PYG{p}{(}\PYG{n}{T}\PYG{p}{)}\PYG{+w}{                    }\PYG{c}{\PYGZsh{}   S\PYGZca{}\PYGZob{}μ\PYGZcb{}T\PYGZca{}\PYGZob{}∘ν\PYGZcb{}\PYGZus{}\PYGZob{}μ∘\PYGZcb{}   }
\PYG{+w}{                                          }\PYG{c}{\PYGZsh{}(μ is a dummy, so untouched)}
\end{Verbatim}

\subsection{\jil{rename_dummies()}}\label{sec:rename-dummies}
In Einstein summation notation, repeated indices in an expression denote summation and are thus ``dummy'' indices: the name of their label is irrelevant. Often in the process of performing algebraic manipulations, being able to recognise terms that are equivalent up to dummy index labels is important. Thus, the built-in function \jil{rename_dummies()} is able to relabel dummy indices in an expression to some canonical ordering, which allows for like terms to be combined. Indices are guaranteed to be taken from the correct \jil{IndexSet}, and will be replaced with the first available indices from the set that are not already used in the expression. For example,
\begin{Verbatim}[commandchars=\\\{\},xleftmargin=\parindent,numbers=left,bgcolor=bg]
\PYG{+w}{    }\PYG{n}{A}\PYG{+w}{ }\PYG{o}{=}\PYG{+w}{ }\PYG{l+s+sa}{T}\PYG{l+s}{\PYGZdq{}}\PYG{l+s}{A\PYGZus{}af\PYGZca{}fc}\PYG{l+s}{\PYGZdq{}}
\PYG{+w}{    }\PYG{n+nd}{@show}\PYG{+w}{ }\PYG{n}{rename\PYGZus{}dummies}\PYG{p}{(}\PYG{n}{A}\PYG{p}{)}
\PYG{+w}{    }\PYG{c}{\PYGZsh{}   rename\PYGZus{}dummies(A) = A\PYGZca{}\PYGZob{}∘∘bc\PYGZcb{}\PYGZus{}\PYGZob{}ab∘∘\PYGZcb{}}
\end{Verbatim}

\subsubsection{\jil{sort()}}\label{sec:sort}

The \jil{sort()} function sorts a \jil{TensorExpression} into some canonical form. By default, this function will sort a \jil{TensorTerm} such that the \jil{Tensor}s are in alphabetical order. However, the user may also define a custom sort order via \jil{custom_sort_order!()}, which takes a \jil{Vector} of strings which \jil{sort()} will prioritise to sort into that specified order. Any name not found in this user-specified order will be sorted alphabetically after the user-specified names. In the event that two tensors have the same name, \jil{sort()} will aim to lexicographically order them by referring to the indices, ignoring dummy indices.

The \jil{sort()} function will also take grading and non-commutativity into account. To do this, the function uses a modified bubble sort, running in $O(n^2)$ time, for a \jil{TensorTerm} with $n$ \jil{Tensor} factors, and thus $O(kn^2)$ for a $k$ term \jil{TensorExpression}. The fact that some terms in the expression may not commute is important, as extra terms may accumulate when performing swaps. This prohibits the usage of faster sorting algorithms. When non-commuting factors are not present in a term, which we check for by a linear scan (and a quadratic check over detected matrices to see if they \jil{can_swap()}), we are able to resort to a faster $O(n\log n)$ merge sort and count the number of inversions to compute the overall sign for fermion inversions (if they exist, which is again determined by a linear scan) in the resulting permutation after the fact, which is $O(m^2)$ in the number of fermions $m$. This allows significant performance improvements for bosonic monomials. In the case where matrices with no commutation relations are present, their order is not changed, but commuting factors are allowed to move freely in the sort.

In addition to the order of tensors in a \jil{TensorTerm}, the indices attached to tensors will also be sorted into a canonical ordering. This is done by exploiting symmetries of the tensor when possible to arrange the indices as close to lexicographical order as possible. 
For example
\begin{Verbatim}[commandchars=\\\{\},xleftmargin=\parindent,numbers=left,bgcolor=bg]
\PYG{+w}{    }\PYG{n+nd}{@show}\PYG{+w}{ }\PYG{n}{sort}\PYG{p}{(}\PYG{l+s+sa}{T}\PYG{l+s}{\PYGZdq{}}\PYG{l+s}{C\PYGZus{}[d\PYGZca{}c]}\PYG{l+s}{\PYGZdq{}}\PYG{o}{*}\PYG{l+s+sa}{T}\PYG{l+s}{\PYGZdq{}}\PYG{l+s}{A\PYGZus{}(cb)a}\PYG{l+s}{\PYGZdq{}}\PYG{p}{)}
\PYG{+w}{    }\PYG{c}{\PYGZsh{}   \PYGZhy{}A\PYGZus{}\PYGZob{}bca\PYGZcb{}C\PYGZca{}\PYGZob{}c∘\PYGZcb{}\PYGZus{}\PYGZob{}∘d\PYGZcb{}}
\end{Verbatim}
and when the objects are fermionic with spinor indices
\begin{Verbatim}[commandchars=\\\{\},xleftmargin=\parindent,numbers=left,bgcolor=bg]
\PYG{+w}{    }\PYG{n+nd}{@show}\PYG{+w}{ }\PYG{n}{sort}\PYG{p}{(}\PYG{l+s+sa}{T}\PYG{l+s}{\PYGZdq{}}\PYG{l+s}{ϕ\PYGZca{}β\PYGZus{}[d\PYGZca{}c]}\PYG{l+s}{\PYGZdq{}}\PYG{o}{*}\PYG{l+s+sa}{T}\PYG{l+s}{\PYGZdq{}}\PYG{l+s}{ψ\PYGZca{}α\PYGZus{}(cb)a}\PYG{l+s}{\PYGZdq{}}\PYG{p}{)}
\PYG{+w}{    }\PYG{c}{\PYGZsh{}   ψ\PYGZca{}\PYGZob{}α∘∘∘\PYGZcb{}\PYGZus{}\PYGZob{}∘bca\PYGZcb{}ϕ\PYGZca{}\PYGZob{}βc∘\PYGZcb{}\PYGZus{}\PYGZob{}∘∘d\PYGZcb{}}
\end{Verbatim}
\subsubsection{\jil{canonicalise()}}\label{sec:canonicalise}
This is a rudimentary, cheap simplification algorithm. It consists of two steps: merge terms of the \jil{TensorExpression}, combining coefficients of summands when the summands are equivalent via simple symmetries, and then \jil{sort()} the resulting expression into canonical ordering. This lightweight algorithm is sufficient for simplifying many basic expressions.

\subsubsection{\jil{collect_terms()}}\label{sec:collect-terms}

\jil{collect_terms()} merges summands that are equal up to the naming of dummy indices and the
order of factors that are free to commute. These are exactly the duplicates that arise from
expanding a derivative or a product, where the same term is generated many times over with
different dummy labels and its factors in different orders.

\begin{Verbatim}[commandchars=\\\{\},xleftmargin=\parindent,numbers=left,bgcolor=bg]
\PYG{+w}{    }\PYG{n+nd}{@show}\PYG{+w}{ }\PYG{n}{length}\PYG{p}{(}\PYG{n}{summands}\PYG{p}{(}\PYG{n}{apply\PYGZus{}derivative}\PYG{p}{(}\PYG{n}{D}\PYG{p}{(}\PYG{n}{mono}\PYG{p}{(}\PYG{l+m+mi}{6}\PYG{p}{)}\PYG{p}{)}\PYG{p}{)}\PYG{p}{)}\PYG{p}{)}\PYG{+w}{ }\PYG{c}{\PYGZsh{}6 terms}
\PYG{+w}{    }\PYG{n+nd}{@show}\PYG{+w}{ }\PYG{n}{length}\PYG{p}{(}\PYG{n}{summands}\PYG{p}{(}\PYG{n}{collect\PYGZus{}terms}\PYG{p}{(}\PYG{n}{ans}\PYG{p}{)}\PYG{p}{)}\PYG{p}{)}\PYG{+w}{  }\PYG{c}{\PYGZsh{}2 terms:\PYGZti{}6δ*mono(5)+6η*mono(5)}
\end{Verbatim}

The merge is done without comparing terms to one another. Each summand is reduced to a canonical signature, its factors are placed in a canonical order, and a key is emitted in which every factor contributes an identity token followed by its index slots; a free index contributing its own identifier and a dummy contributing the position at which it first appears in the term. Operators contribute their name and then recurse into their operand. Two summands that
differ only by dummy relabelling or by a permutation of commuting factors therefore produce
identical keys, and summands are grouped by hashing the key rather than by structural comparison.
The graded sign induced by the reordering is carried alongside the key, so coefficients are combined
with the correct sign. The cost is one signature per summand, so collection is linear in the number
of terms, whereas a pairwise merge such as the one inside \jil{hard_simplify()} is quadratic.

\jil{collect_terms()} is the first step of many of our internal algorithms, which is
deliberate. Most other algorithms are superlinear, involving pairwise comparisons between summands. \jil{collect_terms()} acts as a cheap check to identify summands that are clearly the same and is heavily rewarding for large expressions.

The function is exported and may be called directly. This is worth doing whenever a user assembles their own simplification routine of operations, and particularly before any
expensive stage, for the reason above.

\subsubsection{\jil{simplify()}}\label{sec:simplify}
The \jil{simplify()} function performs a series of other functions in a particular order to simplify a \jil{TensorExpression}.
In particular, this function:
\begin{itemize}
    \item Collects summands that are equal up to dummy relabeling and reordering of commuting factors using \jil{collect_terms()}
    \item Symmetrises any partial derivatives in the expression using \jil{symmetrise_partials()}
    \item Pops all metrics in the expression using \jil{pop_metric()}
    \item Sorts the expression using \jil{sort()}
    \item Eliminates \jil{EpsilonTensor}s with \jil{eliminate_epsilon()}
    \item Sorts the operands of any operators into a canonical order \jil{simp_ops_n_sym()}
    \item Contracts all remaining \jil{Metric}s \jil{contract_metrics()}
    \item Eliminates \jil{KroneckerDelta}s with \jil{eliminate_kronecker}
    \item Renames dummy indices with \jil{rename_dummies!()}
    \item Sorts a final time with \jil{sort()}
\end{itemize}
This has worst case performance $O\Big(kn(n+m+i^2)\Big)$, where $k$ is the number of distinct \jil{TensorTerms} in the collected \jil{TensorExpression} (often much lower than the original input number of terms), $m$ is the maximal number of \jil{Metric}s in a \jil{TensorTerm} in the initial expression, $n$ is the maximal number of non-metric \jil{Tensor}s in a term in the initial expression, and $i$ is the maximal number of indices attached to a \jil{Tensor} in the \jil{TensorExpression}. In practice, the algorithm will run much faster: this asymptotic runtime is for extreme pathological inputs.

This particular routine was chosen as a roughly minimal set of operations to be able to successfully simplify a variety of expressions under different use cases. The user is, of course, free to define their own simplification function with a subset of these operations, or with additional operations, more specific to their use case.

\subsection{\jil{hard_simplify()}}\label{sec:hard-simplify}
In addition to the regular simplification algorithm, we can also perform a more in-depth merge of terms in a \jil{TensorExpression}. The algorithm \jil{hard_simplify()} will perform a potentially more expensive routine that can reduce expressions by recognising equivalences via the intrinsic symmetries of
their tensors and a relabelling of dummy indices. This is done by collecting terms, constructing a graph representing each remaining \jil{TensorTerm}, and checking for graph isomorphisms.

For example, consider
\begin{Verbatim}[commandchars=\\\{\},xleftmargin=\parindent,numbers=left,bgcolor=bg]
\PYG{+w}{    }\PYG{n}{expr}\PYG{o}{=}\PYG{p}{(}\PYG{n}{AntisymmetricTensor}\PYG{p}{(}\PYG{l+s}{\PYGZdq{}}\PYG{l+s}{A}\PYG{l+s}{\PYGZdq{}}\PYG{p}{,}\PYG{p}{[}\PYG{n}{a}\PYG{o}{=\PYGZgt{}}\PYG{n}{upper}\PYG{p}{,}\PYG{n}{b}\PYG{o}{=\PYGZgt{}}\PYG{n}{upper}\PYG{p}{]}\PYG{p}{)}
\PYG{+w}{        }\PYG{o}{*}\PYG{n}{SymmetricTensor}\PYG{p}{(}\PYG{l+s}{\PYGZdq{}}\PYG{l+s}{B}\PYG{l+s}{\PYGZdq{}}\PYG{p}{,}\PYG{p}{[}\PYG{n}{a}\PYG{o}{=\PYGZgt{}}\PYG{n}{lower}\PYG{p}{,}\PYG{n}{b}\PYG{o}{=\PYGZgt{}}\PYG{n}{lower}\PYG{p}{]}\PYG{p}{)}\PYG{p}{)}
\end{Verbatim}
This expression is identically zero, since it is an antisymmetric object contracted against a symmetric object. However, naive simplification cannot recognise this fact:
\begin{Verbatim}[commandchars=\\\{\},xleftmargin=\parindent,numbers=left,bgcolor=bg]
\PYG{+w}{    }\PYG{n+nd}{@show}\PYG{+w}{ }\PYG{n}{simplify}\PYG{p}{(}\PYG{n}{expr}\PYG{p}{)}
\PYG{+w}{    }\PYG{c}{\PYGZsh{}   A\PYGZca{}\PYGZob{}ab\PYGZcb{}B\PYGZus{}\PYGZob{}ab\PYGZcb{}}
\end{Verbatim}
Instead, one must look for patterns across tensors that imply this expression is zero via
\begin{Verbatim}[commandchars=\\\{\},xleftmargin=\parindent,numbers=left,bgcolor=bg]
\PYG{+w}{    }\PYG{n}{hard\PYGZus{}simplify}\PYG{p}{(}\PYG{n}{expr}\PYG{p}{)}
\PYG{+w}{    }\PYG{c}{\PYGZsh{}   0}
\end{Verbatim}
Similarly, two \jil{TensorTerm}s may be equivalent by a nontrivial symmetry or relabelling of dummy indices that may not be obtainable from simple manipulations. We use a graph isomorphism algorithm to detect such equivalences. We defer the technical details to~\ref{sec:hardsimpimp}.

\subsubsection{Muliterm Symmetries}
\label{sec:multiterm}
A multiterm symmetry is a linear identity relating several monomials, each of which is separately non-zero and separately in canonical form. The standard example is the algebraic Bianchi identity $R_{a[bcd]}=0$. This identity is not implied by signed permutation of a single monomial's slots, so neither Butler-Portugal nor graph isomorphism is capable of recognising it. Identities of this sort are however, implied by the Young tableaux of the monomials. As such, these relations can be recovered by the Young projection, briefly mentioned in~\ref{sec:symmetries-and-youngtableau}, which writes a tensor in a form where all of the symmetries are manifest. Our \jil{hard_simplify()} will use this on request:
\begin{Verbatim}[commandchars=\\\{\},xleftmargin=\parindent,numbers=left,bgcolor=bg]
\PYG{+w}{    }\PYG{n}{R}\PYG{p}{(}\PYG{n}{i}\PYG{p}{,}\PYG{n}{j}\PYG{p}{,}\PYG{n}{k}\PYG{p}{,}\PYG{n}{l}\PYG{p}{)}\PYG{+w}{ }\PYG{o}{=}\PYG{+w}{ }\PYG{n}{Riemann}\PYG{p}{(}\PYG{l+s}{\PYGZdq{}}\PYG{l+s}{R}\PYG{l+s}{\PYGZdq{}}\PYG{p}{,}\PYG{+w}{ }\PYG{p}{[}\PYG{n}{i}\PYG{o}{=\PYGZgt{}}\PYG{n}{lower}\PYG{p}{,}\PYG{+w}{ }\PYG{n}{j}\PYG{o}{=\PYGZgt{}}\PYG{n}{lower}\PYG{p}{,}\PYG{+w}{ }\PYG{n}{k}\PYG{o}{=\PYGZgt{}}\PYG{n}{lower}\PYG{p}{,}\PYG{+w}{ }\PYG{n}{l}\PYG{o}{=\PYGZgt{}}\PYG{n}{lower}\PYG{p}{]}\PYG{p}{)}
\PYG{+w}{    }\PYG{n}{bianchi}\PYG{+w}{ }\PYG{o}{=}\PYG{+w}{ }\PYG{n}{R}\PYG{p}{(}\PYG{n}{a}\PYG{p}{,}\PYG{n}{b}\PYG{p}{,}\PYG{n}{c}\PYG{p}{,}\PYG{n}{d}\PYG{p}{)}\PYG{+w}{ }\PYG{o}{+}\PYG{+w}{ }\PYG{n}{R}\PYG{p}{(}\PYG{n}{a}\PYG{p}{,}\PYG{n}{c}\PYG{p}{,}\PYG{n}{d}\PYG{p}{,}\PYG{n}{b}\PYG{p}{)}\PYG{+w}{ }\PYG{o}{+}\PYG{+w}{ }\PYG{n}{R}\PYG{p}{(}\PYG{n}{a}\PYG{p}{,}\PYG{n}{d}\PYG{p}{,}\PYG{n}{b}\PYG{p}{,}\PYG{n}{c}\PYG{p}{)}

\PYG{+w}{    }\PYG{n+nd}{@show}\PYG{+w}{ }\PYG{n}{hard\PYGZus{}simplify}\PYG{p}{(}\PYG{n}{bianchi}\PYG{p}{)}
\PYG{+w}{    }\PYG{c}{\PYGZsh{}   R\PYGZus{}\PYGZob{}adbc\PYGZcb{}+R\PYGZus{}\PYGZob{}abcd\PYGZcb{}\PYGZhy{}R\PYGZus{}\PYGZob{}acbd\PYGZcb{}}
\PYG{+w}{    }\PYG{n+nd}{@show}\PYG{+w}{ }\PYG{n}{hard\PYGZus{}simplify}\PYG{p}{(}\PYG{n}{bianchi}\PYG{p}{;}\PYG{+w}{ }\PYG{n}{multiterm}\PYG{o}{=}\PYG{n+nb}{true}\PYG{p}{)}
\PYG{+w}{    }\PYG{c}{\PYGZsh{}   0}
\end{Verbatim}
The same option reduces contracted consequences of the identity, such as
$R_{abcd}R^{acbd}=\tfrac12 R_{abcd}R^{abcd}$, which reduces to zero under
\jil{multiterm=true} but not otherwise.

Two aspects of this deserve comment. First, multiterm reduction is inherently a global operation on a sum and cannot be folded into the pairwise graph relations. Second, it is off by default because projection is exponential. The Young symmetriser of a tableau
expands into $\prod_i r_i!\cdot\prod_j c_j!$ signed permutations, for row lengths $r_i$ and column
lengths $c_j$, and these multiply across the factors of a term: one Riemann tensor gives $16$, two
give $256$, three give $4096$. Since this count depends only on tableau shape it can be
evaluated before any work is done, which is what \jil{young_projector_size()} reports, and
\jil{hard_simplify()} declines the projection when it exceeds the optional parameter \jil{multiterm_budget=1024}. A term whose
tableaux are trivial has size $1$ and is skipped outright, so ordinary expressions pay nothing for
the option being available. The projected form is kept only when it is strictly smaller by number of terms than the input, and discarded otherwise.

\subsection{\jil{==} and \jil{≅}}\label{sec:equality-and-congruence}
Equality comparison checks are a rudimentary operation. The equality checks in \jil{Alakazam} performed by \jil{==} are simple, naively checking the \jil{name()}, \jil{type()}, \jil{indices()}, \jil{function_of()}, \jil{tableaux()}, and \jil{labels()} of two \jil{Tensor}s match, and likewise for \jil{TensorTerm}s and \jil{TensorExpression}s.

For a deeper equality check, \jil{≅} (also accessible by \jil{cong()}) should be used. This explicitly checks whether two \jil{Tensorial} objects are isomorphic using the aforementioned graph isomorphism algorithm and/or via Young projection. This should be able to recognise many equivalent expressions.

\subsection{Display and \jil{to_latex()}}\label{sec:display-and-to-latex}
Alakazam implements custom pretty printing for all \jil{Tensorial} objects by overloading \jil{Base.show}, as seen throughout this document.
In addition, all \jil{Tensorial} objects can be converted to a LaTeX string using \jil{to_latex()}. For example, we can produce LaTeX markup of the Maxwell Lagrangian
\begin{Verbatim}[commandchars=\\\{\},xleftmargin=\parindent,numbers=left,bgcolor=bg]
\PYG{+w}{    }\PYG{n}{L}\PYG{o}{=}\PYG{p}{(}\PYG{o}{\PYGZhy{}}\PYG{l+m+mi}{1}\PYG{o}{//}\PYG{l+m+mi}{4}\PYG{o}{*}\PYG{l+s+sa}{T}\PYG{l+s}{\PYGZdq{}}\PYG{l+s}{F\PYGZca{}[μν]\PYGZob{}x\PYGZcb{}}\PYG{l+s}{\PYGZdq{}}\PYG{o}{*}\PYG{l+s+sa}{T}\PYG{l+s}{\PYGZdq{}}\PYG{l+s}{F\PYGZus{}[μν]\PYGZob{}x\PYGZcb{}}\PYG{l+s}{\PYGZdq{}}\PYG{o}{+}\PYG{n}{ψ}\PYG{o}{\PYGZsq{}}\PYG{o}{*}\PYG{p}{(}\PYG{n+nb}{im}\PYG{o}{*}\PYG{n}{Gamma}\PYG{p}{(}\PYG{n}{μ}\PYG{o}{=\PYGZgt{}}\PYG{o}{↑}\PYG{p}{)}\PYG{o}{*}\PYG{p}{(}\PYG{n}{PD}\PYG{p}{(}\PYG{l+s}{\PYGZdq{}}\PYG{l+s}{∂}\PYG{l+s}{\PYGZdq{}}\PYG{p}{,}\PYG{+w}{ }\PYG{n}{ψ}\PYG{p}{,}\PYG{+w}{ }\PYG{n}{x}\PYG{p}{,}\PYG{+w}{ }\PYG{n}{μ}\PYG{o}{=\PYGZgt{}}\PYG{o}{↓}\PYG{p}{)}
\PYG{+w}{                                                    }\PYG{o}{+}\PYG{n+nb}{im}\PYG{o}{*}\PYG{n}{e}\PYG{o}{*}\PYG{l+s+sa}{T}\PYG{l+s}{\PYGZdq{}}\PYG{l+s}{A\PYGZus{}μ\PYGZob{}x\PYGZcb{}}\PYG{l+s}{\PYGZdq{}}\PYG{o}{*}\PYG{n}{ψ}\PYG{p}{)}\PYG{o}{\PYGZhy{}}\PYG{n}{m}\PYG{o}{*}\PYG{n}{ψ}\PYG{p}{)}\PYG{p}{)}
\end{Verbatim}
Using \jil{@show}, we get
\begin{Verbatim}[commandchars=\\\{\},xleftmargin=\parindent,numbers=left,bgcolor=bg]
\PYG{+w}{    }\PYG{c}{\PYGZsh{}   L = \PYGZhy{}1//4F\PYGZca{}\PYGZob{}μν\PYGZcb{}F\PYGZus{}\PYGZob{}μν\PYGZcb{}+īψγ\PYGZca{}\PYGZob{}μ\PYGZcb{}∂\PYGZus{}\PYGZob{}μ\PYGZcb{}⟦ψ⟧\PYGZhy{}ēψγ\PYGZca{}\PYGZob{}μ\PYGZcb{}A\PYGZus{}\PYGZob{}μ\PYGZcb{}ψ\PYGZhy{}m̄ψψ}
\end{Verbatim}
while \jil{to_latex()} produces the markup
\begin{verbatim}  
-\frac{1}{4} F { }^{\mu}{ }^{\nu} F { }_{\mu}{ }_{\nu}  
+i \overline{\psi}  \gamma { }^{\mu} \partial_{\mu}\left( \psi    \right)  
- e \overline{\psi}  \gamma { }^{\mu} A { }_{\mu} \psi   
- m \overline{\psi}  \psi   
\end{verbatim}
which renders as
\begin{equation}
-\frac{1}{4} F { }^{\mu}{ }^{\nu} F { }_{\mu}{ }_{\nu}  +i \overline{\psi}  \gamma { }^{\mu} \partial_{\mu}\left( \psi    \right)   - e \overline{\psi}  \gamma { }^{\mu} A { }_{\mu} \psi    - m \overline{\psi}  \psi  . 
\end{equation}

\subsection{Global Variables and Restoring Defaults}\label{sec:clearing}
In Alakazam, things like the \jil{IndexSets}, algebra of covariant derivatives, and commutators, are stored as global variables. In order to establish a ``clean'' workspace after specifying values such as these, \jil{clear_global!()} can be called. This will clear all of the global variables that are considered ``header'' information encoding the current state (see Section~\ref{sec:serialisation-and-saving}). This includes any user-specified global value, which can be added to the \jil{Dict{String,Dict}} registry named \jil{global_values} via the function \jil{add_global_property!}. Note that after clearing, Alakazam will automatically re-establish pre-specified relations (like gamma matrix relations) via \jil{load_defaults!}.

\newpage
\section{Operators}\label{sec:operators}
For a variety of applications, a notion of an \jil{Operator} is useful, which consists of two parts - the operator $\mathcal{O}$ itself, and its operand $T$
\begin{equation}
    \mathcal{O}(T).
\end{equation}
These operators can be endowed with various useful properties and behaviours.
The simplest example would be a derivative, which has a (graded) Leibniz rule action,
\begin{equation}
    D\!\left(\sum_{\alpha} c_\alpha \prod_{i=1}^{n_\alpha} T_{\alpha i}\right)
=
\sum_{\alpha}
c_\alpha
\sum_{i=1}^{n_\alpha}
(-1)^{|D|\sum_{j<i}|T_{\alpha j}|}
\left(\prod_{j<i}T_{\alpha j}\right)
(DT_{\alpha i})
\left(\prod_{j>i}T_{\alpha j}\right),
\end{equation}
where $|\cdot|$ denotes the grading. In addition to explicit implementation of operators of this sort, the package can also handle other types of operators via several reserved abstract types that can be implemented, for example, those that act with a ``homomorphism-like'' action, $O(AB)=O(A)O(B)$, like the determinant. 

All operators descend from the \jil{OperatorSuperType} type, which allows them to share a common set of functions. Users can extend \jil{OperatorSuperType} to create their own custom operators with other unique behaviour. All generic operators extending \jil{OperatorSuperType} are however assumed to possess a set of minimal common parameters,
\begin{Verbatim}[commandchars=\\\{\},xleftmargin=\parindent,numbers=left,bgcolor=bg]
\PYG{+w}{    }\PYG{n}{Operator}\PYG{p}{(}\PYG{n}{name}\PYG{o}{::}\PYG{k+kt}{String}\PYG{p}{,}\PYG{+w}{ }
\PYG{+w}{        }\PYG{n}{operand}\PYG{o}{::}\PYG{k+kt}{TensorExpression}\PYG{p}{,}\PYG{+w}{ }
\PYG{+w}{        }\PYG{n}{operator\PYGZus{}indices}\PYG{o}{::}\PYG{k+kt}{Vector}\PYG{p}{\PYGZob{}}\PYG{k+kt}{Pair}\PYG{p}{\PYGZob{}}\PYG{k+kt}{Index}\PYG{p}{,}\PYG{+w}{ }\PYG{k+kt}{IndexPosition}\PYG{p}{\PYGZcb{}}\PYG{p}{\PYGZcb{}}\PYG{p}{)}
\end{Verbatim}
Here, \jil{name} will be the name of the operator used in pattern matching and display. \jil{operand} is the \jil{TensorExpression} the operator acts on. An operator is free to have its own \jil{operator_indices}, though it should be understood that the full set of indices on an \jil{Operator} contains also those extracted from the \jil{operand} as a union. The properties of any \jil{Operator} will be stored internally in an \jil{OperatorData} structure.

The various properties of an \jil{Operator} can be accessed via the functions
\begin{Verbatim}[commandchars=\\\{\},xleftmargin=\parindent,numbers=left,bgcolor=bg]
\PYG{+w}{    }\PYG{n}{operand}\PYG{p}{(}\PYG{n}{O}\PYG{p}{)}
\PYG{+w}{    }\PYG{n}{op\PYGZus{}data}\PYG{p}{(}\PYG{n}{O}\PYG{p}{)}
\PYG{+w}{    }\PYG{n}{op\PYGZus{}indices}\PYG{p}{(}\PYG{n}{O}\PYG{p}{)}
\PYG{+w}{    }\PYG{n}{op\PYGZus{}function\PYGZus{}of}\PYG{p}{(}\PYG{n}{O}\PYG{p}{)}
\PYG{+w}{    }\PYG{n}{op\PYGZus{}weights}\PYG{p}{(}\PYG{n}{O}\PYG{p}{)}
\PYG{+w}{    }\PYG{n}{op\PYGZus{}tableaux}\PYG{p}{(}\PYG{n}{O}\PYG{p}{)}
\PYG{+w}{    }\PYG{n}{op\PYGZus{}labels}\PYG{p}{(}\PYG{n}{O}\PYG{p}{)}
\PYG{+w}{    }\PYG{n}{operand\PYGZus{}indices}\PYG{p}{(}\PYG{n}{O}\PYG{p}{)}
\end{Verbatim}
in addition to the functions for ordinary \jil{Tensor}s, which in this case access the combined properties of operator and operand.

Note that \jil{OperatorSuperType} is a subtype of \jil{TensorSuperType}, and thus an \jil{Operator} is treated as a single tensor under functions. Some functions (for example simplify algorithms) may check if the tensor in question is an operator, and recursively call the function on its operand.

\subsection{Derivatives}\label{sec:derivatives}
The ability to perform some type of calculus on tensors is essential for field theory. 
In \jil{Alakazam}, derivatives come in two basic types: \jil{Derivative} and \jil{PartialDerivative}, with the difference being that partial derivatives (anti)commute amongst each other, whilst the \jil{Derivative} is assumed to behave like a covariant derivative, and is a subtype of \jil{CovDerivSuperType}. These two concrete types are both subtypes of \jil{LeibnizOperatorSuperType}, and thus automatically obey the graded product rule. Since often we care about covariant differentiation, for which derivatives can have long expressions involving the connection, we are able to use a symbolic form for the derivative, and expand derivatives into a full form later in calculations. We will give an explicit construction shortly.

Derivatives are with respect to (potentially multiple) coordinates. Tensors which are not a \jil{function_of()} these coordinates will be annihilated by such derivatives.

\subsubsection{\jil{PartialDerivative}}\label{sec:partialderivative}
Let us see how to construct a basic derivative operator.

Firstly, we can define a simple partial derivative,
\begin{Verbatim}[commandchars=\\\{\},xleftmargin=\parindent,numbers=left,bgcolor=bg]
\PYG{+w}{    }\PYG{n}{x}\PYG{+w}{ }\PYG{o}{=}\PYG{+w}{ }\PYG{n}{Coordinate}\PYG{p}{(}\PYG{l+s}{\PYGZdq{}}\PYG{l+s}{x}\PYG{l+s}{\PYGZdq{}}\PYG{p}{,}\PYG{+w}{ }\PYG{p}{[}\PYG{n}{spacetime\PYGZus{}indices}\PYG{p}{]}\PYG{p}{)}
\PYG{+w}{    }\PYG{n}{D}\PYG{o}{=}\PYG{n}{PartialDerivative}\PYG{p}{(}\PYG{l+s}{\PYGZdq{}}\PYG{l+s}{∂}\PYG{l+s}{\PYGZdq{}}\PYG{p}{,}\PYG{+w}{ }\PYG{n}{x}\PYG{p}{,}\PYG{n}{μ}\PYG{o}{=\PYGZgt{}}\PYG{n}{lower}\PYG{p}{)}
\end{Verbatim}
Note that this also has shorter aliases \jil{Partial} and \jil{PD}.
In order for a derivative to be useful, we should define its action on some basis of tensors. This can be done through the \jil{@action} macro, for example, 
\begin{Verbatim}[commandchars=\\\{\},xleftmargin=\parindent,numbers=left,bgcolor=bg]
\PYG{+w}{    }\PYG{n}{X}\PYG{o}{=}\PYG{n}{Tensor}\PYG{p}{(}\PYG{l+s}{\PYGZdq{}}\PYG{l+s}{X}\PYG{l+s}{\PYGZdq{}}\PYG{p}{,}\PYG{n}{x}\PYG{p}{,}\PYG{n}{ν}\PYG{o}{=\PYGZgt{}}\PYG{n}{upper}\PYG{p}{)}
\PYG{+w}{    }\PYG{n+nd}{@action}\PYG{+w}{ }\PYG{n}{D}\PYG{p}{(}\PYG{n}{X}\PYG{p}{)}\PYG{o}{=}\PYG{n}{KroneckerDelta}\PYG{p}{(}\PYG{n}{μ}\PYG{o}{=\PYGZgt{}}\PYG{n}{lower}\PYG{p}{,}\PYG{n}{ν}\PYG{o}{=\PYGZgt{}}\PYG{n}{upper}\PYG{p}{)}
\end{Verbatim}
This registers a symbolic rule for how to apply the derivative to tensors named \jil{"X"}. To see how this can be evaluated, consider the polynomial expression
\begin{Verbatim}[commandchars=\\\{\},xleftmargin=\parindent,numbers=left,bgcolor=bg]
\PYG{+w}{    }\PYG{n}{eq}\PYG{o}{=}\PYG{p}{(}\PYG{n}{SymmetricTensor}\PYG{p}{(}\PYG{l+s}{\PYGZdq{}}\PYG{l+s}{A}\PYG{l+s}{\PYGZdq{}}\PYG{p}{,}\PYG{+w}{ }\PYG{p}{[}\PYG{n}{μ}\PYG{+w}{ }\PYG{o}{=\PYGZgt{}}\PYG{+w}{ }\PYG{n}{lower}\PYG{p}{,}\PYG{+w}{ }\PYG{n}{ν}\PYG{+w}{ }\PYG{o}{=\PYGZgt{}}\PYG{+w}{ }\PYG{n}{lower}\PYG{p}{]}\PYG{p}{)}\PYG{o}{*}
\PYG{+w}{    }\PYG{n}{Tensor}\PYG{p}{(}\PYG{l+s}{\PYGZdq{}}\PYG{l+s}{X}\PYG{l+s}{\PYGZdq{}}\PYG{p}{,}\PYG{+w}{ }\PYG{p}{[}\PYG{n}{μ}\PYG{+w}{ }\PYG{o}{=\PYGZgt{}}\PYG{+w}{ }\PYG{n}{upper}\PYG{p}{]}\PYG{p}{,}\PYG{+w}{ }\PYG{n}{x}\PYG{p}{)}\PYG{o}{*}\PYG{n}{Tensor}\PYG{p}{(}\PYG{l+s}{\PYGZdq{}}\PYG{l+s}{X}\PYG{l+s}{\PYGZdq{}}\PYG{p}{,}\PYG{+w}{ }\PYG{p}{[}\PYG{n}{ν}\PYG{+w}{ }\PYG{o}{=\PYGZgt{}}\PYG{+w}{ }\PYG{n}{upper}\PYG{p}{]}\PYG{p}{,}\PYG{+w}{ }\PYG{n}{x}\PYG{p}{)}
\PYG{+w}{    }\PYG{o}{+}\PYG{n}{Tensor}\PYG{p}{(}\PYG{l+s}{\PYGZdq{}}\PYG{l+s}{B}\PYG{l+s}{\PYGZdq{}}\PYG{p}{,}\PYG{+w}{ }\PYG{p}{[}\PYG{n}{μ}\PYG{+w}{ }\PYG{o}{=\PYGZgt{}}\PYG{+w}{ }\PYG{n}{lower}\PYG{p}{]}\PYG{p}{)}\PYG{o}{*}\PYG{n}{Tensor}\PYG{p}{(}\PYG{l+s}{\PYGZdq{}}\PYG{l+s}{X}\PYG{l+s}{\PYGZdq{}}\PYG{p}{,}\PYG{+w}{ }\PYG{p}{[}\PYG{n}{μ}\PYG{+w}{ }\PYG{o}{=\PYGZgt{}}\PYG{+w}{ }\PYG{n}{upper}\PYG{p}{]}\PYG{p}{,}\PYG{+w}{ }\PYG{n}{x}\PYG{p}{)}
\PYG{+w}{    }\PYG{o}{+}\PYG{n}{Tensor}\PYG{p}{(}\PYG{l+s}{\PYGZdq{}}\PYG{l+s}{C}\PYG{l+s}{\PYGZdq{}}\PYG{p}{)}\PYG{p}{)}
\end{Verbatim}
We can create the derivative operator acting on this expression,
\begin{Verbatim}[commandchars=\\\{\},xleftmargin=\parindent,numbers=left,bgcolor=bg]
\PYG{+w}{    }\PYG{n}{D}\PYG{o}{=}\PYG{n}{Partial}\PYG{p}{(}\PYG{l+s}{\PYGZdq{}}\PYG{l+s}{∂}\PYG{l+s}{\PYGZdq{}}\PYG{p}{,}\PYG{+w}{ }\PYG{n}{x}\PYG{p}{,}\PYG{n}{ρ}\PYG{o}{=\PYGZgt{}}\PYG{n}{lower}\PYG{p}{)}
\PYG{+w}{    }\PYG{n+nd}{@show}\PYG{+w}{ }\PYG{n}{exp}\PYG{o}{=}\PYG{n}{D}\PYG{p}{(}\PYG{n}{eq}\PYG{p}{)}
\end{Verbatim}
This will yield \jil{∂_{ρ}⟦C+A_{μν}X^{μ}X^{ν}+B_{μ}X^{μ}⟧}. To apply the action of the derivative, we can use
\begin{Verbatim}[commandchars=\\\{\},xleftmargin=\parindent,numbers=left,bgcolor=bg]
\PYG{+w}{    }\PYG{n+nd}{@show}\PYG{+w}{ }\PYG{n}{derive}\PYG{p}{(}\PYG{n}{exp}\PYG{p}{)}
\end{Verbatim}
which will output \jil{2A^{μ∘}_{∘ρ}X_{μ}+B_{ρ}}, automatically applying the product rule and action of the derivatives.

\subsubsection{Integration by Parts}\label{sec:ibp}
In many cases in physics, we are only interested in expressions modulo total derivatives, which vanish under integration assuming sufficiently fast falloff conditions. For example, actions are defined under an integral, so two integrands differing by a total
derivative describe the same physics.  By discarding boundary terms, integration by parts implies
\begin{equation}
    \left(\partial_\mu X\right)Y \;\simeq\; -X\left(\partial_\mu Y\right), 
\end{equation}
where \(\simeq\) denotes equality modulo total derivatives.  When either the
fields or the derivative are fermionic-graded, the rearrangement costs an extra
sign,
\begin{equation}
    \left(\partial X\right)Y \;\simeq\;
    -(-1)^{|\partial||X|}\,X\left(\partial Y\right),    
\end{equation}
with \(|\cdot|\) denoting the grading, similar to in the Leibniz rule.
As we would like to detect equivalence of expressions modulo total derivatives, Alakazam is able to integrate expressions by parts, respecting the grading automatically.

The basic function to do this is \jil{drop_total_derivatives}.  Products appearing under a
derivative are expanded by the product rule first, so an expression need not be
presented in any particular form:
\begin{Verbatim}[commandchars=\\\{\},xleftmargin=\parindent,numbers=left,bgcolor=bg]
\PYG{+w}{    }\PYG{n}{x}\PYG{+w}{ }\PYG{o}{=}\PYG{+w}{ }\PYG{n}{Coordinate}\PYG{p}{(}\PYG{l+s}{\PYGZdq{}}\PYG{l+s}{x}\PYG{l+s}{\PYGZdq{}}\PYG{p}{,}\PYG{+w}{ }\PYG{n}{st}\PYG{p}{)}
\PYG{+w}{    }\PYG{n}{∂}\PYG{+w}{ }\PYG{o}{=}\PYG{+w}{ }\PYG{n}{PD}\PYG{p}{(}\PYG{l+s}{\PYGZdq{}}\PYG{l+s}{∂}\PYG{l+s}{\PYGZdq{}}\PYG{p}{,}\PYG{+w}{ }\PYG{p}{[}\PYG{n}{x}\PYG{p}{]}\PYG{p}{,}\PYG{+w}{ }\PYG{p}{[}\PYG{n}{μ}\PYG{+w}{ }\PYG{o}{=\PYGZgt{}}\PYG{+w}{ }\PYG{n}{lower}\PYG{p}{]}\PYG{p}{)}
\PYG{+w}{    }\PYG{n}{A}\PYG{+w}{ }\PYG{o}{=}\PYG{+w}{ }\PYG{n}{Tensor}\PYG{p}{(}\PYG{l+s}{\PYGZdq{}}\PYG{l+s}{A}\PYG{l+s}{\PYGZdq{}}\PYG{p}{,}\PYG{+w}{ }\PYG{p}{[}\PYG{n}{ν}\PYG{+w}{ }\PYG{o}{=\PYGZgt{}}\PYG{+w}{ }\PYG{n}{lower}\PYG{p}{]}\PYG{p}{,}\PYG{+w}{ }\PYG{n}{x}\PYG{p}{)}
\PYG{+w}{    }\PYG{n}{B}\PYG{+w}{ }\PYG{o}{=}\PYG{+w}{ }\PYG{n}{Tensor}\PYG{p}{(}\PYG{l+s}{\PYGZdq{}}\PYG{l+s}{B}\PYG{l+s}{\PYGZdq{}}\PYG{p}{,}\PYG{+w}{ }\PYG{p}{[}\PYG{n}{ρ}\PYG{+w}{ }\PYG{o}{=\PYGZgt{}}\PYG{+w}{ }\PYG{n}{lower}\PYG{p}{]}\PYG{p}{,}\PYG{+w}{ }\PYG{n}{x}\PYG{p}{)}

\PYG{+w}{    }\PYG{n}{drop\PYGZus{}total\PYGZus{}derivatives}\PYG{p}{(}\PYG{n}{∂}\PYG{p}{(}\PYG{n}{A}\PYG{p}{)}\PYG{+w}{ }\PYG{o}{*}\PYG{+w}{ }\PYG{n}{B}\PYG{+w}{ }\PYG{o}{+}\PYG{+w}{ }\PYG{n}{A}\PYG{+w}{ }\PYG{o}{*}\PYG{+w}{ }\PYG{n}{∂}\PYG{p}{(}\PYG{n}{B}\PYG{p}{)}\PYG{p}{)}\PYG{+w}{   }\PYG{c}{\PYGZsh{} 0}
\PYG{+w}{    }\PYG{n}{drop\PYGZus{}total\PYGZus{}derivatives}\PYG{p}{(}\PYG{n}{∂}\PYG{p}{(}\PYG{n}{A}\PYG{+w}{ }\PYG{o}{*}\PYG{+w}{ }\PYG{n}{B}\PYG{p}{)}\PYG{p}{)}\PYG{+w}{              }\PYG{c}{\PYGZsh{} 0}
\PYG{+w}{    }\PYG{n}{drop\PYGZus{}total\PYGZus{}derivatives}\PYG{p}{(}\PYG{n}{∂}\PYG{p}{(}\PYG{n}{A}\PYG{p}{)}\PYG{+w}{ }\PYG{o}{*}\PYG{+w}{ }\PYG{n}{B}\PYG{p}{)}\PYG{+w}{              }\PYG{c}{\PYGZsh{} \PYGZhy{}A * ∂(B)}
\PYG{+w}{    }\PYG{n}{drop\PYGZus{}total\PYGZus{}derivatives}\PYG{p}{(}\PYG{n}{A}\PYG{+w}{ }\PYG{o}{*}\PYG{+w}{ }\PYG{n}{∂}\PYG{p}{(}\PYG{n}{B}\PYG{p}{)}\PYG{p}{)}\PYG{+w}{             }\PYG{c}{\PYGZsh{} A * ∂(B)}
\end{Verbatim}
For terms quadratic in the fields, the reduction is a genuine normal form: every
term is driven to carry all of its derivatives on one of the two factors, after
which a total derivative cancels against itself identically.  The result is
therefore zero exactly when the input was a total derivative.
Which factor collects the derivatives is set by the optional \jil{onto} keyword: the
default \jil{:key} selects by a canonical lexicographical ordering of the two base fields (with tie-breakers on the indices), so
that the normal form does not depend on the order in which the term happened to
be written, while \jil{:first} and \jil{:last} force the positional choices.

Beyond quadratic order no such normal form is available, and terms with three or
more factors fall through to \jil{drop_total_derivatives_general}.  This
constructs the space of total derivatives in the relevant sector explicitly and
reduces the target against it by exact row reduction, which is subject to no
restriction on the number of factors but is correspondingly slower.  To keep the
cost bounded it returns its input unchanged if the sector grows past the optional
\jil{budget::Int=512} monomials, or if a coefficient is not of a type admitting exact
division, so a caller is never handed a silently guessed result.

Two lower-level utilities are exposed for building on top of this.
\jil{strip_derivatives} peels a chain \(\partial^k\phi\) into its base field and
the list of operators acting on it, outermost first, and \jil{rebuild_nested} is
its inverse.  A derivative acting on a product is not of this form and is
returned whole, so the two compose safely on arbitrary input.

\subsubsection{\jil{Derivative}}\label{sec:derivative}
Beyond partial derivatives, we may also want to perform covariant differentiation. To illustrate this in a more complicated example for which grading and coordinates are important, let us consider superspace covariant differentiation.

Consider $4D$ $\mathcal{N}=1$ superspace with coordinates $(x^\mu,\theta^\alpha,\bar{\theta}^{\dot{\alpha}})$,
\begin{Verbatim}[commandchars=\\\{\},xleftmargin=\parindent,numbers=left,bgcolor=bg]
\PYG{+w}{    }\PYG{n}{spacetime\PYGZus{}indices}\PYG{+w}{ }\PYG{o}{=}\PYG{+w}{ }\PYG{n}{IndexSet}\PYG{p}{(}\PYG{l+s}{\PYGZdq{}}\PYG{l+s}{spacetime}\PYG{l+s}{\PYGZdq{}}\PYG{p}{,}\PYG{+w}{ }\PYG{l+m+mi}{4}\PYG{p}{,}\PYG{n}{lower}\PYG{p}{)}
\PYG{+w}{    }\PYG{n}{spinor\PYGZus{}indices}\PYG{o}{=}\PYG{n}{IndexSet}\PYG{p}{(}\PYG{l+s}{\PYGZdq{}}\PYG{l+s}{spinor\PYGZus{}indices}\PYG{l+s}{\PYGZdq{}}\PYG{p}{,}\PYG{+w}{ }\PYG{l+m+mi}{2}\PYG{p}{,}\PYG{n}{lower}\PYG{p}{,}
\PYG{+w}{        }\PYG{n}{fermionic}\PYG{p}{,}\PYG{n}{EpsilonTensor}\PYG{p}{)}
\PYG{+w}{    }\PYG{n}{spinor\PYGZus{}conj\PYGZus{}indices}\PYG{o}{=}\PYG{n}{IndexSet}\PYG{p}{(}\PYG{l+s}{\PYGZdq{}}\PYG{l+s}{spinor\PYGZus{}conj\PYGZus{}indices}\PYG{l+s}{\PYGZdq{}}\PYG{p}{,}\PYG{+w}{ }\PYG{l+m+mi}{2}\PYG{p}{,}
\PYG{+w}{        }\PYG{n}{fermionic}\PYG{p}{,}\PYG{n}{lower}\PYG{p}{,}\PYG{n}{EpsilonTensor}\PYG{p}{)}
\PYG{+w}{    }\PYG{n}{α}\PYG{p}{,}\PYG{n}{β}\PYG{p}{,}\PYG{n}{γ}\PYG{p}{,}\PYG{n}{χ}\PYG{+w}{ }\PYG{o}{=}\PYG{+w}{ }\PYG{n}{Index}\PYG{p}{(}\PYG{p}{[}\PYG{l+s}{\PYGZdq{}}\PYG{l+s}{α}\PYG{l+s}{\PYGZdq{}}\PYG{p}{,}\PYG{l+s}{\PYGZdq{}}\PYG{l+s}{β}\PYG{l+s}{\PYGZdq{}}\PYG{p}{,}\PYG{l+s}{\PYGZdq{}}\PYG{l+s}{γ}\PYG{l+s}{\PYGZdq{}}\PYG{p}{,}\PYG{l+s}{\PYGZdq{}}\PYG{l+s}{χ}\PYG{l+s}{\PYGZdq{}}\PYG{p}{]}\PYG{p}{,}\PYG{+w}{ }\PYG{n}{spinor\PYGZus{}indices}\PYG{p}{)}
\PYG{+w}{    }\PYG{n}{αd}\PYG{p}{,}\PYG{n}{βd}\PYG{p}{,}\PYG{n}{γd}\PYG{p}{,}\PYG{n}{χd}\PYG{+w}{ }\PYG{o}{=}\PYG{+w}{ }\PYG{n}{Index}\PYG{p}{(}\PYG{p}{[}\PYG{l+s}{\PYGZdq{}}\PYG{l+s}{̇α}\PYG{l+s}{\PYGZdq{}}\PYG{p}{,}\PYG{l+s}{\PYGZdq{}}\PYG{l+s}{̇β}\PYG{l+s}{\PYGZdq{}}\PYG{p}{,}\PYG{l+s}{\PYGZdq{}}\PYG{l+s}{̇γ}\PYG{l+s}{\PYGZdq{}}\PYG{p}{,}\PYG{l+s}{\PYGZdq{}}\PYG{l+s}{̇χ}\PYG{l+s}{\PYGZdq{}}\PYG{p}{]}\PYG{p}{,}\PYG{+w}{ }\PYG{n}{spinor\PYGZus{}conj\PYGZus{}indices}\PYG{p}{)}
\PYG{+w}{    }\PYG{n}{μ}\PYG{p}{,}\PYG{n}{ν}\PYG{p}{,}\PYG{n}{ρ}\PYG{p}{,}\PYG{n}{τ}\PYG{+w}{ }\PYG{o}{=}\PYG{+w}{ }\PYG{n}{Index}\PYG{p}{(}\PYG{p}{[}\PYG{l+s}{\PYGZdq{}}\PYG{l+s}{μ}\PYG{l+s}{\PYGZdq{}}\PYG{p}{,}\PYG{l+s}{\PYGZdq{}}\PYG{l+s}{ν}\PYG{l+s}{\PYGZdq{}}\PYG{p}{,}\PYG{l+s}{\PYGZdq{}}\PYG{l+s}{ρ}\PYG{l+s}{\PYGZdq{}}\PYG{p}{,}\PYG{l+s}{\PYGZdq{}}\PYG{l+s}{τ}\PYG{l+s}{\PYGZdq{}}\PYG{p}{]}\PYG{p}{,}\PYG{n}{spacetime\PYGZus{}indices}\PYG{p}{)}
\PYG{+w}{    }\PYG{n}{x}\PYG{+w}{ }\PYG{o}{=}\PYG{+w}{ }\PYG{n}{Coordinate}\PYG{p}{(}\PYG{l+s}{\PYGZdq{}}\PYG{l+s}{x}\PYG{l+s}{\PYGZdq{}}\PYG{p}{,}\PYG{+w}{ }\PYG{n}{spacetime\PYGZus{}indices}\PYG{p}{)}
\PYG{+w}{    }\PYG{n}{θ}\PYG{+w}{ }\PYG{o}{=}\PYG{+w}{ }\PYG{n}{Coordinate}\PYG{p}{(}\PYG{l+s}{\PYGZdq{}}\PYG{l+s}{θ}\PYG{l+s}{\PYGZdq{}}\PYG{p}{,}\PYG{+w}{ }\PYG{n}{spinor\PYGZus{}indices}\PYG{p}{)}
\PYG{+w}{    }\PYG{n}{θbar}\PYG{+w}{ }\PYG{o}{=}\PYG{+w}{ }\PYG{n}{Coordinate}\PYG{p}{(}\PYG{l+s}{\PYGZdq{}}\PYG{l+s}{Θ}\PYG{l+s}{\PYGZdq{}}\PYG{p}{,}\PYG{+w}{ }\PYG{n}{spinor\PYGZus{}conj\PYGZus{}indices}\PYG{p}{)}
\PYG{+w}{    }\PYG{n}{custom\PYGZus{}sort\PYGZus{}order!}\PYG{p}{(}\PYG{p}{[}\PYG{l+s}{\PYGZdq{}}\PYG{l+s}{θ}\PYG{l+s}{\PYGZdq{}}\PYG{p}{,}\PYG{l+s}{\PYGZdq{}}\PYG{l+s}{Θ}\PYG{l+s}{\PYGZdq{}}\PYG{p}{,}\PYG{l+s}{\PYGZdq{}}\PYG{l+s}{ϵ}\PYG{l+s}{\PYGZdq{}}\PYG{p}{]}\PYG{p}{)}
\end{Verbatim}
There are associated superspace covariant derivatives
\begin{equation}
    D_\alpha=\del_\alpha+\frac{i}{2}\bar{\theta}^{\dot{\alpha}}\sigma^\mu_{\alpha\dot{\alpha}}\del_\mu,\quad \bar{D}_{\dot{\alpha}}=\del_{\dot{\alpha}}+\frac{i}{2}{\theta}^{{\alpha}}\sigma^\mu_{\alpha\dot{\alpha}}\del_\mu
\end{equation}
which satisfy the anticommutation relations
\begin{equation}
    \{D_{\alpha},\bar{D}_{\dot{\alpha}}\}=i\sigma^\mu_{\alpha\dot{\alpha}}\del_\mu.
\end{equation}
Let us show that we can register covariant derivatives, which can later be expanded and will satisfy this algebra.
We begin by registering the action of partial derivatives on a basis of tensors,
\begin{Verbatim}[commandchars=\\\{\},xleftmargin=\parindent,numbers=left,bgcolor=bg]
\PYG{+w}{    }\PYG{n}{∇}\PYG{o}{=}\PYG{n}{PartialDerivative}\PYG{p}{(}\PYG{l+s}{\PYGZdq{}}\PYG{l+s}{∇}\PYG{l+s}{\PYGZdq{}}\PYG{p}{,}\PYG{+w}{ }\PYG{p}{[}\PYG{n}{x}\PYG{p}{]}\PYG{p}{,}\PYG{+w}{ }\PYG{p}{[}\PYG{n}{μ}\PYG{+w}{ }\PYG{o}{=\PYGZgt{}}\PYG{+w}{ }\PYG{n}{lower}\PYG{p}{]}\PYG{p}{)}
\PYG{+w}{    }\PYG{n}{∂}\PYG{o}{=}\PYG{n}{PartialDerivative}\PYG{p}{(}\PYG{l+s}{\PYGZdq{}}\PYG{l+s}{∂}\PYG{l+s}{\PYGZdq{}}\PYG{p}{,}\PYG{+w}{ }\PYG{p}{[}\PYG{n}{θ}\PYG{p}{]}\PYG{p}{,}\PYG{+w}{ }\PYG{p}{[}\PYG{n}{α}\PYG{+w}{ }\PYG{o}{=\PYGZgt{}}\PYG{+w}{ }\PYG{n}{lower}\PYG{p}{]}\PYG{p}{)}
\PYG{+w}{    }\PYG{n}{𝛛}\PYG{o}{=}\PYG{n}{PartialDerivative}\PYG{p}{(}\PYG{l+s}{\PYGZdq{}}\PYG{l+s}{𝛛}\PYG{l+s}{\PYGZdq{}}\PYG{p}{,}\PYG{+w}{ }\PYG{p}{[}\PYG{n}{θbar}\PYG{p}{]}\PYG{p}{,}\PYG{+w}{ }\PYG{p}{[}\PYG{n}{αd}\PYG{+w}{ }\PYG{o}{=\PYGZgt{}}\PYG{+w}{ }\PYG{n}{lower}\PYG{p}{]}\PYG{p}{)}
\PYG{+w}{    }\PYG{n+nd}{@action}\PYG{+w}{ }\PYG{n}{∂}\PYG{p}{(}\PYG{n}{Tensor}\PYG{p}{(}\PYG{l+s}{\PYGZdq{}}\PYG{l+s}{θ}\PYG{l+s}{\PYGZdq{}}\PYG{p}{,}\PYG{p}{[}\PYG{n}{β}\PYG{o}{=\PYGZgt{}}\PYG{n}{lower}\PYG{p}{]}\PYG{p}{,}\PYG{n}{θ}\PYG{p}{)}\PYG{p}{)}\PYG{o}{=}\PYG{n}{EpsilonTensor}\PYG{p}{(}\PYG{p}{[}\PYG{n}{β}\PYG{+w}{ }\PYG{o}{=\PYGZgt{}}\PYG{n}{lower}\PYG{p}{,}\PYG{+w}{ }\PYG{n}{α}\PYG{o}{=\PYGZgt{}}\PYG{n}{lower}\PYG{p}{]}\PYG{p}{)}
\PYG{+w}{    }\PYG{n+nd}{@action}\PYG{+w}{ }\PYG{n}{∂}\PYG{p}{(}\PYG{n}{Tensor}\PYG{p}{(}\PYG{l+s}{\PYGZdq{}}\PYG{l+s}{θ}\PYG{l+s}{\PYGZdq{}}\PYG{p}{,}\PYG{p}{[}\PYG{n}{β}\PYG{o}{=\PYGZgt{}}\PYG{n}{upper}\PYG{p}{]}\PYG{p}{,}\PYG{n}{θ}\PYG{p}{)}\PYG{p}{)}\PYG{o}{=}\PYG{n}{KroneckerDelta}\PYG{p}{(}\PYG{p}{[}\PYG{n}{α}\PYG{+w}{ }\PYG{o}{=\PYGZgt{}}\PYG{n}{lower}\PYG{p}{,}\PYG{+w}{ }\PYG{n}{β}\PYG{o}{=\PYGZgt{}}\PYG{n}{upper}\PYG{p}{]}\PYG{p}{)}
\PYG{+w}{    }\PYG{n+nd}{@action}\PYG{+w}{ }\PYG{n}{𝛛}\PYG{p}{(}\PYG{n}{Tensor}\PYG{p}{(}\PYG{l+s}{\PYGZdq{}}\PYG{l+s}{Θ}\PYG{l+s}{\PYGZdq{}}\PYG{p}{,}\PYG{p}{[}\PYG{n}{βd}\PYG{o}{=\PYGZgt{}}\PYG{n}{lower}\PYG{p}{]}\PYG{p}{,}\PYG{n}{θbar}\PYG{p}{)}\PYG{p}{)}\PYG{o}{=}\PYG{n}{EpsilonTensor}\PYG{p}{(}\PYG{p}{[}\PYG{n}{βd}\PYG{+w}{ }\PYG{o}{=\PYGZgt{}}\PYG{n}{lower}\PYG{p}{,}\PYG{+w}{ }\PYG{n}{αd}\PYG{o}{=\PYGZgt{}}\PYG{n}{lower}\PYG{p}{]}\PYG{p}{)}
\PYG{+w}{    }\PYG{n+nd}{@action}\PYG{+w}{ }\PYG{n}{𝛛}\PYG{p}{(}\PYG{n}{Tensor}\PYG{p}{(}\PYG{l+s}{\PYGZdq{}}\PYG{l+s}{Θ}\PYG{l+s}{\PYGZdq{}}\PYG{p}{,}\PYG{p}{[}\PYG{n}{βd}\PYG{o}{=\PYGZgt{}}\PYG{n}{upper}\PYG{p}{]}\PYG{p}{,}\PYG{n}{θbar}\PYG{p}{)}\PYG{p}{)}\PYG{o}{=}\PYG{n}{KroneckerDelta}\PYG{p}{(}\PYG{p}{[}\PYG{n}{αd}\PYG{+w}{ }\PYG{o}{=\PYGZgt{}}\PYG{n}{lower}\PYG{p}{,}\PYG{+w}{ }\PYG{n}{βd}\PYG{o}{=\PYGZgt{}}\PYG{n}{upper}\PYG{p}{]}\PYG{p}{)}
\end{Verbatim}
We can register the expansion of a covariant derivative in terms of these using the \jil{@covariant} macro
\begin{Verbatim}[commandchars=\\\{\},xleftmargin=\parindent,numbers=left,bgcolor=bg]
\PYG{+w}{    }\PYG{n}{covar\PYGZus{}bar\PYGZus{}expansion}\PYG{o}{=}\PYG{p}{(}\PYG{n}{𝛛}\PYG{o}{+}\PYG{n+nb}{im}\PYG{o}{//}\PYG{l+m+mi}{2}\PYG{o}{*}\PYG{n}{Tensor}\PYG{p}{(}\PYG{l+s}{\PYGZdq{}}\PYG{l+s}{θ}\PYG{l+s}{\PYGZdq{}}\PYG{p}{,}\PYG{+w}{ }\PYG{p}{[}\PYG{n}{α}\PYG{o}{=\PYGZgt{}}\PYG{n}{upper}\PYG{p}{]}\PYG{p}{,}\PYG{+w}{ }\PYG{n}{θ}\PYG{p}{)}
\PYG{+w}{    }\PYG{o}{*}\PYG{n}{Tensor}\PYG{p}{(}\PYG{l+s}{\PYGZdq{}}\PYG{l+s}{σ}\PYG{l+s}{\PYGZdq{}}\PYG{p}{,}\PYG{+w}{ }\PYG{p}{[}\PYG{n}{μ}\PYG{o}{=\PYGZgt{}}\PYG{n}{upper}\PYG{p}{,}\PYG{+w}{ }\PYG{n}{α}\PYG{o}{=\PYGZgt{}}\PYG{n}{lower}\PYG{p}{,}\PYG{+w}{ }\PYG{n}{αd}\PYG{o}{=\PYGZgt{}}\PYG{n}{lower}\PYG{p}{]}\PYG{p}{)}\PYG{o}{*}\PYG{n}{∇}
\PYG{+w}{    }\PYG{n}{covar\PYGZus{}expansion}\PYG{o}{=}\PYG{n}{∂}\PYG{o}{+}\PYG{n+nb}{im}\PYG{o}{//}\PYG{l+m+mi}{2}\PYG{o}{*}\PYG{n}{Tensor}\PYG{p}{(}\PYG{l+s}{\PYGZdq{}}\PYG{l+s}{σ}\PYG{l+s}{\PYGZdq{}}\PYG{p}{,}\PYG{+w}{ }\PYG{p}{[}\PYG{n}{μ}\PYG{o}{=\PYGZgt{}}\PYG{n}{upper}\PYG{p}{,}\PYG{+w}{ }\PYG{n}{α}\PYG{o}{=\PYGZgt{}}\PYG{n}{lower}\PYG{p}{,}\PYG{+w}{ }\PYG{n}{αd}\PYG{o}{=\PYGZgt{}}\PYG{n}{lower}\PYG{p}{]}\PYG{p}{)}
\PYG{+w}{    }\PYG{o}{*}\PYG{n}{Tensor}\PYG{p}{(}\PYG{l+s}{\PYGZdq{}}\PYG{l+s}{Θ}\PYG{l+s}{\PYGZdq{}}\PYG{p}{,}\PYG{+w}{ }\PYG{p}{[}\PYG{n}{αd}\PYG{o}{=\PYGZgt{}}\PYG{n}{upper}\PYG{p}{]}\PYG{p}{,}\PYG{+w}{ }\PYG{n}{θbar}\PYG{p}{)}\PYG{o}{*}\PYG{n}{∇}\PYG{p}{)}
\PYG{+w}{    }\PYG{n}{Dbar}\PYG{o}{=}\PYG{n}{Derivative}\PYG{p}{(}\PYG{l+s}{\PYGZdq{}}\PYG{l+s}{̄D}\PYG{l+s}{\PYGZdq{}}\PYG{p}{,}\PYG{+w}{ }\PYG{p}{[}\PYG{n}{x}\PYG{p}{,}\PYG{+w}{ }\PYG{n}{θbar}\PYG{p}{]}\PYG{p}{,}\PYG{+w}{ }\PYG{p}{[}\PYG{n}{αd}\PYG{+w}{ }\PYG{o}{=\PYGZgt{}}\PYG{+w}{ }\PYG{n}{lower}\PYG{p}{]}\PYG{p}{)}
\PYG{+w}{    }\PYG{n}{D}\PYG{o}{=}\PYG{n}{Derivative}\PYG{p}{(}\PYG{l+s}{\PYGZdq{}}\PYG{l+s}{D}\PYG{l+s}{\PYGZdq{}}\PYG{p}{,}\PYG{+w}{ }\PYG{p}{[}\PYG{n}{x}\PYG{p}{,}\PYG{+w}{ }\PYG{n}{θ}\PYG{p}{]}\PYG{p}{,}\PYG{+w}{ }\PYG{p}{[}\PYG{n}{α}\PYG{+w}{ }\PYG{o}{=\PYGZgt{}}\PYG{+w}{ }\PYG{n}{lower}\PYG{p}{]}\PYG{p}{)}
\PYG{+w}{    }
\PYG{+w}{    }\PYG{n+nd}{@covariant}\PYG{+w}{ }\PYG{n}{D}\PYG{o}{=}\PYG{n}{covar\PYGZus{}expansion}
\PYG{+w}{    }\PYG{n+nd}{@covariant}\PYG{+w}{ }\PYG{n}{Dbar}\PYG{o}{=}\PYG{n}{covar\PYGZus{}bar\PYGZus{}expansion}
\end{Verbatim}

Then, we can create the two orderings of the derivatives acting on a generic tensor,
\begin{Verbatim}[commandchars=\\\{\},xleftmargin=\parindent,numbers=left,bgcolor=bg]
\PYG{+w}{    }\PYG{n}{a1}\PYG{o}{=}\PYG{n}{Derivative}\PYG{p}{(}\PYG{l+s}{\PYGZdq{}}\PYG{l+s}{D}\PYG{l+s}{\PYGZdq{}}\PYG{p}{,}\PYG{n}{Derivative}\PYG{p}{(}\PYG{l+s}{\PYGZdq{}}\PYG{l+s}{̄D}\PYG{l+s}{\PYGZdq{}}\PYG{p}{,}\PYG{n}{Tensor}\PYG{p}{(}\PYG{l+s}{\PYGZdq{}}\PYG{l+s}{X}\PYG{l+s}{\PYGZdq{}}\PYG{p}{,}\PYG{+w}{ }\PYG{p}{[}\PYG{n}{x}\PYG{p}{,}\PYG{n}{θ}\PYG{p}{,}\PYG{n}{θbar}\PYG{p}{]}\PYG{p}{)}\PYG{p}{,}\PYG{+w}{ }
\PYG{+w}{        }\PYG{p}{[}\PYG{n}{x}\PYG{p}{,}\PYG{n}{θbar}\PYG{p}{]}\PYG{p}{,}\PYG{+w}{ }\PYG{p}{[}\PYG{n}{γd}\PYG{o}{=\PYGZgt{}}\PYG{n}{lower}\PYG{p}{]}\PYG{p}{)}\PYG{p}{,}\PYG{+w}{ }\PYG{p}{[}\PYG{n}{x}\PYG{p}{,}\PYG{n}{θ}\PYG{p}{]}\PYG{p}{,}\PYG{+w}{ }\PYG{p}{[}\PYG{n}{γ}\PYG{o}{=\PYGZgt{}}\PYG{n}{lower}\PYG{p}{]}\PYG{p}{)}
\PYG{+w}{    }\PYG{n}{b1}\PYG{o}{=}\PYG{n}{Derivative}\PYG{p}{(}\PYG{l+s}{\PYGZdq{}}\PYG{l+s}{̄D}\PYG{l+s}{\PYGZdq{}}\PYG{p}{,}\PYG{n}{Derivative}\PYG{p}{(}\PYG{l+s}{\PYGZdq{}}\PYG{l+s}{D}\PYG{l+s}{\PYGZdq{}}\PYG{p}{,}\PYG{n}{Tensor}\PYG{p}{(}\PYG{l+s}{\PYGZdq{}}\PYG{l+s}{X}\PYG{l+s}{\PYGZdq{}}\PYG{p}{,}\PYG{+w}{ }\PYG{p}{[}\PYG{n}{x}\PYG{p}{,}\PYG{n}{θ}\PYG{p}{,}\PYG{n}{θbar}\PYG{p}{]}\PYG{p}{)}\PYG{p}{,}\PYG{+w}{ }
\PYG{+w}{        }\PYG{p}{[}\PYG{n}{x}\PYG{p}{,}\PYG{n}{θ}\PYG{p}{]}\PYG{p}{,}\PYG{+w}{ }\PYG{p}{[}\PYG{n}{γ}\PYG{o}{=\PYGZgt{}}\PYG{n}{lower}\PYG{p}{]}\PYG{p}{)}\PYG{p}{,}\PYG{+w}{ }\PYG{p}{[}\PYG{n}{x}\PYG{p}{,}\PYG{n}{θbar}\PYG{p}{]}\PYG{p}{,}\PYG{+w}{ }\PYG{p}{[}\PYG{n}{γd}\PYG{o}{=\PYGZgt{}}\PYG{n}{lower}\PYG{p}{]}\PYG{p}{)}
\end{Verbatim}
and then apply the covariant expansion then perform the action of the derivatives (including the product rule), or simply do all these steps via \jil{derive()}
\begin{Verbatim}[commandchars=\\\{\},xleftmargin=\parindent,numbers=left,bgcolor=bg]
\PYG{+w}{    }\PYG{n+nd}{@show}\PYG{+w}{ }\PYG{n}{simplify}\PYG{p}{(}\PYG{n}{apply\PYGZus{}derivative}\PYG{p}{(}\PYG{n}{expand\PYGZus{}derivative}\PYG{p}{(}\PYG{n}{a1}\PYG{o}{+}\PYG{n}{b1}\PYG{p}{)}\PYG{p}{)}\PYG{p}{)}
\PYG{+w}{    }\PYG{n+nd}{@show}\PYG{+w}{ }\PYG{n}{derive}\PYG{p}{(}\PYG{n}{a1}\PYG{o}{+}\PYG{n}{b1}\PYG{p}{)}
\end{Verbatim}
which will yield the output \jil{iσ^{μ∘∘}_{∘γ̇γ}∇_{μ}⟦X⟧}, as expected from the known anticommutation relation.

The algebra for generic covariant derivatives can also be encoded explicitly via the \jil{@commutator} macro
\begin{Verbatim}[commandchars=\\\{\},xleftmargin=\parindent,numbers=left,bgcolor=bg]
\PYG{+w}{    }\PYG{n+nd}{@commutator}\PYG{+w}{ }\PYG{p}{\PYGZob{}}\PYG{k+kt}{D}\PYG{p}{,}\PYG{k+kt}{̄D}\PYG{p}{\PYGZcb{}}\PYG{o}{=}\PYG{n+nb}{im}\PYG{o}{*}\PYG{n}{∇}
\PYG{+w}{    }\PYG{n+nd}{@commutator}\PYG{+w}{ }\PYG{p}{[}\PYG{n}{D}\PYG{p}{,}\PYG{n}{∇}\PYG{p}{]}\PYG{o}{=}\PYG{l+m+mi}{0}
\PYG{+w}{    }\PYG{n+nd}{@commutator}\PYG{+w}{ }\PYG{p}{[}\PYG{n}{̄D}\PYG{p}{,}\PYG{n}{∇}\PYG{p}{]}\PYG{o}{=}\PYG{l+m+mi}{0}
\end{Verbatim}
These will be obeyed when rearranging the depth of nested operators.

\subsection{Lie Brackets}\label{sec:lie-brackets}
In addition to registering commutation relations for \jil{MatrixSuperType} tensors, one may desire to directly work with the Lie bracket of two matrices instead, and manipulate such expressions. This can be expressed as a \jil{Lie} operator.
To do so, the user can use the \jil{@Lie} macro to create an (anti-)commutator of matrices. For the generic matrix structure \jil{Mat}, we can do
\begin{Verbatim}[commandchars=\\\{\},xleftmargin=\parindent,numbers=left,bgcolor=bg]
\PYG{+w}{    }\PYG{n}{X}\PYG{o}{=}\PYG{n}{Mat}\PYG{p}{(}\PYG{l+s}{\PYGZdq{}}\PYG{l+s}{x}\PYG{l+s}{\PYGZdq{}}\PYG{p}{,}\PYG{+w}{ }\PYG{p}{[}\PYG{n}{μ}\PYG{o}{=\PYGZgt{}}\PYG{n}{upper}\PYG{p}{]}\PYG{p}{)}
\PYG{+w}{    }\PYG{n}{P}\PYG{o}{=}\PYG{n}{Mat}\PYG{p}{(}\PYG{l+s}{\PYGZdq{}}\PYG{l+s}{p}\PYG{l+s}{\PYGZdq{}}\PYG{p}{,}\PYG{+w}{ }\PYG{p}{[}\PYG{n}{ν}\PYG{o}{=\PYGZgt{}}\PYG{n}{upper}\PYG{p}{]}\PYG{p}{)}
\PYG{+w}{    }\PYG{n}{br}\PYG{+w}{ }\PYG{o}{=}\PYG{+w}{ }\PYG{n+nd}{@Lie}\PYG{+w}{ }\PYG{p}{[}\PYG{n}{X}\PYG{p}{,}\PYG{+w}{ }\PYG{n}{P}\PYG{p}{]}
\PYG{+w}{    }\PYG{n}{abr}\PYG{o}{=}\PYG{+w}{ }\PYG{n+nd}{@Lie}\PYG{+w}{ }\PYG{p}{\PYGZob{}}\PYG{k+kt}{X}\PYG{p}{,}\PYG{k+kt}{P}\PYG{p}{\PYGZcb{}}
\end{Verbatim}
These macros return a \jil{Lie} operator object, which is a subtype of \jil{LieBracketSuperType} and in turn \jil{LeibnizOperatorSuperType}. This macro construction works for arbitrary nested brackets.
There are various functions defined on Lie brackets, for example a basic swap function
\begin{Verbatim}[commandchars=\\\{\},xleftmargin=\parindent,numbers=left,bgcolor=bg]
\PYG{+w}{    }\PYG{n+nd}{@show}\PYG{+w}{ }\PYG{n}{swap}\PYG{p}{(}\PYG{n}{br}\PYG{p}{)}
\PYG{+w}{    }\PYG{c}{\PYGZsh{}   \PYGZhy{}[p\PYGZca{}\PYGZob{}ν\PYGZcb{},x\PYGZca{}\PYGZob{}μ\PYGZcb{}]}
\end{Verbatim}
Thanks to subtyping \jil{LeibnizOperatorSuperType}, given
\begin{Verbatim}[commandchars=\\\{\},xleftmargin=\parindent,numbers=left,bgcolor=bg]
\PYG{+w}{    }\PYG{n}{Q}\PYG{o}{=}\PYG{n}{Mat}\PYG{p}{(}\PYG{l+s}{\PYGZdq{}}\PYG{l+s}{q}\PYG{l+s}{\PYGZdq{}}\PYG{p}{,}\PYG{+w}{ }\PYG{p}{[}\PYG{n}{ρ}\PYG{o}{=\PYGZgt{}}\PYG{n}{upper}\PYG{p}{]}\PYG{p}{)}\PYG{+w}{    }
\PYG{+w}{    }\PYG{n}{br2}\PYG{+w}{ }\PYG{o}{=}\PYG{+w}{ }\PYG{n+nd}{@Lie}\PYG{+w}{ }\PYG{p}{[}\PYG{n}{X}\PYG{p}{,}\PYG{+w}{ }\PYG{n}{P}\PYG{o}{*}\PYG{n}{Q}\PYG{p}{]}
\end{Verbatim}
we can perform
\begin{Verbatim}[commandchars=\\\{\},xleftmargin=\parindent,numbers=left,bgcolor=bg]
\PYG{+w}{    }\PYG{n+nd}{@show}\PYG{+w}{ }\PYG{n}{product\PYGZus{}rule}\PYG{p}{(}\PYG{n}{br2}\PYG{p}{)}
\PYG{+w}{    }\PYG{c}{\PYGZsh{}   [x\PYGZca{}\PYGZob{}μ\PYGZcb{},p\PYGZca{}\PYGZob{}ν\PYGZcb{}]q\PYGZca{}\PYGZob{}ρ\PYGZcb{}+p\PYGZca{}\PYGZob{}ν\PYGZcb{}[x\PYGZca{}\PYGZob{}μ\PYGZcb{},q\PYGZca{}\PYGZob{}ρ\PYGZcb{}]}
\end{Verbatim}
to apply the usual graded derivation identity of the Lie bracket. Given nested brackets, we can also expanded commutators into explicit products of matrices,
\begin{Verbatim}[commandchars=\\\{\},xleftmargin=\parindent,numbers=left,bgcolor=bg]
\PYG{+w}{    }\PYG{n}{br3}\PYG{+w}{ }\PYG{o}{=}\PYG{+w}{ }\PYG{n+nd}{@Lie}\PYG{+w}{ }\PYG{p}{[}\PYG{n}{X}\PYG{p}{,}\PYG{+w}{ }\PYG{p}{\PYGZob{}}\PYG{k+kt}{P}\PYG{p}{,}\PYG{+w}{ }\PYG{k+kt}{Q}\PYG{p}{\PYGZcb{}}\PYG{p}{]}
\PYG{+w}{    }\PYG{n+nd}{@show}\PYG{+w}{ }\PYG{n}{expand\PYGZus{}commutators}\PYG{p}{(}\PYG{n}{br3}\PYG{p}{)}
\PYG{+w}{    }\PYG{c}{\PYGZsh{}   x\PYGZca{}\PYGZob{}μ\PYGZcb{}p\PYGZca{}\PYGZob{}ν\PYGZcb{}q\PYGZca{}\PYGZob{}ρ\PYGZcb{}+x\PYGZca{}\PYGZob{}μ\PYGZcb{}q\PYGZca{}\PYGZob{}ρ\PYGZcb{}p\PYGZca{}\PYGZob{}ν\PYGZcb{}\PYGZhy{}p\PYGZca{}\PYGZob{}ν\PYGZcb{}q\PYGZca{}\PYGZob{}ρ\PYGZcb{}x\PYGZca{}\PYGZob{}μ\PYGZcb{}\PYGZhy{}q\PYGZca{}\PYGZob{}ρ\PYGZcb{}p\PYGZca{}\PYGZob{}ν\PYGZcb{}x\PYGZca{}\PYGZob{}μ\PYGZcb{}}
\end{Verbatim}
Manifestly, the (graded) Jacobi identity is satisfied
\begin{Verbatim}[commandchars=\\\{\},xleftmargin=\parindent,numbers=left,bgcolor=bg]
\PYG{+w}{    }\PYG{n}{jac}\PYG{o}{=}\PYG{n+nd}{@Lie}\PYG{p}{[}\PYG{n}{X}\PYG{p}{,}\PYG{+w}{ }\PYG{p}{[}\PYG{n}{P}\PYG{p}{,}\PYG{+w}{ }\PYG{n}{Q}\PYG{p}{]}\PYG{p}{]}\PYG{+w}{ }\PYG{o}{+}\PYG{+w}{ }\PYG{n+nd}{@Lie}\PYG{p}{[}\PYG{n}{P}\PYG{p}{,}\PYG{+w}{ }\PYG{p}{[}\PYG{n}{Q}\PYG{p}{,}\PYG{+w}{ }\PYG{n}{X}\PYG{p}{]}\PYG{p}{]}\PYG{+w}{ }\PYG{o}{+}\PYG{+w}{ }\PYG{n+nd}{@Lie}\PYG{p}{[}\PYG{n}{Q}\PYG{p}{,}\PYG{+w}{ }\PYG{p}{[}\PYG{n}{X}\PYG{p}{,}\PYG{+w}{ }\PYG{n}{P}\PYG{p}{]}\PYG{p}{]}
\PYG{+w}{    }\PYG{n+nd}{@show}\PYG{+w}{ }\PYG{n}{expand\PYGZus{}commutators}\PYG{p}{(}\PYG{n}{jac}\PYG{p}{)}
\PYG{+w}{    }\PYG{c}{\PYGZsh{}   expand\PYGZus{}commutators(jac) = 0}
\end{Verbatim}
If there is a registered commutation relation for the matrices,
\begin{Verbatim}[commandchars=\\\{\},xleftmargin=\parindent,numbers=left,bgcolor=bg]
\PYG{+w}{    }\PYG{n+nd}{@syms}\PYG{+w}{ }\PYG{n}{ħ}
\PYG{+w}{    }\PYG{n}{result}\PYG{o}{=}\PYG{n+nb}{im}\PYG{o}{*}\PYG{n}{ħ}\PYG{o}{*}\PYG{n}{Metric}\PYG{p}{(}\PYG{p}{[}\PYG{n}{μ}\PYG{o}{=\PYGZgt{}}\PYG{n}{upper}\PYG{p}{,}\PYG{+w}{ }\PYG{n}{ν}\PYG{o}{=\PYGZgt{}}\PYG{n}{upper}\PYG{p}{]}\PYG{p}{)}
\PYG{+w}{    }\PYG{n+nd}{@commutator}\PYG{+w}{ }\PYG{p}{[}\PYG{n}{X}\PYG{p}{,}\PYG{n}{P}\PYG{p}{]}\PYG{o}{=}\PYG{n}{result}
\end{Verbatim}
then \jil{expand_commutators()} will expand in terms of this commutation relation instead of the matrix products,
\begin{Verbatim}[commandchars=\\\{\},xleftmargin=\parindent,numbers=left,bgcolor=bg]
\PYG{+w}{    }\PYG{n+nd}{@show}\PYG{+w}{ }\PYG{n}{expand\PYGZus{}commutators}\PYG{p}{(}\PYG{n+nd}{@Lie}\PYG{p}{[}\PYG{n}{X}\PYG{p}{,}\PYG{+w}{ }\PYG{n}{P}\PYG{p}{]}\PYG{p}{)}
\PYG{+w}{    }\PYG{c}{\PYGZsh{}   iħη\PYGZca{}\PYGZob{}μν\PYGZcb{}}
\PYG{+w}{    }\PYG{n+nd}{@show}\PYG{+w}{ }\PYG{n}{expand\PYGZus{}commutators}\PYG{p}{(}\PYG{n+nd}{@Lie}\PYG{p}{[}\PYG{n}{P}\PYG{p}{,}\PYG{+w}{ }\PYG{n}{X}\PYG{p}{]}\PYG{p}{)}
\PYG{+w}{    }\PYG{c}{\PYGZsh{}   \PYGZhy{}iħη\PYGZca{}\PYGZob{}μν\PYGZcb{}}
\end{Verbatim}

\subsection{Lie Algebra Generators}\label{sec:lie-algebra-generators}
The generators of Lie algebras can naturally act on tensors via some representation. We can encode this via the \jil{LieGenerator} operator, which is a subtype of \jil{GeneratorDerivSuperType} of \jil{DerivativeSuperType}. These operators are generically assumed not to commute, similar to a covariant derivative. The commutation relations between these operators can be encoded via the \jil{@commutator} macro.

As a simple example, consider the conformal algebra, given by generators
\begin{Verbatim}[commandchars=\\\{\},xleftmargin=\parindent,numbers=left,bgcolor=bg]
\PYG{+w}{    }\PYG{n}{P}\PYG{o}{=}\PYG{n}{LieGenerator}\PYG{p}{(}\PYG{l+s}{\PYGZdq{}}\PYG{l+s}{P}\PYG{l+s}{\PYGZdq{}}\PYG{p}{,}\PYG{n}{x}\PYG{p}{,}\PYG{p}{[}\PYG{n}{μ}\PYG{o}{=\PYGZgt{}}\PYG{n}{lower}\PYG{p}{]}\PYG{p}{)}
\PYG{+w}{    }\PYG{n}{M}\PYG{o}{=}\PYG{n}{LieGenerator}\PYG{p}{(}\PYG{l+s}{\PYGZdq{}}\PYG{l+s}{M}\PYG{l+s}{\PYGZdq{}}\PYG{p}{,}\PYG{n}{x}\PYG{p}{,}\PYG{p}{[}\PYG{n}{μ}\PYG{o}{=\PYGZgt{}}\PYG{n}{lower}\PYG{p}{,}\PYG{n}{ν}\PYG{o}{=\PYGZgt{}}\PYG{n}{lower}\PYG{p}{]}\PYG{p}{,}\PYG{n}{generate\PYGZus{}antisymmetric\PYGZus{}tableau}\PYG{p}{(}\PYG{p}{[}\PYG{l+m+mi}{1}\PYG{p}{,}\PYG{l+m+mi}{2}\PYG{p}{]}\PYG{p}{)}\PYG{p}{)}
\PYG{+w}{    }\PYG{n}{D}\PYG{o}{=}\PYG{n}{LieGenerator}\PYG{p}{(}\PYG{l+s}{\PYGZdq{}}\PYG{l+s}{D}\PYG{l+s}{\PYGZdq{}}\PYG{p}{,}\PYG{n}{x}\PYG{p}{)}
\PYG{+w}{    }\PYG{n}{K}\PYG{o}{=}\PYG{n}{LieGenerator}\PYG{p}{(}\PYG{l+s}{\PYGZdq{}}\PYG{l+s}{K}\PYG{l+s}{\PYGZdq{}}\PYG{p}{,}\PYG{n}{x}\PYG{p}{,}\PYG{p}{[}\PYG{n}{μ}\PYG{o}{=\PYGZgt{}}\PYG{n}{lower}\PYG{p}{]}\PYG{p}{)}
\end{Verbatim}
which satisfy
\begin{align}
[M_{\mu\nu},M_{\rho\sigma}]
&=
i(\eta_{\mu\rho}M_{\nu\sigma}
-\eta_{\mu\sigma}M_{\nu\rho}
-\eta_{\nu\rho}M_{\mu\sigma}
+\eta_{\nu\sigma}M_{\mu\rho}),
\\
[M_{\mu\nu},P_\rho]
&=
i(\eta_{\mu\rho}P_\nu
-\eta_{\nu\rho}P_\mu),
\\
[M_{\mu\nu},K_\rho]
&=
i(\eta_{\mu\rho}K_\nu
-\eta_{\nu\rho}K_\mu),
\\
[D,P_\mu]
&=iP_\mu,
\\
[D,K_\mu]
&=-iK_\mu,
\\
[K_\mu,P_\nu]
&=
2i\left(\eta_{\mu\nu}D-M_{\mu\nu}\right),
\\
[D,M_{\mu\nu}]
&=0,
\\
[P_\mu,P_\nu]
&=0,
\\
[K_\mu,K_\nu]
&=0,
\end{align}
which one can encode for each, for example using the syntax
\begin{Verbatim}[commandchars=\\\{\},xleftmargin=\parindent,numbers=left,bgcolor=bg]
\PYG{+w}{    }\PYG{n+nd}{@commutator}\PYG{+w}{ }\PYG{p}{[}\PYG{n}{D}\PYG{p}{,}\PYG{n}{P}\PYG{p}{]}\PYG{o}{=}\PYG{n+nb}{im}\PYG{o}{*}\PYG{n}{P}
\end{Verbatim}
For ease, these generators and the full algebra can be generated from scratch via the function (in 4D)
\begin{Verbatim}[commandchars=\\\{\},xleftmargin=\parindent,numbers=left,bgcolor=bg]
\PYG{+w}{    }\PYG{n}{P}\PYG{p}{,}\PYG{+w}{ }\PYG{n}{M}\PYG{p}{,}\PYG{+w}{ }\PYG{n}{D}\PYG{p}{,}\PYG{+w}{ }\PYG{n}{K}\PYG{o}{=}\PYG{n}{generate\PYGZus{}conformal\PYGZus{}algebra}\PYG{p}{(}\PYG{l+m+mi}{4}\PYG{p}{)}
\end{Verbatim}
Now, suppose we want to consider a primary operator $X$ with weight $\Delta$
\begin{equation}
    DX=i\Delta X.
\end{equation}
We can encode the action of this generator on the tensor via
\begin{Verbatim}[commandchars=\\\{\},xleftmargin=\parindent,numbers=left,bgcolor=bg]
\PYG{+w}{    }\PYG{n+nd}{@syms}\PYG{+w}{ }\PYG{n}{Δ}
\PYG{+w}{    }\PYG{n}{X}\PYG{o}{=}\PYG{n}{Tensor}\PYG{p}{(}\PYG{l+s}{\PYGZdq{}}\PYG{l+s}{X}\PYG{l+s}{\PYGZdq{}}\PYG{p}{,}\PYG{+w}{ }\PYG{n}{x}\PYG{p}{)}
\PYG{+w}{    }\PYG{n+nd}{@action}\PYG{+w}{ }\PYG{n}{D}\PYG{p}{(}\PYG{n}{X}\PYG{p}{)}\PYG{+w}{ }\PYG{o}{=}\PYG{+w}{ }\PYG{n+nb}{im}\PYG{o}{*}\PYG{n}{Δ}\PYG{o}{*}\PYG{n}{X}
\end{Verbatim}
Now, consider a descendant
\begin{Verbatim}[commandchars=\\\{\},xleftmargin=\parindent,numbers=left,bgcolor=bg]
\PYG{+w}{    }\PYG{n}{desc}\PYG{o}{=}\PYG{n}{P}\PYG{p}{(}\PYG{n}{X}\PYG{p}{)}
\end{Verbatim}
We can compute the weight of this operator simply via
\begin{Verbatim}[commandchars=\\\{\},xleftmargin=\parindent,numbers=left,bgcolor=bg]
\PYG{+w}{    }\PYG{n+nd}{@show}\PYG{+w}{ }\PYG{n}{derive}\PYG{p}{(}\PYG{n}{commute}\PYG{p}{(}\PYG{n}{D}\PYG{p}{(}\PYG{n}{desc}\PYG{p}{)}\PYG{p}{)}\PYG{p}{)}
\PYG{+w}{    }\PYG{c}{\PYGZsh{}   (i + iΔ)P\PYGZus{}\PYGZob{}μ\PYGZcb{}∘X}
\end{Verbatim}
where \jil{commute()} commutes the two highest-lying nested operators in the expression, producing the associated additional terms from the commutation relations.
We can also consider higher descendants, for example
\begin{Verbatim}[commandchars=\\\{\},xleftmargin=\parindent,numbers=left,bgcolor=bg]
\PYG{+w}{    }\PYG{n}{Pρ}\PYG{o}{=}\PYG{n}{set\PYGZus{}op\PYGZus{}indices!}\PYG{p}{(}\PYG{n}{copy}\PYG{p}{(}\PYG{n}{P}\PYG{p}{)}\PYG{p}{,}\PYG{+w}{ }\PYG{p}{[}\PYG{n}{ρ}\PYG{o}{=\PYGZgt{}}\PYG{n}{lower}\PYG{p}{]}\PYG{p}{)}
\PYG{+w}{    }\PYG{n}{Pν}\PYG{o}{=}\PYG{n}{set\PYGZus{}op\PYGZus{}indices!}\PYG{p}{(}\PYG{n}{copy}\PYG{p}{(}\PYG{n}{P}\PYG{p}{)}\PYG{p}{,}\PYG{+w}{ }\PYG{p}{[}\PYG{n}{ν}\PYG{o}{=\PYGZgt{}}\PYG{n}{lower}\PYG{p}{]}\PYG{p}{)}
\PYG{+w}{    }\PYG{n}{desc}\PYG{o}{=}\PYG{n}{Pρ}\PYG{p}{(}\PYG{n}{Pν}\PYG{p}{(}\PYG{n}{P}\PYG{p}{(}\PYG{n}{X}\PYG{p}{)}\PYG{p}{)}\PYG{p}{)}
\end{Verbatim}
Then, the weight is determined by
\begin{Verbatim}[commandchars=\\\{\},xleftmargin=\parindent,numbers=left,bgcolor=bg]
\PYG{+w}{    }\PYG{n+nd}{@show}\PYG{+w}{ }\PYG{n}{derive}\PYG{p}{(}\PYG{n}{commute\PYGZus{}right}\PYG{p}{(}\PYG{n}{D}\PYG{p}{(}\PYG{n}{desc}\PYG{p}{)}\PYG{p}{,}\PYG{l+s}{\PYGZdq{}}\PYG{l+s}{D}\PYG{l+s}{\PYGZdq{}}\PYG{p}{)}\PYG{p}{)}
\PYG{+w}{    }\PYG{c}{\PYGZsh{}   (3i + iΔ)P\PYGZus{}\PYGZob{}ρ\PYGZcb{}∘P\PYGZus{}\PYGZob{}ν\PYGZcb{}∘P\PYGZus{}\PYGZob{}μ\PYGZcb{}∘X}
\end{Verbatim}
which shows that the conformal weight has increased by 3. Here, \jil{commute_right()} drags the operator with name ``D'' as deep as possible in a tree of nested operators. This is just one example, hopefully illustrative of the potential for these kinds of operators.

\subsection{Traces}\label{sec:traces}
\jil{Trace}s exist in Alakazam as an operator that subtypes \jil{<:TraceSuperType} \jil{<:CyclicSuperType} \jil{<:OperatorSuperType}.
They obey the expected properties: scalar tensors can be factored out and sums can be split by linearity, as well as obeying cyclicity:
\begin{Verbatim}[commandchars=\\\{\},xleftmargin=\parindent,numbers=left,bgcolor=bg]
\PYG{+w}{    }\PYG{n}{A}\PYG{+w}{ }\PYG{o}{=}\PYG{+w}{ }\PYG{n}{Mat}\PYG{p}{(}\PYG{l+s}{\PYGZdq{}}\PYG{l+s}{A}\PYG{l+s}{\PYGZdq{}}\PYG{p}{)}\PYG{p}{;}\PYG{+w}{ }\PYG{n}{B}\PYG{+w}{ }\PYG{o}{=}\PYG{+w}{ }\PYG{n}{Mat}\PYG{p}{(}\PYG{l+s}{\PYGZdq{}}\PYG{l+s}{B}\PYG{l+s}{\PYGZdq{}}\PYG{p}{)}\PYG{p}{;}\PYG{+w}{ }\PYG{n}{C}\PYG{+w}{ }\PYG{o}{=}\PYG{+w}{ }\PYG{n}{Mat}\PYG{p}{(}\PYG{l+s}{\PYGZdq{}}\PYG{l+s}{C}\PYG{l+s}{\PYGZdq{}}\PYG{p}{)}
\PYG{+w}{    }\PYG{n}{V}\PYG{p}{(}\PYG{n}{i}\PYG{p}{)}\PYG{+w}{ }\PYG{o}{=}\PYG{+w}{ }\PYG{n}{Tensor}\PYG{p}{(}\PYG{l+s}{\PYGZdq{}}\PYG{l+s}{V}\PYG{l+s}{\PYGZdq{}}\PYG{p}{,}\PYG{+w}{ }\PYG{p}{[}\PYG{n}{i}\PYG{+w}{ }\PYG{o}{=\PYGZgt{}}\PYG{+w}{ }\PYG{n}{lower}\PYG{p}{]}\PYG{p}{)}

\PYG{+w}{    }\PYG{n}{factor\PYGZus{}trace}\PYG{p}{(}\PYG{n}{Trace}\PYG{p}{(}\PYG{l+m+mi}{2}\PYG{+w}{ }\PYG{o}{*}\PYG{+w}{ }\PYG{n}{A}\PYG{+w}{ }\PYG{o}{*}\PYG{+w}{ }\PYG{n}{B}\PYG{p}{)}\PYG{p}{)}\PYG{+w}{          }\PYG{c}{\PYGZsh{} 2Tr⟦AB⟧}
\PYG{+w}{    }\PYG{n}{factor\PYGZus{}trace}\PYG{p}{(}\PYG{n}{Trace}\PYG{p}{(}\PYG{n}{V}\PYG{p}{(}\PYG{n}{μ}\PYG{p}{)}\PYG{+w}{ }\PYG{o}{*}\PYG{+w}{ }\PYG{n}{A}\PYG{+w}{ }\PYG{o}{*}\PYG{+w}{ }\PYG{n}{B}\PYG{p}{)}\PYG{p}{)}\PYG{+w}{       }\PYG{c}{\PYGZsh{} V\PYGZus{}\PYGZob{}μ\PYGZcb{}Tr⟦AB⟧}
\PYG{+w}{    }\PYG{n}{factor\PYGZus{}trace}\PYG{p}{(}\PYG{n}{Trace}\PYG{p}{(}\PYG{n}{A}\PYG{+w}{ }\PYG{o}{*}\PYG{+w}{ }\PYG{n}{B}\PYG{+w}{ }\PYG{o}{+}\PYG{+w}{ }\PYG{n}{c}\PYG{+w}{ }\PYG{o}{*}\PYG{+w}{ }\PYG{n}{B}\PYG{+w}{ }\PYG{o}{*}\PYG{+w}{ }\PYG{n}{C}\PYG{p}{)}\PYG{p}{)}\PYG{+w}{  }\PYG{c}{\PYGZsh{} Tr⟦AB⟧+cTr⟦BC⟧}

\PYG{+w}{    }\PYG{n}{Trace}\PYG{p}{(}\PYG{n}{A}\PYG{+w}{ }\PYG{o}{*}\PYG{+w}{ }\PYG{n}{B}\PYG{p}{)}\PYG{+w}{ }\PYG{o}{==}\PYG{+w}{ }\PYG{n}{Trace}\PYG{p}{(}\PYG{n}{B}\PYG{+w}{ }\PYG{o}{*}\PYG{+w}{ }\PYG{n}{A}\PYG{p}{)}\PYG{+w}{            }\PYG{c}{\PYGZsh{} true}
\PYG{+w}{    }\PYG{n}{Trace}\PYG{p}{(}\PYG{n}{A}\PYG{+w}{ }\PYG{o}{*}\PYG{+w}{ }\PYG{n}{B}\PYG{+w}{ }\PYG{o}{*}\PYG{+w}{ }\PYG{n}{C}\PYG{p}{)}\PYG{+w}{ }\PYG{o}{==}\PYG{+w}{ }\PYG{n}{Trace}\PYG{p}{(}\PYG{n}{C}\PYG{+w}{ }\PYG{o}{*}\PYG{+w}{ }\PYG{n}{A}\PYG{+w}{ }\PYG{o}{*}\PYG{+w}{ }\PYG{n}{B}\PYG{p}{)}\PYG{+w}{    }\PYG{c}{\PYGZsh{} true}
\PYG{+w}{    }\PYG{n}{Trace}\PYG{p}{(}\PYG{n}{A}\PYG{+w}{ }\PYG{o}{*}\PYG{+w}{ }\PYG{n}{B}\PYG{+w}{ }\PYG{o}{*}\PYG{+w}{ }\PYG{n}{C}\PYG{p}{)}\PYG{+w}{ }\PYG{o}{==}\PYG{+w}{ }\PYG{n}{Trace}\PYG{p}{(}\PYG{n}{B}\PYG{+w}{ }\PYG{o}{*}\PYG{+w}{ }\PYG{n}{A}\PYG{+w}{ }\PYG{o}{*}\PYG{+w}{ }\PYG{n}{C}\PYG{p}{)}\PYG{+w}{    }\PYG{c}{\PYGZsh{} false, not a rotation}
\end{Verbatim}
Any tensor that is not a matrix is brought outside the trace by \jil{factor_trace}, leaving only matrices within it, and cyclicity is recognised by \jil{==} itself. A permutation which is not a rotation is correctly not equal to the original trace. Note that \jil{Trace} is implemented as a supertrace - it can change by a sign when cycling fermions:
\begin{equation}
    \mathrm{str}(XY)=(-1)^{|X||Y|}\mathrm{str}(YX). 
\end{equation} 
For example,
\begin{Verbatim}[commandchars=\\\{\},xleftmargin=\parindent,numbers=left,bgcolor=bg]
\PYG{+w}{    }\PYG{n}{spinor\PYGZus{}indices}\PYG{o}{=}\PYG{n}{IndexSet}\PYG{p}{(}\PYG{l+s}{\PYGZdq{}}\PYG{l+s}{spinor\PYGZus{}indices}\PYG{l+s}{\PYGZdq{}}\PYG{p}{,}\PYG{l+m+mi}{2}\PYG{p}{,}\PYG{n}{lower}\PYG{p}{,}\PYG{n}{fermionic}\PYG{p}{)}
\PYG{+w}{    }\PYG{n}{α}\PYG{p}{,}\PYG{n}{β}\PYG{p}{,}\PYG{n}{γ}\PYG{p}{,}\PYG{n}{χ}\PYG{+w}{ }\PYG{o}{=}\PYG{+w}{ }\PYG{n}{Index}\PYG{p}{(}\PYG{p}{[}\PYG{l+s}{\PYGZdq{}}\PYG{l+s}{α}\PYG{l+s}{\PYGZdq{}}\PYG{p}{,}\PYG{l+s}{\PYGZdq{}}\PYG{l+s}{β}\PYG{l+s}{\PYGZdq{}}\PYG{p}{,}\PYG{l+s}{\PYGZdq{}}\PYG{l+s}{γ}\PYG{l+s}{\PYGZdq{}}\PYG{p}{,}\PYG{l+s}{\PYGZdq{}}\PYG{l+s}{χ}\PYG{l+s}{\PYGZdq{}}\PYG{p}{]}\PYG{p}{,}\PYG{+w}{ }\PYG{n}{spinor\PYGZus{}indices}\PYG{p}{)}
\PYG{+w}{    }\PYG{n}{Af}\PYG{+w}{ }\PYG{o}{=}\PYG{+w}{ }\PYG{n}{Mat}\PYG{p}{(}\PYG{l+s}{\PYGZdq{}}\PYG{l+s}{A}\PYG{l+s}{\PYGZdq{}}\PYG{p}{,}\PYG{+w}{ }\PYG{p}{[}\PYG{n}{α}\PYG{+w}{ }\PYG{o}{=\PYGZgt{}}\PYG{+w}{ }\PYG{n}{lower}\PYG{p}{]}\PYG{p}{)}\PYG{+w}{              }\PYG{c}{\PYGZsh{} odd grading}
\PYG{+w}{    }\PYG{n}{Bf}\PYG{+w}{ }\PYG{o}{=}\PYG{+w}{ }\PYG{n}{Mat}\PYG{p}{(}\PYG{l+s}{\PYGZdq{}}\PYG{l+s}{B}\PYG{l+s}{\PYGZdq{}}\PYG{p}{,}\PYG{+w}{ }\PYG{p}{[}\PYG{n}{β}\PYG{+w}{ }\PYG{o}{=\PYGZgt{}}\PYG{+w}{ }\PYG{n}{lower}\PYG{p}{]}\PYG{p}{)}\PYG{+w}{              }\PYG{c}{\PYGZsh{} odd grading}

\PYG{+w}{    }\PYG{n}{grading}\PYG{p}{(}\PYG{n}{Af}\PYG{p}{)}\PYG{+w}{                             }\PYG{c}{\PYGZsh{} 1}
\PYG{+w}{    }\PYG{n}{Trace}\PYG{p}{(}\PYG{n}{Af}\PYG{+w}{ }\PYG{o}{*}\PYG{+w}{ }\PYG{n}{Bf}\PYG{p}{)}\PYG{+w}{ }\PYG{o}{==}\PYG{+w}{ }\PYG{n}{Trace}\PYG{p}{(}\PYG{n}{Bf}\PYG{+w}{ }\PYG{o}{*}\PYG{+w}{ }\PYG{n}{Af}\PYG{p}{)}\PYG{+w}{        }\PYG{c}{\PYGZsh{} false}
\PYG{+w}{    }\PYG{n}{signed\PYGZus{}cyclic\PYGZus{}equals}\PYG{p}{(}\PYG{n}{Trace}\PYG{p}{(}\PYG{n}{Af}\PYG{+w}{ }\PYG{o}{*}\PYG{+w}{ }\PYG{n}{Bf}\PYG{p}{)}\PYG{p}{,}\PYG{+w}{ }\PYG{n}{Trace}\PYG{p}{(}\PYG{n}{Bf}\PYG{+w}{ }\PYG{o}{*}\PYG{+w}{ }\PYG{n}{Af}\PYG{p}{)}\PYG{p}{)}\PYG{+w}{   }\PYG{c}{\PYGZsh{} (true, \PYGZhy{}1)}
\PYG{+w}{    }\PYG{n}{Trace}\PYG{p}{(}\PYG{n}{Af}\PYG{+w}{ }\PYG{o}{*}\PYG{+w}{ }\PYG{n}{B}\PYG{p}{)}\PYG{+w}{ }\PYG{o}{==}\PYG{+w}{ }\PYG{n}{Trace}\PYG{p}{(}\PYG{n}{B}\PYG{+w}{ }\PYG{o}{*}\PYG{+w}{ }\PYG{n}{Af}\PYG{p}{)}\PYG{+w}{          }\PYG{c}{\PYGZsh{} true, only one factor is odd}
\end{Verbatim}

In field theory, traces of certain matrices obey special identities. The key example are gamma matrices, whose product generally reduces when inside a trace. For gamma matrices, these laws are generated by the package automatically, and take the dimension into account:
\begin{Verbatim}[commandchars=\\\{\},xleftmargin=\parindent,numbers=left,bgcolor=bg]
\PYG{+w}{    }\PYG{n}{tr}\PYG{p}{(}\PYG{n}{x}\PYG{p}{)}\PYG{+w}{ }\PYG{o}{=}\PYG{+w}{ }\PYG{n}{simplify}\PYG{p}{(}\PYG{n}{eliminate\PYGZus{}trace}\PYG{p}{(}\PYG{n}{Trace}\PYG{p}{(}\PYG{n}{TensorExpression}\PYG{p}{(}\PYG{n}{x}\PYG{p}{)}\PYG{p}{)}\PYG{p}{)}\PYG{p}{)}

\PYG{+w}{    }\PYG{n}{tr}\PYG{p}{(}\PYG{n}{γ}\PYG{p}{(}\PYG{n}{μ}\PYG{p}{)}\PYG{+w}{ }\PYG{o}{*}\PYG{+w}{ }\PYG{n}{γ}\PYG{p}{(}\PYG{n}{ν}\PYG{p}{)}\PYG{p}{)}\PYG{+w}{                   }\PYG{c}{\PYGZsh{} 4η\PYGZus{}\PYGZob{}μν\PYGZcb{}}
\PYG{+w}{    }\PYG{n}{tr}\PYG{p}{(}\PYG{n}{γ}\PYG{p}{(}\PYG{n}{μ}\PYG{p}{)}\PYG{+w}{ }\PYG{o}{*}\PYG{+w}{ }\PYG{n}{γ}\PYG{p}{(}\PYG{n}{ν}\PYG{p}{)}\PYG{+w}{ }\PYG{o}{*}\PYG{+w}{ }\PYG{n}{γ}\PYG{p}{(}\PYG{n}{ρ}\PYG{p}{)}\PYG{p}{)}\PYG{+w}{            }\PYG{c}{\PYGZsh{} 0}
\PYG{+w}{    }\PYG{n}{tr}\PYG{p}{(}\PYG{n}{γ}\PYG{p}{(}\PYG{n}{μ}\PYG{p}{)}\PYG{+w}{ }\PYG{o}{*}\PYG{+w}{ }\PYG{n}{γ}\PYG{p}{(}\PYG{n}{ν}\PYG{p}{)}\PYG{+w}{ }\PYG{o}{*}\PYG{+w}{ }\PYG{n}{γ}\PYG{p}{(}\PYG{n}{ρ}\PYG{p}{)}\PYG{+w}{ }\PYG{o}{*}\PYG{+w}{ }\PYG{n}{γ}\PYG{p}{(}\PYG{n}{σ}\PYG{p}{)}\PYG{p}{)}\PYG{+w}{     }\PYG{c}{\PYGZsh{} 4η\PYGZus{}\PYGZob{}μν\PYGZcb{}η\PYGZus{}\PYGZob{}ρσ\PYGZcb{}\PYGZhy{}4η\PYGZus{}\PYGZob{}μρ\PYGZcb{}η\PYGZus{}\PYGZob{}νσ\PYGZcb{}+4η\PYGZus{}\PYGZob{}μσ\PYGZcb{}η\PYGZus{}\PYGZob{}νρ\PYGZcb{}}
\PYG{+w}{    }\PYG{n}{tr}\PYG{p}{(}\PYG{n}{Gamma}\PYG{p}{(}\PYG{n}{μ}\PYG{+w}{ }\PYG{o}{=\PYGZgt{}}\PYG{+w}{ }\PYG{n}{upper}\PYG{p}{)}\PYG{+w}{ }\PYG{o}{*}\PYG{+w}{ }\PYG{n}{γ}\PYG{p}{(}\PYG{n}{μ}\PYG{p}{)}\PYG{p}{)}\PYG{+w}{      }\PYG{c}{\PYGZsh{} 16}
\end{Verbatim}
The identity matrix is a related case, since \(\mathrm{tr}(\mathbb{1})\) is the
size of the space being traced over and an \jil{IdMatrix} carries no index of
its own.  It is therefore built over the space it is the identity on. This is important for gamma matrices, since identities generated here will be for the spinor space rather than the spacetime:
\begin{Verbatim}[commandchars=\\\{\},xleftmargin=\parindent,numbers=left,bgcolor=bg]
\PYG{+w}{    }\PYG{n}{tr}\PYG{p}{(}\PYG{n}{IdMatrix}\PYG{p}{(}\PYG{n}{spinor\PYGZus{}indexset}\PYG{p}{(}\PYG{n}{st4}\PYG{p}{)}\PYG{p}{)}\PYG{p}{)}\PYG{+w}{       }\PYG{c}{\PYGZsh{} 4, for st of dimension 4}
\PYG{+w}{    }\PYG{n}{tr}\PYG{p}{(}\PYG{n}{IdMatrix}\PYG{p}{(}\PYG{n}{spinor\PYGZus{}indexset}\PYG{p}{(}\PYG{n}{st10}\PYG{p}{)}\PYG{p}{)}\PYG{p}{)}\PYG{+w}{       }\PYG{c}{\PYGZsh{} 32, for st of dimension 10}
\PYG{+w}{    }\PYG{n}{tr}\PYG{p}{(}\PYG{n}{IdMatrix}\PYG{p}{(}\PYG{p}{)}\PYG{p}{)}\PYG{+w}{                          }\PYG{c}{\PYGZsh{} Tr⟦1⟧}
\end{Verbatim}
An identity built with no space is left unreduced rather than guessed at.

For general matrices of a concrete type, reduction rules when inside a trace can be registered by the \jil{@mrule} macro, similar to reduction rules for matrix products:
\begin{Verbatim}[commandchars=\\\{\},xleftmargin=\parindent,numbers=left,bgcolor=bg]
\PYG{+w}{    }\PYG{n+nd}{@matrix}\PYG{+w}{ }\PYG{n}{Gx}
\PYG{+w}{    }\PYG{n+nd}{@matrix}\PYG{+w}{ }\PYG{n}{Px}
\PYG{+w}{    }\PYG{n}{G}\PYG{p}{(}\PYG{n}{i}\PYG{p}{)}\PYG{+w}{ }\PYG{o}{=}\PYG{+w}{ }\PYG{n}{Gx}\PYG{p}{(}\PYG{l+s}{\PYGZdq{}}\PYG{l+s}{G}\PYG{l+s}{\PYGZdq{}}\PYG{p}{,}\PYG{+w}{ }\PYG{p}{[}\PYG{n}{i}\PYG{+w}{ }\PYG{o}{=\PYGZgt{}}\PYG{+w}{ }\PYG{n}{lower}\PYG{p}{]}\PYG{p}{)}\PYG{p}{;}\PYG{+w}{  }\PYG{n}{P}\PYG{+w}{ }\PYG{o}{=}\PYG{+w}{ }\PYG{n}{Px}\PYG{p}{(}\PYG{l+s}{\PYGZdq{}}\PYG{l+s}{P}\PYG{l+s}{\PYGZdq{}}\PYG{p}{)}
\PYG{+w}{    }\PYG{n+nd}{@mrule}\PYG{+w}{ }\PYG{n}{Trace}\PYG{p}{(}\PYG{n}{Gx}\PYG{p}{(}\PYG{l+s}{\PYGZdq{}}\PYG{l+s}{G}\PYG{l+s}{\PYGZdq{}}\PYG{p}{,}\PYG{p}{[}\PYG{n}{μ}\PYG{+w}{ }\PYG{o}{=\PYGZgt{}}\PYG{+w}{ }\PYG{n}{lower}\PYG{p}{]}\PYG{p}{)}\PYG{+w}{ }\PYG{o}{*}\PYG{+w}{ }\PYG{n}{Gx}\PYG{p}{(}\PYG{l+s}{\PYGZdq{}}\PYG{l+s}{G}\PYG{l+s}{\PYGZdq{}}\PYG{p}{,}\PYG{p}{[}\PYG{n}{ν}\PYG{+w}{ }\PYG{o}{=\PYGZgt{}}\PYG{+w}{ }\PYG{n}{lower}\PYG{p}{]}\PYG{p}{)}\PYG{p}{)}\PYG{+w}{ }\PYG{o}{=}
\PYG{+w}{        }\PYG{n}{k}\PYG{+w}{ }\PYG{o}{*}\PYG{+w}{ }\PYG{n}{Metric}\PYG{p}{(}\PYG{p}{[}\PYG{n}{μ}\PYG{+w}{ }\PYG{o}{=\PYGZgt{}}\PYG{+w}{ }\PYG{n}{lower}\PYG{p}{,}\PYG{+w}{ }\PYG{n}{ν}\PYG{+w}{ }\PYG{o}{=\PYGZgt{}}\PYG{+w}{ }\PYG{n}{lower}\PYG{p}{]}\PYG{p}{)}

\PYG{+w}{    }\PYG{n}{reduce\PYGZus{}trace}\PYG{p}{(}\PYG{n}{Trace}\PYG{p}{(}\PYG{n}{G}\PYG{p}{(}\PYG{n}{μ}\PYG{p}{)}\PYG{+w}{ }\PYG{o}{*}\PYG{+w}{ }\PYG{n}{G}\PYG{p}{(}\PYG{n}{ν}\PYG{p}{)}\PYG{p}{)}\PYG{p}{)}\PYG{+w}{        }\PYG{c}{\PYGZsh{} kη\PYGZus{}\PYGZob{}μν\PYGZcb{}}
\PYG{+w}{    }\PYG{n}{reduce\PYGZus{}trace}\PYG{p}{(}\PYG{n}{Trace}\PYG{p}{(}\PYG{n}{G}\PYG{p}{(}\PYG{n}{ρ}\PYG{p}{)}\PYG{+w}{ }\PYG{o}{*}\PYG{+w}{ }\PYG{n}{G}\PYG{p}{(}\PYG{n}{σ}\PYG{p}{)}\PYG{p}{)}\PYG{p}{)}\PYG{+w}{        }\PYG{c}{\PYGZsh{} kη\PYGZus{}\PYGZob{}ρσ\PYGZcb{}, matched by index}
\PYG{+w}{    }\PYG{n}{reduce\PYGZus{}trace}\PYG{p}{(}\PYG{n}{Trace}\PYG{p}{(}\PYG{n}{P}\PYG{+w}{ }\PYG{o}{*}\PYG{+w}{ }\PYG{n}{P}\PYG{p}{)}\PYG{p}{)}\PYG{+w}{              }\PYG{c}{\PYGZsh{} Tr⟦PP⟧, no rule registered}
\end{Verbatim}
Rules are dispatched on the types of the factors and matched by index, exactly as the \jil{reduce_matrix} rules are. The rule will also reduce products that are equivalent up to cyclic order.

The two halves are combined by \jil{eliminate_trace}, which factors scalars out
and then applies the reduction rules.

\subsection{\jil{Dagger} and Complex Conjugation}\label{sec:dagger-and-complex-conjugation}
Basic complex conjugation behaviour is implemented in Alakazam with associated \jil{Dagger} operator. This operation can be performed using Julia's adjoint \jil{'} postfixed to a \jil{TensorExpression}. It can also be reached via \jil{^✝} as a synonym that reads closer to the physics notation. Note that this character is the Unicode Latin cross (U+271D) rather than the Unicode dagger, as the latter is not legal syntax in Julia.

Either spelling carries out the conjugation immediately, and applies the \jil{Dagger} object for the case where the operand's transformation is not yet known.

Conjugation is an antimorphism, so a product reverses,
\begin{equation}
    (AB)^\dagger = B^\dagger A^\dagger,
\end{equation}
and this reversal is implemented once, on \jil{AntimorphismSuperType}. This allows the implementation to be easily extended to custom antimorphisms in future. Sums distribute with complex-conjugated coefficients by default, and a dagger applied to a dagger collapses by involution, $O(O(A)) = A$. The collapse tests the operator's concrete type, so a dagger never cancels against a different antimorphism operator that happens to sit beneath it.

The complex conjugation behaviour is determined primarily by the indices of the object, via Lie algebra, metric, and conjugate \jil{IndexSet}. This allows for a variety of conjugation behaviours to be defined by an implied Frobenius-Schur class of ``fixed'' (real-orthogonal), ``complex'', or ``quaternionic'' (pseudo-real).

First, we check for the conjugate \jil{IndexSet}. When this is another set, this implies the behaviour is complex. For a self-conjugate set, the behaviour is inferred through the Lie algebra by considering its division algebra, which implies whether the indices should be fixed or quaternionic. If no Lie algebra is specified, then we fall back to the ``fixed'' behaviour.

Each class implies how the indices change, and what sign is accumulated. This is fixed through \jil{conj_convention}, which determines the behaviour based on a reference position at which conjugation costs nothing (by default the \jil{default_position}), and a sign charged to the opposite slot, so that one value describes both levels. This can be seen in Table~\ref{tab:conjclass}.
\begin{table}[]
\begin{center}
\begin{tabular}{@{}llll@{}}
\toprule
class & index set & position & sign \\
\midrule
fixed & unchanged & unchanged & $+1$ at the reference, $s=1$ away from it \\
complex & partner set & unchanged & $+1$ at the reference, $s=1$ away from it \\
quaternionic & unchanged & flipped & $+1$ at the reference, $s=-1$ away from it \\
\bottomrule
\end{tabular}

\end{center}
    \caption{Conjugation Properties}
    \label{tab:conjclass}
\end{table}
By default $s$ is $-1$ for the quaternionic class and $+1$ otherwise, which reproduces our own convention. This can be overridden with a custom antisymmetric symbol using \jil{conj_convention} and need not be dependent on the class, as we will demonstrate in one of the following examples.

This is quite technical as it stands, and not very transparent how the behaviour is in practice. Thus, let us consider some key examples which demonstrate the versatility of this implementation.

The basic spinor one may like to work with is the complex Weyl spinor in $3+1D$. These have a conjugate pair of indices, for example $\alpha$ and $\dot\alpha$. By declaring the \jil{IndexSet}s as conjugates, we get behaviour
\begin{Verbatim}[commandchars=\\\{\},xleftmargin=\parindent,numbers=left,bgcolor=bg]
\PYG{+w}{    }\PYG{n}{weyl}\PYG{+w}{ }\PYG{o}{=}\PYG{+w}{ }\PYG{n}{IndexSet}\PYG{p}{(}\PYG{l+s}{\PYGZdq{}}\PYG{l+s}{weyl}\PYG{l+s}{\PYGZdq{}}\PYG{p}{,}\PYG{+w}{ }\PYG{n}{Spin}\PYG{p}{(}\PYG{l+m+mi}{4}\PYG{p}{)}\PYG{p}{)}
\PYG{+w}{    }\PYG{n}{weyldot}\PYG{+w}{ }\PYG{o}{=}\PYG{+w}{ }\PYG{n}{IndexSet}\PYG{p}{(}\PYG{l+s}{\PYGZdq{}}\PYG{l+s}{weyldot}\PYG{l+s}{\PYGZdq{}}\PYG{p}{,}\PYG{+w}{ }\PYG{n}{Spin}\PYG{p}{(}\PYG{l+m+mi}{4}\PYG{p}{)}\PYG{p}{)}
\PYG{+w}{    }\PYG{n+nd}{@conjugates}\PYG{+w}{ }\PYG{n}{weyl}\PYG{p}{,}\PYG{+w}{ }\PYG{n}{weyldot}
\PYG{+w}{    }\PYG{n+nd}{@indices}\PYG{+w}{ }\PYG{n}{α}\PYG{o}{:}\PYG{n}{β∈weyl}
\PYG{+w}{    }\PYG{n+nd}{@indices}\PYG{+w}{ }\PYG{n}{αdot}\PYG{o}{:}\PYG{n}{βdot∈weyldot}
\PYG{+w}{    }\PYG{n}{ψ}\PYG{o}{=}\PYG{l+s+sa}{T}\PYG{l+s}{\PYGZdq{}}\PYG{l+s}{ψ\PYGZca{}α}\PYG{l+s}{\PYGZdq{}}
\PYG{+w}{    }\PYG{n+nd}{@show}\PYG{+w}{ }\PYG{n}{ψ}\PYG{o}{\PYGZsq{}}
\PYG{+w}{    }\PYG{c}{\PYGZsh{}   ψ\PYGZsq{} = ̄ψ\PYGZca{}\PYGZob{}̇α\PYGZcb{}}
\PYG{+w}{    }\PYG{n}{χbar}\PYG{o}{=}\PYG{l+s+sa}{T}\PYG{l+s}{\PYGZdq{}}\PYG{l+s}{χbar\PYGZus{}αdot}\PYG{l+s}{\PYGZdq{}}
\PYG{+w}{    }\PYG{n+nd}{@show}\PYG{+w}{ }\PYG{n}{χbar}\PYG{o}{\PYGZsq{}}
\PYG{+w}{    }\PYG{c}{\PYGZsh{}   ̄χ\PYGZsq{} = χ\PYGZus{}\PYGZob{}α\PYGZcb{}}
\end{Verbatim}
Alakazam will find the equivalent (un-)barred/dotted/hatted etc. index in the conjugate \jil{IndexSet} automatically, and toggle the bar in the name of the \jil{Tensor}. Only when the partner set holds no match is a name
constructed, and then the decoration is inferred from the indices already declared in that set rather than fixed in advance; a dot is used as the last resort for a fermionic set that is still empty and so offers nothing to infer from. A user who labels conjugate spinors indices with a bar for example therefore gets barred indices back, not dotted ones.

For $SU(2)$ indices, these are quaternionic. They are self-conjugate and also change level under conjugation. Furthermore, one of the levels picks up a sign on conjugation.
\begin{Verbatim}[commandchars=\\\{\},xleftmargin=\parindent,numbers=left,bgcolor=bg]
\PYG{+w}{    }\PYG{n}{su2inds}\PYG{+w}{ }\PYG{o}{=}\PYG{+w}{ }\PYG{n}{IndexSet}\PYG{p}{(}\PYG{l+s}{\PYGZdq{}}\PYG{l+s}{su2}\PYG{l+s}{\PYGZdq{}}\PYG{p}{,}\PYG{+w}{ }\PYG{n}{lower}\PYG{p}{,}\PYG{+w}{ }\PYG{n}{SUAlgebra}\PYG{p}{(}\PYG{l+m+mi}{2}\PYG{p}{)}\PYG{p}{)}
\PYG{+w}{    }\PYG{n+nd}{@conjugates}\PYG{+w}{ }\PYG{n}{su2inds}
\PYG{+w}{    }\PYG{n+nd}{@indices}\PYG{+w}{ }\PYG{n}{i}\PYG{o}{:}\PYG{n}{j∈su2inds}
\PYG{+w}{    }\PYG{n}{λ}\PYG{o}{=}\PYG{l+s+sa}{T}\PYG{l+s}{\PYGZdq{}}\PYG{l+s}{λ\PYGZus{}i}\PYG{l+s}{\PYGZdq{}}
\PYG{+w}{    }\PYG{n+nd}{@show}\PYG{+w}{ }\PYG{n}{λ}\PYG{o}{\PYGZsq{}}
\PYG{+w}{    }\PYG{c}{\PYGZsh{}   λ\PYGZsq{} = ̄λ\PYGZca{}\PYGZob{}i\PYGZcb{}}
\PYG{+w}{    }\PYG{n}{μ}\PYG{o}{=}\PYG{l+s+sa}{T}\PYG{l+s}{\PYGZdq{}}\PYG{l+s}{μ\PYGZca{}i}\PYG{l+s}{\PYGZdq{}}
\PYG{+w}{    }\PYG{n+nd}{@show}\PYG{+w}{ }\PYG{n}{μ}\PYG{o}{\PYGZsq{}}
\PYG{+w}{    }\PYG{c}{\PYGZsh{}   μ\PYGZsq{} = \PYGZhy{}̄μ\PYGZus{}\PYGZob{}i\PYGZcb{}}
\end{Verbatim}
For regular spacetime fields, we can expect different complex conjugation behaviour depending on whether the field is real, imaginary, or complex.
\begin{Verbatim}[commandchars=\\\{\},xleftmargin=\parindent,numbers=left,bgcolor=bg]
\PYG{+w}{    }\PYG{n}{spacetime}\PYG{+w}{ }\PYG{o}{=}\PYG{+w}{ }\PYG{n}{IndexSet}\PYG{p}{(}\PYG{l+s}{\PYGZdq{}}\PYG{l+s}{spacetime}\PYG{l+s}{\PYGZdq{}}\PYG{p}{,}\PYG{+w}{ }\PYG{l+m+mi}{4}\PYG{p}{)}
\PYG{+w}{    }\PYG{n+nd}{@indices}\PYG{+w}{ }\PYG{n}{μ}\PYG{o}{:}\PYG{n}{ν∈spacetime}
\PYG{+w}{    }\PYG{n}{A}\PYG{o}{=}\PYG{n}{Tensor}\PYG{p}{(}\PYG{l+s}{\PYGZdq{}}\PYG{l+s}{A}\PYG{l+s}{\PYGZdq{}}\PYG{p}{,}\PYG{+w}{ }\PYG{n}{μ}\PYG{o}{=\PYGZgt{}}\PYG{n}{lower}\PYG{p}{,}\PYG{+w}{ }\PYG{n}{ℝ}\PYG{p}{)}\PYG{p}{;}
\PYG{+w}{    }\PYG{n+nd}{@show}\PYG{+w}{ }\PYG{n}{A}\PYG{o}{\PYGZsq{}}
\PYG{+w}{    }\PYG{c}{\PYGZsh{}   A\PYGZsq{} = A\PYGZus{}\PYGZob{}μ\PYGZcb{}}
\PYG{+w}{    }\PYG{n}{B}\PYG{o}{=}\PYG{n}{Tensor}\PYG{p}{(}\PYG{l+s}{\PYGZdq{}}\PYG{l+s}{B}\PYG{l+s}{\PYGZdq{}}\PYG{p}{,}\PYG{+w}{ }\PYG{n}{μ}\PYG{o}{=\PYGZgt{}}\PYG{n}{lower}\PYG{p}{,}\PYG{+w}{ }\PYG{n}{iℝ}\PYG{p}{)}\PYG{p}{;}
\PYG{+w}{    }\PYG{n+nd}{@show}\PYG{+w}{ }\PYG{n}{B}\PYG{o}{\PYGZsq{}}
\PYG{+w}{    }\PYG{c}{\PYGZsh{}   B\PYGZsq{} = \PYGZhy{}B\PYGZus{}\PYGZob{}μ\PYGZcb{}}
\PYG{+w}{    }\PYG{n}{c}\PYG{o}{=}\PYG{n}{Tensor}\PYG{p}{(}\PYG{l+s}{\PYGZdq{}}\PYG{l+s}{c}\PYG{l+s}{\PYGZdq{}}\PYG{p}{,}\PYG{+w}{ }\PYG{n}{μ}\PYG{o}{=\PYGZgt{}}\PYG{n}{lower}\PYG{p}{,}\PYG{+w}{ }\PYG{n}{ℂ}\PYG{p}{)}\PYG{p}{;}
\PYG{+w}{    }\PYG{n+nd}{@show}\PYG{+w}{ }\PYG{n}{c}\PYG{o}{\PYGZsq{}}
\PYG{+w}{    }\PYG{c}{\PYGZsh{}   c\PYGZsq{} = ̄c\PYGZus{}\PYGZob{}μ\PYGZcb{}}
\end{Verbatim}
If no number field is specified, then the complex conjugation behaviour is inferred from the attached indices.

In $2+1D$, spinors admit the Majorana condition. The appropriate properties are implied by specifying that the indices transform under \jil{Spin(3)} and are self-conjugate.
\begin{Verbatim}[commandchars=\\\{\},xleftmargin=\parindent,numbers=left,bgcolor=bg]
\PYG{+w}{    }\PYG{n}{majorana\PYGZus{}indices}\PYG{+w}{ }\PYG{o}{=}\PYG{+w}{ }\PYG{n}{IndexSet}\PYG{p}{(}\PYG{l+s}{\PYGZdq{}}\PYG{l+s}{majorana}\PYG{l+s}{\PYGZdq{}}\PYG{p}{,}\PYG{n}{Spin}\PYG{p}{(}\PYG{l+m+mi}{3}\PYG{p}{)}\PYG{p}{)}
\PYG{+w}{    }\PYG{n+nd}{@indices}\PYG{+w}{ }\PYG{n}{α}\PYG{o}{:}\PYG{n}{β∈majorana\PYGZus{}indices}
\PYG{+w}{    }\PYG{n+nd}{@conjugates}\PYG{+w}{ }\PYG{n}{majorana\PYGZus{}indices}
\PYG{+w}{    }\PYG{n}{ψ}\PYG{o}{=}\PYG{l+s+sa}{T}\PYG{l+s}{\PYGZdq{}}\PYG{l+s}{ψ\PYGZca{}α}\PYG{l+s}{\PYGZdq{}}
\PYG{+w}{    }\PYG{n+nd}{@show}\PYG{+w}{ }\PYG{n}{ψ}\PYG{o}{\PYGZsq{}}
\PYG{+w}{        }\PYG{c}{\PYGZsh{}   ψ\PYGZsq{} = ψ\PYGZca{}\PYGZob{}α\PYGZcb{}}
\PYG{+w}{    }\PYG{n}{χ}\PYG{o}{=}\PYG{l+s+sa}{T}\PYG{l+s}{\PYGZdq{}}\PYG{l+s}{χ\PYGZus{}β}\PYG{l+s}{\PYGZdq{}}
\PYG{+w}{    }\PYG{n+nd}{@show}\PYG{+w}{ }\PYG{n}{χ}\PYG{o}{\PYGZsq{}}
\PYG{+w}{        }\PYG{c}{\PYGZsh{}   χ\PYGZus{}\PYGZob{}β\PYGZcb{}}
\end{Verbatim}
However, following the convention of Gates~\cite{gates2001superspace}, the antisymmetric symbol is imaginary. This implies that a lower-index Majorana spinor is imaginary. This convention can be enforced by using a custom antisymmetric-metric structure
\begin{Verbatim}[commandchars=\\\{\},xleftmargin=\parindent,numbers=left,bgcolor=bg]
\PYG{k}{struct} \PYG{k+kt}{GatesC}\PYG{+w}{ }\PYG{o}{\PYGZlt{}:}\PYG{+w}{ }\PYG{k+kt}{AntisymMetricSuperType}
\PYG{+w}{    }\PYG{n}{data}\PYG{o}{::}\PYG{k+kt}{TensorData}
\PYG{+w}{    }\PYG{n}{indices}\PYG{o}{::}\PYG{k+kt}{Vector}\PYG{p}{\PYGZob{}}\PYG{k+kt}{IndexPair}\PYG{p}{\PYGZcb{}}
\PYG{k}{end}
\PYG{n}{Alakazam}\PYG{o}{.}\PYG{n}{conj\PYGZus{}convention}\PYG{p}{(}\PYG{o}{::}\PYG{k+kt}{Type}\PYG{p}{\PYGZob{}}\PYG{k+kt}{GatesC}\PYG{p}{\PYGZcb{}}\PYG{p}{,}\PYG{+w}{ }\PYG{n}{S}\PYG{o}{::}\PYG{k+kt}{IndexSet}\PYG{p}{)}\PYG{+w}{ }\PYG{o}{=}\PYG{+w}{ }\PYG{p}{(}\PYG{n}{upper}\PYG{p}{,}\PYG{+w}{ }\PYG{o}{\PYGZhy{}}\PYG{l+m+mi}{1}\PYG{p}{)}
\end{Verbatim}
which overrides the complex-conjugation convention, in which case giving
\begin{Verbatim}[commandchars=\\\{\},xleftmargin=\parindent,numbers=left,bgcolor=bg]
\PYG{+w}{    }\PYG{n}{gates\PYGZus{}majorana\PYGZus{}indices}\PYG{+w}{ }\PYG{o}{=}\PYG{+w}{ }\PYG{n}{IndexSet}\PYG{p}{(}\PYG{l+s}{\PYGZdq{}}\PYG{l+s}{gates\PYGZus{}majorana}\PYG{l+s}{\PYGZdq{}}\PYG{p}{,}\PYG{+w}{ }\PYG{n}{GatesC}\PYG{p}{,}\PYG{+w}{ }\PYG{n}{Spin}\PYG{p}{(}\PYG{l+m+mi}{3}\PYG{p}{)}\PYG{p}{)}
\PYG{+w}{    }\PYG{n+nd}{@conjugates}\PYG{+w}{ }\PYG{n}{gates\PYGZus{}majorana\PYGZus{}indices}
\PYG{+w}{    }\PYG{n+nd}{@indices}\PYG{+w}{ }\PYG{n}{γ}\PYG{o}{:}\PYG{n}{δ∈gates\PYGZus{}majorana\PYGZus{}indices}
\PYG{+w}{    }\PYG{n}{ψ}\PYG{o}{=}\PYG{l+s+sa}{T}\PYG{l+s}{\PYGZdq{}}\PYG{l+s}{ψ\PYGZca{}γ}\PYG{l+s}{\PYGZdq{}}
\PYG{+w}{    }\PYG{n+nd}{@show}\PYG{+w}{ }\PYG{n}{ψ}\PYG{o}{\PYGZsq{}}
\PYG{+w}{    }\PYG{c}{\PYGZsh{}   ψ\PYGZsq{} = ψ\PYGZca{}\PYGZob{}γ\PYGZcb{}}
\PYG{+w}{    }\PYG{n}{χ}\PYG{o}{=}\PYG{l+s+sa}{T}\PYG{l+s}{\PYGZdq{}}\PYG{l+s}{χ\PYGZus{}δ}\PYG{l+s}{\PYGZdq{}}
\PYG{+w}{    }\PYG{n+nd}{@show}\PYG{+w}{ }\PYG{n}{χ}\PYG{o}{\PYGZsq{}}
\PYG{+w}{    }\PYG{c}{\PYGZsh{}   \PYGZhy{}χ\PYGZus{}\PYGZob{}δ\PYGZcb{}}
\end{Verbatim}
Specifying \jil{Spin(3)} here is a convenience rather than a necessity, as it just implies the metric, fermionic grading, and index range (which is inferred from the minimal spinor). This can be useful in cases that are non-trivial. Consider for example $6D$. Here, Majorana–Weyl spinors are symplectic and behave quaternionically under conjugation
\begin{Verbatim}[commandchars=\\\{\},xleftmargin=\parindent,numbers=left,bgcolor=bg]
\PYG{+w}{    }\PYG{n}{sympmaj}\PYG{+w}{ }\PYG{o}{=}\PYG{+w}{ }\PYG{n}{IndexSet}\PYG{p}{(}\PYG{l+s}{\PYGZdq{}}\PYG{l+s}{sympmaj}\PYG{l+s}{\PYGZdq{}}\PYG{p}{,}\PYG{n}{lower}\PYG{p}{,}\PYG{+w}{ }\PYG{n}{Spin}\PYG{p}{(}\PYG{l+m+mi}{6}\PYG{p}{)}\PYG{p}{)}
\PYG{+w}{    }\PYG{n+nd}{@conjugates}\PYG{+w}{ }\PYG{n}{sympmaj}
\PYG{+w}{    }\PYG{n+nd}{@indices}\PYG{+w}{ }\PYG{n}{m}\PYG{o}{:}\PYG{n}{n∈sympmaj}
\PYG{+w}{    }\PYG{n}{λ}\PYG{o}{=}\PYG{l+s+sa}{T}\PYG{l+s}{\PYGZdq{}}\PYG{l+s}{λ\PYGZus{}m}\PYG{l+s}{\PYGZdq{}}
\PYG{+w}{    }\PYG{n+nd}{@show}\PYG{+w}{ }\PYG{n}{λ}\PYG{o}{\PYGZsq{}}
\PYG{+w}{    }\PYG{c}{\PYGZsh{}   λ\PYGZsq{} = ̄λ\PYGZca{}\PYGZob{}m\PYGZcb{}}
\PYG{+w}{    }\PYG{n}{μ}\PYG{o}{=}\PYG{l+s+sa}{T}\PYG{l+s}{\PYGZdq{}}\PYG{l+s}{μ\PYGZca{}m}\PYG{l+s}{\PYGZdq{}}
\PYG{+w}{    }\PYG{n+nd}{@show}\PYG{+w}{ }\PYG{n}{μ}\PYG{o}{\PYGZsq{}}
\PYG{+w}{    }\PYG{c}{\PYGZsh{}   μ\PYGZsq{} = \PYGZhy{}̄μ\PYGZus{}\PYGZob{}m\PYGZcb{}}
\end{Verbatim}

We note that if the declared parameters defining complex conjugation are contradictory (for example, declaring the tensor as real but with a conjugate index set), an error will be emitted. As an example, in $4D$ there are no Majorana-Weyl spinors,
\begin{Verbatim}[commandchars=\\\{\},xleftmargin=\parindent,numbers=left,bgcolor=bg]
\PYG{+w}{    }\PYG{n}{weyl}\PYG{+w}{ }\PYG{o}{=}\PYG{+w}{ }\PYG{n}{IndexSet}\PYG{p}{(}\PYG{l+s}{\PYGZdq{}}\PYG{l+s}{weyl}\PYG{l+s}{\PYGZdq{}}\PYG{p}{,}\PYG{+w}{ }\PYG{n}{Spin}\PYG{p}{(}\PYG{l+m+mi}{4}\PYG{p}{)}\PYG{p}{)}
\PYG{+w}{    }\PYG{n+nd}{@conjugates}\PYG{+w}{ }\PYG{n}{weyl}
\PYG{+w}{        }\PYG{c}{\PYGZsh{}   ERROR: index set weyl carries an algebra whose defining representation is}
\PYG{+w}{        }\PYG{c}{\PYGZsh{}   complex, so its conjugate is a different representation, but the set names}
\PYG{+w}{        }\PYG{c}{\PYGZsh{}   itself as its own conjugate; declare the partner set and point the two at}
\PYG{+w}{        }\PYG{c}{\PYGZsh{}   each other.}
\end{Verbatim}

\newpage

\section{Pattern Matching and Replacements}\label{sec:pattern-matching-and-replacements}
No computer algebra software would be complete without the ability to perform symbolic substitutions in expressions. Naturally, \jil{Alakazam} supports replacement patterns including wildcard matching. This is probably one of the most sophisticated features of the package.

\subsection{Basic Replacement Objects}\label{sec:basic-replacement-objects}
It is worth distinguishing outright the objects involved in a replacement. These are constructed from \jil{Tensor}s, and \jil{Index} and \jil{DummyPatternIndex} index types. Firstly, it is important to distinguish between the concrete types \jil{Index} and \jil{DummyPatternIndex}. Why do we need these two types? An index is a wildcard exactly when it is a \jil{DummyPatternIndex}: the \jil{?} written
in a pattern string is consumed by the parser, so \jil{?x} becomes a \jil{DummyPatternIndex} named
\jil{x}, and the type is the only thing marking it. If this were not the case, a new \jil{Index} with name $?x$ would need to be created. This increases the number of \jil{Index} objects in the \jil{IndexSet}, and any loop checking indices would have to examine the literal string name for the character ``?'' to distinguish a dummy from a regular \jil{Index} - much slower than detecting on a type due to dispatch used in our methods.

A tensor, by contrast, has a wildcard name if it is written with a leading ``?''. A \jil{Tensor} named ``?A'' for example matches a tensor of any name, and the name of the old term will be written into the \jil{Tensor}, maintaining \jil{Tensor}'s other properties. For a double leading ``??'', the \jil{Tensor} will inherit the tensor type and data from the matching.

\subsection{Basic Usage}\label{sec:pattern-matching-basic-usage}
In Alakazam, we are able to match literal terms in a \jil{TensorTerm}, or tensors with a fixed shape but a wildcard name or index names, via strings. Wildcards in patterns are represented by \jil{?}, to distinguish them from concrete names, followed by a character to distinguish between wildcards. This allows for reasonably sophisticated replacement patterns to be constructed. \jil{replace_pattern(exp, pattern, replace)}
accepts each of its three arguments in whatever form is convenient: a bare tensor, a
\jil{TensorTerm} or a \jil{TensorExpression}; and for the \jil{pattern} and the \jil{replacement}, also a
pattern string. A scalar, symbolic or numeric, is accepted as a replacement too. A tensor name in a pattern string should be one
character, and a wildcard is \jil{?} plus one character, so \jil{"k4^{?w}"} and
\jil{"?abc"} match nothing rather than erroring. This is a robust rule when using the macro to parse a string to a tensor.

Consider a basic expression
\begin{Verbatim}[commandchars=\\\{\},xleftmargin=\parindent,numbers=left,bgcolor=bg]
\PYG{+w}{    }\PYG{n}{T}\PYG{+w}{ }\PYG{o}{=}\PYG{+w}{ }\PYG{n}{Tensor}\PYG{p}{(}\PYG{l+s}{\PYGZdq{}}\PYG{l+s}{T}\PYG{l+s}{\PYGZdq{}}\PYG{p}{,}\PYG{+w}{ }\PYG{p}{[}\PYG{n}{a}\PYG{+w}{ }\PYG{o}{=\PYGZgt{}}\PYG{+w}{ }\PYG{n}{lower}\PYG{p}{,}\PYG{+w}{ }\PYG{n}{b}\PYG{+w}{ }\PYG{o}{=\PYGZgt{}}\PYG{+w}{ }\PYG{n}{lower}\PYG{p}{,}\PYG{+w}{ }\PYG{n}{c}\PYG{+w}{ }\PYG{o}{=\PYGZgt{}}\PYG{+w}{ }\PYG{n}{lower}\PYG{p}{]}\PYG{p}{,}\PYG{+w}{ }\PYG{n}{x}\PYG{p}{)}
\PYG{+w}{    }\PYG{n}{S}\PYG{+w}{ }\PYG{o}{=}\PYG{+w}{ }\PYG{n}{Tensor}\PYG{p}{(}\PYG{l+s}{\PYGZdq{}}\PYG{l+s}{S}\PYG{l+s}{\PYGZdq{}}\PYG{p}{,}\PYG{+w}{ }\PYG{p}{[}\PYG{n}{s}\PYG{+w}{ }\PYG{o}{=\PYGZgt{}}\PYG{+w}{ }\PYG{n}{lower}\PYG{p}{,}\PYG{+w}{ }\PYG{n}{t}\PYG{+w}{ }\PYG{o}{=\PYGZgt{}}\PYG{+w}{ }\PYG{n}{upper}\PYG{p}{,}\PYG{+w}{ }\PYG{n}{u}\PYG{+w}{ }\PYG{o}{=\PYGZgt{}}\PYG{+w}{ }\PYG{n}{lower}\PYG{p}{]}\PYG{p}{,}\PYG{+w}{ }\PYG{n}{x}\PYG{p}{)}
\PYG{+w}{    }\PYG{n}{U}\PYG{+w}{ }\PYG{o}{=}\PYG{+w}{ }\PYG{n}{Tensor}\PYG{p}{(}\PYG{l+s}{\PYGZdq{}}\PYG{l+s}{U}\PYG{l+s}{\PYGZdq{}}\PYG{p}{,}\PYG{+w}{ }\PYG{p}{[}\PYG{n}{d}\PYG{+w}{ }\PYG{o}{=\PYGZgt{}}\PYG{+w}{ }\PYG{n}{lower}\PYG{p}{,}\PYG{+w}{ }\PYG{n}{e}\PYG{+w}{ }\PYG{o}{=\PYGZgt{}}\PYG{+w}{ }\PYG{n}{upper}\PYG{p}{,}\PYG{+w}{ }\PYG{n}{t}\PYG{+w}{ }\PYG{o}{=\PYGZgt{}}\PYG{+w}{ }\PYG{n}{lower}\PYG{p}{]}\PYG{p}{,}\PYG{+w}{ }\PYG{n}{x}\PYG{p}{)}
\PYG{+w}{    }\PYG{n}{term}\PYG{o}{=}\PYG{n}{T}\PYG{o}{*}\PYG{n}{S}\PYG{o}{*}\PYG{n}{U}
\end{Verbatim}
Suppose we would like to replace one of the tensors with a tensor $R$
\begin{Verbatim}[commandchars=\\\{\},xleftmargin=\parindent,numbers=left,bgcolor=bg]
\PYG{+w}{    }\PYG{n}{R}\PYG{+w}{ }\PYG{o}{=}\PYG{+w}{ }\PYG{n}{Tensor}\PYG{p}{(}\PYG{l+s}{\PYGZdq{}}\PYG{l+s}{R}\PYG{l+s}{\PYGZdq{}}\PYG{p}{,}\PYG{+w}{ }\PYG{p}{[}\PYG{n}{i}\PYG{+w}{ }\PYG{o}{=\PYGZgt{}}\PYG{+w}{ }\PYG{n}{lower}\PYG{p}{,}\PYG{+w}{ }\PYG{n}{j}\PYG{+w}{ }\PYG{o}{=\PYGZgt{}}\PYG{+w}{ }\PYG{n}{lower}\PYG{p}{,}\PYG{+w}{ }\PYG{n}{k}\PYG{+w}{ }\PYG{o}{=\PYGZgt{}}\PYG{+w}{ }\PYG{n}{lower}\PYG{p}{]}\PYG{p}{,}\PYG{+w}{ }\PYG{n}{x}\PYG{p}{)}
\end{Verbatim}
Of course, we should only replace a tensor in this expression, if it has index placement consistent  with $R$. In this case we would like to match a tensor with only lower indices. If we would like to find a match, ignoring the names of the tensor and indices, we can use wildcard patterns signified by \jil{?} before a placeholder name. For example,
\begin{Verbatim}[commandchars=\\\{\},xleftmargin=\parindent,numbers=left,bgcolor=bg]
\PYG{+w}{    }\PYG{n+nd}{@show}\PYG{+w}{ }\PYG{n}{replace\PYGZus{}pattern}\PYG{p}{(}\PYG{n}{term}\PYG{p}{,}\PYG{+w}{ }\PYG{l+s}{\PYGZdq{}}\PYG{l+s}{?A\PYGZus{}\PYGZob{}?f?g?h\PYGZcb{}}\PYG{l+s}{\PYGZdq{}}\PYG{p}{,}\PYG{+w}{ }\PYG{n}{R}\PYG{p}{)}\PYG{+w}{ }
\PYG{+w}{     }\PYG{c}{\PYGZsh{}   R\PYGZus{}\PYGZob{}ijk\PYGZcb{}S\PYGZca{}\PYGZob{}∘t∘\PYGZcb{}\PYGZus{}\PYGZob{}s∘u\PYGZcb{}U\PYGZca{}\PYGZob{}∘e∘\PYGZcb{}\PYGZus{}\PYGZob{}d∘t\PYGZcb{}}
\end{Verbatim}
Note that the final result has both the name and indices of the replacement tensor. This is a fairly basic instance of replacing one tensor with another specific tensor, with specific indices. 

In a different scenario, we perhaps would instead like to copy the indices of the original expression onto the replacement tensor. To do this, we should use \jil{DummyPatternIndex} in the replacement expression. For example, consider the Maxwell Lagrangian
\begin{Verbatim}[commandchars=\\\{\},xleftmargin=\parindent,numbers=left,bgcolor=bg]
\PYG{+w}{    }\PYG{n}{f1}\PYG{+w}{ }\PYG{o}{=}\PYG{+w}{ }\PYG{n}{AntisymmetricTensor}\PYG{p}{(}\PYG{l+s}{\PYGZdq{}}\PYG{l+s}{F}\PYG{l+s}{\PYGZdq{}}\PYG{p}{,}\PYG{+w}{ }\PYG{p}{[}\PYG{n}{μ}\PYG{+w}{ }\PYG{o}{=\PYGZgt{}}\PYG{+w}{ }\PYG{n}{lower}\PYG{p}{,}\PYG{n}{ν}\PYG{o}{=\PYGZgt{}}\PYG{n}{lower}\PYG{p}{]}\PYG{p}{,}\PYG{+w}{ }\PYG{n}{x}\PYG{p}{)}
\PYG{+w}{    }\PYG{n}{f2}\PYG{+w}{ }\PYG{o}{=}\PYG{+w}{ }\PYG{n}{AntisymmetricTensor}\PYG{p}{(}\PYG{l+s}{\PYGZdq{}}\PYG{l+s}{F}\PYG{l+s}{\PYGZdq{}}\PYG{p}{,}\PYG{+w}{ }\PYG{p}{[}\PYG{n}{μ}\PYG{+w}{ }\PYG{o}{=\PYGZgt{}}\PYG{+w}{ }\PYG{n}{upper}\PYG{p}{,}\PYG{n}{ν}\PYG{o}{=\PYGZgt{}}\PYG{n}{upper}\PYG{p}{]}\PYG{p}{,}\PYG{+w}{ }\PYG{n}{x}\PYG{p}{)}
\PYG{+w}{    }\PYG{n}{L}\PYG{o}{=}\PYG{o}{\PYGZhy{}}\PYG{l+m+mi}{1}\PYG{o}{//}\PYG{l+m+mi}{4}\PYG{o}{*}\PYG{n}{f1}\PYG{o}{*}\PYG{n}{f2}
\end{Verbatim}
Suppose we would like to express this in terms of the gauge field using replacements. By creating some dummy indices and the terms
\begin{Verbatim}[commandchars=\\\{\},xleftmargin=\parindent,numbers=left,bgcolor=bg]
\PYG{+w}{    }\PYG{n}{a}\PYG{o}{=}\PYG{n}{DummyPatternIndex}\PYG{p}{(}\PYG{l+s}{\PYGZdq{}}\PYG{l+s}{a}\PYG{l+s}{\PYGZdq{}}\PYG{p}{,}\PYG{+w}{ }\PYG{n}{spacetime\PYGZus{}indices}\PYG{p}{)}
\PYG{+w}{    }\PYG{n}{b}\PYG{o}{=}\PYG{n}{DummyPatternIndex}\PYG{p}{(}\PYG{l+s}{\PYGZdq{}}\PYG{l+s}{b}\PYG{l+s}{\PYGZdq{}}\PYG{p}{,}\PYG{+w}{ }\PYG{n}{spacetime\PYGZus{}indices}\PYG{p}{)}
\PYG{+w}{    }\PYG{n}{da}\PYG{o}{=}\PYG{n}{PartialDerivative}\PYG{p}{(}\PYG{l+s}{\PYGZdq{}}\PYG{l+s}{∂}\PYG{l+s}{\PYGZdq{}}\PYG{p}{,}\PYG{n}{Tensor}\PYG{p}{(}\PYG{l+s}{\PYGZdq{}}\PYG{l+s}{A}\PYG{l+s}{\PYGZdq{}}\PYG{p}{,}\PYG{p}{[}\PYG{n}{b}\PYG{o}{=\PYGZgt{}}\PYG{n}{lower}\PYG{p}{]}\PYG{p}{,}\PYG{n}{x}\PYG{p}{)}\PYG{p}{,}\PYG{+w}{ }\PYG{n}{x}\PYG{p}{,}\PYG{n}{a}\PYG{o}{=\PYGZgt{}}\PYG{n}{lower}\PYG{p}{)}
\PYG{+w}{    }\PYG{n}{db}\PYG{o}{=}\PYG{n}{PartialDerivative}\PYG{p}{(}\PYG{l+s}{\PYGZdq{}}\PYG{l+s}{∂}\PYG{l+s}{\PYGZdq{}}\PYG{p}{,}\PYG{n}{Tensor}\PYG{p}{(}\PYG{l+s}{\PYGZdq{}}\PYG{l+s}{A}\PYG{l+s}{\PYGZdq{}}\PYG{p}{,}\PYG{p}{[}\PYG{n}{a}\PYG{o}{=\PYGZgt{}}\PYG{n}{lower}\PYG{p}{]}\PYG{p}{,}\PYG{n}{x}\PYG{p}{)}\PYG{p}{,}\PYG{+w}{ }\PYG{n}{x}\PYG{p}{,}\PYG{n}{b}\PYG{o}{=\PYGZgt{}}\PYG{n}{lower}\PYG{p}{)}
\PYG{+w}{    }\PYG{n}{dau}\PYG{o}{=}\PYG{n}{PartialDerivative}\PYG{p}{(}\PYG{l+s}{\PYGZdq{}}\PYG{l+s}{∂}\PYG{l+s}{\PYGZdq{}}\PYG{p}{,}\PYG{n}{Tensor}\PYG{p}{(}\PYG{l+s}{\PYGZdq{}}\PYG{l+s}{A}\PYG{l+s}{\PYGZdq{}}\PYG{p}{,}\PYG{p}{[}\PYG{n}{b}\PYG{o}{=\PYGZgt{}}\PYG{n}{upper}\PYG{p}{]}\PYG{p}{,}\PYG{n}{x}\PYG{p}{)}\PYG{p}{,}\PYG{+w}{ }\PYG{n}{x}\PYG{p}{,}\PYG{n}{a}\PYG{o}{=\PYGZgt{}}\PYG{n}{upper}\PYG{p}{)}
\PYG{+w}{    }\PYG{n}{dbu}\PYG{o}{=}\PYG{n}{PartialDerivative}\PYG{p}{(}\PYG{l+s}{\PYGZdq{}}\PYG{l+s}{∂}\PYG{l+s}{\PYGZdq{}}\PYG{p}{,}\PYG{n}{Tensor}\PYG{p}{(}\PYG{l+s}{\PYGZdq{}}\PYG{l+s}{A}\PYG{l+s}{\PYGZdq{}}\PYG{p}{,}\PYG{p}{[}\PYG{n}{a}\PYG{o}{=\PYGZgt{}}\PYG{n}{upper}\PYG{p}{]}\PYG{p}{,}\PYG{n}{x}\PYG{p}{)}\PYG{p}{,}\PYG{+w}{ }\PYG{n}{x}\PYG{p}{,}\PYG{n}{b}\PYG{o}{=\PYGZgt{}}\PYG{n}{upper}\PYG{p}{)}
\end{Verbatim}
we can perform the replacement
\begin{Verbatim}[commandchars=\\\{\},xleftmargin=\parindent,numbers=left,bgcolor=bg]
\PYG{+w}{    }\PYG{n}{simplify}\PYG{p}{(}\PYG{n}{replace\PYGZus{}pattern}\PYG{p}{(}\PYG{n}{replace\PYGZus{}pattern}\PYG{p}{(}\PYG{n}{L}\PYG{p}{,}
\PYG{+w}{        }\PYG{l+s}{\PYGZdq{}}\PYG{l+s}{?T\PYGZus{}\PYGZob{}?a?b\PYGZcb{}}\PYG{l+s}{\PYGZdq{}}\PYG{p}{,}\PYG{n}{da}\PYG{o}{\PYGZhy{}}\PYG{n}{db}\PYG{p}{)}\PYG{p}{,}\PYG{+w}{ }\PYG{l+s}{\PYGZdq{}}\PYG{l+s}{?T\PYGZca{}\PYGZob{}?a?b\PYGZcb{}}\PYG{l+s}{\PYGZdq{}}\PYG{p}{,}\PYG{n}{dau}\PYG{o}{\PYGZhy{}}\PYG{n}{dbu}\PYG{p}{)}\PYG{p}{)}
\end{Verbatim}
which yields \jil{-1//2∂^{μ}⟦A^{ν}⟧∂_{μ}⟦A_{ν}⟧+1//2∂^{μ}⟦A^{ν}⟧∂_{ν}⟦A_{μ}⟧}, where each \jil{DummyPatternIndex} has been replaced with corresponding \jil{Index} in the original expression, found by wildcard matching.

A similar manipulation can for example be performed to verify the derived Bianchi identity of the Riemann tensor in normal coordinates.
\begin{Verbatim}[commandchars=\\\{\},xleftmargin=\parindent,numbers=left,bgcolor=bg]
\PYG{+w}{    }\PYG{n}{R1}\PYG{+w}{ }\PYG{o}{=}\PYG{+w}{ }\PYG{n}{RiemannTensor}\PYG{p}{(}\PYG{p}{[}\PYG{n}{μ}\PYG{+w}{ }\PYG{o}{=\PYGZgt{}}\PYG{+w}{ }\PYG{n}{lower}\PYG{p}{,}\PYG{+w}{ }\PYG{n}{ν}\PYG{+w}{ }\PYG{o}{=\PYGZgt{}}\PYG{+w}{ }\PYG{n}{lower}\PYG{p}{,}\PYG{+w}{ }\PYG{n}{σ}\PYG{+w}{ }\PYG{o}{=\PYGZgt{}}\PYG{+w}{ }\PYG{n}{lower}\PYG{p}{,}\PYG{+w}{ }\PYG{n}{τ}\PYG{+w}{ }\PYG{o}{=\PYGZgt{}}\PYG{+w}{ }\PYG{n}{lower}\PYG{p}{]}\PYG{p}{,}\PYG{+w}{ }\PYG{n}{x}\PYG{p}{)}
\PYG{+w}{    }\PYG{n}{D1}\PYG{+w}{ }\PYG{o}{=}\PYG{+w}{ }\PYG{n}{Partial}\PYG{p}{(}\PYG{l+s}{\PYGZdq{}}\PYG{l+s}{∇}\PYG{l+s}{\PYGZdq{}}\PYG{p}{,}\PYG{+w}{ }\PYG{n}{R1}\PYG{p}{,}\PYG{+w}{ }\PYG{n}{x}\PYG{p}{,}\PYG{+w}{ }\PYG{n}{ρ}\PYG{+w}{ }\PYG{o}{=\PYGZgt{}}\PYG{+w}{ }\PYG{n}{lower}\PYG{p}{)}\PYG{+w}{   }
\PYG{+w}{    }\PYG{n}{D2}\PYG{+w}{ }\PYG{o}{=}\PYG{+w}{ }\PYG{n}{Partial}\PYG{p}{(}\PYG{l+s}{\PYGZdq{}}\PYG{l+s}{∇}\PYG{l+s}{\PYGZdq{}}\PYG{p}{,}\PYG{+w}{ }\PYG{n}{RiemannTensor}\PYG{p}{(}\PYG{p}{[}\PYG{n}{ν}\PYG{+w}{ }\PYG{o}{=\PYGZgt{}}\PYG{+w}{ }\PYG{n}{lower}\PYG{p}{,}\PYG{+w}{ }\PYG{n}{ρ}\PYG{+w}{ }\PYG{o}{=\PYGZgt{}}\PYG{+w}{ }\PYG{n}{lower}\PYG{p}{,}
\PYG{+w}{        }\PYG{n}{σ}\PYG{+w}{ }\PYG{o}{=\PYGZgt{}}\PYG{+w}{ }\PYG{n}{lower}\PYG{p}{,}\PYG{+w}{ }\PYG{n}{τ}\PYG{+w}{ }\PYG{o}{=\PYGZgt{}}\PYG{+w}{ }\PYG{n}{lower}\PYG{p}{]}\PYG{p}{,}\PYG{+w}{ }\PYG{n}{x}\PYG{p}{)}\PYG{p}{,}\PYG{+w}{ }\PYG{n}{x}\PYG{p}{,}\PYG{+w}{ }\PYG{n}{μ}\PYG{+w}{ }\PYG{o}{=\PYGZgt{}}\PYG{+w}{ }\PYG{n}{lower}\PYG{p}{)}\PYG{+w}{   }
\PYG{+w}{    }\PYG{n}{D3}\PYG{+w}{ }\PYG{o}{=}\PYG{+w}{ }\PYG{n}{Partial}\PYG{p}{(}\PYG{l+s}{\PYGZdq{}}\PYG{l+s}{∇}\PYG{l+s}{\PYGZdq{}}\PYG{p}{,}\PYG{+w}{ }\PYG{n}{RiemannTensor}\PYG{p}{(}\PYG{p}{[}\PYG{n}{ρ}\PYG{+w}{ }\PYG{o}{=\PYGZgt{}}\PYG{+w}{ }\PYG{n}{lower}\PYG{p}{,}\PYG{+w}{ }\PYG{n}{μ}\PYG{+w}{ }\PYG{o}{=\PYGZgt{}}\PYG{+w}{ }\PYG{n}{lower}\PYG{p}{,}
\PYG{+w}{        }\PYG{n}{σ}\PYG{+w}{ }\PYG{o}{=\PYGZgt{}}\PYG{+w}{ }\PYG{n}{lower}\PYG{p}{,}\PYG{+w}{ }\PYG{n}{τ}\PYG{+w}{ }\PYG{o}{=\PYGZgt{}}\PYG{+w}{ }\PYG{n}{lower}\PYG{p}{]}\PYG{p}{,}\PYG{+w}{ }\PYG{n}{x}\PYG{p}{)}\PYG{p}{,}\PYG{+w}{ }\PYG{n}{x}\PYG{p}{,}\PYG{+w}{ }\PYG{n}{ν}\PYG{+w}{ }\PYG{o}{=\PYGZgt{}}\PYG{+w}{ }\PYG{n}{lower}\PYG{p}{)}
\PYG{+w}{    }\PYG{n}{c}\PYG{o}{=}\PYG{n}{DummyPatternIndex}\PYG{p}{(}\PYG{l+s}{\PYGZdq{}}\PYG{l+s}{c}\PYG{l+s}{\PYGZdq{}}\PYG{p}{,}\PYG{+w}{ }\PYG{n}{spacetime\PYGZus{}indices}\PYG{p}{)}
\PYG{+w}{    }\PYG{n}{d}\PYG{o}{=}\PYG{n}{DummyPatternIndex}\PYG{p}{(}\PYG{l+s}{\PYGZdq{}}\PYG{l+s}{d}\PYG{l+s}{\PYGZdq{}}\PYG{p}{,}\PYG{+w}{ }\PYG{n}{spacetime\PYGZus{}indices}\PYG{p}{)}
\PYG{+w}{    }\PYG{n}{ddg}\PYG{o}{=}\PYG{l+m+mi}{1}\PYG{o}{//}\PYG{l+m+mi}{2}\PYG{o}{*}\PYG{p}{(}\PYG{n}{PD}\PYG{p}{(}\PYG{l+s}{\PYGZdq{}}\PYG{l+s}{∇}\PYG{l+s}{\PYGZdq{}}\PYG{p}{,}\PYG{p}{(}\PYG{n}{PD}\PYG{p}{(}\PYG{l+s}{\PYGZdq{}}\PYG{l+s}{∇}\PYG{l+s}{\PYGZdq{}}\PYG{p}{,}\PYG{n}{SymmetricTensor}\PYG{p}{(}\PYG{l+s}{\PYGZdq{}}\PYG{l+s}{g}\PYG{l+s}{\PYGZdq{}}\PYG{p}{,}
\PYG{+w}{        }\PYG{p}{[}\PYG{n}{a}\PYG{o}{=\PYGZgt{}}\PYG{n}{lower}\PYG{p}{,}\PYG{n}{d}\PYG{o}{=\PYGZgt{}}\PYG{n}{lower}\PYG{p}{]}\PYG{p}{,}\PYG{n}{x}\PYG{p}{)}\PYG{p}{,}\PYG{+w}{ }\PYG{n}{x}\PYG{p}{,}\PYG{n}{c}\PYG{o}{=\PYGZgt{}}\PYG{n}{lower}\PYG{p}{)}\PYG{p}{)}\PYG{p}{,}\PYG{n}{x}\PYG{p}{,}\PYG{n}{b}\PYG{o}{=\PYGZgt{}}\PYG{n}{lower}\PYG{p}{)}
\PYG{+w}{        }\PYG{o}{+}\PYG{n}{PD}\PYG{p}{(}\PYG{l+s}{\PYGZdq{}}\PYG{l+s}{∇}\PYG{l+s}{\PYGZdq{}}\PYG{p}{,}\PYG{p}{(}\PYG{n}{PD}\PYG{p}{(}\PYG{l+s}{\PYGZdq{}}\PYG{l+s}{∇}\PYG{l+s}{\PYGZdq{}}\PYG{p}{,}\PYG{n}{SymmetricTensor}\PYG{p}{(}\PYG{l+s}{\PYGZdq{}}\PYG{l+s}{g}\PYG{l+s}{\PYGZdq{}}\PYG{p}{,}\PYG{p}{[}\PYG{n}{b}\PYG{o}{=\PYGZgt{}}\PYG{n}{lower}\PYG{p}{,}\PYG{n}{c}\PYG{o}{=\PYGZgt{}}\PYG{n}{lower}\PYG{p}{]}\PYG{p}{,}\PYG{n}{x}\PYG{p}{)}\PYG{p}{,}\PYG{+w}{ }\PYG{n}{x}\PYG{p}{,}\PYG{n}{d}\PYG{o}{=\PYGZgt{}}\PYG{n}{lower}\PYG{p}{)}\PYG{p}{)}\PYG{p}{,}\PYG{n}{x}\PYG{p}{,}
\PYG{+w}{            }\PYG{n}{a}\PYG{o}{=\PYGZgt{}}\PYG{n}{lower}\PYG{p}{)}
\PYG{+w}{        }\PYG{o}{\PYGZhy{}}\PYG{n}{PD}\PYG{p}{(}\PYG{l+s}{\PYGZdq{}}\PYG{l+s}{∇}\PYG{l+s}{\PYGZdq{}}\PYG{p}{,}\PYG{p}{(}\PYG{n}{PD}\PYG{p}{(}\PYG{l+s}{\PYGZdq{}}\PYG{l+s}{∇}\PYG{l+s}{\PYGZdq{}}\PYG{p}{,}\PYG{n}{SymmetricTensor}\PYG{p}{(}\PYG{l+s}{\PYGZdq{}}\PYG{l+s}{g}\PYG{l+s}{\PYGZdq{}}\PYG{p}{,}\PYG{p}{[}\PYG{n}{a}\PYG{o}{=\PYGZgt{}}\PYG{n}{lower}\PYG{p}{,}\PYG{n}{c}\PYG{o}{=\PYGZgt{}}\PYG{n}{lower}\PYG{p}{]}\PYG{p}{,}\PYG{n}{x}\PYG{p}{)}\PYG{p}{,}\PYG{+w}{ }\PYG{n}{x}\PYG{p}{,}\PYG{n}{d}\PYG{o}{=\PYGZgt{}}\PYG{n}{lower}\PYG{p}{)}\PYG{p}{)}\PYG{p}{,}
\PYG{+w}{            }\PYG{n}{x}\PYG{p}{,}\PYG{n}{b}\PYG{o}{=\PYGZgt{}}\PYG{n}{lower}\PYG{p}{)}
\PYG{+w}{        }\PYG{o}{\PYGZhy{}}\PYG{n}{PD}\PYG{p}{(}\PYG{l+s}{\PYGZdq{}}\PYG{l+s}{∇}\PYG{l+s}{\PYGZdq{}}\PYG{p}{,}\PYG{p}{(}\PYG{n}{PD}\PYG{p}{(}\PYG{l+s}{\PYGZdq{}}\PYG{l+s}{∇}\PYG{l+s}{\PYGZdq{}}\PYG{p}{,}\PYG{n}{SymmetricTensor}\PYG{p}{(}\PYG{l+s}{\PYGZdq{}}\PYG{l+s}{g}\PYG{l+s}{\PYGZdq{}}\PYG{p}{,}\PYG{p}{[}\PYG{n}{b}\PYG{o}{=\PYGZgt{}}\PYG{n}{lower}\PYG{p}{,}\PYG{n}{d}\PYG{o}{=\PYGZgt{}}\PYG{n}{lower}\PYG{p}{]}\PYG{p}{,}\PYG{n}{x}\PYG{p}{)}\PYG{p}{,}\PYG{+w}{ }\PYG{n}{x}\PYG{p}{,}
\PYG{+w}{            }\PYG{n}{c}\PYG{o}{=\PYGZgt{}}\PYG{n}{lower}\PYG{p}{)}\PYG{p}{)}\PYG{p}{,}\PYG{n}{x}\PYG{p}{,}\PYG{n}{a}\PYG{o}{=\PYGZgt{}}\PYG{n}{lower}\PYG{p}{)}\PYG{p}{)}
\PYG{+w}{    }\PYG{c}{\PYGZsh{} @show ddg}
\PYG{+w}{    }\PYG{n}{bia}\PYG{o}{=}\PYG{+w}{ }\PYG{n}{replace\PYGZus{}pattern}\PYG{p}{(}\PYG{n}{D1}\PYG{o}{+}\PYG{n}{D2}\PYG{o}{+}\PYG{n}{D3}\PYG{p}{,}\PYG{+w}{ }\PYG{l+s}{\PYGZdq{}}\PYG{l+s}{R\PYGZus{}\PYGZob{}?a?b?c?d\PYGZcb{}}\PYG{l+s}{\PYGZdq{}}\PYG{p}{,}\PYG{n}{ddg}\PYG{p}{)}
\PYG{+w}{    }\PYG{n}{hard\PYGZus{}simplify}\PYG{p}{(}\PYG{n}{apply\PYGZus{}derivative}\PYG{p}{(}\PYG{n}{bia}\PYG{p}{)}\PYG{p}{)}
\end{Verbatim}
which will evaluate to \jil{zero}.

\subsection{Multi Tensor Patterns}\label{sec:multi-tensor-patterns}
The previous examples have replaced a single tensor at a time.
A pattern string parses into a \jil{TensorTerm}, not a single tensor, so it may name several tensors
at once. This is the most capable part of \jil{replace_pattern()}, and forms the basis for more sophisticated manipulations of expressions. As an example, we can replace a pair of contracted terms with a single scalar
\begin{Verbatim}[commandchars=\\\{\},xleftmargin=\parindent,numbers=left,bgcolor=bg]
\PYG{+w}{    }\PYG{n}{A}\PYG{p}{(}\PYG{n}{q}\PYG{p}{)}\PYG{+w}{ }\PYG{o}{=}\PYG{+w}{ }\PYG{n}{Tensor}\PYG{p}{(}\PYG{l+s}{\PYGZdq{}}\PYG{l+s}{A}\PYG{l+s}{\PYGZdq{}}\PYG{p}{,}\PYG{+w}{ }\PYG{p}{[}\PYG{n}{μ}\PYG{+w}{ }\PYG{o}{=\PYGZgt{}}\PYG{+w}{ }\PYG{n}{q}\PYG{p}{]}\PYG{p}{)}\PYG{p}{;}\PYG{+w}{ }\PYG{n}{B}\PYG{p}{(}\PYG{n}{q}\PYG{p}{)}\PYG{+w}{ }\PYG{o}{=}\PYG{+w}{ }\PYG{n}{Tensor}\PYG{p}{(}\PYG{l+s}{\PYGZdq{}}\PYG{l+s}{B}\PYG{l+s}{\PYGZdq{}}\PYG{p}{,}\PYG{+w}{ }\PYG{p}{[}\PYG{n}{μ}\PYG{+w}{ }\PYG{o}{=\PYGZgt{}}\PYG{+w}{ }\PYG{n}{q}\PYG{p}{]}\PYG{p}{)}
\PYG{+w}{    }\PYG{n}{C}\PYG{+w}{ }\PYG{o}{=}\PYG{+w}{ }\PYG{n}{Tensor}\PYG{p}{(}\PYG{l+s}{\PYGZdq{}}\PYG{l+s}{C}\PYG{l+s}{\PYGZdq{}}\PYG{p}{,}\PYG{+w}{ }\PYG{p}{[}\PYG{n}{ν}\PYG{+w}{ }\PYG{o}{=\PYGZgt{}}\PYG{+w}{ }\PYG{n}{upper}\PYG{p}{]}\PYG{p}{)}\PYG{p}{;}\PYG{+w}{ }\PYG{n}{K}\PYG{p}{(}\PYG{n}{q}\PYG{p}{)}\PYG{+w}{ }\PYG{o}{=}\PYG{+w}{ }\PYG{n}{Tensor}\PYG{p}{(}\PYG{l+s}{\PYGZdq{}}\PYG{l+s}{K}\PYG{l+s}{\PYGZdq{}}\PYG{p}{,}\PYG{+w}{ }\PYG{p}{[}\PYG{n}{μ}\PYG{+w}{ }\PYG{o}{=\PYGZgt{}}\PYG{+w}{ }\PYG{n}{q}\PYG{p}{]}\PYG{p}{)}\PYG{+w}{ }
\PYG{+w}{    }\PYG{n}{Z}\PYG{+w}{ }\PYG{o}{=}\PYG{+w}{ }\PYG{n}{Tensor}\PYG{p}{(}\PYG{l+s}{\PYGZdq{}}\PYG{l+s}{Z}\PYG{l+s}{\PYGZdq{}}\PYG{p}{)}

\PYG{+w}{    }\PYG{n+nd}{@show}\PYG{+w}{ }\PYG{n}{replace\PYGZus{}pattern}\PYG{p}{(}\PYG{n}{A}\PYG{p}{(}\PYG{n}{lower}\PYG{p}{)}\PYG{o}{*}\PYG{n}{B}\PYG{p}{(}\PYG{n}{upper}\PYG{p}{)}\PYG{p}{,}\PYG{+w}{ }\PYG{l+s}{\PYGZdq{}}\PYG{l+s}{A\PYGZus{}\PYGZob{}?x\PYGZcb{}B\PYGZca{}\PYGZob{}?x\PYGZcb{}}\PYG{l+s}{\PYGZdq{}}\PYG{p}{,}\PYG{+w}{ }\PYG{n}{Z}\PYG{p}{)}
\PYG{+w}{    }\PYG{c}{\PYGZsh{}   Z}
\PYG{+w}{    }\PYG{n+nd}{@show}\PYG{+w}{ }\PYG{n}{replace\PYGZus{}pattern}\PYG{p}{(}\PYG{n}{B}\PYG{p}{(}\PYG{n}{upper}\PYG{p}{)}\PYG{o}{*}\PYG{n}{A}\PYG{p}{(}\PYG{n}{lower}\PYG{p}{)}\PYG{p}{,}\PYG{+w}{ }\PYG{l+s}{\PYGZdq{}}\PYG{l+s}{A\PYGZus{}\PYGZob{}?x\PYGZcb{}B\PYGZca{}\PYGZob{}?x\PYGZcb{}}\PYG{l+s}{\PYGZdq{}}\PYG{p}{,}\PYG{+w}{ }\PYG{n}{Z}\PYG{p}{)}
\PYG{+w}{    }\PYG{c}{\PYGZsh{}   Z}
\PYG{+w}{    }\PYG{n+nd}{@show}\PYG{+w}{ }\PYG{n}{replace\PYGZus{}pattern}\PYG{p}{(}\PYG{n}{A}\PYG{p}{(}\PYG{n}{lower}\PYG{p}{)}\PYG{o}{*}\PYG{n}{B}\PYG{p}{(}\PYG{n}{upper}\PYG{p}{)}\PYG{o}{*}\PYG{n}{C}\PYG{p}{,}\PYG{+w}{ }\PYG{l+s}{\PYGZdq{}}\PYG{l+s}{A\PYGZus{}\PYGZob{}?x\PYGZcb{}B\PYGZca{}\PYGZob{}?x\PYGZcb{}}\PYG{l+s}{\PYGZdq{}}\PYG{p}{,}\PYG{+w}{ }\PYG{n}{Z}\PYG{p}{)}
\PYG{+w}{    }\PYG{c}{\PYGZsh{}   ZC\PYGZca{}\PYGZob{}ν\PYGZcb{}}
\end{Verbatim}
Matching is order-independent, and the matched factors need not be adjacent: the second line finds the pattern with its factors reversed, and the third leaves the unmatched $C$ in place. This respects grading and noncommutativity, generating new terms if swaps are needed to bring terms adjacent before replacement. Tensor names
may themselves be wildcards within a multi-tensor pattern, so in this case \jil{"?P_{?x}?Q^{?x}"} matches any contracted pair. An on-shell condition, for example, could be expressed by sending a self-contraction to zero,
\begin{Verbatim}[commandchars=\\\{\},xleftmargin=\parindent,numbers=left,bgcolor=bg]
\PYG{+w}{    }\PYG{n+nd}{@show}\PYG{+w}{ }\PYG{n}{replace\PYGZus{}pattern}\PYG{p}{(}\PYG{n}{K}\PYG{p}{(}\PYG{n}{lower}\PYG{p}{)}\PYG{o}{*}\PYG{n}{K}\PYG{p}{(}\PYG{n}{upper}\PYG{p}{)}\PYG{+w}{ }\PYG{o}{+}\PYG{+w}{ }\PYG{n}{A}\PYG{p}{(}\PYG{n}{lower}\PYG{p}{)}\PYG{o}{*}\PYG{n}{A}\PYG{p}{(}\PYG{n}{upper}\PYG{p}{)}\PYG{p}{,}\PYG{+w}{ }\PYG{l+s}{\PYGZdq{}}\PYG{l+s}{K\PYGZus{}\PYGZob{}?x\PYGZcb{}K\PYGZca{}\PYGZob{}?x\PYGZcb{}}\PYG{l+s}{\PYGZdq{}}\PYG{p}{,}\PYG{+w}{ }\PYG{n}{zero}\PYG{p}{(}\PYG{p}{)}\PYG{p}{)}
\PYG{+w}{    }\PYG{c}{\PYGZsh{}   A\PYGZus{}\PYGZob{}μ\PYGZcb{}A\PYGZca{}\PYGZob{}μ\PYGZcb{}}
\end{Verbatim}
which is applied to every summand of the expression independently.

Note that a wildcard index binds consistently across the whole pattern. A repeated \jil{?x} demands a contraction and will not match two unrelated indices
\begin{Verbatim}[commandchars=\\\{\},xleftmargin=\parindent,numbers=left,bgcolor=bg]
\PYG{+w}{    }\PYG{n}{W}\PYG{p}{(}\PYG{n}{i}\PYG{p}{,}\PYG{n}{q}\PYG{p}{)}\PYG{+w}{ }\PYG{o}{=}\PYG{+w}{ }\PYG{n}{Tensor}\PYG{p}{(}\PYG{l+s}{\PYGZdq{}}\PYG{l+s}{W}\PYG{l+s}{\PYGZdq{}}\PYG{p}{,}\PYG{+w}{ }\PYG{p}{[}\PYG{n}{i}\PYG{+w}{ }\PYG{o}{=\PYGZgt{}}\PYG{+w}{ }\PYG{n}{q}\PYG{p}{]}\PYG{p}{)}
\PYG{+w}{    }\PYG{n+nd}{@show}\PYG{+w}{ }\PYG{n}{replace\PYGZus{}pattern}\PYG{p}{(}\PYG{n}{W}\PYG{p}{(}\PYG{n}{μ}\PYG{p}{,}\PYG{n}{lower}\PYG{p}{)}\PYG{o}{*}\PYG{n}{W}\PYG{p}{(}\PYG{n}{μ}\PYG{p}{,}\PYG{n}{upper}\PYG{p}{)}\PYG{p}{,}\PYG{+w}{ }\PYG{l+s}{\PYGZdq{}}\PYG{l+s}{W\PYGZus{}\PYGZob{}?x\PYGZcb{}W\PYGZca{}\PYGZob{}?x\PYGZcb{}}\PYG{l+s}{\PYGZdq{}}\PYG{p}{,}\PYG{+w}{ }\PYG{n}{Z}\PYG{p}{)}\PYG{+w}{   }
\PYG{+w}{        }\PYG{c}{\PYGZsh{} Z}
\PYG{+w}{    }\PYG{n+nd}{@show}\PYG{+w}{ }\PYG{n}{replace\PYGZus{}pattern}\PYG{p}{(}\PYG{n}{W}\PYG{p}{(}\PYG{n}{μ}\PYG{p}{,}\PYG{n}{lower}\PYG{p}{)}\PYG{o}{*}\PYG{n}{W}\PYG{p}{(}\PYG{n}{ν}\PYG{p}{,}\PYG{n}{upper}\PYG{p}{)}\PYG{p}{,}\PYG{+w}{ }\PYG{l+s}{\PYGZdq{}}\PYG{l+s}{W\PYGZus{}\PYGZob{}?x\PYGZcb{}W\PYGZca{}\PYGZob{}?x\PYGZcb{}}\PYG{l+s}{\PYGZdq{}}\PYG{p}{,}\PYG{+w}{ }\PYG{n}{Z}\PYG{p}{)}\PYG{+w}{   }
\PYG{+w}{        }\PYG{c}{\PYGZsh{} W\PYGZus{}\PYGZob{}μ\PYGZcb{}W\PYGZca{}\PYGZob{}ν\PYGZcb{} \PYGZhy{} no match, since ?x binds to μ}
\PYG{+w}{    }\PYG{n+nd}{@show}\PYG{+w}{ }\PYG{n}{replace\PYGZus{}pattern}\PYG{p}{(}\PYG{n}{W}\PYG{p}{(}\PYG{n}{μ}\PYG{p}{,}\PYG{n}{lower}\PYG{p}{)}\PYG{o}{*}\PYG{n}{W}\PYG{p}{(}\PYG{n}{ν}\PYG{p}{,}\PYG{n}{upper}\PYG{p}{)}\PYG{p}{,}\PYG{+w}{ }\PYG{l+s}{\PYGZdq{}}\PYG{l+s}{W\PYGZus{}\PYGZob{}?x\PYGZcb{}W\PYGZca{}\PYGZob{}?y\PYGZcb{}}\PYG{l+s}{\PYGZdq{}}\PYG{p}{,}\PYG{+w}{ }\PYG{n}{Z}\PYG{p}{)}\PYG{+w}{   }
\PYG{+w}{        }\PYG{c}{\PYGZsh{} Z}
\end{Verbatim}
This is what makes patterns such as \jil{"k_{?x}k^{?x}"} safe to apply to an expression in which
several distinct momenta appear. Note also that a pattern is sensitive to index position: \jil{"q_{?x}q^{?x}"} matches
$q_{a}q^{a}$ and not $q^{a}q_{a}$. This is intentional, to be compatible with spinors.

\subsection{Labels in patterns}
\label{sec:pat-labels}
Labels participate in matching, and do so as wildcards by default: a pattern that includes no label
matches a tensor carrying any label for convenience, while a pattern that names one must match it exactly. For example, if we construct $k_1$ and $k_2$, where $1$ and $2$ are automatically parsed as labels by the macro,
\begin{Verbatim}[commandchars=\\\{\},xleftmargin=\parindent,numbers=left,bgcolor=bg]
\PYG{+w}{    }\PYG{n}{k1}\PYG{+w}{ }\PYG{o}{=}\PYG{+w}{ }\PYG{l+s+sa}{T}\PYG{l+s}{\PYGZdq{}}\PYG{l+s}{k\PYGZus{}1\PYGZca{}μ}\PYG{l+s}{\PYGZdq{}}\PYG{+w}{ }\PYG{o}{*}\PYG{+w}{ }\PYG{n}{Tensor}\PYG{p}{(}\PYG{l+s}{\PYGZdq{}}\PYG{l+s}{m}\PYG{l+s}{\PYGZdq{}}\PYG{p}{,}\PYG{+w}{ }\PYG{p}{[}\PYG{n}{μ}\PYG{+w}{ }\PYG{o}{=\PYGZgt{}}\PYG{+w}{ }\PYG{n}{lower}\PYG{p}{]}\PYG{p}{)}
\PYG{+w}{    }\PYG{n}{k2}\PYG{+w}{ }\PYG{o}{=}\PYG{+w}{ }\PYG{l+s+sa}{T}\PYG{l+s}{\PYGZdq{}}\PYG{l+s}{k\PYGZus{}2\PYGZca{}μ}\PYG{l+s}{\PYGZdq{}}\PYG{+w}{ }\PYG{o}{*}\PYG{+w}{ }\PYG{n}{Tensor}\PYG{p}{(}\PYG{l+s}{\PYGZdq{}}\PYG{l+s}{m}\PYG{l+s}{\PYGZdq{}}\PYG{p}{,}\PYG{+w}{ }\PYG{p}{[}\PYG{n}{μ}\PYG{+w}{ }\PYG{o}{=\PYGZgt{}}\PYG{+w}{ }\PYG{n}{lower}\PYG{p}{]}\PYG{p}{)}

\PYG{+w}{    }\PYG{n}{replace\PYGZus{}pattern}\PYG{p}{(}\PYG{n}{k1}\PYG{p}{,}\PYG{+w}{ }\PYG{n}{TensorTerm}\PYG{p}{(}\PYG{l+s+sa}{T}\PYG{l+s}{\PYGZdq{}}\PYG{l+s}{k\PYGZus{}1\PYGZca{}?x}\PYG{l+s}{\PYGZdq{}}\PYG{p}{)}\PYG{p}{,}\PYG{+w}{ }\PYG{n}{Z}\PYG{p}{)}\PYG{+w}{   }\PYG{c}{\PYGZsh{}  Zm\PYGZus{}\PYGZob{}μ\PYGZcb{}}
\PYG{+w}{    }\PYG{n}{replace\PYGZus{}pattern}\PYG{p}{(}\PYG{n}{k2}\PYG{p}{,}\PYG{+w}{ }\PYG{n}{TensorTerm}\PYG{p}{(}\PYG{l+s+sa}{T}\PYG{l+s}{\PYGZdq{}}\PYG{l+s}{k\PYGZus{}1\PYGZca{}?x}\PYG{l+s}{\PYGZdq{}}\PYG{p}{)}\PYG{p}{,}\PYG{+w}{ }\PYG{n}{Z}\PYG{p}{)}\PYG{+w}{   }\PYG{c}{\PYGZsh{}  k\PYGZca{}\PYGZob{}μ\PYGZcb{}\PYGZus{}\PYGZob{}2\PYGZcb{}m\PYGZus{}\PYGZob{}μ\PYGZcb{}, no label match}
\PYG{+w}{    }\PYG{n}{replace\PYGZus{}pattern}\PYG{p}{(}\PYG{n}{k2}\PYG{p}{,}\PYG{+w}{ }\PYG{n}{TensorTerm}\PYG{p}{(}\PYG{l+s+sa}{T}\PYG{l+s}{\PYGZdq{}}\PYG{l+s}{k\PYGZca{}?x}\PYG{l+s}{\PYGZdq{}}\PYG{p}{)}\PYG{p}{,}\PYG{+w}{   }\PYG{n}{Z}\PYG{p}{)}\PYG{+w}{   }\PYG{c}{\PYGZsh{}  Zm\PYGZus{}\PYGZob{}μ\PYGZcb{}, label wildcard}
\end{Verbatim}
Since names in pattern strings are single characters, labels are also how one distinguishes several
objects of the same kind: $k_1$ and $k_2$ are the tensor \jil{k} bearing different labels, and are
matched apart.

\subsection{Patterns as Objects}\label{sec:patterns-as-objects}
Patterns need not be written as strings. The \jil{T"..."} macro can parse strings to tensors, including those with wildcards on the name and indices. Thus, a pattern can be built from a tensor created via the ordinary macro syntax and
inherits labels, index positions, and coordinate dependence from it.
\begin{Verbatim}[commandchars=\\\{\},xleftmargin=\parindent,numbers=left,bgcolor=bg]
\PYG{+w}{    }\PYG{n+nd}{@show}\PYG{+w}{ }\PYG{l+s+sa}{T}\PYG{l+s}{\PYGZdq{}}\PYG{l+s}{?A\PYGZus{}?x?y}\PYG{l+s}{\PYGZdq{}}\PYG{+w}{               }\PYG{c}{\PYGZsh{} ?A\PYGZus{}\PYGZob{}?x?y\PYGZcb{}}
\PYG{+w}{    }\PYG{n+nd}{@show}\PYG{+w}{ }\PYG{n}{name}\PYG{p}{(}\PYG{l+s+sa}{T}\PYG{l+s}{\PYGZdq{}}\PYG{l+s}{?A\PYGZus{}?x?y}\PYG{l+s}{\PYGZdq{}}\PYG{p}{)}\PYG{+w}{         }\PYG{c}{\PYGZsh{} ?A}
\PYG{+w}{    }\PYG{n+nd}{@show}\PYG{+w}{ }\PYG{n}{all}\PYG{p}{(}\PYG{n}{p}\PYG{p}{[}\PYG{l+m+mi}{1}\PYG{p}{]}\PYG{+w}{ }\PYG{k}{isa}\PYG{+w}{ }\PYG{n}{DummyPatternIndex}\PYG{+w}{ }\PYG{k}{for}\PYG{+w}{ }\PYG{n}{p}\PYG{+w}{ }\PYG{k}{in}\PYG{+w}{ }\PYG{n}{indices}\PYG{p}{(}\PYG{l+s+sa}{T}\PYG{l+s}{\PYGZdq{}}\PYG{l+s}{?A\PYGZus{}?x?y}\PYG{l+s}{\PYGZdq{}}\PYG{p}{)}\PYG{p}{)}\PYG{+w}{   }\PYG{c}{\PYGZsh{} true}
\end{Verbatim}
Such patterns may be used directly, and multiplied together to form multi-tensor patterns:
\begin{Verbatim}[commandchars=\\\{\},xleftmargin=\parindent,numbers=left,bgcolor=bg]
\PYG{+w}{    }\PYG{n}{replace\PYGZus{}pattern}\PYG{p}{(}\PYG{n}{P}\PYG{p}{(}\PYG{n}{lower}\PYG{p}{)}\PYG{o}{*}\PYG{n}{Q}\PYG{p}{(}\PYG{n}{upper}\PYG{p}{)}\PYG{o}{*}\PYG{n}{R}\PYG{p}{,}\PYG{+w}{ }\PYG{l+s+sa}{T}\PYG{l+s}{\PYGZdq{}}\PYG{l+s}{P\PYGZus{}?x}\PYG{l+s}{\PYGZdq{}}\PYG{+w}{ }\PYG{o}{*}\PYG{+w}{ }\PYG{l+s+sa}{T}\PYG{l+s}{\PYGZdq{}}\PYG{l+s}{Q\PYGZca{}?x}\PYG{l+s}{\PYGZdq{}}\PYG{p}{,}\PYG{+w}{ }\PYG{n}{Z}\PYG{p}{)}\PYG{+w}{   }\PYG{c}{\PYGZsh{}  ZR\PYGZca{}\PYGZob{}ν\PYGZcb{}}
\end{Verbatim}
The string form \jil{parse_pattern2tensor()} remains available, and the two are interchangeable; the
macro is preferable whenever the pattern needs anything the pattern-string grammar cannot express.

\subsection{Results of Replacement}
\label{sec:pat-position}
For commuting factors, the position of the replacement within a term is immaterial. For matrices and fermions, it is important. Therefore, for consistency the replacement is spliced in at the position where the match began:
\begin{Verbatim}[commandchars=\\\{\},xleftmargin=\parindent,numbers=left,bgcolor=bg]
\PYG{+w}{    }\PYG{n+nd}{@matrix}\PYG{+w}{ }\PYG{n}{Mx}
\PYG{+w}{    }\PYG{n}{A}\PYG{+w}{ }\PYG{o}{=}\PYG{+w}{ }\PYG{n}{Mx}\PYG{p}{(}\PYG{l+s}{\PYGZdq{}}\PYG{l+s}{A}\PYG{l+s}{\PYGZdq{}}\PYG{p}{)}\PYG{p}{;}\PYG{+w}{ }\PYG{n}{B}\PYG{+w}{ }\PYG{o}{=}\PYG{+w}{ }\PYG{n}{Mx}\PYG{p}{(}\PYG{l+s}{\PYGZdq{}}\PYG{l+s}{B}\PYG{l+s}{\PYGZdq{}}\PYG{p}{)}\PYG{p}{;}\PYG{+w}{ }\PYG{n}{C}\PYG{+w}{ }\PYG{o}{=}\PYG{+w}{ }\PYG{n}{Mx}\PYG{p}{(}\PYG{l+s}{\PYGZdq{}}\PYG{l+s}{C}\PYG{l+s}{\PYGZdq{}}\PYG{p}{)}\PYG{p}{;}\PYG{+w}{ }\PYG{n}{D}\PYG{+w}{ }\PYG{o}{=}\PYG{+w}{ }\PYG{n}{Mx}\PYG{p}{(}\PYG{l+s}{\PYGZdq{}}\PYG{l+s}{D}\PYG{l+s}{\PYGZdq{}}\PYG{p}{)}\PYG{p}{;}\PYG{+w}{ }\PYG{n}{X}\PYG{+w}{ }\PYG{o}{=}\PYG{+w}{ }\PYG{n}{Mx}\PYG{p}{(}\PYG{l+s}{\PYGZdq{}}\PYG{l+s}{X}\PYG{l+s}{\PYGZdq{}}\PYG{p}{)}

\PYG{+w}{    }\PYG{n}{replace\PYGZus{}pattern}\PYG{p}{(}\PYG{n}{A}\PYG{o}{*}\PYG{n}{B}\PYG{o}{*}\PYG{n}{C}\PYG{p}{,}\PYG{+w}{   }\PYG{l+s}{\PYGZdq{}}\PYG{l+s}{B}\PYG{l+s}{\PYGZdq{}}\PYG{p}{,}\PYG{+w}{  }\PYG{n}{X}\PYG{p}{)}\PYG{+w}{   }\PYG{c}{\PYGZsh{}  AXC}
\PYG{+w}{    }\PYG{n}{replace\PYGZus{}pattern}\PYG{p}{(}\PYG{n}{A}\PYG{o}{*}\PYG{n}{B}\PYG{o}{*}\PYG{n}{C}\PYG{o}{*}\PYG{n}{D}\PYG{p}{,}\PYG{+w}{ }\PYG{l+s}{\PYGZdq{}}\PYG{l+s}{BC}\PYG{l+s}{\PYGZdq{}}\PYG{p}{,}\PYG{+w}{ }\PYG{n}{X}\PYG{p}{)}\PYG{+w}{   }\PYG{c}{\PYGZsh{}  AXD}
\PYG{+w}{    }\PYG{n}{replace\PYGZus{}pattern}\PYG{p}{(}\PYG{n}{A}\PYG{o}{*}\PYG{n}{B}\PYG{o}{*}\PYG{n}{C}\PYG{o}{*}\PYG{n}{D}\PYG{p}{,}\PYG{+w}{ }\PYG{l+s}{\PYGZdq{}}\PYG{l+s}{BD}\PYG{l+s}{\PYGZdq{}}\PYG{p}{,}\PYG{+w}{ }\PYG{n}{X}\PYG{p}{)}\PYG{+w}{   }\PYG{c}{\PYGZsh{}  AXC}
\end{Verbatim}
In the last line the matched factors are separated by $C$; both are consumed, the replacement lands
at the position of the first, and $C$ stays where it was.

When the factors lying between the matched ones cannot be crossed freely, they are moved using the
registered commutation relations rather than a bare sign, and the commutator remainders
appear as additional summands:
\begin{Verbatim}[commandchars=\\\{\},xleftmargin=\parindent,numbers=left,bgcolor=bg]
\PYG{+w}{    }\PYG{n+nd}{@commutator}\PYG{+w}{ }\PYG{p}{[}\PYG{n}{C}\PYG{p}{,}\PYG{+w}{ }\PYG{n}{B}\PYG{p}{]}\PYG{+w}{ }\PYG{o}{=}\PYG{+w}{ }\PYG{n}{h}\PYG{o}{*}\PYG{n}{E}

\PYG{+w}{    }\PYG{n}{replace\PYGZus{}pattern}\PYG{p}{(}\PYG{n}{A}\PYG{o}{*}\PYG{n}{B}\PYG{o}{*}\PYG{n}{C}\PYG{p}{,}\PYG{+w}{ }\PYG{l+s}{\PYGZdq{}}\PYG{l+s}{AB}\PYG{l+s}{\PYGZdq{}}\PYG{p}{,}\PYG{+w}{ }\PYG{n}{X}\PYG{p}{)}\PYG{+w}{   }\PYG{c}{\PYGZsh{}  XC             \PYGZhy{} adjacent, nothing moves}
\PYG{+w}{    }\PYG{n}{replace\PYGZus{}pattern}\PYG{p}{(}\PYG{n}{A}\PYG{o}{*}\PYG{n}{C}\PYG{o}{*}\PYG{n}{B}\PYG{p}{,}\PYG{+w}{ }\PYG{l+s}{\PYGZdq{}}\PYG{l+s}{AB}\PYG{l+s}{\PYGZdq{}}\PYG{p}{,}\PYG{+w}{ }\PYG{n}{X}\PYG{p}{)}\PYG{+w}{   }\PYG{c}{\PYGZsh{}  XC + h*A*E     \PYGZhy{} ACB = ABC + A[C,B]}
\end{Verbatim}
For graded factors that do commute, the move contributes only the graded sign:
\begin{Verbatim}[commandchars=\\\{\},xleftmargin=\parindent,numbers=left,bgcolor=bg]
\PYG{+w}{    }\PYG{n}{replace\PYGZus{}pattern}\PYG{p}{(}\PYG{n}{F}\PYG{p}{(}\PYG{n}{lower}\PYG{p}{)}\PYG{o}{*}\PYG{n}{G}\PYG{p}{(}\PYG{n}{upper}\PYG{p}{)}\PYG{o}{*}\PYG{n}{H}\PYG{p}{,}\PYG{+w}{ }\PYG{l+s}{\PYGZdq{}}\PYG{l+s}{F\PYGZus{}\PYGZob{}?x\PYGZcb{}G\PYGZca{}\PYGZob{}?x\PYGZcb{}}\PYG{l+s}{\PYGZdq{}}\PYG{p}{,}\PYG{+w}{ }\PYG{n}{Z}\PYG{p}{)}\PYG{+w}{   }\PYG{c}{\PYGZsh{}   ZH\PYGZca{}\PYGZob{}f\PYGZcb{}}
\PYG{+w}{    }\PYG{n}{replace\PYGZus{}pattern}\PYG{p}{(}\PYG{n}{F}\PYG{p}{(}\PYG{n}{lower}\PYG{p}{)}\PYG{o}{*}\PYG{n}{H}\PYG{o}{*}\PYG{n}{G}\PYG{p}{(}\PYG{n}{upper}\PYG{p}{)}\PYG{p}{,}\PYG{+w}{ }\PYG{l+s}{\PYGZdq{}}\PYG{l+s}{F\PYGZus{}\PYGZob{}?x\PYGZcb{}G\PYGZca{}\PYGZob{}?x\PYGZcb{}}\PYG{l+s}{\PYGZdq{}}\PYG{p}{,}\PYG{+w}{ }\PYG{n}{Z}\PYG{p}{)}\PYG{+w}{   }\PYG{c}{\PYGZsh{}  \PYGZhy{}ZH\PYGZca{}\PYGZob{}f\PYGZcb{}}
\end{Verbatim}

Note again that a wildcard appearing in the replacement refers back to what the pattern bound. \jil{?A} supplies
the matched tensor's name and nothing else, so coordinates, tableaux, labels and the concrete
type come from the replacement as written. This is useful for stripping or restating them. \jil{??A}
supplies the matched object: its data and its type, with only the index placement taken from
the replacement. Consequently subsequent replacements \jil{??A -> ??A} is the identity, whereas \jil{?A -> ?A} is not.
Both refer to the same binding, so a pattern always writes \jil{?A}.

Note that replacement may introduce an index that is already free in the term, which silently creates a
contraction - for example replacing \(A_\mu\) by \(B_\nu\) inside \(A_\mu C^\nu\) gives \(B_\nu C^\nu\). Also, an unbound wildcard in a replacement, or an empty pattern, raises
\jil{InvalidReplacementPattern} listing the bindings that were available.

\subsection{Reducing Expressions: Replacements and Derivatives}\label{sec:reducing-expressions-replacements-and}
Replacement patterns have nice interplay with both solving linear equations, and integration by parts, which lets us reduce many complicated expressions.

Recall that we can isolate a tensor in a \jil{TensorExpression} using \jil{solve_linear}. This returns the pairing of the isolated tensor, and the expression it is equal to. This is exactly the pair \jil{replace_pattern} can consume. Rules therefore need not be written out by hand, but can be generated from the equations of the theory itself, for example via constraints, equations of motion or conservation laws. By solving each for a particular tensor, the quantity can be eliminated from another expression.  The \jil{nothing} return branch of \jil{solve_linear} is what makes this practical in bulk: when sweeping a list of equations against a list of candidate targets, those that isolate the tensor successfully yield rules and those that do not are discarded, without the caller inspecting each case. We use this explicitly in the extended example in~\ref{sec:ttbarexample}.

Substitution and integration by parts interact.  A rule keyed on a derivative of
a field, such as a conservation law \(\partial^\mu T_{\mu\nu}=0\) or a relation
expressing \(\partial\phi\) in terms of another field (possibly also containing derivatives), can only match when that derivative is actually sitting on \(\phi\).  If one attempted to perform a reduction modulo total derivatives, the derivative may move off the field to elsewhere, so a later applied rule is left with nothing to match against.  Neither
operation on its own reaches a fixed point: substituting first may create new
total derivatives, while reducing first may hide the pattern inside derivatives that the rules were intending to match.

\jil{reduce_bp_with} interleaves the two, iterating until neither step changes
anything. The essential point is that it does not insist on the canonical
representative. For each term it tries all three placements (\jil{:key},
\jil{:first} and \jil{:last}) applies the rules to each, and keeps whichever
one a rule actually fired on, breaking ties by the number of terms produced.  A
final reduction modulo total derivatives is applied at the end, so the result is
returned in normal form even though non-canonical placements were used along the
way.
\begin{Verbatim}[commandchars=\\\{\},xleftmargin=\parindent,numbers=left,bgcolor=bg]
\PYG{+w}{    }\PYG{n}{rules}\PYG{+w}{ }\PYG{o}{=}\PYG{+w}{ }\PYG{p}{[}\PYG{p}{(}\PYG{n}{∂}\PYG{p}{(}\PYG{n}{A}\PYG{p}{)}\PYG{p}{,}\PYG{+w}{ }\PYG{n}{C}\PYG{p}{)}\PYG{p}{]}\PYG{+w}{                  }\PYG{c}{\PYGZsh{} a relation of the form ∂A = C}

\PYG{+w}{    }\PYG{n}{drop\PYGZus{}total\PYGZus{}derivatives}\PYG{p}{(}\PYG{n}{∂}\PYG{p}{(}\PYG{n}{A}\PYG{p}{)}\PYG{+w}{ }\PYG{o}{*}\PYG{+w}{ }\PYG{n}{B}\PYG{p}{)}\PYG{+w}{     }\PYG{c}{\PYGZsh{} \PYGZhy{}A\PYGZus{}\PYGZob{}ν\PYGZcb{}∂\PYGZus{}\PYGZob{}μ\PYGZcb{}⟦B\PYGZus{}\PYGZob{}ρ\PYGZcb{}⟧, rule cannot match}
\PYG{+w}{    }\PYG{n}{reduce\PYGZus{}bp\PYGZus{}with}\PYG{p}{(}\PYG{n}{∂}\PYG{p}{(}\PYG{n}{A}\PYG{p}{)}\PYG{+w}{ }\PYG{o}{*}\PYG{+w}{ }\PYG{n}{B}\PYG{p}{,}\PYG{+w}{ }\PYG{n}{rules}\PYG{p}{)}\PYG{+w}{      }\PYG{c}{\PYGZsh{} B\PYGZus{}\PYGZob{}ρ\PYGZcb{}C\PYGZus{}\PYGZob{}μν\PYGZcb{}}
\PYG{+w}{    }\PYG{n}{reduce\PYGZus{}bp\PYGZus{}with}\PYG{p}{(}\PYG{n}{A}\PYG{+w}{ }\PYG{o}{*}\PYG{+w}{ }\PYG{n}{∂}\PYG{p}{(}\PYG{n}{B}\PYG{p}{)}\PYG{p}{,}\PYG{+w}{ }\PYG{n}{rules}\PYG{p}{)}\PYG{+w}{      }\PYG{c}{\PYGZsh{} \PYGZhy{}B\PYGZus{}\PYGZob{}ρ\PYGZcb{}C\PYGZus{}\PYGZob{}μν\PYGZcb{}}
\end{Verbatim}
The first line shows the difficulty: the normal form has carried the derivative
onto \(B\), and no rule written for \(\partial A\) will ever see it.  The second
recovers the substitution by performing a substitution first.  The third starts
from the opposite representative of the same equivalence class, automatically integrates by
parts to expose \(\partial A\) again, and replaces it using the substitution rule. This is a quite powerful feature that can reduce many expressions.

The number of sweeps is capped by the optional \jil{passes} keyword, which defaults to
twelve. The loop exits early as soon as a sweep leaves the expression unchanged.

\newpage

\section{Variational Calculus}\label{sec:variational-calculus}
Variational calculus allows one to compute the Euler-Lagrange equations associated with a Lagrangian or action. Given its ubiquity in field theories, \jil{Alakazam} is able to compute the Euler-Lagrange equations via the \jil{EL()} function. This takes as input a \jil{TensorExpression}, the \jil{Tensor} to take the variation with respect to, and the type of derivative with which to take the variation with respect to. These variational variables should use indices not already used in the expression.

Consider for example, the QED Lagrangian,
\begin{equation}
    \mathcal{L}_{\mathrm{QED}}
=-\frac{1}{4} F_{\mu\nu}F^{\mu\nu}+
\bar{\psi}(i\slashed{\partial}-e\slashed{A}-m)\psi.
\end{equation}
We can write this in \jil{Alakazam}
\begin{Verbatim}[commandchars=\\\{\},xleftmargin=\parindent,numbers=left,bgcolor=bg]
\PYG{+w}{    }\PYG{n+nd}{@syms}\PYG{+w}{ }\PYG{n}{m}\PYG{+w}{ }\PYG{n}{e}
\PYG{+w}{    }\PYG{n}{ψ}\PYG{o}{=}\PYG{n}{Spinor}\PYG{p}{(}\PYG{l+s}{\PYGZdq{}}\PYG{l+s}{ψ}\PYG{l+s}{\PYGZdq{}}\PYG{p}{,}\PYG{n}{x}\PYG{p}{)}
\PYG{+w}{    }\PYG{n}{L}\PYG{o}{=}\PYG{p}{(}\PYG{o}{\PYGZhy{}}\PYG{l+m+mi}{1}\PYG{o}{//}\PYG{l+m+mi}{4}\PYG{o}{*}\PYG{l+s+sa}{T}\PYG{l+s}{\PYGZdq{}}\PYG{l+s}{F\PYGZca{}[μν]\PYGZob{}x\PYGZcb{}}\PYG{l+s}{\PYGZdq{}}\PYG{o}{*}\PYG{l+s+sa}{T}\PYG{l+s}{\PYGZdq{}}\PYG{l+s}{F\PYGZus{}[μν]\PYGZob{}x\PYGZcb{}}\PYG{l+s}{\PYGZdq{}}
\PYG{+w}{       }\PYG{o}{+}\PYG{n}{ψ}\PYG{o}{\PYGZsq{}}\PYG{o}{*}\PYG{p}{(}\PYG{n+nb}{im}\PYG{o}{*}\PYG{n}{Gamma}\PYG{p}{(}\PYG{n}{μ}\PYG{o}{=\PYGZgt{}}\PYG{o}{↑}\PYG{p}{)}\PYG{o}{*}\PYG{p}{(}\PYG{n}{PD}\PYG{p}{(}\PYG{l+s}{\PYGZdq{}}\PYG{l+s}{∂}\PYG{l+s}{\PYGZdq{}}\PYG{p}{,}\PYG{+w}{ }\PYG{n}{ψ}\PYG{p}{,}\PYG{+w}{ }\PYG{n}{x}\PYG{p}{,}\PYG{+w}{ }\PYG{n}{μ}\PYG{o}{=\PYGZgt{}}\PYG{o}{↓}\PYG{p}{)}\PYG{o}{+}\PYG{n+nb}{im}\PYG{o}{*}\PYG{n}{e}\PYG{o}{*}\PYG{l+s+sa}{T}\PYG{l+s}{\PYGZdq{}}\PYG{l+s}{A\PYGZus{}μ\PYGZob{}x\PYGZcb{}}\PYG{l+s}{\PYGZdq{}}\PYG{o}{*}\PYG{n}{ψ}\PYG{p}{)}\PYG{o}{\PYGZhy{}}\PYG{n}{m}\PYG{o}{*}\PYG{n}{ψ}\PYG{p}{)}\PYG{p}{)}
\end{Verbatim}
and explicitly expand the Maxwell tensor using replacements
\begin{Verbatim}[commandchars=\\\{\},xleftmargin=\parindent,numbers=left,bgcolor=bg]
\PYG{+w}{    }\PYG{n}{a}\PYG{p}{,}\PYG{+w}{ }\PYG{n}{b}\PYG{o}{=}\PYG{n}{DummyPatternIndex}\PYG{p}{(}\PYG{p}{[}\PYG{l+s}{\PYGZdq{}}\PYG{l+s}{a}\PYG{l+s}{\PYGZdq{}}\PYG{p}{,}\PYG{+w}{ }\PYG{l+s}{\PYGZdq{}}\PYG{l+s}{b}\PYG{l+s}{\PYGZdq{}}\PYG{p}{]}\PYG{p}{,}\PYG{+w}{ }\PYG{n}{st}\PYG{p}{)}
\PYG{+w}{    }\PYG{n}{f1}\PYG{o}{=}\PYG{n}{PD}\PYG{p}{(}\PYG{l+s}{\PYGZdq{}}\PYG{l+s}{∂}\PYG{l+s}{\PYGZdq{}}\PYG{p}{,}\PYG{+w}{ }\PYG{l+s+sa}{T}\PYG{l+s}{\PYGZdq{}}\PYG{l+s}{A\PYGZus{}b\PYGZob{}x\PYGZcb{}}\PYG{l+s}{\PYGZdq{}}\PYG{p}{,}\PYG{+w}{ }\PYG{n}{x}\PYG{p}{,}\PYG{+w}{ }\PYG{n}{a}\PYG{o}{=\PYGZgt{}}\PYG{o}{↓}\PYG{p}{)}\PYG{o}{\PYGZhy{}}\PYG{n}{PD}\PYG{p}{(}\PYG{l+s}{\PYGZdq{}}\PYG{l+s}{∂}\PYG{l+s}{\PYGZdq{}}\PYG{p}{,}\PYG{+w}{ }\PYG{l+s+sa}{T}\PYG{l+s}{\PYGZdq{}}\PYG{l+s}{A\PYGZus{}a\PYGZob{}x\PYGZcb{}}\PYG{l+s}{\PYGZdq{}}\PYG{p}{,}\PYG{+w}{ }\PYG{n}{x}\PYG{p}{,}\PYG{+w}{ }\PYG{n}{b}\PYG{o}{=\PYGZgt{}}\PYG{o}{↓}\PYG{p}{)}
\PYG{+w}{    }\PYG{n}{f2}\PYG{o}{=}\PYG{n}{PD}\PYG{p}{(}\PYG{l+s}{\PYGZdq{}}\PYG{l+s}{∂}\PYG{l+s}{\PYGZdq{}}\PYG{p}{,}\PYG{+w}{ }\PYG{l+s+sa}{T}\PYG{l+s}{\PYGZdq{}}\PYG{l+s}{A\PYGZca{}b\PYGZob{}x\PYGZcb{}}\PYG{l+s}{\PYGZdq{}}\PYG{p}{,}\PYG{+w}{ }\PYG{n}{x}\PYG{p}{,}\PYG{+w}{ }\PYG{n}{a}\PYG{o}{=\PYGZgt{}}\PYG{o}{↑}\PYG{p}{)}\PYG{o}{\PYGZhy{}}\PYG{n}{PD}\PYG{p}{(}\PYG{l+s}{\PYGZdq{}}\PYG{l+s}{∂}\PYG{l+s}{\PYGZdq{}}\PYG{p}{,}\PYG{+w}{ }\PYG{l+s+sa}{T}\PYG{l+s}{\PYGZdq{}}\PYG{l+s}{A\PYGZca{}a\PYGZob{}x\PYGZcb{}}\PYG{l+s}{\PYGZdq{}}\PYG{p}{,}\PYG{+w}{ }\PYG{n}{x}\PYG{p}{,}\PYG{+w}{ }\PYG{n}{b}\PYG{o}{=\PYGZgt{}}\PYG{o}{↑}\PYG{p}{)}
\PYG{+w}{    }
\PYG{+w}{    }\PYG{n}{LA}\PYG{o}{=}\PYG{n}{simplify}\PYG{p}{(}\PYG{n}{replace\PYGZus{}pattern}\PYG{p}{(}\PYG{n}{replace\PYGZus{}pattern}\PYG{p}{(}\PYG{n}{L}\PYG{p}{,}
\PYG{+w}{            }\PYG{l+s}{\PYGZdq{}}\PYG{l+s}{F\PYGZus{}\PYGZob{}?a?b\PYGZcb{}}\PYG{l+s}{\PYGZdq{}}\PYG{p}{,}\PYG{+w}{ }\PYG{n}{f1}\PYG{p}{)}\PYG{p}{,}\PYG{+w}{ }\PYG{l+s}{\PYGZdq{}}\PYG{l+s}{F\PYGZca{}\PYGZob{}?a?b\PYGZcb{}}\PYG{l+s}{\PYGZdq{}}\PYG{p}{,}\PYG{+w}{ }\PYG{n}{f2}\PYG{p}{)}\PYG{p}{)}
\end{Verbatim}
We can then derive the Maxwell equation,
\begin{Verbatim}[commandchars=\\\{\},xleftmargin=\parindent,numbers=left,bgcolor=bg]
\PYG{+w}{    }\PYG{n+nd}{@show}\PYG{+w}{ }\PYG{n}{hard\PYGZus{}simplify}\PYG{p}{(}\PYG{n}{EL}\PYG{p}{(}\PYG{n}{LA}\PYG{p}{,}\PYG{l+s+sa}{T}\PYG{l+s}{\PYGZdq{}}\PYG{l+s}{A\PYGZca{}ρ\PYGZob{}x\PYGZcb{}}\PYG{l+s}{\PYGZdq{}}\PYG{p}{,}\PYG{+w}{ }\PYG{n}{PD}\PYG{p}{(}\PYG{l+s}{\PYGZdq{}}\PYG{l+s}{∂}\PYG{l+s}{\PYGZdq{}}\PYG{p}{,}\PYG{+w}{ }\PYG{n}{x}\PYG{p}{,}\PYG{+w}{ }\PYG{n}{τ}\PYG{o}{=\PYGZgt{}}\PYG{o}{↑}\PYG{p}{)}\PYG{p}{)}\PYG{p}{)}
\PYG{+w}{    }\PYG{c}{\PYGZsh{}   \PYGZhy{}ēψγ\PYGZus{}\PYGZob{}ρ\PYGZcb{}ψ+∂\PYGZus{}\PYGZob{}μ\PYGZcb{}⟦∂\PYGZca{}\PYGZob{}μ\PYGZcb{}⟦A\PYGZus{}\PYGZob{}ρ\PYGZcb{}⟧⟧\PYGZhy{}∂\PYGZus{}\PYGZob{}ρ\PYGZcb{}⟦∂\PYGZca{}\PYGZob{}μ\PYGZcb{}⟦A\PYGZus{}\PYGZob{}μ\PYGZcb{}⟧⟧}
\end{Verbatim}
or the Dirac equations
\begin{Verbatim}[commandchars=\\\{\},xleftmargin=\parindent,numbers=left,bgcolor=bg]
\PYG{+w}{    }\PYG{n+nd}{@show}\PYG{+w}{ }\PYG{n}{hard\PYGZus{}simplify}\PYG{p}{(}\PYG{n}{EL}\PYG{p}{(}\PYG{n}{LA}\PYG{p}{,}\PYG{+w}{ }\PYG{n}{ψ}\PYG{p}{,}\PYG{+w}{ }\PYG{n}{PD}\PYG{p}{(}\PYG{l+s}{\PYGZdq{}}\PYG{l+s}{∂}\PYG{l+s}{\PYGZdq{}}\PYG{p}{,}\PYG{+w}{ }\PYG{n}{x}\PYG{p}{,}\PYG{+w}{ }\PYG{n}{ρ}\PYG{o}{=\PYGZgt{}}\PYG{o}{↑}\PYG{p}{)}\PYG{p}{)}\PYG{p}{)}
\PYG{+w}{    }\PYG{c}{\PYGZsh{}   eA\PYGZus{}\PYGZob{}μ\PYGZcb{}̄ψγ\PYGZca{}\PYGZob{}μ\PYGZcb{}+m̄ψ+iγ\PYGZus{}\PYGZob{}μ\PYGZcb{}∂\PYGZca{}\PYGZob{}μ\PYGZcb{}⟦̄ψ⟧}
\PYG{+w}{    }\PYG{n+nd}{@show}\PYG{+w}{ }\PYG{n}{hard\PYGZus{}simplify}\PYG{p}{(}\PYG{n}{EL}\PYG{p}{(}\PYG{n}{LA}\PYG{p}{,}\PYG{+w}{ }\PYG{n}{ψ}\PYG{o}{\PYGZsq{}}\PYG{p}{,}\PYG{+w}{ }\PYG{n}{PD}\PYG{p}{(}\PYG{l+s}{\PYGZdq{}}\PYG{l+s}{∂}\PYG{l+s}{\PYGZdq{}}\PYG{p}{,}\PYG{+w}{ }\PYG{n}{x}\PYG{p}{,}\PYG{+w}{ }\PYG{n}{ρ}\PYG{o}{=\PYGZgt{}}\PYG{o}{↑}\PYG{p}{)}\PYG{p}{)}\PYG{p}{)}
\PYG{+w}{    }\PYG{c}{\PYGZsh{}   iγ\PYGZca{}\PYGZob{}μ\PYGZcb{}∂\PYGZus{}\PYGZob{}μ\PYGZcb{}⟦ψ⟧\PYGZhy{}eA\PYGZus{}\PYGZob{}μ\PYGZcb{}γ\PYGZca{}\PYGZob{}μ\PYGZcb{}ψ\PYGZhy{}mψ}
\end{Verbatim}
Note that the variational derivative has also taken into account the fermionic grading on the Dirac spinors to obtain the correct relative signs. Internally, the algorithm uses a \jil{VariationalDerivative<:DerivativeSuperType} to do functional calculus, however, this is not exposed at the user-level.

\newpage

\section{Differential Geometry}\label{sec:differential-geometry}
By default, a few rudimentary Differential Geometry operations are available. These currently all use the component form of tensors. Basis-free differential forms in some capacity will be implemented in future versions.

\subsection{\jil{⋆} (Hodge Star)}\label{sec:hodge-star}
The Hodge star is implemented to find the Hodge dual of a differential form. Given an antisymmetric tensor (possibly with other indices, such as flavour), one can easily compute the dual
\begin{Verbatim}[commandchars=\\\{\},xleftmargin=\parindent,numbers=left,bgcolor=bg]
\PYG{+w}{    }\PYG{n}{F}\PYG{o}{=}\PYG{n}{AntisymmetricTensor}\PYG{p}{(}\PYG{l+s}{\PYGZdq{}}\PYG{l+s}{F}\PYG{l+s}{\PYGZdq{}}\PYG{p}{,}\PYG{p}{[}\PYG{n}{a}\PYG{o}{=\PYGZgt{}}\PYG{n}{upper}\PYG{p}{,}\PYG{n}{μ}\PYG{o}{=\PYGZgt{}}\PYG{n}{lower}\PYG{p}{,}\PYG{n}{ν}\PYG{o}{=\PYGZgt{}}\PYG{n}{lower}\PYG{p}{]}\PYG{p}{)}
\PYG{+w}{    }\PYG{n+nd}{@show}\PYG{+w}{ }\PYG{o}{⋆}\PYG{p}{(}\PYG{n}{F}\PYG{p}{)}
\PYG{+w}{    }\PYG{c}{\PYGZsh{}   (⋆)(F) = 1//2ϵ\PYGZca{}\PYGZob{}μντλ\PYGZcb{}F\PYGZca{}\PYGZob{}a∘∘\PYGZcb{}\PYGZus{}\PYGZob{}∘τλ\PYGZcb{}}
\end{Verbatim}
This obeys the convention
\begin{equation}
  (\star\omega)^{\mu_1\ldots\mu_{d-p}}
    \;=\;\frac{1}{p!}\,
    \epsilon^{\mu_1\ldots\mu_{d-p}\nu_1\ldots\nu_p}\,
    \omega_{\nu_1\ldots\nu_p},
\end{equation}

and 
\begin{equation}\label{eq:hodge-squared}
    \star\star\,\omega \;=\; (-1)^{p(d-p)}\,\operatorname{sgn}(\det\eta)\;\omega ,
\end{equation}
for $p$-forms.

\subsection{\jil{⨼} (Interior Product)}\label{sec:interior-product}
One may compute the interior product between a vector and a form using the function \jil{⨼} or \jil{ι}. These can be used with either infix notation, or by bracketing the input arguments. One can also use the function \jil{contraction()}.
\begin{Verbatim}[commandchars=\\\{\},xleftmargin=\parindent,numbers=left,bgcolor=bg]
\PYG{+w}{    }\PYG{n}{V}\PYG{o}{=}\PYG{n}{Tensor}\PYG{p}{(}\PYG{l+s}{\PYGZdq{}}\PYG{l+s}{v}\PYG{l+s}{\PYGZdq{}}\PYG{p}{,}\PYG{p}{[}\PYG{n}{μ}\PYG{o}{=\PYGZgt{}}\PYG{n}{lower}\PYG{p}{]}\PYG{p}{)}
\PYG{+w}{    }\PYG{n}{ω}\PYG{o}{=}\PYG{n}{AntisymmetricTensor}\PYG{p}{(}\PYG{l+s}{\PYGZdq{}}\PYG{l+s}{ω}\PYG{l+s}{\PYGZdq{}}\PYG{p}{,}\PYG{p}{[}\PYG{n}{τ}\PYG{o}{=\PYGZgt{}}\PYG{n}{lower}\PYG{p}{,}\PYG{n}{ν}\PYG{o}{=\PYGZgt{}}\PYG{n}{lower}\PYG{p}{,}\PYG{n}{ρ}\PYG{o}{=\PYGZgt{}}\PYG{n}{lower}\PYG{p}{]}\PYG{p}{)}
\PYG{+w}{    }\PYG{n+nd}{@show}\PYG{+w}{ }\PYG{n}{V⨼ω}
\PYG{+w}{    }\PYG{n+nd}{@show}\PYG{+w}{ }\PYG{n}{ι}\PYG{p}{(}\PYG{n}{V}\PYG{p}{,}\PYG{n}{ω}\PYG{p}{)}
\PYG{+w}{    }\PYG{n+nd}{@show}\PYG{+w}{ }\PYG{n}{contraction}\PYG{p}{(}\PYG{n}{V}\PYG{p}{,}\PYG{n}{ω}\PYG{p}{)}
\PYG{+w}{        }\PYG{c}{\PYGZsh{}V ⨼ ω = v\PYGZus{}\PYGZob{}μ\PYGZcb{}ω\PYGZca{}\PYGZob{}μ∘∘\PYGZcb{}\PYGZus{}\PYGZob{}∘νρ\PYGZcb{}}
\PYG{+w}{        }\PYG{c}{\PYGZsh{}ι(V, ω) = v\PYGZus{}\PYGZob{}μ\PYGZcb{}ω\PYGZca{}\PYGZob{}μ∘∘\PYGZcb{}\PYGZus{}\PYGZob{}∘νρ\PYGZcb{}}
\PYG{+w}{        }\PYG{c}{\PYGZsh{}contraction(V,ω) = v\PYGZus{}\PYGZob{}μ\PYGZcb{}ω\PYGZca{}\PYGZob{}μ∘∘\PYGZcb{}\PYGZus{}\PYGZob{}∘νρ\PYGZcb{}}
\end{Verbatim}
The vector will automatically be contracted with the first slot, as is convention:
\begin{equation}\label{eq:dg-interior}
    (\iota_V\omega)_{\mu_2\ldots\mu_p}\;=\;V^{\mu_1}\,\omega_{\mu_1\mu_2\ldots\mu_p}.
\end{equation}

\subsection{\jil{∧} (Wedge Product)}\label{sec:wedge-product}
We can also take the wedge product between forms. For example, for a $2$-form $F$ and a $3$-form $H$, in $6$D we obtain
\begin{Verbatim}[commandchars=\\\{\},xleftmargin=\parindent,numbers=left,bgcolor=bg]
\PYG{+w}{    }\PYG{n}{F}\PYG{o}{=}\PYG{n}{AntisymmetricTensor}\PYG{p}{(}\PYG{l+s}{\PYGZdq{}}\PYG{l+s}{F}\PYG{l+s}{\PYGZdq{}}\PYG{p}{,}\PYG{p}{[}\PYG{n}{μ}\PYG{o}{=\PYGZgt{}}\PYG{n}{lower}\PYG{p}{,}\PYG{n}{ν}\PYG{o}{=\PYGZgt{}}\PYG{n}{lower}\PYG{p}{]}\PYG{p}{)}
\PYG{+w}{    }\PYG{n}{H}\PYG{o}{=}\PYG{n}{AntisymmetricTensor}\PYG{p}{(}\PYG{l+s}{\PYGZdq{}}\PYG{l+s}{H}\PYG{l+s}{\PYGZdq{}}\PYG{p}{,}\PYG{p}{[}\PYG{n}{λ}\PYG{o}{=\PYGZgt{}}\PYG{n}{lower}\PYG{p}{,}\PYG{n}{ρ}\PYG{o}{=\PYGZgt{}}\PYG{n}{lower}\PYG{p}{,}\PYG{n}{σ}\PYG{o}{=\PYGZgt{}}\PYG{n}{lower}\PYG{p}{]}\PYG{p}{)}
\PYG{+w}{    }\PYG{n+nd}{@show}\PYG{+w}{ }\PYG{n}{F∧H}
\PYG{+w}{        }\PYG{c}{\PYGZsh{}F ∧ H = \PYGZhy{}1//12ϵ\PYGZus{}\PYGZob{}μνλρσχ\PYGZcb{}ϵ\PYGZca{}\PYGZob{}ταβγδχ\PYGZcb{}F\PYGZus{}\PYGZob{}τα\PYGZcb{}H\PYGZus{}\PYGZob{}βγδ\PYGZcb{}}
\end{Verbatim}
Here, we obey the convention
\begin{equation}\label{eq:dg-wedge-eps}
    (\alpha\wedge\beta)_{\mu_1\ldots\mu_{r}}
    \;=\;\frac{\operatorname{sgn}(\det\eta)}{p!\;q!\;m!}\;
    \epsilon_{\mu_1\ldots\mu_{r}\kappa_1\ldots\kappa_{m}}\;
    \epsilon^{\nu_1\ldots\nu_{r}\kappa_1\ldots\kappa_{m}}\;
    \alpha_{\nu_1\ldots\nu_p}\,\beta_{\nu_{p+1}\ldots\nu_{r}}\,,
\end{equation}
where $m=d-r$, $r=p+q$. Note that the metric factor here compensates for the metric factor used in the contraction of epsilons, so that the result is metric-independent.

\subsection{\jilg{\sfd} (Exterior Derivative)}\label{sec:ext-deriv}
A basic implementation of the exterior derivative is also included, satisfying nilpotency $d^2=0$. This is a function denoted by the character \jilg{\sfd} (\jil{\sfd}), not to be confused with the usual character for d. Alternatively, one can call the method directly from \jil{exterior_derivative()}.
\begin{Verbatim}[commandchars=\\\{\},xleftmargin=\parindent,numbers=left,bgcolor=bg]
\PYG{+w}{    }\PYG{n}{f}\PYG{+w}{ }\PYG{o}{=}\PYG{+w}{ }\PYG{n}{Tensor}\PYG{p}{(}\PYG{l+s}{\PYGZdq{}}\PYG{l+s}{f}\PYG{l+s}{\PYGZdq{}}\PYG{p}{,}\PYG{+w}{ }\PYG{n}{x}\PYG{p}{)}\PYG{+w}{                                      }\PYG{c}{\PYGZsh{} a 0\PYGZhy{}form}
\PYG{+w}{    }\PYG{n}{A}\PYG{+w}{ }\PYG{o}{=}\PYG{+w}{ }\PYG{n}{Tensor}\PYG{p}{(}\PYG{l+s}{\PYGZdq{}}\PYG{l+s}{A}\PYG{l+s}{\PYGZdq{}}\PYG{p}{,}\PYG{+w}{ }\PYG{n}{b}\PYG{o}{=\PYGZgt{}}\PYG{n}{lower}\PYG{p}{,}\PYG{+w}{ }\PYG{n}{x}\PYG{p}{)}\PYG{+w}{                            }\PYG{c}{\PYGZsh{} a 1\PYGZhy{}form}
\PYG{+w}{    }\PYG{n}{F}\PYG{+w}{ }\PYG{o}{=}\PYG{+w}{ }\PYG{n}{AntisymmetricTensor}\PYG{p}{(}\PYG{l+s}{\PYGZdq{}}\PYG{l+s}{F}\PYG{l+s}{\PYGZdq{}}\PYG{p}{,}\PYG{+w}{ }\PYG{n}{c}\PYG{o}{=\PYGZgt{}}\PYG{n}{lower}\PYG{p}{,}\PYG{+w}{ }\PYG{n}{d}\PYG{o}{=\PYGZgt{}}\PYG{n}{lower}\PYG{p}{,}\PYG{+w}{ }\PYG{n}{x}\PYG{p}{)}\PYG{+w}{     }\PYG{c}{\PYGZsh{} a 2\PYGZhy{}form}

\PYG{+w}{    }\PYG{n+nd}{@show}\PYG{+w}{ }\PYG{n}{exterior\PYGZus{}derivative}\PYG{p}{(}\PYG{n}{f}\PYG{p}{,}\PYG{+w}{ }\PYG{n}{x}\PYG{p}{,}\PYG{+w}{ }\PYG{n}{a}\PYG{p}{)}
\PYG{+w}{    }\PYG{c}{\PYGZsh{}   ∂\PYGZus{}\PYGZob{}a\PYGZcb{}⟦f⟧    the gradient}
\PYG{+w}{    }\PYG{n+nd}{@show}\PYG{+w}{ }\PYG{n}{exterior\PYGZus{}derivative}\PYG{p}{(}\PYG{n}{A}\PYG{p}{,}\PYG{+w}{ }\PYG{n}{x}\PYG{p}{,}\PYG{+w}{ }\PYG{n}{a}\PYG{p}{)}
\PYG{+w}{    }\PYG{c}{\PYGZsh{}   ∂\PYGZus{}\PYGZob{}a\PYGZcb{}⟦A\PYGZus{}\PYGZob{}b\PYGZcb{}⟧\PYGZhy{}∂\PYGZus{}\PYGZob{}b\PYGZcb{}⟦A\PYGZus{}\PYGZob{}a\PYGZcb{}⟧    the curl, two terms of opposite sign}
\PYG{+w}{    }\PYG{n+nd}{@show}\PYG{+w}{ }\PYG{n}{exterior\PYGZus{}derivative}\PYG{p}{(}\PYG{n}{F}\PYG{p}{,}\PYG{+w}{ }\PYG{n}{x}\PYG{p}{,}\PYG{+w}{ }\PYG{n}{a}\PYG{p}{)}
\PYG{+w}{    }\PYG{c}{\PYGZsh{}   ∂\PYGZus{}\PYGZob{}a\PYGZcb{}⟦F\PYGZus{}\PYGZob{}cd\PYGZcb{}⟧\PYGZhy{}∂\PYGZus{}\PYGZob{}c\PYGZcb{}⟦F\PYGZus{}\PYGZob{}ad\PYGZcb{}⟧+∂\PYGZus{}\PYGZob{}d\PYGZcb{}⟦F\PYGZus{}\PYGZob{}ac\PYGZcb{}⟧    alternating}
\end{Verbatim}
The index for the new slot need not be supplied, and one is generated when it is omitted. A form whose degree exceeds the dimension of the space vanishes automatically.
\begin{Verbatim}[commandchars=\\\{\},xleftmargin=\parindent,numbers=left,bgcolor=bg]
\PYG{+w}{    }\PYG{n}{top}\PYG{+w}{ }\PYG{o}{=}\PYG{+w}{ }\PYG{n}{AntisymmetricTensor}\PYG{p}{(}\PYG{l+s}{\PYGZdq{}}\PYG{l+s}{T}\PYG{l+s}{\PYGZdq{}}\PYG{p}{,}\PYG{+w}{ }\PYG{n}{a}\PYG{o}{=\PYGZgt{}}\PYG{n}{lower}\PYG{p}{,}\PYG{+w}{ }\PYG{n}{b}\PYG{o}{=\PYGZgt{}}\PYG{n}{lower}\PYG{p}{,}\PYG{+w}{ }\PYG{n}{c}\PYG{o}{=\PYGZgt{}}\PYG{n}{lower}\PYG{p}{,}\PYG{+w}{ }\PYG{n}{d}\PYG{o}{=\PYGZgt{}}\PYG{n}{lower}\PYG{p}{,}\PYG{+w}{ }\PYG{n}{x}\PYG{p}{)}
\PYG{+w}{    }\PYG{n+nd}{@show}\PYG{+w}{ }\PYG{n}{exterior\PYGZus{}derivative}\PYG{p}{(}\PYG{n}{top}\PYG{p}{,}\PYG{+w}{ }\PYG{n}{x}\PYG{p}{)}
\PYG{+w}{    }\PYG{c}{\PYGZsh{}   0}
\end{Verbatim}
Nilpotency is an automatic property by construction. For example, writing $F=dA$, then $dF=0$ is the Bianchi identity of a gauge field.
\begin{Verbatim}[commandchars=\\\{\},xleftmargin=\parindent,numbers=left,bgcolor=bg]
\PYG{+w}{    }\PYG{n}{FA}\PYG{+w}{ }\PYG{o}{=}\PYG{+w}{ }\PYG{n}{exterior\PYGZus{}derivative}\PYG{p}{(}\PYG{n}{A}\PYG{p}{,}\PYG{+w}{ }\PYG{n}{x}\PYG{p}{,}\PYG{+w}{ }\PYG{n}{a}\PYG{p}{)}\PYG{+w}{                       }\PYG{c}{\PYGZsh{} F = dA}
\PYG{+w}{    }\PYG{n+nd}{@show}\PYG{+w}{ }\PYG{n}{simplify}\PYG{p}{(}\PYG{n}{apply\PYGZus{}derivative}\PYG{p}{(}\PYG{n}{𝖽}\PYG{p}{(}\PYG{n}{FA}\PYG{p}{,}\PYG{+w}{ }\PYG{n}{x}\PYG{p}{)}\PYG{p}{)}\PYG{p}{)}
\PYG{+w}{    }\PYG{c}{\PYGZsh{}   0                                                   dF = 0}
\end{Verbatim}
A different derivative name may also be given, rather than the default partial
\begin{Verbatim}[commandchars=\\\{\},xleftmargin=\parindent,numbers=left,bgcolor=bg]
\PYG{+w}{    }\PYG{n+nd}{@show}\PYG{+w}{ }\PYG{n}{exterior\PYGZus{}derivative}\PYG{p}{(}\PYG{n}{f}\PYG{p}{,}\PYG{+w}{ }\PYG{n}{x}\PYG{p}{,}\PYG{+w}{ }\PYG{n}{a}\PYG{p}{;}\PYG{+w}{ }\PYG{n}{dname}\PYG{o}{=}\PYG{l+s}{\PYGZdq{}}\PYG{l+s}{D}\PYG{l+s}{\PYGZdq{}}\PYG{p}{)}
\PYG{+w}{    }\PYG{c}{\PYGZsh{}   D\PYGZus{}\PYGZob{}a\PYGZcb{}⟦f⟧}
\end{Verbatim}

The exterior derivative obeys convention
\begin{equation}\label{eq:dg-d}
    (d\omega)_{\mu_0\mu_1\ldots\mu_p}
    \;=\;(p+1)\,\partial_{[\mu_0}\omega_{\mu_1\ldots\mu_p]}
    \;=\;\sum_{i=0}^{p}(-1)^{i}\,
    \partial_{\mu_i}\,\omega_{\mu_0\ldots\widehat{\mu_i}\ldots\mu_p}.
\end{equation}

Together with the interior product, this is sufficient to implement more advanced operations like the Lie derivative, if the user so desires.

\newpage

\section{Serialisation and Saving}\label{sec:serialisation-and-saving}
When working with complicated expressions, it is useful to be able to write them to disk. \jil{Alakazam} includes functions \jil{save()} and \jil{load()} for serialising \jil{TensorExpression}s. The serialisation layer is designed to preserve not only the numerical values of tensor expressions, but also the symbolic structure, type information, and global registries required for exact reconstruction. The expressions are encoded and saved to JSON format, in order to be somewhat readable to the user, and so that the raw JSON may be edited manually, or by other software. In addition, one may also save and load ``header'' data, in addition to an expression. This header information contains other information about the global state of the workspace, such as registered derivative actions and commutators. This should allow the user to seamlessly restore a workspace from file.

The central design principle is that objects are serialised into a type-annotated tree representation rather than relying on Julia's native object serialisation. Each serialised object is represented as a JSON dictionary containing its fully qualified type name and recursively serialised fields. This approach makes the output platform-independent and allows objects to be reconstructed in a fresh Julia session without access to the original runtime state.

The custom JSON-based format provides several advantages:

\begin{itemize}
    \item Human readability: the serialised files can be inspected and version-controlled.
\item Portability: objects can be reconstructed independently of the original Julia process.
\item Symbolic fidelity: symbolic expressions retain their exact tree structure.
\item Extensibility: new tensor object types can be supported by adding specialised serialisation or reconstruction methods, in the case they are not automatically supported
\item State preservation: global state information required for consistent tensor manipulation is stored alongside the expression.
\end{itemize}
The resulting framework provides a reproducible storage mechanism for large symbolic tensor calculations, allowing intermediate results, tensor identities, and complete symbolic environments to be archived and restored across computational sessions.

There are a few subtleties to this procedure. A particular challenge arises for symbolic coefficients represented using \jil{SymbolicUtils}. Symbolic expressions cannot be reliably serialised by storing their printed form alone, since this loses structural information and may depend on the active parsing environment. Therefore, symbolic objects are converted into a structural expression tree. Composite symbolic expressions are stored recursively as operation nodes containing the qualified name of the operation and a list of serialised arguments. During deserialisation, the operation is resolved dynamically from its qualified module path and the expression tree is reconstructed recursively. This allows symbolic coefficients involving arbitrary functions and user-defined operations to be restored without converting them to strings or relying on expression parsing.

Every serialised object contains its Julia type name, allowing the deserialiser to dynamically identify the appropriate reconstruction procedure. Types are resolved from their fully qualified names by searching loaded modules and resolving nested type expressions, including parameterised types. Reconstruction is then performed by recursively rebuilding the fields of the original object.

For primitive objects, arrays, tuples, dictionaries, and numerical values, generic recursive rules are sufficient. More complicated structures use specialised reconstruction routines. These include the \jil{Tensor} objects, \jil{Index} objects, \jil{Coordinate}s, etc.

A primary design objective of the serialisation framework is extensibility. Rather than hard-coding all globally defined symbolic objects into the serialisation process, the framework maintains a registry of global objects that are required to reconstruct the symbolic environment. This registry, \jil{global_values}, stores references to the data structures that define the global state of the tensor algebra, including index sets, coordinates, derivative expansions, commutation relations, derivative actions, and user-defined tensor ordering rules.
At serialisation time, the contents of this registry are written to a dedicated header section of the JSON document, while the tensor expression itself is stored separately. During deserialisation, the header is imported before the tensor expression is reconstructed, ensuring that all globally referenced objects already exist when individual tensor objects are instantiated.
The registry-based approach makes the framework readily extensible. Users may introduce additional globally maintained objects - for example, custom symmetry tables, simplification rules, or application-specific metadata - without modifying the core serialisation routines. Registering a new global property simply requires adding it to the global registry via \jil{add_global_property!(key,value)}, after which it is automatically included in the serialised header.
Supporting reconstruction of a new global property requires only a small amount of additional code. Specifically, the user provides a corresponding reconstruction method of the form
\begin{Verbatim}[commandchars=\\\{\},xleftmargin=\parindent,numbers=left,bgcolor=bg]
\PYG{+w}{    }\PYG{n}{reconstruct\PYGZus{}header}\PYG{p}{(}\PYG{n}{data}\PYG{p}{,}\PYG{+w}{ }\PYG{o}{::}\PYG{k+kt}{Val}\PYG{p}{\PYGZob{}}\PYG{l+s+ss}{:my\PYGZus{}property}\PYG{p}{\PYGZcb{}}\PYG{p}{)}\PYG{+w}{ }\PYG{o}{=}\PYG{+w}{ }\PYG{o}{...}
\end{Verbatim}
which specifies how the serialised data should be interpreted and incorporated into the global symbolic state during deserialisation. Since \jil{import_header()} dispatches reconstruction according to the header key, no modifications to the core loading algorithm are required. Likewise, if a new object is represented by a user-defined Julia type, the user need only implement the corresponding \jil{to_dict()} and \jil{reconstruct()} methods for that type (in the event that the user-defined structures are too specialised for the serialiser to handle automatically), following the same multiple-dispatch interface used throughout the framework.

The core framework remains unchanged as new algebraic structures are introduced, while users extend the system by implementing only the serialisation and reconstruction logic relevant to their own types. Consequently, the serialisation layer is open to extension without requiring modification of the underlying infrastructure, making it suitable for research code in which new symbolic objects and algebraic constructs are continually introduced.

The extensibility described above does induce potential security vulnerabilities. Deserialisation reaches Julia's evaluator at four points, and at
three of them the string being evaluated originates in the file. A document crafted by a third party can therefore potentially execute arbitrary code at load time, and a user should treat a serialised expression exactly as they would treat a
script from the same source. Though unlikely, users should exercise caution. We implement some basic safeguards which should protect against this, and we will continue to add more protection in future versions.

\newpage

\section{Supersymmetry and Superspace}\label{sec:supersymmetry-and-superspace}
One of the main motivations for the creation of \jil{Alakazam} is to manipulate supersymmetric expressions in superspace\cite{Salam:1976ib,gates2001superspace}. Given this, many useful operations are available to the user to study supersymmetric field theory. Note that supersymmetric grading is automatically taken into account in all appropriate methods involving swaps of tensors. The grading property is determined by the \jil{IndexSet} during creation, for example
\begin{Verbatim}[commandchars=\\\{\},xleftmargin=\parindent,numbers=left,bgcolor=bg]
\PYG{+w}{    }\PYG{n}{spacetime\PYGZus{}indices}\PYG{+w}{ }\PYG{o}{=}\PYG{+w}{ }\PYG{n}{IndexSet}\PYG{p}{(}\PYG{l+s}{\PYGZdq{}}\PYG{l+s}{spacetime}\PYG{l+s}{\PYGZdq{}}\PYG{p}{,}\PYG{+w}{ }\PYG{l+m+mi}{4}\PYG{p}{,}\PYG{n}{lower}\PYG{p}{)}
\PYG{+w}{    }\PYG{n}{spinor\PYGZus{}indices}\PYG{o}{=}\PYG{n}{IndexSet}\PYG{p}{(}\PYG{l+s}{\PYGZdq{}}\PYG{l+s}{spinor\PYGZus{}indices}\PYG{l+s}{\PYGZdq{}}\PYG{p}{,}\PYG{+w}{ }\PYG{l+m+mi}{2}\PYG{p}{,}\PYG{n}{lower}\PYG{p}{,}
\PYG{+w}{        }\PYG{n}{fermionic}\PYG{p}{,}\PYG{n}{EpsilonTensor}\PYG{p}{)}
\PYG{+w}{    }\PYG{n}{spinor\PYGZus{}conj\PYGZus{}indices}\PYG{o}{=}\PYG{n}{IndexSet}\PYG{p}{(}\PYG{l+s}{\PYGZdq{}}\PYG{l+s}{spinor\PYGZus{}conj\PYGZus{}indices}\PYG{l+s}{\PYGZdq{}}\PYG{p}{,}\PYG{+w}{ }\PYG{l+m+mi}{2}\PYG{p}{,}
\PYG{+w}{        }\PYG{n}{fermionic}\PYG{p}{,}\PYG{n}{lower}\PYG{p}{,}\PYG{n}{EpsilonTensor}\PYG{p}{)}
\end{Verbatim}
The grading of any tensor is thus determined by the sum of the grading of its indices mod 2, 0 for bosonic tensors and 1 for fermionic ones. For specialised scenarios (for example, perhaps ghosts), this can be overridden by creating \jil{Boson} or \jil{Fermion} tensors, which behave identically to the usual \jil{Tensor} object aside from forcing the grading to be bosonic or fermionic respectively, regardless of the attached indices.

\subsection{Superspace and Superfields}\label{sec:superspace-and-superfields}
Supersymmetry can conveniently be studied in superspace, which extends the spacetime coordinates $x$ with additional anticommuting Grassmann coordinates $\theta$ (and $\bar{\theta})$. In extended supersymmetry, these coordinates may carry extra indices corresponding to the $R$-symmetries of the supercharges. It is often convenient to write the field content of a theory in terms of superfields $\Psi(x,\theta,\bar{\theta})$, which can be Taylor expanded in the Grassmann basis, which truncates at finite order due to the anticommuting nature of the numbers. Generically however, the number of terms in the expansion grows exponentially with the size of superspace, corresponding to the number of supercharges in extended supersymmetry. This makes writing out the expansions by hand tedious. Since these are a foundational object, \jil{Alakazam} implements functions to generate superfields automatically. 

One way to do this is via the function
\begin{Verbatim}[commandchars=\\\{\},xleftmargin=\parindent,numbers=left,bgcolor=bg]
\PYG{+w}{    }\PYG{n}{generate\PYGZus{}superfield\PYGZus{}naive}\PYG{p}{(}\PYG{n}{name}\PYG{p}{,}\PYG{+w}{ }\PYG{n}{inds}\PYG{p}{,}\PYG{+w}{ }\PYG{n}{grassmann\PYGZus{}basis}\PYG{p}{,}\PYG{+w}{ }\PYG{n}{coord}\PYG{p}{)}
\end{Verbatim}
which takes arguments corresponding to the name of the superfield, the external indices attached to the superfield (to indicate a spinor superfield, a vector superfield and so on), a vector of Grassmann number tensors over which to expand, and the spacetime coordinate of which the coefficient fields should be a function of.

For a simple example, consider the chiral part of $4D$ $\mathcal{N}=2$ superspace, spanned by a single Grassmann number $\theta_\alpha{}^i$, where $\alpha=1,2$ is the spinor index, and $i=1,2$ is the flavour index. Then we can generate a superfield expansion
\begin{Verbatim}[commandchars=\\\{\},xleftmargin=\parindent,numbers=left,bgcolor=bg]
\PYG{+w}{    }\PYG{n}{generate\PYGZus{}superfield\PYGZus{}naive}\PYG{p}{(}\PYG{l+s}{\PYGZdq{}}\PYG{l+s}{X}\PYG{l+s}{\PYGZdq{}}\PYG{p}{,}\PYG{+w}{ }\PYG{p}{[}\PYG{n}{Tensor}\PYG{p}{(}\PYG{l+s}{\PYGZdq{}}\PYG{l+s}{θ}\PYG{l+s}{\PYGZdq{}}\PYG{p}{,}\PYG{+w}{ }\PYG{p}{[}\PYG{n}{α}\PYG{o}{=\PYGZgt{}}\PYG{n}{lower}\PYG{p}{,}\PYG{+w}{ }\PYG{n}{i}\PYG{o}{=\PYGZgt{}}\PYG{n}{upper}\PYG{p}{]}\PYG{p}{,}\PYG{+w}{ }\PYG{n}{θ}\PYG{p}{)}\PYG{p}{]}\PYG{p}{,}\PYG{+w}{ }\PYG{n}{x}\PYG{p}{)}
\end{Verbatim}
which yields
\[
\begin{aligned}
X^{\alpha}{}_{i\,(1)}(x)\,\theta_{\alpha}{}^{i}
&+ X^{\alpha\beta\gamma}{}_{ijk\,(3)}(x)\,
\theta_{\alpha}{}^{i}\theta_{\beta}{}^{j}\theta_{\gamma}{}^{k} \\
&+ X^{\alpha\beta}{}_{ij\,(2)}(x)\,
\theta_{\alpha}{}^{i}\theta_{\beta}{}^{j}+ X^{\alpha\beta\gamma\chi}{}_{ijkl\,(4)}(x)\,
\theta_{\alpha}{}^{i}\theta_{\beta}{}^{j}
\theta_{\gamma}{}^{k}\theta_{\chi}{}^{l}+ X_{(0)}(x).
\end{aligned}
\]

Alternatively, we can also generate the superfield from an explicit \jil{Superspace} structure, which can be constructed from a basis of Grassmann tensors and their associated covariant \jil{Derivative} basis,
\begin{Verbatim}[commandchars=\\\{\},xleftmargin=\parindent,numbers=left,bgcolor=bg]
\PYG{+w}{    }\PYG{n}{Superspace}\PYG{p}{(}\PYG{n}{grassmann\PYGZus{}basis}\PYG{o}{::}\PYG{k+kt}{Vector}\PYG{p}{\PYGZob{}}\PYG{k+kt}{TensorSuperType}\PYG{p}{\PYGZcb{}}\PYG{p}{,}
\PYG{+w}{        }\PYG{n}{covariant\PYGZus{}basis}\PYG{o}{::}\PYG{k+kt}{Vector}\PYG{p}{\PYGZob{}}\PYG{k+kt}{CovDerivSuperType}\PYG{p}{\PYGZcb{}}\PYG{p}{)}
\end{Verbatim}
This structure exists mostly for convenience for shorter input for functions.
For example, for $4D$ $\mathcal{N}=1$ supersymmetry we could create
\begin{Verbatim}[commandchars=\\\{\},xleftmargin=\parindent,numbers=left,bgcolor=bg]
\PYG{+w}{    }\PYG{n}{Dbar}\PYG{o}{=}\PYG{n}{Derivative}\PYG{p}{(}\PYG{l+s}{\PYGZdq{}}\PYG{l+s}{̄D}\PYG{l+s}{\PYGZdq{}}\PYG{p}{,}\PYG{+w}{ }\PYG{p}{[}\PYG{n}{x}\PYG{p}{,}\PYG{+w}{ }\PYG{n}{θbar}\PYG{p}{]}\PYG{p}{,}\PYG{+w}{ }\PYG{p}{[}\PYG{n}{αd}\PYG{+w}{ }\PYG{o}{=\PYGZgt{}}\PYG{+w}{ }\PYG{n}{lower}\PYG{p}{]}\PYG{p}{)}
\PYG{+w}{    }\PYG{n}{D}\PYG{o}{=}\PYG{n}{Derivative}\PYG{p}{(}\PYG{l+s}{\PYGZdq{}}\PYG{l+s}{D}\PYG{l+s}{\PYGZdq{}}\PYG{p}{,}\PYG{+w}{ }\PYG{p}{[}\PYG{n}{x}\PYG{p}{,}\PYG{+w}{ }\PYG{n}{θ}\PYG{p}{]}\PYG{p}{,}\PYG{+w}{ }\PYG{p}{[}\PYG{n}{α}\PYG{+w}{ }\PYG{o}{=\PYGZgt{}}\PYG{+w}{ }\PYG{n}{lower}\PYG{p}{]}\PYG{p}{)}

\PYG{+w}{    }\PYG{n}{grassmann}\PYG{o}{=}\PYG{p}{[}\PYG{n}{Tensor}\PYG{p}{(}\PYG{l+s}{\PYGZdq{}}\PYG{l+s}{θ}\PYG{l+s}{\PYGZdq{}}\PYG{p}{,}\PYG{+w}{ }\PYG{p}{[}\PYG{n}{α}\PYG{o}{=\PYGZgt{}}\PYG{n}{lower}\PYG{p}{]}\PYG{p}{,}\PYG{+w}{ }\PYG{n}{θ}\PYG{p}{)}\PYG{p}{,}\PYG{n}{Tensor}\PYG{p}{(}\PYG{l+s}{\PYGZdq{}}\PYG{l+s}{Θ}\PYG{l+s}{\PYGZdq{}}\PYG{p}{,}\PYG{+w}{ }\PYG{p}{[}\PYG{n}{αd}\PYG{o}{=\PYGZgt{}}\PYG{n}{upper}\PYG{p}{]}\PYG{p}{,}\PYG{+w}{ }\PYG{n}{θbar}\PYG{p}{)}\PYG{p}{]}
\PYG{+w}{    }\PYG{n}{sspace}\PYG{o}{=}\PYG{n}{Superspace}\PYG{p}{(}\PYG{n}{grassmann}\PYG{p}{,}\PYG{p}{[}\PYG{n}{D}\PYG{p}{,}\PYG{n}{Dbar}\PYG{p}{]}\PYG{p}{)}
\end{Verbatim}
We can then generate a superfield expansion directly from this superspace,
\begin{Verbatim}[commandchars=\\\{\},xleftmargin=\parindent,numbers=left,bgcolor=bg]
\PYG{+w}{    }\PYG{n}{generate\PYGZus{}superfield\PYGZus{}naive}\PYG{p}{(}\PYG{l+s}{\PYGZdq{}}\PYG{l+s}{Z}\PYG{l+s}{\PYGZdq{}}\PYG{p}{,}\PYG{n}{sspace}\PYG{p}{,}\PYG{n}{x}\PYG{p}{)}
\end{Verbatim}
which yields
\[
\begin{aligned}
Z(x,\theta,\bar{\theta})=\;&
Z_{(0,0)}(x)
+Z_{(1,0)}^{\alpha}(x)\,\theta_{\alpha}
+Z_{(0,1)\dot{\alpha}}(x)\,\bar{\theta}^{\dot{\alpha}}
+Z_{(2,0)}^{\alpha\beta}(x)\,
\theta_{\alpha}\theta_{\beta}
+Z_{(0,2)\dot{\alpha}\dot{\beta}}(x)\,
\bar{\theta}^{\dot{\alpha}}\bar{\theta}^{\dot{\beta}}
\\[1mm]
&+Z_{(1,1)}^{\alpha}{}_{\dot{\alpha}}(x)\,
\theta_{\alpha}\bar{\theta}^{\dot{\alpha}}
+Z_{(2,1)}^{\alpha\beta}{}_{\dot{\alpha}}(x)\,
\theta_{\alpha}\theta_{\beta}\bar{\theta}^{\dot{\alpha}}
\\[1mm]
&+Z_{(1,2)}^{\alpha}{}_{\dot{\alpha}\dot{\beta}}(x)\,
\theta_{\alpha}\bar{\theta}^{\dot{\alpha}}\bar{\theta}^{\dot{\beta}}
+Z_{(2,2)}^{\alpha\beta}{}_{\dot{\alpha}\dot{\beta}}(x)\,
\theta_{\alpha}\theta_{\beta}
\bar{\theta}^{\dot{\alpha}}\bar{\theta}^{\dot{\beta}},
\end{aligned}
\]
the full 9 terms of the naive expansion.
This is one choice of superfield expansion. In practice, it is more useful to expand the superfield in terms of independent structures constructed from group invariants. For example, consider again the chiral part of $4D$ $\mathcal{N}=2$ superspace. At $\mathcal{O}(\theta^2)$ in the expansion, we have the term
\begin{equation}
     X^{\alpha\beta}{}_{ij\,(2)}(x)\,
\theta_{\alpha}{}^{i}\theta_{\beta}{}^{j}.
\end{equation}
Due to the fact that there exists a totally antisymmetric symbol $\epsilon$ for both index types, this term can be reduced to
\begin{equation}
     X^{\alpha\beta}(x) \theta_{\alpha\beta}+X_{ij}(x)\theta^{ij},
\end{equation}
where
\begin{equation}
    \theta_{\alpha\beta}=\epsilon_{ij}\theta_{\alpha}{}^{i}\theta_{\beta}{}^{j}, \quad \theta^{ij}=\epsilon^{\alpha\beta}\theta_{\alpha}{}^{i}\theta_{\beta}{}^{j}.
\end{equation}
This allows the field content to be seen more transparently: the 6 components of this superfield correspond to a self-dual two form $X^{\alpha\beta}(x)$ and scalar triplet $X_{ij}(x)$. Additionally, at $\mathcal{O}(\theta^4)$, we should expect the coefficient field to be a scalar, since each index must take a different value in the product due to the anticommutativity of the Grassmann variables
\begin{equation}
      X^{\alpha\beta\gamma\delta}{}_{ijkl\,(4)}(x)\,
\theta_{\alpha}{}^{i}\theta_{\beta}{}^{j}\theta_{\gamma}{}^{k}\theta_{\delta}{}^{l}= X^{1212}{}_{1122\,(4)}(x)\,
\theta_{1}{}^{1}\theta_{2}{}^{1}\theta_{1}{}^{2}\theta_{2}{}^{2},
\end{equation}
and so the true field should take a much simpler form.
Thus, it can be desirable to work with a basis of independent Grassmann structures when performing the superfield expansion, where the structures are constructed from the appropriate invariant tensors of the groups. Note that this comes at the expense of having more terms in the superfield expansion. \jil{Alakazam} implements this decomposition - a full explanation of the algorithm, which is entirely self-contained, is in the companion paper~\cite{woods2026decomposing}. This allows the package in its current iteration to decompose superfields in $D\leq 6$ spacetime dimensions in terms of the group invariants.
Applied to the chiral $4D$ $\mathcal{N}=2$ example, we can run
\begin{Verbatim}[commandchars=\\\{\},xleftmargin=\parindent,numbers=left,bgcolor=bg]
\PYG{+w}{    }\PYG{n}{generate\PYGZus{}superfield}\PYG{p}{(}\PYG{l+s}{\PYGZdq{}}\PYG{l+s}{X}\PYG{l+s}{\PYGZdq{}}\PYG{p}{,}\PYG{+w}{ }\PYG{p}{[}\PYG{n}{Tensor}\PYG{p}{(}\PYG{l+s}{\PYGZdq{}}\PYG{l+s}{θ}\PYG{l+s}{\PYGZdq{}}\PYG{p}{,}\PYG{+w}{ }\PYG{p}{[}\PYG{n}{α}\PYG{o}{=\PYGZgt{}}\PYG{n}{lower}\PYG{p}{,}\PYG{+w}{ }\PYG{n}{i}\PYG{o}{=\PYGZgt{}}\PYG{n}{upper}\PYG{p}{]}\PYG{p}{,}\PYG{+w}{ }\PYG{n}{θ}\PYG{p}{)}\PYG{p}{]}\PYG{p}{,}\PYG{+w}{ }\PYG{n}{x}\PYG{p}{)}
\end{Verbatim}
which yields the 6-term expansion
\[
\begin{aligned}
X(x,\theta)=\;&
X_{(0)}(x)
+X^{\alpha}{}_{i\,(1)}(x)\,\theta_{\alpha}{}^{i}-X^{\alpha\beta}{}_{(2)}(x)\,
\theta_{\alpha j}\theta_{\beta}{}^{j}
-X_{ij\,(2)}(x)\,
\theta^{\beta i}\theta_{\beta}{}^{j}
\\[1mm]
&+X^{\beta}{}_{k\,(3)}(x)\,
\theta^{\gamma}{}_{j}\theta_{\beta}{}^{j}
\theta_{\gamma}{}^{k}+X_{(4)}(x)\,
\theta^{\gamma}{}_{j}\theta^{\chi j}
\theta_{\gamma l}\theta_{\chi}{}^{l},
\end{aligned}
\]
where contractions of the indices using the \jil{EpsilonTensor} have been performed to reduce the number of indices attached to the coefficient fields. The remaining coefficient fields will additionally be equipped with the appropriate Young tableaux. This algorithm can also be applied to theories where the $R$-symmetry is different, for example $Sp(2n)$ or $SO(n)$, where the invariant tensors are the symplectic form and its wedge products; or, the symmetric metric and volume form, respectively. This changes the form of the superfield expansion. For example, for the $R$-symmetry group $Sp(4)$, this expansion would have $22$ terms due to the independent structures that can be produced using the symplectic form.

\subsection{Superspace Integration}\label{sec:superspace-integration}
One of the key manipulations of superspace is to integrate over the Grassmann numbers to return a function over only spacetime. For Grassmann numbers, integration is equivalent to differentiation. Thus, superspace integration is supported by \jil{Alakazam}. The only subtlety is the choice of measure. By convention, we take
\begin{equation}
    \int \mathcal{X}(x,\theta)d\theta^n=\frac{1}{n!}\epsilon^{\alpha_1\alpha_2...\alpha_n}\int\mathcal{X}(x,\theta)d\theta_{\alpha_1}d\theta_{\alpha_2}...d\theta_{\alpha_n}
\end{equation}
where the innermost integral is taken first by differentiating the integrand.
As an example, we can consider again the $4D$ $\mathcal{N}=1$ superspace and superfield
\begin{Verbatim}[commandchars=\\\{\},xleftmargin=\parindent,numbers=left,bgcolor=bg]
\PYG{+w}{    }\PYG{n}{grassmann}\PYG{o}{=}\PYG{p}{[}\PYG{n}{Tensor}\PYG{p}{(}\PYG{l+s}{\PYGZdq{}}\PYG{l+s}{θ}\PYG{l+s}{\PYGZdq{}}\PYG{p}{,}\PYG{+w}{ }\PYG{p}{[}\PYG{n}{α}\PYG{o}{=\PYGZgt{}}\PYG{n}{lower}\PYG{p}{]}\PYG{p}{,}\PYG{+w}{ }\PYG{n}{θ}\PYG{p}{)}\PYG{p}{,}\PYG{n}{Tensor}\PYG{p}{(}\PYG{l+s}{\PYGZdq{}}\PYG{l+s}{Θ}\PYG{l+s}{\PYGZdq{}}\PYG{p}{,}\PYG{+w}{ }\PYG{p}{[}\PYG{n}{αd}\PYG{o}{=\PYGZgt{}}\PYG{n}{lower}\PYG{p}{]}\PYG{p}{,}\PYG{+w}{ }\PYG{n}{θbar}\PYG{p}{)}\PYG{p}{]}
\PYG{+w}{    }\PYG{n}{sspace}\PYG{o}{=}\PYG{n}{Superspace}\PYG{p}{(}\PYG{n}{grassmann}\PYG{p}{,}\PYG{p}{[}\PYG{n}{D}\PYG{p}{,}\PYG{n}{Dbar}\PYG{p}{]}\PYG{p}{)}
\PYG{+w}{    }\PYG{n}{X}\PYG{o}{=}\PYG{n}{generate\PYGZus{}superfield}\PYG{p}{(}\PYG{l+s}{\PYGZdq{}}\PYG{l+s}{X}\PYG{l+s}{\PYGZdq{}}\PYG{p}{,}\PYG{p}{[}\PYG{n}{i}\PYG{o}{=\PYGZgt{}}\PYG{n}{upper}\PYG{p}{]}\PYG{p}{,}\PYG{+w}{ }\PYG{n}{sspace}\PYG{p}{,}\PYG{+w}{ }\PYG{n}{x}\PYG{p}{)}
\end{Verbatim}
Upon performing integration, 
\begin{Verbatim}[commandchars=\\\{\},xleftmargin=\parindent,numbers=left,bgcolor=bg]
\PYG{+w}{    }\PYG{n}{hard\PYGZus{}simplify}\PYG{p}{(}\PYG{n}{integrate}\PYG{p}{(}\PYG{n}{X}\PYG{p}{,}\PYG{n}{sspace}\PYG{p}{)}\PYG{p}{)}
\end{Verbatim}
we obtain
\[
\begin{aligned}
4 X^i_{(2,2)}(x)
+\frac{i}{2}\,
\sigma^{\mu\,\alpha}{}_{\dot{\alpha}}\,
\nabla_{\mu}
X^i_{\alpha}{}^{\dot{\alpha}}{}_{(1,1)}(x)
+\frac{1}{8}\,
\sigma^{\mu\,\alpha\dot{\alpha}}
\sigma^{\nu}{}_{\alpha\dot{\alpha}}\,
\nabla_{\nu}
\nabla_{\mu}
X^i_{(0,0)}(x).
\end{aligned}
\]

In addition to projection, we also have \jil{grassmann_coefficient}, which naively extracts a component using the weight property.

\subsection{Superspace Projection}\label{sec:superspace-projection}
Superspace projection integrates over only a subset of full superspace. This is equivalent to differentiating, and setting $\theta=0$ on the result.
To do this here, we can specify the covariant derivative(s) associated with the coordinate(s) we would like project onto, as well as the desired weight/level of the projection - for example, to extract the $(\theta=1,\bar{\theta}=2)$ part of an expression in superspace. This can be done with
\begin{Verbatim}[commandchars=\\\{\},xleftmargin=\parindent,numbers=left,bgcolor=bg]
\PYG{+w}{    }\PYG{n}{project\PYGZus{}to\PYGZus{}component}\PYG{p}{(}\PYG{n}{Tex}\PYG{o}{::}\PYG{k+kt}{TensorExpression}\PYG{p}{,}\PYG{+w}{ }\PYG{n}{covariant}\PYG{o}{::}\PYG{k+kt}{Vector}\PYG{p}{\PYGZob{}}\PYG{k+kt}{D}\PYG{p}{\PYGZcb{}}\PYG{p}{,}\PYG{+w}{ }
\PYG{+w}{        }\PYG{n}{theta}\PYG{o}{::}\PYG{k+kt}{Int}\PYG{p}{,}\PYG{+w}{ }\PYG{n}{theta\PYGZus{}bar}\PYG{o}{::}\PYG{k+kt}{Int}\PYG{p}{)}
\end{Verbatim}
where \jil{D<:CovDerivSuperType} is the superspace covariant derivative. Algorithmically, this functions the same as the full superspace integration.

\subsection{Lightcone}\label{sec:lightcone}

In 2D theories, it is often convenient to work in terms of lightcone coordinates given by
\begin{equation}
    x^{\pm \pm}=\frac{1}{\sqrt{2}}(x^0\pm x^1),
\end{equation}
where $(x^0,x^1)$ are the usual space and time coordinates. These coordinates describe the position of a particle on the lightcone, and thus are natural for describing massless particles in a 2D theory. They are the real analogue of holomorphic and antiholomorphic complex coordinates of the Euclidean plane. These coordinates are termed either left-moving ($x^{--}$), or right-moving ($x^{++}$), as a massless particle moving left ($-x^1$ direction) will be described entirely by the $x^{--}$ coordinate with $x^{++}=0$ along its trajectory, and likewise for a right-moving massless particle. The number of $\pm$ indices on objects in this coordinate system describes how the object transforms under Lorentz transformations. For example, spinorial objects will have only a single $\pm$ index, while vectors will have two $\pm$ indices, and so-on for higher spin objects.

In usual operation of Alakazam, repeated indices denote summation. When dealing with lightcone coordinates, the indices $\pm$ can be attached an arbitrary number of times to an object, and the repetitions do not denote summation, but do dictate grading. This means special treatment is needed for lightcone coordinates. These special indices come prepackaged in the submodule \jil{Alakazam.Lightcone}, which can be imported into the workspace via
\begin{Verbatim}[commandchars=\\\{\},xleftmargin=\parindent,numbers=left,bgcolor=bg]
\PYG{+w}{    }\PYG{k}{using}\PYG{+w}{ }\PYG{n}{Alakazam}\PYG{o}{.}\PYG{n}{Lightcone}
\end{Verbatim}
This brings into scope the constant variables
\begin{Verbatim}[commandchars=\\\{\},xleftmargin=\parindent,numbers=left,bgcolor=bg]
\PYG{+w}{    }\PYG{n}{lightcone\PYGZus{}indices}\PYG{o}{::}\PYG{k+kt}{IndexSet}
\end{Verbatim}
and lightcone index variables
\begin{Verbatim}[commandchars=\\\{\},xleftmargin=\parindent,numbers=left,bgcolor=bg]
\PYG{+w}{    }\PYG{n}{ℓ⁺⁻}\PYG{p}{,}\PYG{+w}{ }\PYG{n}{ℓ⁻⁺}\PYG{p}{,}\PYG{+w}{ }\PYG{n}{ℓ⁺}\PYG{p}{,}\PYG{+w}{ }\PYG{n}{ℓ⁻}\PYG{+w}{ }\PYG{o}{∈}\PYG{n}{lightcone\PYGZus{}indices}
\end{Verbatim}
corresponding to ``$\pm$'', ``$\mp$'', ``$+$'', and ``$-$''. These handles can be used explicitly to create tensors
\begin{Verbatim}[commandchars=\\\{\},xleftmargin=\parindent,numbers=left,bgcolor=bg]
\PYG{+w}{    }\PYG{n}{T}\PYG{o}{=}\PYG{n}{Tensor}\PYG{p}{(}\PYG{l+s}{\PYGZdq{}}\PYG{l+s}{T}\PYG{l+s}{\PYGZdq{}}\PYG{p}{,}\PYG{+w}{ }\PYG{p}{[}\PYG{n}{ℓ⁺}\PYG{o}{=\PYGZgt{}}\PYG{n}{upper}\PYG{p}{,}\PYG{+w}{ }\PYG{n}{ℓ⁺}\PYG{o}{=\PYGZgt{}}\PYG{n}{upper}\PYG{p}{]}\PYG{p}{)}
\PYG{+w}{    }\PYG{n}{S}\PYG{o}{=}\PYG{n}{Tensor}\PYG{p}{(}\PYG{l+s}{\PYGZdq{}}\PYG{l+s}{S}\PYG{l+s}{\PYGZdq{}}\PYG{p}{,}\PYG{+w}{ }\PYG{p}{[}\PYG{n}{ℓ⁺⁻}\PYG{+w}{ }\PYG{o}{=\PYGZgt{}}\PYG{+w}{ }\PYG{n}{upper}\PYG{p}{,}\PYG{+w}{ }\PYG{n}{ℓ⁺⁻}\PYG{+w}{ }\PYG{o}{=\PYGZgt{}}\PYG{+w}{ }\PYG{n}{upper}\PYG{p}{]}\PYG{p}{)}
\end{Verbatim}
though it is likely more convenient to use macros like
\begin{Verbatim}[commandchars=\\\{\},xleftmargin=\parindent,numbers=left,bgcolor=bg]
\PYG{+w}{    }\PYG{n}{T}\PYG{o}{=}\PYG{l+s+sa}{T}\PYG{l+s}{\PYGZdq{}}\PYG{l+s}{T\PYGZca{}++}\PYG{l+s}{\PYGZdq{}}
\PYG{+w}{    }\PYG{n}{S}\PYG{o}{=}\PYG{l+s+sa}{T}\PYG{l+s}{\PYGZdq{}}\PYG{l+s}{S\PYGZca{}±±}\PYG{l+s}{\PYGZdq{}}
\end{Verbatim}
than explicitly typing out the characters \jil{\ell\^+}, for example. The submodule also includes a function \jil{expand_lc()}, which expands terms with $\pm$ and $\mp$ indices into sums of $+$ and $-$ indices:
\begin{Verbatim}[commandchars=\\\{\},xleftmargin=\parindent,numbers=left,bgcolor=bg]
\PYG{+w}{    }\PYG{n}{expr}\PYG{o}{=}\PYG{l+s+sa}{T}\PYG{l+s}{\PYGZdq{}}\PYG{l+s}{S\PYGZca{}±±}\PYG{l+s}{\PYGZdq{}}\PYG{o}{*}\PYG{l+s+sa}{T}\PYG{l+s}{\PYGZdq{}}\PYG{l+s}{S\PYGZca{}±∓}\PYG{l+s}{\PYGZdq{}}
\PYG{+w}{    }\PYG{n+nd}{@show}\PYG{+w}{ }\PYG{n}{expand\PYGZus{}lc}\PYG{p}{(}\PYG{n}{expr}\PYG{p}{)}
\PYG{+w}{    }\PYG{c}{\PYGZsh{}   S\PYGZca{}\PYGZob{}++\PYGZcb{}S\PYGZca{}\PYGZob{}+\PYGZhy{}\PYGZcb{}+S\PYGZca{}\PYGZob{}\PYGZhy{}\PYGZhy{}\PYGZcb{}S\PYGZca{}\PYGZob{}\PYGZhy{}+\PYGZcb{}}
\end{Verbatim}

\section{Extended Examples}\label{sec:extended-examples}
In this section, we provide some extended examples to illustrate some of the features of Alakazam. We have aimed to keep the code somewhat compact, while retaining as much readability as possible. We provide the .jl scripts as supplementary material. The first example is a scattering amplitude calculation. It uses mainly trace reduction rules and pattern matching. The second example computes an operator in superspace. It uses mainly superspace integration and integration by parts functionality.

\subsection{Extended Example: Compton Scattering}\label{sec:extended-example-compton-scattering}
Consider the tree-level Compton scattering process in spinor QED,
\[
    \gamma(k_1)\,e^-(k_3)\rightarrow \gamma(k_2)\,e^-(k_4),
\]
where the electron is coupled to the photon through the covariant derivative
\(D_\mu=\partial_\mu+ieA_\mu\) acting on a Dirac field.  The minimal coupling is
linear in \(A_\mu\), so the amplitude receives contributions from exactly two
diagrams, corresponding to electron propagators in the \(s\)- and \(u\)-channels.
Each carries two vertices, and \(e\) therefore enters as an overall factor of
\(e^4\) in the squared amplitude. We label the four-momenta \(k_1,\dots,k_4\) throughout, with
\(k_1,k_2\) corresponding to the photons and \(k_3,k_4\) to the electrons, and
we retain the electron mass \(m\).

\begin{figure}[H]
    \centering
\begin{tikzpicture}[scale=0.8]

  \begin{scope}[xshift=0cm]
    \begin{feynman}
      \vertex (a1) at (-2, 1.1) {\(\gamma(k_1)\)};
      \vertex (a2) at (-2,-1.1) {\(e^-(k_3)\)};
      \vertex (v1) at ( 0, 0);
      \vertex (v2) at ( 2, 0);
      \vertex (b1) at ( 4, 1.1) {\(\gamma(k_2)\)};
      \vertex (b2) at ( 4,-1.1) {\(e^-(k_4)\)};

      \diagram* {
        (a1) -- [photon] (v1),
        (a2) -- [fermion] (v1),
        (v1) -- [fermion, edge label=\(e^-\)] (v2),
        (v2) -- [photon] (b1),
        (v2) -- [fermion] (b2),
      };

      \node at (1, -1.9) {\((k_1+k_3)^2-m^2 = 2s_1\)};
      \node at (1,  1.9) {\textbf{(a) $s$-channel}};
    \end{feynman}
  \end{scope}

  \begin{scope}[xshift=7.5cm]
    \begin{feynman}
      \vertex (a1) at (-2, 1.1) {\(\gamma(k_2)\)};
      \vertex (a2) at (-2,-1.1) {\(e^-(k_3)\)};
      \vertex (v1) at ( 0, 0);
      \vertex (v2) at ( 2, 0);
      \vertex (b1) at ( 4, 1.1) {\(\gamma(k_1)\)};
      \vertex (b2) at ( 4,-1.1) {\(e^-(k_4)\)};

      \diagram* {
        (v1) -- [photon] (a1),
        (a2) -- [fermion] (v1),
        (v1) -- [fermion, edge label=\(e^-\)] (v2),
        (b1) -- [photon] (v2),
        (v2) -- [fermion] (b2),
      };

      \node at (1, -1.9) {\((k_3-k_2)^2-m^2 = -2s_2\)};
      \node at (1,  1.9) {\textbf{(b) $u$-channel (crossed)}};
    \end{feynman}
  \end{scope}
\end{tikzpicture}
\caption{Tree level contributions to Compton scattering in spinor QED.}
\label{figure:fdiag}
\end{figure}
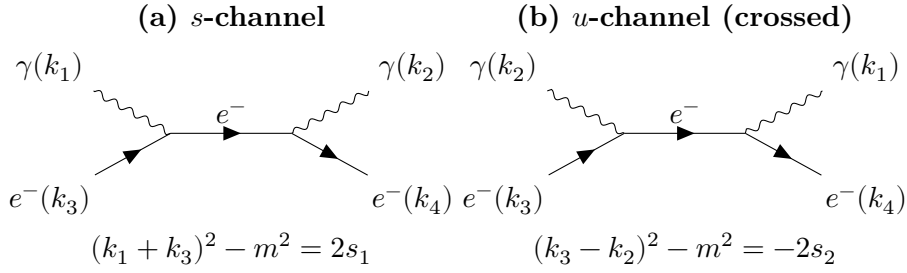

Using the QED vertex
\[
    V^\mu=-ie\gamma^\mu
\]
and the fermion propagator \(S(p)=i(\slashed{p}+m)/(p^2-m^2)\), the amplitude
factorises as
\(\mathcal{M}=\varepsilon_\mu(k_1)\varepsilon^\ast_\nu(k_2)\,
\bar{u}(k_4)\,\Gamma^{\mu\nu}\,u(k_3)\), with the vertex structure
\[
    \Gamma^{\mu\nu}
    =-ie^2\left[
      \frac{\gamma^\nu(\slashed{k}_3+\slashed{k}_1+m)\gamma^\mu}{2s_1}
     -\frac{\gamma^\mu(\slashed{k}_3-\slashed{k}_2+m)\gamma^\nu}{2s_2}
    \right].
\]
Summing over the electron spins with
\(\sum_s u(p)\bar{u}(p)=\slashed{p}+m\) and over the photon polarisations in
Feynman gauge with \(\sum_\lambda\varepsilon_\mu\varepsilon^\ast_{\mu'}\to
-g_{\mu\mu'}\), the squared matrix element collapses to a single Dirac trace,
\[
    \sum|\mathcal{M}|^2
    =\mathrm{tr}\!\left[(\slashed{k}_4+m)\,\Gamma^{\mu\nu}\,
      (\slashed{k}_3+m)\,\bar{\Gamma}_{\mu\nu}\right],
    \qquad
    \bar{\Gamma}=\gamma^0\Gamma^\dagger\gamma^0,
\]
where the Dirac conjugate \(\bar{\Gamma}\) reverses the order of the gamma
matrices, and the two polarisation sums contribute \((-1)^2=+1\).  Expanding the
product gives traces of up to eight gamma matrices.

Only two scalar products are independent, and we write them as
\[
    s_1=k_1\cdot k_3 ,\qquad s_2=k_2\cdot k_3 ,
\]
with a single subscript to keep them distinct from the Mandelstam variables, to
which they are related by \(s-m^2=2s_1\), \(u-m^2=-2s_2\) and
\(t=-2(s_1-s_2)\).  The external legs satisfy the on-shell conditions
\(k_1^2=k_2^2=0\) and \(k_3^2=k_4^2=m^2\), together with momentum conservation
\(k_4=k_1+k_3-k_2\); imposing \(k_4^2=m^2\) on the latter fixes the remaining
product as \(k_1\cdot k_2=s_1-s_2\).  The following script constructs the
amplitude, evaluates the trace, and applies these relations to obtain an expression equal to the averaged squared matrix element
\[
    \frac{1}{4}\sum|\mathcal{M}|^2
    =2e^4\left[
      \frac{s_1}{s_2}+\frac{s_2}{s_1}
      +2m^2\!\left(\frac{1}{s_1}-\frac{1}{s_2}\right)
      +m^4\!\left(\frac{1}{s_1}-\frac{1}{s_2}\right)^{\!2}
    \right],
\]
which is the Klein--Nishina result; dividing by the four initial spin and
polarisation states reproduces the standard averaged expression.

\begin{Verbatim}[commandchars=\\\{\},xleftmargin=\parindent,numbers=left,bgcolor=bg]
\PYG{k}{using}\PYG{+w}{ }\PYG{n}{Alakazam}\PYG{p}{,}\PYG{+w}{ }\PYG{n}{Symbolics}
\PYG{n}{st}\PYG{+w}{ }\PYG{o}{=}\PYG{+w}{ }\PYG{n}{IndexSet}\PYG{p}{(}\PYG{l+s}{\PYGZdq{}}\PYG{l+s}{spacetime}\PYG{l+s}{\PYGZdq{}}\PYG{p}{,}\PYG{+w}{ }\PYG{l+m+mi}{4}\PYG{p}{,}\PYG{+w}{ }\PYG{n}{upper}\PYG{p}{)}
\PYG{n+nd}{@indices}\PYG{+w}{ }\PYG{n}{μ}\PYG{o}{:}\PYG{n}{ν}\PYG{+w}{ }\PYG{o}{∈}\PYG{+w}{ }\PYG{n}{st}\PYG{+w}{ }\PYG{p}{;}\PYG{+w}{ }\PYG{n+nd}{@indices}\PYG{+w}{ }\PYG{n}{a}\PYG{o}{:}\PYG{n}{d}\PYG{+w}{ }\PYG{o}{∈}\PYG{+w}{ }\PYG{n}{st}\PYG{+w}{ }\PYG{p}{;}\PYG{+w}{ }\PYG{n+nd}{@indices}\PYG{+w}{ }\PYG{n}{f}\PYG{o}{:}\PYG{n}{k}\PYG{+w}{ }\PYG{o}{∈}\PYG{+w}{ }\PYG{n}{st}
\PYG{n+nd}{@syms}\PYG{+w}{ }\PYG{n}{m}\PYG{+w}{ }\PYG{n}{e}\PYG{+w}{ }\PYG{n}{s1}\PYG{+w}{ }\PYG{n}{s2}\PYG{+w}{   }\PYG{c}{\PYGZsh{}s1 = k₁·k₃, s2 = k₂·k₃ (scalar invars)}
\PYG{n}{γ}\PYG{p}{(}\PYG{n}{i}\PYG{p}{,}\PYG{+w}{ }\PYG{n}{p}\PYG{o}{=}\PYG{n}{upper}\PYG{p}{)}\PYG{+w}{ }\PYG{o}{=}\PYG{+w}{ }\PYG{n}{Gamma}\PYG{p}{(}\PYG{n}{i}\PYG{+w}{ }\PYG{o}{=\PYGZgt{}}\PYG{+w}{ }\PYG{n}{p}\PYG{p}{)}
\PYG{n}{slash}\PYG{p}{(}\PYG{n}{n}\PYG{p}{,}\PYG{n}{i}\PYG{p}{)}\PYG{+w}{ }\PYG{o}{=}\PYG{+w}{ }\PYG{n}{Tensor}\PYG{p}{(}\PYG{l+s}{\PYGZdq{}}\PYG{l+s}{k}\PYG{l+s}{\PYGZdq{}}\PYG{p}{,}\PYG{n}{i}\PYG{o}{=\PYGZgt{}}\PYG{n}{upper}\PYG{p}{,}\PYG{+w}{ }\PYG{l+s}{\PYGZdq{}}\PYG{l+s+si}{\PYGZdl{}n}\PYG{l+s}{\PYGZdq{}}\PYG{+w}{ }\PYG{o}{=\PYGZgt{}}\PYG{+w}{ }\PYG{n}{lower}\PYG{p}{)}\PYG{o}{*}\PYG{n}{γ}\PYG{p}{(}\PYG{n}{i}\PYG{p}{,}\PYG{n}{lower}\PYG{p}{)}\PYG{c}{\PYGZsh{} k̸ₙ}
\PYG{n}{Id}\PYG{o}{=}\PYG{n}{IdMatrix}\PYG{p}{(}\PYG{n}{spinor\PYGZus{}indexset}\PYG{p}{(}\PYG{n}{st}\PYG{p}{)}\PYG{p}{)}\PYG{c}{\PYGZsh{} the spinor space tr runs over, not st itself}
\PYG{c}{\PYGZsh{} s and u channel vertex structs, denoms (k₃+k₁)²\PYGZhy{}m² = 2s1 and (k₃\PYGZhy{}k₂)²\PYGZhy{}m² = \PYGZhy{}2s2}
\PYG{n}{Γ}\PYG{o}{=}\PYG{o}{\PYGZhy{}}\PYG{n+nb}{im}\PYG{o}{*}\PYG{n}{e}\PYG{o}{\PYGZca{}}\PYG{l+m+mi}{2}\PYG{o}{*}\PYG{p}{(}\PYG{n}{γ}\PYG{p}{(}\PYG{n}{ν}\PYG{p}{)}\PYG{o}{*}\PYG{p}{(}\PYG{n}{slash}\PYG{p}{(}\PYG{l+m+mi}{3}\PYG{p}{,}\PYG{+w}{ }\PYG{n}{a}\PYG{p}{)}\PYG{o}{+}\PYG{n}{slash}\PYG{p}{(}\PYG{l+m+mi}{1}\PYG{p}{,}\PYG{+w}{ }\PYG{n}{b}\PYG{p}{)}\PYG{+w}{ }\PYG{o}{+}\PYG{+w}{ }\PYG{n}{m}\PYG{o}{*}\PYG{n}{Id}\PYG{p}{)}\PYG{o}{*}\PYG{n}{γ}\PYG{p}{(}\PYG{n}{μ}\PYG{p}{)}\PYG{o}{*}\PYG{p}{(}\PYG{l+m+mi}{1}\PYG{o}{//}\PYG{l+m+mi}{2}\PYG{p}{)}\PYG{+w}{ }\PYG{o}{/}\PYG{+w}{ }\PYG{n}{s1}\PYG{+w}{ }\PYG{o}{\PYGZhy{}}
\PYG{+w}{              }\PYG{n}{γ}\PYG{p}{(}\PYG{n}{μ}\PYG{p}{)}\PYG{o}{*}\PYG{p}{(}\PYG{n}{slash}\PYG{p}{(}\PYG{l+m+mi}{3}\PYG{p}{,}\PYG{+w}{ }\PYG{n}{c}\PYG{p}{)}\PYG{o}{\PYGZhy{}}\PYG{n}{slash}\PYG{p}{(}\PYG{l+m+mi}{2}\PYG{p}{,}\PYG{+w}{ }\PYG{n}{d}\PYG{p}{)}\PYG{+w}{ }\PYG{o}{+}\PYG{+w}{ }\PYG{n}{m}\PYG{o}{*}\PYG{n}{Id}\PYG{p}{)}\PYG{o}{*}\PYG{n}{γ}\PYG{p}{(}\PYG{n}{ν}\PYG{p}{)}\PYG{o}{*}\PYG{p}{(}\PYG{l+m+mi}{1}\PYG{o}{//}\PYG{l+m+mi}{2}\PYG{p}{)}\PYG{+w}{ }\PYG{o}{/}\PYG{+w}{ }\PYG{n}{s2}\PYG{p}{)}
\PYG{n}{Γbar}\PYG{o}{=}\PYG{n+nb}{im}\PYG{o}{*}\PYG{n}{e}\PYG{o}{\PYGZca{}}\PYG{l+m+mi}{2}\PYG{o}{*}\PYG{p}{(}\PYG{n}{γ}\PYG{p}{(}\PYG{n}{μ}\PYG{p}{,}\PYG{n}{lower}\PYG{p}{)}\PYG{o}{*}\PYG{p}{(}\PYG{n}{slash}\PYG{p}{(}\PYG{l+m+mi}{3}\PYG{p}{,}\PYG{+w}{ }\PYG{n}{f}\PYG{p}{)}\PYG{o}{+}\PYG{n}{slash}\PYG{p}{(}\PYG{l+m+mi}{1}\PYG{p}{,}\PYG{+w}{ }\PYG{n}{g}\PYG{p}{)}\PYG{o}{+}\PYG{n}{m}\PYG{o}{*}\PYG{n}{Id}\PYG{p}{)}\PYG{o}{*}\PYG{n}{γ}\PYG{p}{(}\PYG{n}{ν}\PYG{p}{,}\PYG{n}{lower}\PYG{p}{)}\PYG{o}{*}\PYG{p}{(}\PYG{l+m+mi}{1}\PYG{o}{//}\PYG{l+m+mi}{2}\PYG{p}{)}\PYG{+w}{ }\PYG{o}{/}\PYG{+w}{ }\PYG{n}{s1}\PYG{+w}{ }\PYG{o}{\PYGZhy{}}
\PYG{+w}{              }\PYG{n}{γ}\PYG{p}{(}\PYG{n}{ν}\PYG{p}{,}\PYG{n}{lower}\PYG{p}{)}\PYG{o}{*}\PYG{p}{(}\PYG{n}{slash}\PYG{p}{(}\PYG{l+m+mi}{3}\PYG{p}{,}\PYG{+w}{ }\PYG{n}{h}\PYG{p}{)}\PYG{o}{\PYGZhy{}}\PYG{n}{slash}\PYG{p}{(}\PYG{l+m+mi}{2}\PYG{p}{,}\PYG{+w}{ }\PYG{n}{i}\PYG{p}{)}\PYG{o}{+}\PYG{n}{m}\PYG{o}{*}\PYG{n}{Id}\PYG{p}{)}\PYG{o}{*}\PYG{n}{γ}\PYG{p}{(}\PYG{n}{μ}\PYG{p}{,}\PYG{n}{lower}\PYG{p}{)}\PYG{o}{*}\PYG{p}{(}\PYG{l+m+mi}{1}\PYG{o}{//}\PYG{l+m+mi}{2}\PYG{p}{)}\PYG{+w}{ }\PYG{o}{/}\PYG{+w}{ }\PYG{n}{s2}\PYG{p}{)}

\PYG{n}{kernel}\PYG{+w}{ }\PYG{o}{=}\PYG{+w}{ }\PYG{n}{Trace}\PYG{p}{(}\PYG{p}{(}\PYG{n}{slash}\PYG{p}{(}\PYG{l+m+mi}{4}\PYG{p}{,}\PYG{+w}{ }\PYG{n}{j}\PYG{p}{)}\PYG{+w}{ }\PYG{o}{+}\PYG{+w}{ }\PYG{n}{m}\PYG{+w}{ }\PYG{o}{*}\PYG{+w}{ }\PYG{n}{Id}\PYG{p}{)}\PYG{+w}{ }\PYG{o}{*}\PYG{+w}{ }\PYG{n}{Γ}\PYG{+w}{ }\PYG{o}{*}\PYG{+w}{ }\PYG{p}{(}\PYG{n}{slash}\PYG{p}{(}\PYG{l+m+mi}{3}\PYG{p}{,}\PYG{+w}{ }\PYG{n}{k}\PYG{p}{)}\PYG{+w}{ }\PYG{o}{+}\PYG{+w}{ }\PYG{n}{m}\PYG{+w}{ }\PYG{o}{*}\PYG{+w}{ }\PYG{n}{Id}\PYG{p}{)}\PYG{+w}{ }\PYG{o}{*}\PYG{+w}{ }\PYG{n}{Γbar}\PYG{p}{)}
\PYG{n}{traced}\PYG{+w}{ }\PYG{o}{=}\PYG{+w}{ }\PYG{n}{simplify}\PYG{p}{(}\PYG{n}{eliminate\PYGZus{}trace}\PYG{p}{(}\PYG{n}{kernel}\PYG{p}{)}\PYG{p}{)}

\PYG{c}{\PYGZsh{} k₄ = k₁ + k₃ \PYGZhy{} k₂, in both index positions}
\PYG{n}{onshell}\PYG{+w}{ }\PYG{o}{=}\PYG{+w}{ }\PYG{n}{replace\PYGZus{}pattern}\PYG{p}{(}\PYG{n}{traced}\PYG{p}{,}\PYG{+w}{  }\PYG{l+s+sa}{T}\PYG{l+s}{\PYGZdq{}}\PYG{l+s}{k\PYGZus{}4\PYGZca{}?w}\PYG{l+s}{\PYGZdq{}}\PYG{p}{,}\PYG{+w}{ }\PYG{l+s+sa}{T}\PYG{l+s}{\PYGZdq{}}\PYG{l+s}{k\PYGZus{}1\PYGZca{}?w}\PYG{l+s}{\PYGZdq{}}\PYG{+w}{ }\PYG{o}{+}\PYG{+w}{ }\PYG{l+s+sa}{T}\PYG{l+s}{\PYGZdq{}}\PYG{l+s}{k\PYGZus{}3\PYGZca{}?w}\PYG{l+s}{\PYGZdq{}}\PYG{+w}{ }\PYG{o}{\PYGZhy{}}\PYG{+w}{ }\PYG{l+s+sa}{T}\PYG{l+s}{\PYGZdq{}}\PYG{l+s}{k\PYGZus{}2\PYGZca{}?w}\PYG{l+s}{\PYGZdq{}}\PYG{p}{)}
\PYG{n}{onshell}\PYG{+w}{ }\PYG{o}{=}\PYG{+w}{ }\PYG{n}{replace\PYGZus{}pattern}\PYG{p}{(}\PYG{n}{onshell}\PYG{p}{,}\PYG{+w}{ }\PYG{l+s+sa}{T}\PYG{l+s}{\PYGZdq{}}\PYG{l+s}{k\PYGZus{}4\PYGZus{}?w}\PYG{l+s}{\PYGZdq{}}\PYG{p}{,}\PYG{+w}{ }\PYG{l+s+sa}{T}\PYG{l+s}{\PYGZdq{}}\PYG{l+s}{k\PYGZus{}1\PYGZus{}?w}\PYG{l+s}{\PYGZdq{}}\PYG{+w}{ }\PYG{o}{+}\PYG{+w}{ }\PYG{l+s+sa}{T}\PYG{l+s}{\PYGZdq{}}\PYG{l+s}{k\PYGZus{}3\PYGZus{}?w}\PYG{l+s}{\PYGZdq{}}\PYG{+w}{ }\PYG{o}{\PYGZhy{}}\PYG{+w}{ }\PYG{l+s+sa}{T}\PYG{l+s}{\PYGZdq{}}\PYG{l+s}{k\PYGZus{}2\PYGZus{}?w}\PYG{l+s}{\PYGZdq{}}\PYG{p}{)}

\PYG{c}{\PYGZsh{} replacement rules for the scalar invariants \PYGZsh{} kₙ, wildcard index helper}
\PYG{n}{kx}\PYG{p}{(}\PYG{n}{n}\PYG{p}{,}\PYG{+w}{ }\PYG{n}{p}\PYG{p}{)}\PYG{+w}{ }\PYG{o}{=}\PYG{+w}{ }\PYG{n}{Tensor}\PYG{p}{(}\PYG{l+s}{\PYGZdq{}}\PYG{l+s}{k}\PYG{l+s}{\PYGZdq{}}\PYG{p}{,}\PYG{+w}{ }\PYG{p}{[}\PYG{n}{DummyPatternIndex}\PYG{p}{(}\PYG{l+s}{\PYGZdq{}}\PYG{l+s}{x}\PYG{l+s}{\PYGZdq{}}\PYG{p}{,}\PYG{+w}{ }\PYG{n}{st}\PYG{p}{)}\PYG{+w}{ }\PYG{o}{=\PYGZgt{}}\PYG{+w}{ }\PYG{n}{p}\PYG{p}{]}\PYG{p}{,}\PYG{+w}{ }\PYG{l+s}{\PYGZdq{}}\PYG{l+s+si}{\PYGZdl{}n}\PYG{l+s}{\PYGZdq{}}\PYG{+w}{ }\PYG{o}{=\PYGZgt{}}\PYG{+w}{ }\PYG{n}{lower}\PYG{p}{)}
\PYG{n}{prods}\PYG{+w}{ }\PYG{o}{=}\PYG{+w}{ }\PYG{p}{(}\PYG{p}{(}\PYG{l+m+mi}{1}\PYG{p}{,}\PYG{l+m+mi}{1}\PYG{p}{,}\PYG{l+m+mi}{0}\PYG{p}{)}\PYG{p}{,}\PYG{+w}{ }\PYG{p}{(}\PYG{l+m+mi}{2}\PYG{p}{,}\PYG{l+m+mi}{2}\PYG{p}{,}\PYG{l+m+mi}{0}\PYG{p}{)}\PYG{p}{,}\PYG{+w}{ }\PYG{p}{(}\PYG{l+m+mi}{3}\PYG{p}{,}\PYG{l+m+mi}{3}\PYG{p}{,}\PYG{n}{m}\PYG{o}{\PYGZca{}}\PYG{l+m+mi}{2}\PYG{p}{)}\PYG{p}{,}\PYG{c}{\PYGZsh{} k₁² = k₂² = 0,  k₃² = m²}
\PYG{+w}{       }\PYG{p}{(}\PYG{l+m+mi}{1}\PYG{p}{,}\PYG{l+m+mi}{3}\PYG{p}{,}\PYG{n}{s1}\PYG{p}{)}\PYG{p}{,}\PYG{+w}{ }\PYG{p}{(}\PYG{l+m+mi}{2}\PYG{p}{,}\PYG{l+m+mi}{3}\PYG{p}{,}\PYG{n}{s2}\PYG{p}{)}\PYG{p}{,}\PYG{+w}{ }\PYG{p}{(}\PYG{l+m+mi}{1}\PYG{p}{,}\PYG{l+m+mi}{2}\PYG{p}{,}\PYG{n}{s1}\PYG{o}{\PYGZhy{}}\PYG{n}{s2}\PYG{p}{)}\PYG{p}{)}\PYG{c}{\PYGZsh{} k₁·k₃ = s1, k₂·k₃ = s2, k₁·k₂ = s1\PYGZhy{}s2}
\PYG{n}{rules}\PYG{+w}{ }\PYG{o}{=}\PYG{+w}{ }\PYG{p}{[}\PYG{p}{(}\PYG{n}{kx}\PYG{p}{(}\PYG{n}{x}\PYG{p}{,}\PYG{n}{p}\PYG{p}{)}\PYG{o}{*}\PYG{n}{kx}\PYG{p}{(}\PYG{n}{y}\PYG{p}{,}\PYG{n}{op\PYGZus{}pos}\PYG{p}{(}\PYG{n}{p}\PYG{p}{)}\PYG{p}{)}\PYG{p}{,}\PYG{+w}{ }\PYG{n}{r}\PYG{p}{)}\PYG{+w}{ }\PYG{k}{for}\PYG{+w}{ }\PYG{p}{(}\PYG{n}{x}\PYG{p}{,}\PYG{n}{y}\PYG{p}{,}\PYG{n}{r}\PYG{p}{)}\PYG{o}{∈}\PYG{n}{prods}\PYG{+w}{ }\PYG{k}{for}\PYG{+w}{ }\PYG{n}{p∈}\PYG{p}{(}\PYG{n}{lower}\PYG{p}{,}\PYG{n}{upper}\PYG{p}{)}\PYG{p}{]}
\PYG{c}{\PYGZsh{} replace momenta by the scalar invariants}
\PYG{k}{for}\PYG{+w}{  }\PYG{p}{(}\PYG{n}{pat}\PYG{p}{,}\PYG{+w}{ }\PYG{n}{rep}\PYG{p}{)}\PYG{+w}{ }\PYG{k}{in}\PYG{+w}{ }\PYG{n}{rules}
\PYG{+w}{    }\PYG{k}{global}\PYG{+w}{ }\PYG{n}{onshell}\PYG{+w}{ }\PYG{o}{=}\PYG{+w}{ }\PYG{n}{replace\PYGZus{}pattern}\PYG{p}{(}\PYG{n}{onshell}\PYG{p}{,}\PYG{+w}{ }\PYG{n}{pat}\PYG{p}{,}\PYG{+w}{ }\PYG{n}{rep}\PYG{p}{)}
\PYG{k}{end}
\PYG{n}{length}\PYG{p}{(}\PYG{n}{summands}\PYG{p}{(}\PYG{n}{onshell}\PYG{p}{)}\PYG{p}{)}\PYG{o}{==}\PYG{l+m+mi}{1}\PYG{+w}{ }\PYG{o}{||}\PYG{+w}{ }\PYG{n}{error}\PYG{p}{(}\PYG{l+s}{\PYGZdq{}}\PYG{l+s}{Simplify failed.}\PYG{l+s}{\PYGZdq{}}\PYG{p}{)}\PYG{+w}{ }\PYG{p}{;}\PYG{+w}{ }\PYG{n}{M2}\PYG{o}{=}\PYG{n}{simplify}\PYG{p}{(}\PYG{n}{onshell}\PYG{p}{)}
\PYG{n}{println}\PYG{p}{(}\PYG{l+s}{\PYGZdq{}}\PYG{l+s}{¼Σ|M|²      = }\PYG{l+s}{\PYGZdq{}}\PYG{p}{,}\PYG{+w}{ }\PYG{n}{simplify}\PYG{p}{(}\PYG{n}{expand}\PYG{p}{(}\PYG{n}{first\PYGZus{}coef}\PYG{p}{(}\PYG{l+m+mi}{1}\PYG{o}{/}\PYG{l+m+mi}{4}\PYG{o}{*}\PYG{n}{M2}\PYG{p}{)}\PYG{p}{)}\PYG{p}{)}\PYG{p}{)}
\PYG{n}{kn}\PYG{+w}{ }\PYG{o}{=}\PYG{+w}{ }\PYG{l+m+mi}{2}\PYG{n}{e}\PYG{o}{\PYGZca{}}\PYG{l+m+mi}{4}\PYG{o}{*}\PYG{p}{(}\PYG{n}{s1}\PYG{o}{/}\PYG{n}{s2}\PYG{+w}{ }\PYG{o}{+}\PYG{+w}{ }\PYG{n}{s2}\PYG{o}{/}\PYG{n}{s1}\PYG{+w}{ }\PYG{o}{+}\PYG{+w}{ }\PYG{l+m+mi}{2}\PYG{n}{m}\PYG{o}{\PYGZca{}}\PYG{l+m+mi}{2}\PYG{o}{*}\PYG{p}{(}\PYG{l+m+mi}{1}\PYG{o}{/}\PYG{n}{s1}\PYG{o}{\PYGZhy{}}\PYG{l+m+mi}{1}\PYG{o}{/}\PYG{n}{s2}\PYG{p}{)}\PYG{+w}{ }\PYG{o}{+}\PYG{+w}{ }\PYG{n}{m}\PYG{o}{\PYGZca{}}\PYG{l+m+mi}{4}\PYG{o}{*}\PYG{p}{(}\PYG{l+m+mi}{1}\PYG{o}{/}\PYG{n}{s1}\PYG{o}{\PYGZhy{}}\PYG{l+m+mi}{1}\PYG{o}{/}\PYG{n}{s2}\PYG{p}{)}\PYG{o}{\PYGZca{}}\PYG{l+m+mi}{2}\PYG{p}{)}
\PYG{n}{println}\PYG{p}{(}\PYG{l+s}{\PYGZdq{}}\PYG{l+s}{Our form\PYGZhy{} Klein\PYGZhy{}Nishina = }\PYG{l+s}{\PYGZdq{}}\PYG{p}{,}\PYG{+w}{ }\PYG{n}{simplify}\PYG{p}{(}\PYG{n}{expand}\PYG{p}{(}\PYG{n}{first\PYGZus{}coef}\PYG{p}{(}\PYG{l+m+mi}{1}\PYG{o}{/}\PYG{l+m+mi}{4}\PYG{o}{*}\PYG{n}{M2}\PYG{p}{)}\PYG{+w}{ }\PYG{o}{\PYGZhy{}}\PYG{+w}{ }\PYG{n}{kn}\PYG{p}{)}\PYG{p}{)}\PYG{p}{)}
\end{Verbatim}

A few details of the construction are worth noting for anyone adapting the script. The momenta are the single tensor \jil{k} carrying \(1,\dots,4\) as labels, so
\jil{T"k_1^i"} reads as \(k_1^{\,i}\) does on paper. Labels take part in matching, and a pattern naming one matches only that label. The substitution rules are built with the constructor in line 22, rather than the macro, because they are generated in a loop over the momenta and over index position, and a string macro takes a literal. Note also that both index orders are required: a pattern fixes which slot is raised, so \(k_{1\,i}k_2^{\,i}\) and \(k_1^{\,i}k_{2\,i}\) are distinct patterns even though they denote the same scalar product, which is what is seen in the construction of \jil{rules} in line 25.

\subsection{Extended Example: Supersymmetric TTBar deformations}\label{sec:ttbarexample}
In two dimensional Quantum Field Theory, there exists a distinguished, one-parameter class of deformations generated by the stress-energy tensor
\begin{equation}
    \frac{\del \mathcal{L}^{(\lambda)}}{\del \lambda}=-\frac{1}{\pi^2}\ttb^{(\lambda)}.
\end{equation}
where
\begin{equation}
      \ttb(z)\equiv-\pi^2\det T_{\mu\nu}=4\pi^2(T_{zz}T_{\Bar{z}\Bar{z}}-T_{z\Bar{z}}^2)\equiv T(z)\Bar{T}(z)-\Theta(z)^2
\end{equation}
This deforming operator is unique in the fact that it is irrelevant, solvable, and well-defined at the quantum level. Physical quantities in the deformed theory, like the $S$ matrix and energy spectrum, can all be computed exactly from the undeformed theory~\cite{zamolodchikov2004ttbar,bonelli2018ttdeformations,jiang2021pedagogical}.
In supersymmetric theories, the stress-energy tensor lives as one current inside a supercurrent multiplet. It is natural to ask whether supersymmetric theories possess a more general ``supercurrent-squared'' deformation, which reduces to $\ttb$ on-shell, up to total derivatives. This has been answered in the affirmative for  $\mathcal{N}=(0,1)$, $\mathcal{N}=(1,1)$, $\mathcal{N}=(0,2)$ , and $\mathcal{N}=(2,2)$ supersymmetry~\cite{Baggio_2019,Jiang2dn02,ferko2019susy4Dand2D,Chang_2020,ferko2021ttbarSUSY}. 
As an illustration we can use Alakazam to construct the known operator for $\mathcal{N}=(0,2)$ supersymmetry. In this case, we consider the theory in superspace parametrised by $\{x^{\pm\pm},\theta^+,\bar{\theta}^+\}$. These have associated covariant derivatives
\begin{equation}
{D}_{+}=\frac{\partial}{\partial \theta^{+}}-\frac{\mathrm{i}}{2} \bar{\theta}^{+} \partial_{++}, \quad \overline{{D}}_{+}=-\frac{\partial}{\partial \bar{\theta}^{+}}+\frac{\mathrm{i}}{2} \theta^{+} \partial_{++}, \quad   \left\{{D}_{+}, \overline{{D}}_{+}\right\}=\mathrm{i} \partial_{++}.
\end{equation}
In this case, we have three supercurrent superfields: a real vector, real spin-2, and a complex spinor superfield
\begin{equation}
    \mathcal{S}_{++}(x,\theta^+,\overline{\theta}^+),\quad \mathcal{T}_{----}(x,\theta^+,\overline{\theta}^+),\quad \mathcal{W}_-(x,\theta^+,\overline{\theta}^+),
\end{equation}
which are subject to the constraints~\cite{Jiang2dn02}
\begin{equation}\label{eq::2dn2noncondS}
    \del_{--}\mathcal{S}_{++}=D_+\mathcal{W}_--\overline{D}_+\overline{\mathcal{W}}_-,
\end{equation}
\begin{equation}\label{eq::2dn2nonconDT}
    D_+\mathcal{T}_{----}=\frac{1}{2}\del_{--}\overline{\mathcal{W}}_{-},
\quad  \overline{D}_+\mathcal{T}_{----}=\frac{1}{2}\del_{--}\mathcal{W}_{-},
\end{equation}
\begin{equation}
    \overline{D}_+\mathcal{W}_{-}=0, \quad
D_+\overline{\mathcal{W}}_{-}=0.
\end{equation}

Solving these equations will yield constraints on the components of the superfields. It is known that in superspace, the supercurrent-squared term is then given by~\cite{Jiang2dn02}
\begin{equation}
    \mathcal{O}_{T^2}=\mathcal{T}_{----}\mathcal{S}_{++}-\overline{\mathcal{W}}_-\mathcal{W}_-,
\end{equation}
which can then be projected to spacetime
\begin{equation}
    \mathrm{O}_{T^2}=\int d\theta d\bar{\theta}\mathcal{O}_{T^2}\Big\vert_{\theta=\bar{\theta}=0}.
\end{equation}
The question then is: is $\mathrm{O}_{T^2}=\ttb$ on-shell, modulo total derivatives?
In Alakazam, we can answer this with a reasonably simple script.
\begin{Verbatim}[commandchars=\\\{\},xleftmargin=\parindent,numbers=left,bgcolor=bg]
\PYG{k}{using}\PYG{+w}{ }\PYG{n}{Alakazam}\PYG{p}{,}\PYG{+w}{ }\PYG{n}{Alakazam}\PYG{o}{.}\PYG{n}{Lightcone}
\PYG{c}{\PYGZsh{}\PYGZsh{}\PYGZsh{}\PYGZsh{}\PYGZsh{}\PYGZsh{}\PYGZsh{}\PYGZsh{}\PYGZsh{}\PYGZsh{}\PYGZsh{}\PYGZsh{}\PYGZsh{}\PYGZsh{}\PYGZsh{}\PYGZsh{}\PYGZsh{}\PYGZsh{}\PYGZsh{} Setup \PYGZsh{}\PYGZsh{}\PYGZsh{}\PYGZsh{}\PYGZsh{}\PYGZsh{}\PYGZsh{}\PYGZsh{}\PYGZsh{}\PYGZsh{}\PYGZsh{}\PYGZsh{}\PYGZsh{}\PYGZsh{}\PYGZsh{}\PYGZsh{}\PYGZsh{}\PYGZsh{}\PYGZsh{}}
\PYG{c}{\PYGZsh{} coordinates}
\PYG{n}{x}\PYG{+w}{ }\PYG{o}{=}\PYG{+w}{ }\PYG{n}{Coordinate}\PYG{p}{(}\PYG{l+s}{\PYGZdq{}}\PYG{l+s}{x}\PYG{l+s}{\PYGZdq{}}\PYG{p}{,}\PYG{+w}{ }\PYG{p}{[}\PYG{n}{lightcone\PYGZus{}indices}\PYG{p}{,}\PYG{n}{lightcone\PYGZus{}indices}\PYG{p}{]}\PYG{p}{)}
\PYG{n}{θ}\PYG{p}{,}\PYG{n}{Θ}\PYG{o}{=}\PYG{n}{Coordinate}\PYG{p}{(}\PYG{l+s}{\PYGZdq{}}\PYG{l+s}{θ}\PYG{l+s}{\PYGZdq{}}\PYG{p}{,}\PYG{+w}{ }\PYG{n}{lightcone\PYGZus{}indices}\PYG{p}{)}\PYG{p}{,}\PYG{n}{Coordinate}\PYG{p}{(}\PYG{l+s}{\PYGZdq{}}\PYG{l+s}{̄θ}\PYG{l+s}{\PYGZdq{}}\PYG{p}{,}\PYG{+w}{ }\PYG{n}{lightcone\PYGZus{}indices}\PYG{p}{)}
\PYG{n}{θ⁺}\PYG{p}{,}\PYG{n}{Θ⁺}\PYG{o}{=}\PYG{n}{Tensor}\PYG{p}{(}\PYG{l+s}{\PYGZdq{}}\PYG{l+s}{θ}\PYG{l+s}{\PYGZdq{}}\PYG{p}{,}\PYG{n}{ℓ⁺}\PYG{o}{=\PYGZgt{}}\PYG{o}{↑}\PYG{p}{,}\PYG{n}{θ}\PYG{p}{)}\PYG{p}{,}\PYG{n}{Tensor}\PYG{p}{(}\PYG{l+s}{\PYGZdq{}}\PYG{l+s}{Θ}\PYG{l+s}{\PYGZdq{}}\PYG{p}{,}\PYG{n}{ℓ⁺}\PYG{o}{=\PYGZgt{}}\PYG{o}{↑}\PYG{p}{,}\PYG{n}{Θ}\PYG{p}{)}
\PYG{c}{\PYGZsh{} basic partial derivatives}
\PYG{n}{∂₊₊}\PYG{p}{,}\PYG{n}{∂₋₋}\PYG{o}{=}\PYG{+w}{ }\PYG{n}{PD}\PYG{p}{(}\PYG{l+s}{\PYGZdq{}}\PYG{l+s}{∇}\PYG{l+s}{\PYGZdq{}}\PYG{p}{,}\PYG{n}{x}\PYG{p}{,}\PYG{n}{ℓ⁺}\PYG{o}{=\PYGZgt{}}\PYG{o}{↓}\PYG{p}{,}\PYG{n}{ℓ⁺}\PYG{o}{=\PYGZgt{}}\PYG{o}{↓}\PYG{p}{)}\PYG{p}{,}\PYG{n}{PD}\PYG{p}{(}\PYG{l+s}{\PYGZdq{}}\PYG{l+s}{∇}\PYG{l+s}{\PYGZdq{}}\PYG{p}{,}\PYG{n}{x}\PYG{p}{,}\PYG{n}{ℓ⁻}\PYG{o}{=\PYGZgt{}}\PYG{o}{↓}\PYG{p}{,}\PYG{n}{ℓ⁻}\PYG{o}{=\PYGZgt{}}\PYG{o}{↓}\PYG{p}{)}
\PYG{n}{∂θ}\PYG{p}{,}\PYG{n}{∂Θ}\PYG{o}{=}\PYG{n}{PD}\PYG{p}{(}\PYG{l+s}{\PYGZdq{}}\PYG{l+s}{∂}\PYG{l+s}{\PYGZdq{}}\PYG{p}{,}\PYG{n}{θ}\PYG{p}{,}\PYG{n}{ℓ⁺}\PYG{o}{=\PYGZgt{}}\PYG{o}{↓}\PYG{p}{)}\PYG{p}{,}\PYG{n}{PD}\PYG{p}{(}\PYG{l+s}{\PYGZdq{}}\PYG{l+s}{𝛛}\PYG{l+s}{\PYGZdq{}}\PYG{p}{,}\PYG{n}{Θ}\PYG{p}{,}\PYG{n}{ℓ⁺}\PYG{o}{=\PYGZgt{}}\PYG{o}{↓}\PYG{p}{)}
\PYG{n+nd}{@action}\PYG{+w}{ }\PYG{n}{∂θ}\PYG{p}{(}\PYG{n}{θ⁺}\PYG{p}{)}\PYG{+w}{ }\PYG{o}{=}\PYG{+w}{ }\PYG{l+m+mi}{1}\PYG{p}{;}\PYG{+w}{ }\PYG{n+nd}{@action}\PYG{+w}{ }\PYG{n}{∂Θ}\PYG{p}{(}\PYG{n}{Θ⁺}\PYG{p}{)}\PYG{+w}{ }\PYG{o}{=}\PYG{+w}{ }\PYG{l+m+mi}{1}
\PYG{c}{\PYGZsh{} covariant derivatives}
\PYG{n}{D}\PYG{p}{,}\PYG{n}{Dbar}\PYG{o}{=}\PYG{n}{Derivative}\PYG{p}{(}\PYG{l+s}{\PYGZdq{}}\PYG{l+s}{D}\PYG{l+s}{\PYGZdq{}}\PYG{p}{,}\PYG{p}{[}\PYG{n}{x}\PYG{p}{,}\PYG{+w}{ }\PYG{n}{θ}\PYG{p}{]}\PYG{p}{,}\PYG{n}{ℓ⁺}\PYG{o}{=\PYGZgt{}}\PYG{o}{↓}\PYG{p}{)}\PYG{p}{,}\PYG{n}{Derivative}\PYG{p}{(}\PYG{l+s}{\PYGZdq{}}\PYG{l+s}{̄D}\PYG{l+s}{\PYGZdq{}}\PYG{p}{,}\PYG{p}{[}\PYG{n}{x}\PYG{p}{,}\PYG{+w}{ }\PYG{n}{Θ}\PYG{p}{]}\PYG{p}{,}\PYG{n}{ℓ⁺}\PYG{o}{=\PYGZgt{}}\PYG{o}{↓}\PYG{p}{)}
\PYG{n+nd}{@covariant}\PYG{+w}{ }\PYG{n}{D}\PYG{o}{=}\PYG{n}{∂θ}\PYG{o}{\PYGZhy{}}\PYG{n+nb}{im}\PYG{o}{//}\PYG{l+m+mi}{2}\PYG{o}{*}\PYG{n}{Θ⁺}\PYG{o}{*}\PYG{n}{∂₊₊}\PYG{p}{;}\PYG{n+nd}{@covariant}\PYG{+w}{ }\PYG{n}{Dbar}\PYG{o}{=}\PYG{o}{\PYGZhy{}}\PYG{n}{∂Θ}\PYG{o}{+}\PYG{n+nb}{im}\PYG{o}{//}\PYG{l+m+mi}{2}\PYG{o}{*}\PYG{n}{θ⁺}\PYG{o}{*}\PYG{n}{∂₊₊}
\PYG{n}{superspace}\PYG{+w}{ }\PYG{o}{=}\PYG{+w}{ }\PYG{n}{Superspace}\PYG{p}{(}\PYG{p}{[}\PYG{n}{θ⁺}\PYG{p}{,}\PYG{+w}{ }\PYG{n}{Θ⁺}\PYG{p}{]}\PYG{p}{,}\PYG{+w}{ }\PYG{p}{[}\PYG{n}{D}\PYG{p}{,}\PYG{+w}{ }\PYG{n}{Dbar}\PYG{p}{]}\PYG{p}{)}
\PYG{c}{\PYGZsh{} generate the superfields}
\PYG{n}{sf}\PYG{p}{(}\PYG{n}{n}\PYG{p}{,}\PYG{n}{inds}\PYG{p}{)}\PYG{o}{=}\PYG{n}{generate\PYGZus{}superfield\PYGZus{}naive}\PYG{p}{(}\PYG{n}{n}\PYG{p}{,}\PYG{n}{inds}\PYG{p}{,}\PYG{n}{superspace}\PYG{p}{,}\PYG{n}{x}\PYG{p}{)}
\PYG{n}{S}\PYG{p}{,}\PYG{n}{T}\PYG{o}{=}\PYG{n}{sf}\PYG{p}{(}\PYG{l+s}{\PYGZdq{}}\PYG{l+s}{𝒮}\PYG{l+s}{\PYGZdq{}}\PYG{p}{,}\PYG{+w}{ }\PYG{p}{[}\PYG{n}{ℓ⁺}\PYG{o}{=\PYGZgt{}}\PYG{o}{↓}\PYG{p}{,}\PYG{n}{ℓ⁺}\PYG{o}{=\PYGZgt{}}\PYG{o}{↓}\PYG{p}{]}\PYG{p}{)}\PYG{p}{,}\PYG{n}{sf}\PYG{p}{(}\PYG{l+s}{\PYGZdq{}}\PYG{l+s}{𝒯}\PYG{l+s}{\PYGZdq{}}\PYG{p}{,}\PYG{+w}{ }\PYG{p}{[}\PYG{p}{(}\PYG{n}{ℓ⁻}\PYG{+w}{ }\PYG{o}{=\PYGZgt{}}\PYG{+w}{ }\PYG{o}{↓}\PYG{p}{)}\PYG{+w}{ }\PYG{k}{for}\PYG{+w}{ }\PYG{n}{\PYGZus{}}\PYG{+w}{ }\PYG{k}{in}\PYG{+w}{ }\PYG{l+m+mi}{1}\PYG{o}{:}\PYG{l+m+mi}{4}\PYG{p}{]}\PYG{p}{)}
\PYG{n}{W}\PYG{p}{,}\PYG{n}{Wb}\PYG{+w}{ }\PYG{o}{=}\PYG{+w}{ }\PYG{n}{sf}\PYG{p}{(}\PYG{l+s}{\PYGZdq{}}\PYG{l+s}{𝒲}\PYG{l+s}{\PYGZdq{}}\PYG{p}{,}\PYG{+w}{ }\PYG{p}{[}\PYG{n}{ℓ⁻}\PYG{o}{=\PYGZgt{}}\PYG{o}{↓}\PYG{p}{]}\PYG{p}{)}\PYG{p}{,}\PYG{n}{sf}\PYG{p}{(}\PYG{l+s}{\PYGZdq{}}\PYG{l+s}{𝒲̄}\PYG{l+s}{\PYGZdq{}}\PYG{p}{,}\PYG{+w}{ }\PYG{p}{[}\PYG{n}{ℓ⁻}\PYG{o}{=\PYGZgt{}}\PYG{o}{↓}\PYG{p}{]}\PYG{p}{)}
\PYG{c}{\PYGZsh{} coefficient of θᵃΘᵇ in the expansion of the superfield F}
\PYG{n}{sfcoef}\PYG{p}{(}\PYG{n}{F}\PYG{p}{,}\PYG{n}{a}\PYG{p}{,}\PYG{n}{b}\PYG{p}{)}\PYG{o}{=}\PYG{n}{isTensor}\PYG{p}{(}\PYG{n}{grassmann\PYGZus{}coefficient}\PYG{p}{(}\PYG{n}{F}\PYG{p}{,}\PYG{n}{superspace}\PYG{p}{,}\PYG{p}{[}\PYG{n}{a}\PYG{p}{,}\PYG{n}{b}\PYG{p}{]}\PYG{p}{)}\PYG{p}{)}
\PYG{c}{\PYGZsh{} substitute target =\PYGZgt{} value rules via our replacement pattern}
\PYG{n}{apply\PYGZus{}rules}\PYG{p}{(}\PYG{n}{e}\PYG{p}{,}\PYG{+w}{ }\PYG{n}{rules}\PYG{p}{)}\PYG{+w}{ }\PYG{o}{=}\PYG{+w}{ }\PYG{n}{foldl}\PYG{p}{(}\PYG{p}{(}\PYG{n}{a}\PYG{p}{,}\PYG{+w}{ }\PYG{n}{r}\PYG{p}{)}\PYG{+w}{ }\PYG{o}{\PYGZhy{}\PYGZgt{}}\PYG{+w}{ }\PYG{n}{replace\PYGZus{}pattern}\PYG{p}{(}\PYG{n}{a}\PYG{p}{,}
\PYG{+w}{    }\PYG{n}{TensorTerm}\PYG{p}{(}\PYG{n}{r}\PYG{p}{[}\PYG{l+m+mi}{1}\PYG{p}{]}\PYG{p}{)}\PYG{p}{,}\PYG{+w}{ }\PYG{n}{r}\PYG{p}{[}\PYG{l+m+mi}{2}\PYG{p}{]}\PYG{p}{)}\PYG{p}{,}\PYG{n}{rules}\PYG{p}{;}\PYG{+w}{ }\PYG{n}{init}\PYG{+w}{ }\PYG{o}{=}\PYG{+w}{ }\PYG{n}{TensorExpression}\PYG{p}{(}\PYG{n}{e}\PYG{p}{)}\PYG{p}{)}

\PYG{c}{\PYGZsh{}\PYGZsh{}\PYGZsh{}\PYGZsh{}\PYGZsh{}\PYGZsh{}\PYGZsh{}\PYGZsh{}\PYGZsh{}\PYGZsh{}\PYGZsh{}\PYGZsh{}\PYGZsh{}\PYGZsh{}\PYGZsh{} Defining the Problem \PYGZsh{}\PYGZsh{}\PYGZsh{}\PYGZsh{}\PYGZsh{}\PYGZsh{}\PYGZsh{}\PYGZsh{}\PYGZsh{}\PYGZsh{}\PYGZsh{}\PYGZsh{}\PYGZsh{}\PYGZsh{}\PYGZsh{}}
\PYG{c}{\PYGZsh{} The constraints, each = 0, on the set F = \PYGZob{}𝒮, 𝒯, 𝒲, ̄𝒲\PYGZcb{}}
\PYG{n}{constraints}\PYG{+w}{ }\PYG{o}{=}\PYG{+w}{ }\PYG{p}{[}\PYG{p}{(}\PYG{n}{F}\PYG{p}{)}\PYG{o}{\PYGZhy{}\PYGZgt{}}\PYG{n}{Dbar}\PYG{p}{(}\PYG{n}{F}\PYG{p}{[}\PYG{l+m+mi}{3}\PYG{p}{]}\PYG{p}{)}\PYG{p}{,}\PYG{p}{(}\PYG{n}{F}\PYG{p}{)}\PYG{o}{\PYGZhy{}\PYGZgt{}}\PYG{n}{D}\PYG{p}{(}\PYG{n}{F}\PYG{p}{[}\PYG{l+m+mi}{4}\PYG{p}{]}\PYG{p}{)}\PYG{p}{,}\PYG{p}{(}\PYG{n}{F}\PYG{p}{)}\PYG{o}{\PYGZhy{}\PYGZgt{}}\PYG{n}{D}\PYG{p}{(}\PYG{n}{F}\PYG{p}{[}\PYG{l+m+mi}{2}\PYG{p}{]}\PYG{p}{)}\PYG{o}{\PYGZhy{}}\PYG{l+m+mi}{1}\PYG{o}{//}\PYG{l+m+mi}{2}\PYG{o}{*}\PYG{n}{∂₋₋}\PYG{p}{(}\PYG{n}{F}\PYG{p}{[}\PYG{l+m+mi}{4}\PYG{p}{]}\PYG{p}{)}\PYG{p}{,}
\PYG{+w}{    }\PYG{p}{(}\PYG{n}{F}\PYG{p}{)}\PYG{o}{\PYGZhy{}\PYGZgt{}}\PYG{n}{Dbar}\PYG{p}{(}\PYG{n}{F}\PYG{p}{[}\PYG{l+m+mi}{2}\PYG{p}{]}\PYG{p}{)}\PYG{o}{\PYGZhy{}}\PYG{l+m+mi}{1}\PYG{o}{//}\PYG{l+m+mi}{2}\PYG{o}{*}\PYG{n}{∂₋₋}\PYG{p}{(}\PYG{n}{F}\PYG{p}{[}\PYG{l+m+mi}{3}\PYG{p}{]}\PYG{p}{)}\PYG{p}{,}\PYG{p}{(}\PYG{n}{F}\PYG{p}{)}\PYG{o}{\PYGZhy{}\PYGZgt{}}\PYG{n}{∂₋₋}\PYG{p}{(}\PYG{n}{F}\PYG{p}{[}\PYG{l+m+mi}{1}\PYG{p}{]}\PYG{p}{)}\PYG{o}{\PYGZhy{}}\PYG{n}{D}\PYG{p}{(}\PYG{n}{F}\PYG{p}{[}\PYG{l+m+mi}{3}\PYG{p}{]}\PYG{p}{)}\PYG{o}{+}\PYG{n}{Dbar}\PYG{p}{(}\PYG{n}{F}\PYG{p}{[}\PYG{l+m+mi}{4}\PYG{p}{]}\PYG{p}{)}\PYG{p}{,}\PYG{p}{]}
\PYG{c}{\PYGZsh{} Helper for ith constraint for each superfield at order θʲΘᵏ}
\PYG{n}{expand}\PYG{p}{(}\PYG{n}{i}\PYG{p}{,}\PYG{n}{j}\PYG{p}{,}\PYG{n}{k}\PYG{p}{)}\PYG{o}{=}\PYG{n}{simplify}\PYG{p}{(}\PYG{n}{derive}\PYG{p}{(}\PYG{n}{apply\PYGZus{}rules}\PYG{p}{(}\PYG{n}{grassmann\PYGZus{}coefficient}\PYG{p}{(}
\PYG{+w}{    }\PYG{n}{derive}\PYG{p}{(}\PYG{n}{constraints}\PYG{p}{[}\PYG{n}{i}\PYG{p}{]}\PYG{p}{(}\PYG{p}{[}\PYG{n}{apply\PYGZus{}rules}\PYG{p}{(}\PYG{n}{F}\PYG{p}{,}\PYG{+w}{ }\PYG{n}{sols}\PYG{p}{)}\PYG{+w}{ }
\PYG{+w}{    }\PYG{k}{for}\PYG{+w}{ }\PYG{n}{F}\PYG{+w}{ }\PYG{k}{in}\PYG{+w}{ }\PYG{p}{(}\PYG{n}{S}\PYG{p}{,}\PYG{+w}{ }\PYG{n}{T}\PYG{p}{,}\PYG{+w}{ }\PYG{n}{W}\PYG{p}{,}\PYG{+w}{ }\PYG{n}{Wb}\PYG{p}{)}\PYG{p}{]}\PYG{p}{)}\PYG{p}{,}\PYG{n}{superspace}\PYG{p}{)}\PYG{p}{,}\PYG{n}{superspace}\PYG{p}{,}\PYG{+w}{ }\PYG{n}{j}\PYG{p}{,}\PYG{+w}{ }\PYG{n}{k}\PYG{p}{)}\PYG{p}{,}\PYG{n}{diffs}\PYG{p}{)}\PYG{p}{)}\PYG{p}{)}

\PYG{c}{\PYGZsh{}\PYGZsh{}\PYGZsh{}\PYGZsh{}\PYGZsh{}\PYGZsh{}\PYGZsh{}\PYGZsh{}\PYGZsh{}\PYGZsh{}\PYGZsh{}\PYGZsh{}\PYGZsh{}\PYGZsh{}\PYGZsh{}\PYGZsh{}\PYGZsh{}\PYGZsh{}\PYGZsh{} Solving \PYGZsh{}\PYGZsh{}\PYGZsh{}\PYGZsh{}\PYGZsh{}\PYGZsh{}\PYGZsh{}\PYGZsh{}\PYGZsh{}\PYGZsh{}\PYGZsh{}\PYGZsh{}\PYGZsh{}\PYGZsh{}\PYGZsh{}\PYGZsh{}\PYGZsh{}\PYGZsh{}\PYGZsh{}}
\PYG{c}{\PYGZsh{} Brute force solve the constraints wherever a component appears}
\PYG{n}{sols}\PYG{p}{,}\PYG{+w}{ }\PYG{n}{diffs}\PYG{+w}{ }\PYG{o}{=}\PYG{+w}{ }\PYG{k+kt}{Pair}\PYG{p}{[}\PYG{p}{]}\PYG{p}{,}\PYG{+w}{ }\PYG{k+kt}{Pair}\PYG{p}{[}\PYG{p}{]}
\PYG{k}{for}\PYG{+w}{ }\PYG{n}{F∈}\PYG{p}{(}\PYG{n}{S}\PYG{p}{,}\PYG{n}{T}\PYG{p}{,}\PYG{n}{W}\PYG{p}{,}\PYG{n}{Wb}\PYG{p}{)}\PYG{p}{,}\PYG{n}{i∈1}\PYG{o}{:}\PYG{n}{length}\PYG{p}{(}\PYG{n}{constraints}\PYG{p}{)}\PYG{p}{,}\PYG{n}{j∈0}\PYG{o}{:}\PYG{l+m+mi}{1}\PYG{p}{,}\PYG{n}{k∈0}\PYG{o}{:}\PYG{l+m+mi}{1}\PYG{p}{,}
\PYG{+w}{    }\PYG{n}{∂∈}\PYG{p}{(}\PYG{n+nb}{nothing}\PYG{p}{,}\PYG{+w}{ }\PYG{n}{∂₊₊}\PYG{p}{,}\PYG{+w}{ }\PYG{n}{∂₋₋}\PYG{p}{)}\PYG{p}{,}\PYG{n}{a}\PYG{+w}{ }\PYG{o}{∈}\PYG{+w}{ }\PYG{l+m+mi}{0}\PYG{o}{:}\PYG{l+m+mi}{1}\PYG{p}{,}\PYG{n}{b∈0}\PYG{o}{:}\PYG{l+m+mi}{1}
\PYG{+w}{        }\PYG{n}{target}\PYG{+w}{ }\PYG{o}{=}\PYG{+w}{ }\PYG{n}{∂}\PYG{o}{===}\PYG{n+nb}{nothing}\PYG{+w}{ }\PYG{o}{?}\PYG{+w}{ }\PYG{n}{sfcoef}\PYG{p}{(}\PYG{n}{F}\PYG{p}{,}\PYG{n}{a}\PYG{p}{,}\PYG{n}{b}\PYG{p}{)}\PYG{+w}{ }\PYG{o}{:}\PYG{+w}{ }\PYG{n}{∂}\PYG{p}{(}\PYG{n}{sfcoef}\PYG{p}{(}\PYG{n}{F}\PYG{p}{,}\PYG{n}{a}\PYG{p}{,}\PYG{n}{b}\PYG{p}{)}\PYG{p}{)}
\PYG{+w}{        }\PYG{p}{(}\PYG{n}{comp}\PYG{p}{,}\PYG{n}{val}\PYG{p}{)}\PYG{o}{=}\PYG{n}{solve\PYGZus{}linear}\PYG{p}{(}\PYG{n}{expand}\PYG{p}{(}\PYG{n}{i}\PYG{p}{,}\PYG{+w}{ }\PYG{n}{j}\PYG{p}{,}\PYG{+w}{ }\PYG{n}{k}\PYG{p}{)}\PYG{p}{,}\PYG{+w}{ }\PYG{n}{target}\PYG{p}{)}
\PYG{+w}{        }\PYG{n}{comp}\PYG{o}{===}\PYG{n+nb}{nothing}\PYG{+w}{ }\PYG{o}{||}\PYG{+w}{ }\PYG{n}{push!}\PYG{p}{(}\PYG{n}{∂}\PYG{o}{===}\PYG{n+nb}{nothing}\PYG{+w}{ }\PYG{o}{?}\PYG{+w}{ }\PYG{n}{sols}\PYG{+w}{ }\PYG{o}{:}\PYG{+w}{ }\PYG{n}{diffs}\PYG{p}{,}\PYG{n}{comp}\PYG{o}{=\PYGZgt{}}\PYG{n}{val}\PYG{p}{)}
\PYG{k}{end}
\PYG{c}{\PYGZsh{} simplify constraints, drop redundant ones}
\PYG{n}{diffs}\PYG{o}{=}\PYG{p}{[}\PYG{n}{r}\PYG{+w}{ }\PYG{k}{for}\PYG{+w}{ }\PYG{n}{r∈}\PYG{p}{(}\PYG{n}{t}\PYG{o}{=\PYGZgt{}}\PYG{n}{simplify}\PYG{p}{(}\PYG{n}{derive}\PYG{p}{(}\PYG{n}{apply\PYGZus{}rules}\PYG{p}{(}\PYG{n}{v}\PYG{p}{,}\PYG{n}{sols}\PYG{p}{)}\PYG{p}{)}\PYG{p}{)}\PYG{+w}{ }
\PYG{+w}{    }\PYG{k}{for}\PYG{+w}{ }\PYG{p}{(}\PYG{n}{t}\PYG{p}{,}\PYG{n}{v}\PYG{p}{)}\PYG{o}{∈}\PYG{n}{diffs}\PYG{p}{)}\PYG{+w}{ }\PYG{k}{if}\PYG{+w}{ }\PYG{o}{!}\PYG{n}{isZeroTens}\PYG{p}{(}\PYG{n}{simplify}\PYG{p}{(}\PYG{n}{r}\PYG{p}{[}\PYG{l+m+mi}{2}\PYG{p}{]}\PYG{o}{\PYGZhy{}}\PYG{n}{r}\PYG{p}{[}\PYG{l+m+mi}{1}\PYG{p}{]}\PYG{p}{)}\PYG{p}{)}\PYG{p}{]}
\PYG{c}{\PYGZsh{} apply the constraints, then integrate over superspace}
\PYG{n}{Ss}\PYG{p}{,}\PYG{+w}{ }\PYG{n}{Ts}\PYG{p}{,}\PYG{+w}{ }\PYG{n}{Ws}\PYG{p}{,}\PYG{+w}{ }\PYG{n}{Wbs}\PYG{+w}{ }\PYG{o}{=}\PYG{+w}{ }\PYG{p}{(}\PYG{n}{apply\PYGZus{}rules}\PYG{p}{(}\PYG{n}{F}\PYG{p}{,}\PYG{+w}{ }\PYG{n}{sols}\PYG{p}{)}\PYG{+w}{ }\PYG{k}{for}\PYG{+w}{ }\PYG{n}{F}\PYG{+w}{ }\PYG{o}{∈}\PYG{+w}{ }\PYG{p}{(}\PYG{n}{S}\PYG{p}{,}\PYG{+w}{ }\PYG{n}{T}\PYG{p}{,}\PYG{+w}{ }\PYG{n}{W}\PYG{p}{,}\PYG{+w}{ }\PYG{n}{Wb}\PYG{p}{)}\PYG{p}{)}
\PYG{n}{𝒪}\PYG{+w}{ }\PYG{o}{=}\PYG{+w}{ }\PYG{n}{∫}\PYG{p}{(}\PYG{n}{Ts}\PYG{+w}{ }\PYG{o}{*}\PYG{+w}{ }\PYG{n}{Ss}\PYG{+w}{ }\PYG{o}{\PYGZhy{}}\PYG{+w}{ }\PYG{n}{Wbs}\PYG{+w}{ }\PYG{o}{*}\PYG{+w}{ }\PYG{n}{Ws}\PYG{p}{,}\PYG{+w}{ }\PYG{n}{superspace}\PYG{p}{)}
\PYG{n}{O}\PYG{+w}{ }\PYG{o}{=}\PYG{+w}{ }\PYG{n}{reduce\PYGZus{}bp\PYGZus{}with}\PYG{p}{(}\PYG{n}{𝒪}\PYG{p}{,}\PYG{+w}{ }\PYG{n}{diffs}\PYG{p}{)}\PYG{+w}{   }\PYG{c}{\PYGZsh{} by parts, drop total derivs}

\PYG{c}{\PYGZsh{}\PYGZsh{}\PYGZsh{}\PYGZsh{}\PYGZsh{}\PYGZsh{}\PYGZsh{}\PYGZsh{}\PYGZsh{}\PYGZsh{}\PYGZsh{}\PYGZsh{}\PYGZsh{}\PYGZsh{}\PYGZsh{}\PYGZsh{}\PYGZsh{}\PYGZsh{}\PYGZsh{} Display \PYGZsh{}\PYGZsh{}\PYGZsh{}\PYGZsh{}\PYGZsh{}\PYGZsh{}\PYGZsh{}\PYGZsh{}\PYGZsh{}\PYGZsh{}\PYGZsh{}\PYGZsh{}\PYGZsh{}\PYGZsh{}\PYGZsh{}\PYGZsh{}\PYGZsh{}\PYGZsh{}\PYGZsh{}}
\PYG{n}{println}\PYG{p}{(}\PYG{l+s}{\PYGZdq{}}\PYG{l+s}{solved components (}\PYG{l+s}{\PYGZdq{}}\PYG{p}{,}\PYG{+w}{ }\PYG{n}{length}\PYG{p}{(}\PYG{n}{sols}\PYG{p}{)}\PYG{p}{,}\PYG{+w}{ }\PYG{l+s}{\PYGZdq{}}\PYG{l+s}{)}\PYG{l+s}{\PYGZdq{}}\PYG{p}{)}
\PYG{n}{foreach}\PYG{p}{(}\PYG{n}{r}\PYG{+w}{ }\PYG{o}{\PYGZhy{}\PYGZgt{}}\PYG{+w}{ }\PYG{n}{println}\PYG{p}{(}\PYG{l+s}{\PYGZdq{}}\PYG{l+s}{   }\PYG{l+s}{\PYGZdq{}}\PYG{p}{,}\PYG{+w}{ }\PYG{n}{r}\PYG{p}{[}\PYG{l+m+mi}{1}\PYG{p}{]}\PYG{p}{,}\PYG{+w}{ }\PYG{l+s}{\PYGZdq{}}\PYG{l+s}{ = }\PYG{l+s}{\PYGZdq{}}\PYG{p}{,}\PYG{+w}{ }\PYG{n}{r}\PYG{p}{[}\PYG{l+m+mi}{2}\PYG{p}{]}\PYG{p}{)}\PYG{p}{,}\PYG{+w}{ }\PYG{n}{sols}\PYG{p}{)}
\PYG{n}{println}\PYG{p}{(}\PYG{l+s}{\PYGZdq{}}\PYG{l+s}{relations (}\PYG{l+s}{\PYGZdq{}}\PYG{p}{,}\PYG{+w}{ }\PYG{n}{length}\PYG{p}{(}\PYG{n}{diffs}\PYG{p}{)}\PYG{p}{,}\PYG{+w}{ }\PYG{l+s}{\PYGZdq{}}\PYG{l+s}{)}\PYG{l+s}{\PYGZdq{}}\PYG{p}{)}
\PYG{n}{foreach}\PYG{p}{(}\PYG{n}{r}\PYG{+w}{ }\PYG{o}{\PYGZhy{}\PYGZgt{}}\PYG{+w}{ }\PYG{n}{println}\PYG{p}{(}\PYG{l+s}{\PYGZdq{}}\PYG{l+s}{   }\PYG{l+s}{\PYGZdq{}}\PYG{p}{,}\PYG{+w}{ }\PYG{n}{r}\PYG{p}{[}\PYG{l+m+mi}{1}\PYG{p}{]}\PYG{p}{,}\PYG{+w}{ }\PYG{l+s}{\PYGZdq{}}\PYG{l+s}{ = }\PYG{l+s}{\PYGZdq{}}\PYG{p}{,}\PYG{+w}{ }\PYG{n}{r}\PYG{p}{[}\PYG{l+m+mi}{2}\PYG{p}{]}\PYG{p}{)}\PYG{p}{,}\PYG{+w}{ }\PYG{n}{diffs}\PYG{p}{)}
\PYG{n}{println}\PYG{p}{(}\PYG{l+s}{\PYGZdq{}}\PYG{l+s+se}{\PYGZbs{}n}\PYG{l+s}{constrained superfields}\PYG{l+s}{\PYGZdq{}}\PYG{p}{)}
\PYG{n}{foreach}\PYG{p}{(}\PYG{p}{(}\PYG{n}{nm}\PYG{p}{,}\PYG{+w}{ }\PYG{n}{F}\PYG{p}{)}\PYG{+w}{ }\PYG{o}{\PYGZhy{}\PYGZgt{}}\PYG{+w}{ }\PYG{n}{println}\PYG{p}{(}\PYG{l+s}{\PYGZdq{}}\PYG{l+s}{  }\PYG{l+s}{\PYGZdq{}}\PYG{p}{,}\PYG{+w}{ }\PYG{n}{rpad}\PYG{p}{(}\PYG{n}{nm}\PYG{p}{,}\PYG{+w}{ }\PYG{l+m+mi}{6}\PYG{p}{)}\PYG{p}{,}\PYG{+w}{ }\PYG{l+s}{\PYGZdq{}}\PYG{l+s}{ = }\PYG{l+s}{\PYGZdq{}}\PYG{p}{,}\PYG{+w}{ }\PYG{n}{F}\PYG{p}{)}\PYG{p}{,}
\PYG{+w}{    }\PYG{p}{(}\PYG{l+s}{\PYGZdq{}}\PYG{l+s}{𝒮₊₊}\PYG{l+s}{\PYGZdq{}}\PYG{p}{,}\PYG{+w}{ }\PYG{l+s}{\PYGZdq{}}\PYG{l+s}{𝒯₋₋₋₋}\PYG{l+s}{\PYGZdq{}}\PYG{p}{,}\PYG{+w}{ }\PYG{l+s}{\PYGZdq{}}\PYG{l+s}{𝒲₋}\PYG{l+s}{\PYGZdq{}}\PYG{p}{,}\PYG{+w}{ }\PYG{l+s}{\PYGZdq{}}\PYG{l+s}{𝒲̄₋}\PYG{l+s}{\PYGZdq{}}\PYG{p}{)}\PYG{p}{,}\PYG{+w}{ }\PYG{p}{(}\PYG{n}{Ss}\PYG{p}{,}\PYG{+w}{ }\PYG{n}{Ts}\PYG{p}{,}\PYG{+w}{ }\PYG{n}{Ws}\PYG{p}{,}\PYG{+w}{ }\PYG{n}{Wbs}\PYG{p}{)}\PYG{p}{)}
\PYG{n+nd}{@show}\PYG{+w}{ }\PYG{n}{𝒪}
\PYG{n+nd}{@show}\PYG{+w}{ }\PYG{n}{O}
\end{Verbatim}

This yields the answer as the operator (after factorising manually)
\begin{equation}
    \mathrm{O}_{T^2}=- \mathscr{S} { }_{+}{ }_{+}{ }_{+}{ }_{+}{ }_{(1,1)} \mathscr{T} { }_{-}{ }_{-}{ }_{-}{ }_{-}{ }_{(0,0)} +\frac{1}{4} \left(\mathscr{W} { }_{-}{ }_{+}{ }_{(1,0)}-\overline{\mathscr{W}} { }_{-}{ }_{+}{ }_{(0,1)} \right)^2
\end{equation}
for which we identify
\begin{equation}
    \mathscr{S} { }_{+}{ }_{+}{ }_{+}{ }_{+}{ }_{(1,1)}=-T_{++++},\quad \mathscr{T} { }_{-}{ }_{-}{ }_{-}{ }_{-}{ }_{(0,0)}=T_{----},\quad \Theta=\textrm{Im}\left(\mathscr{W} { }_{-}{ }_{+}{ }_{(1,0)}\right)
\end{equation}
to see that this is of the form
\begin{equation}
    \mathrm{O}_{T^2}=T_{++++}T_{----}-\Theta^2,
\end{equation}
matching the known form in~\cite{Jiang2dn02}.

\newpage
\section{Implementation Details}\label{sec:implementation}
In this section, we touch on some of the design considerations and principles for Alakazam, and provide some indicative numbers for speed of basic structures and operations.
We mention throughout some of the limitations of our implementation. We also overview the more technical side of the \jil{hard_simplify} and pattern matching algorithms.

\subsection{Efficiency Considerations}\label{sec:efficiency-considerations}
\subsubsection{Stable Typing}
Julia has many optimisations performed by the compiler in order to help code achieve very fast speeds. The performance rests mainly on type inference and specialisation. Whenever the compiler can determine concrete types that are used in a function, it produces machine code specialised to those types which has direct memory access and no runtime dispatch. When the type cannot be inferred, the performance tends to degrade substantially. In the creation of Alakazam, we have been careful to ensure compliance where possible with typing to make the most of Julia.

In many cases, we work with \jil{Vector{T}} where \jil{T} is an abstract type, rather than concrete. Our \jil{TensorTerm}s  for example necessarily hold many concrete types, like \jil{Tensor} and \jil{Operator}, so this necessitates abstract vectors. Where such a container is iterated and each element used in several operations, we sometimes interpose a function barrier: the loop body is placed in a separate function so that Julia can specialise on the concrete type and make all work inside statically typed. This allows for optimised runtime in some cases.

Due to these optimisations tied to typing, it is desirable for any function to always return a certain fixed type. To be precise, a function should have a return type determined by the types of the inputs, rather than their values. Sometimes, when the type cannot be inferred, we can include an explicit return annotation to recover stability. This has been tested and included selectively where helpful to performance - for example, \jil{::Vector{IndexPair}} annotations increased allocations in some locations by $\sim 26\%$ and were thus removed. Annotations can pessimise when the asserted type forces a conversion at each call, which is why they
are applied only where measured to help.

This return typing constraint shapes the arithmetic interface. The sum of two tensors may vanish identically, but returning a numerical zero would make the return type depend on the value rather than the arguments, destabilising every caller. We therefore return an empty \jil{TensorExpression}, which represents zero while preserving the type. Since zero is a common object, we can store a dedicated \jil{_ZERO} constant to avoid reconstruction and have fast comparison. However, this object is exposed through an accessor \jil{zero()} rather than as a bare constant: a \jil{const} binding is tracked by the compiler as referring to that particular instance, which in our case caused worse performance when used directly. The accessor participates in dispatch. Note that this method extends \jil{Base.zero} to avoid the user having to specify \jil{Alakazam.zero()}, which would be inconvenient as such a frequently used object. As \jil{Base} has no zero-argument \jil{zero()} function, the signature is not ambiguous, and should hopefully be benign despite type piracy. Note that the cached zero is never mutated within the library, however, functionality is contingent on users also not modifying this.

In a similar vein, \jil{simplify()} overloads \jil{Symbolics.simplify()} and dispatches on \jil{Tensorial} types owned by Alakazam, and we name our basic matrix object \jil{Mat} to not collide with \jil{Base.Matrix}. 

\subsubsection{Inlining}
Another compiler optimisation is inlining. This means replacing a function call with the function's body at the call site. Rather than jumping to a separate block of code, passing arguments and returning, the compiler substitutes the instructions directly where the call appeared. The immediate saving is the call overhead itself which is small, but adds up for a function called millions of times. This can also sometimes allow the optimiser to resolve more complicated functions. Julia inlines small functions automatically, but the heuristic is conservative and an inlining attempt can be requested explicitly with \jil{@inline}. We apply it to short accessors and predicates that appear inside hot loops to try to avoid the call overhead that would otherwise dominate.

\subsubsection{Global State}
Another important observation is that Alakazam has a mutable global state. Index sets, coordinates, commutators, registered derivative
actions, and the custom sort order all live in module-level mutable dictionaries. This makes results depend on session
history and makes the package potentially unsafe to use from multiple threads. However, these global objects make sense mathematically, as these are shared properties of a particular problem, rather than individual tensors. Note that \jil{clear_global!()} exists in order to aid with reproducibility, and the global variables can be exported and imported using the JSON serialisation to realise a particular global state.

\subsubsection{\jil{TensorData}}
The main limiting factor on speed for our basic operations comes from performing excessive memory allocations - for example, creating unnecessary arrays. Thus, at the base level, Alakazam is designed to be as minimal as possible in this regard. 

To ensure adherence, all basic \jil{Tensor}s carry only an \jil{indices::Vector{IndexPair}} and a \jil{TensorData} struct. The fields of this object are
\begin{Verbatim}[commandchars=\\\{\},xleftmargin=\parindent,numbers=left,bgcolor=bg]
\PYG{+w}{    }\PYG{k}{struct} \PYG{k+kt}{TensorData}\PYG{+w}{ }\PYG{o}{\PYGZlt{}:}\PYG{+w}{ }\PYG{k+kt}{TensorDataSuperType}
\PYG{+w}{        }\PYG{n}{name}\PYG{o}{::}\PYG{k+kt}{String}
\PYG{+w}{        }\PYG{n}{function\PYGZus{}of}\PYG{o}{::}\PYG{k+kt}{Vector}\PYG{p}{\PYGZob{}}\PYG{k+kt}{Coordinate}\PYG{p}{\PYGZcb{}}
\PYG{+w}{        }\PYG{n}{weights}\PYG{o}{::}\PYG{k+kt}{Dict}\PYG{p}{\PYGZob{}}\PYG{k+kt}{String}\PYG{p}{,}\PYG{k+kt}{Number}\PYG{p}{\PYGZcb{}}
\PYG{+w}{        }\PYG{n}{tableaux}\PYG{o}{::}\PYG{k+kt}{Set}\PYG{p}{\PYGZob{}}\PYG{k+kt}{YoungTableau}\PYG{p}{\PYGZcb{}}
\PYG{+w}{        }\PYG{n}{labels}\PYG{o}{::}\PYG{k+kt}{Vector}\PYG{p}{\PYGZob{}}\PYG{k+kt}{LabelPair}\PYG{p}{\PYGZcb{}}
\PYG{+w}{        }\PYG{n}{numberfield}\PYG{o}{::}\PYG{k+kt}{NumberSet}
\PYG{+w}{        }\PYG{n}{uid}\PYG{o}{::}\PYG{k+kt}{UInt64}
\PYG{+w}{    }\PYG{k}{end}
\end{Verbatim}
Since the fields of \jil{TensorData} rarely need to change after construction (indeed, it's only usually indices that change across objects, so we store these separately), this means the data can be an immutable object, shared by copied tensors, saving on overhead. It also saves time on comparison, as these objects can be compared via identity rather than fields. Indeed, we include a \jil{uid} field to each \jil{TensorData} instance which hashes the data. Comparing these UIDs is much faster than comparing the literal objectids, about $\sim$1\,ns vs $\sim$75\,ns in rudimentary tests. Since it's a hash, there may be collisions in which case we resort to a proper equality check, though in most cases tensors are not equal so the fast rejection is a big time save.

We choose \jil{indices} as a \jil{Vector} since ordering is important, while it takes only 1 allocation vs 5 for an \jil{OrderedSet}. We can enforce uniqueness manually for indices, and specific element access is rare since we usually have to iterate over indices anyway to check properties, so there is no real loss to not have constant time access. An \jil{OrderedSet} would also be problematic for lightcone indices which can be repeated.
For \jil{function_of}, we also use a vector, as even though order is not important, the \jil{Set} overhead outweighs the cost of manually sorting as these contain usually only $0-3$ coordinates.

In many applications, a \jil{Tensor} could have an empty \jil{function_of}, indeed, if we're not doing calculus. In this case, \jil{function_of} shares a reference to a global empty set. This is safe provided that functions downstream strictly do not mutate this set. The code is designed in a way to avoid this, but this is a genuine fragility that the user must ultimately respect. Similar logic holds for the other fields, where we refer to generic empty instances when not needed.
\subsubsection{\jil{@inbounds} and Allocation}
For the creation of a \jil{TensorTerm}, which has fields
\begin{Verbatim}[commandchars=\\\{\},xleftmargin=\parindent,numbers=left,bgcolor=bg]
\PYG{+w}{  }\PYG{k}{mutable}\PYG{+w}{ }\PYG{k}{struct} \PYG{k+kt}{TensorTerm}\PYG{+w}{ }\PYG{o}{\PYGZlt{}:}\PYG{+w}{ }\PYG{k+kt}{Tensorial}
\PYG{+w}{    }\PYG{n}{terms}\PYG{o}{::}\PYG{k+kt}{Vector}\PYG{p}{\PYGZob{}}\PYG{k+kt}{TensorSuperType}\PYG{p}{\PYGZcb{}}
\PYG{+w}{    }\PYG{n}{indices}\PYG{o}{::}\PYG{k+kt}{Vector}\PYG{p}{\PYGZob{}}\PYG{k+kt}{IndexPair}\PYG{p}{\PYGZcb{}}
\end{Verbatim}
we perform checks on the lengths needed for the vectors, depending on the constructor used. In these constructions, and likewise many other cases in our basic structures, we know the vectors are of a fixed length so we can bypass explicit array bounds checking with \jil{@inbounds}, saving a little when the bounds are provably within the array. We apply this as much as possible where it is safe to do so, especially in hot functions.

Additionally, in many places in the code we know in advance the size of the vector needed. In these cases, we preallocate space for $n$ objects in the data structure, for example, \jil{Vector{T}(undef, n)}. This saves time on operations where not specifying the size would lead to multiple resizings.
We occasionally give a \jil{sizehint!} to some objects to save on repeated allocations where the size is less certain.

\subsubsection{Coefficients}

For \jil{TensorExpression}s, we store
\begin{Verbatim}[commandchars=\\\{\},xleftmargin=\parindent,numbers=left,bgcolor=bg]
\PYG{+w}{    }\PYG{k}{mutable}\PYG{+w}{ }\PYG{k}{struct} \PYG{k+kt}{TensorExpression}\PYG{+w}{ }\PYG{o}{\PYGZlt{}:}\PYG{+w}{ }\PYG{k+kt}{Tensorial}
\PYG{+w}{    }\PYG{n}{summands}\PYG{o}{::}\PYG{k+kt}{Vector}\PYG{p}{\PYGZob{}}\PYG{k+kt}{CoeffPair}\PYG{p}{\PYGZcb{}}
\end{Verbatim}
Here, \jil{CoeffPair} is
\begin{Verbatim}[commandchars=\\\{\},xleftmargin=\parindent,numbers=left,bgcolor=bg]
\PYG{+w}{    }\PYG{k}{const}\PYG{+w}{ }\PYG{n}{CoeffPair}\PYG{+w}{ }\PYG{o}{=}\PYG{+w}{ }\PYG{k+kt}{Pair}\PYG{p}{\PYGZob{}}\PYG{k+kt}{SymorNum}\PYG{p}{,}\PYG{k+kt}{TensorTerm}\PYG{p}{\PYGZcb{}}
\end{Verbatim}
were
\begin{Verbatim}[commandchars=\\\{\},xleftmargin=\parindent,numbers=left,bgcolor=bg]
\PYG{+w}{    }\PYG{k}{const}\PYG{+w}{ }\PYG{n}{SymorNum}\PYG{+w}{ }\PYG{o}{=}\PYG{+w}{ }\PYG{k+kt}{Union}\PYG{p}{\PYGZob{}}
\PYG{+w}{    }\PYG{k+kt}{SymbolicUtils}\PYG{o}{.}\PYG{k+kt}{BasicSymbolic}\PYG{p}{\PYGZob{}}\PYG{k+kt}{SymbolicUtils}\PYG{o}{.}\PYG{k+kt}{SymReal}\PYG{p}{\PYGZcb{}}\PYG{p}{,}
\PYG{+w}{    }\PYG{k+kt}{Number}\PYG{p}{\PYGZcb{}}
\end{Verbatim}
However, this is a union of a concrete type, and an abstract type for \jil{Number}. This appears to be suboptimal, given the above discussion on concrete typing. However, in practice, after testing using just a concrete subtype of \jil{Number} (for example, \jil{Complex{Rational{Int}}}) this was deemed imperceptible in performance - while having slightly fewer allocations (3-5$\%$), there were slightly worse median timings (1-5$\%$) for the concrete type due to the fact that many operations now had additional conversions at every place a coefficient entered. Thus, the union with \jil{Number} was retained for the flexibility here of allowing for generic coefficients.

\subsubsection{Copy-On-Write}
One design choice that we have made is that functions seldom manipulate \jil{TensorExpression}s in place. What is meant by this is that calling, for example, \jil{simplify()} on an expression will return a new simplified expression, rather than editing the input. For example,
\begin{Verbatim}[commandchars=\\\{\},xleftmargin=\parindent,numbers=left,bgcolor=bg]
\PYG{+w}{    }\PYG{n}{expr}\PYG{o}{=}\PYG{l+s+sa}{T}\PYG{l+s}{\PYGZdq{}}\PYG{l+s}{A\PYGZca{}(μν)}\PYG{l+s}{\PYGZdq{}}\PYG{o}{+}\PYG{l+s+sa}{T}\PYG{l+s}{\PYGZdq{}}\PYG{l+s}{A\PYGZca{}(νμ)}\PYG{l+s}{\PYGZdq{}}
\PYG{+w}{    }\PYG{n+nd}{@show}\PYG{+w}{ }\PYG{n}{expr}
\PYG{+w}{    }\PYG{n+nd}{@show}\PYG{+w}{ }\PYG{n}{canonicalise}\PYG{p}{(}\PYG{n}{expr}\PYG{p}{)}
\PYG{+w}{    }\PYG{n+nd}{@show}\PYG{+w}{ }\PYG{n}{expr}
\end{Verbatim}
will yield
\begin{Verbatim}[commandchars=\\\{\},xleftmargin=\parindent,numbers=left,bgcolor=bg]
\PYG{+w}{    }\PYG{c}{\PYGZsh{}   expr = A\PYGZca{}\PYGZob{}μν\PYGZcb{}+A\PYGZca{}\PYGZob{}νμ\PYGZcb{}}
\PYG{+w}{    }\PYG{c}{\PYGZsh{}   canonicalise(expr) = 2A\PYGZca{}\PYGZob{}μν\PYGZcb{}}
\PYG{+w}{    }\PYG{c}{\PYGZsh{}   expr = A\PYGZca{}\PYGZob{}μν\PYGZcb{}+A\PYGZca{}\PYGZob{}νμ\PYGZcb{}}
\end{Verbatim}
This allows one to retain original expressions for later comparison or operations.

The result of this is that symbolic manipulation copies aggressively. If the result of multiplication of two tensors refers to the original operands, a subsequent index substitution on the product would alter the original factors, or vice versa. A naive implementation would therefore deep-copy every field of each tensor at every step. This is redundant however, since the index vector is by convention the only mutable part of a \jil{Tensor}. The naive copy dominated early benchmarks. 

\begin{figure}
    \centering
    \includegraphics[width=0.95\linewidth]{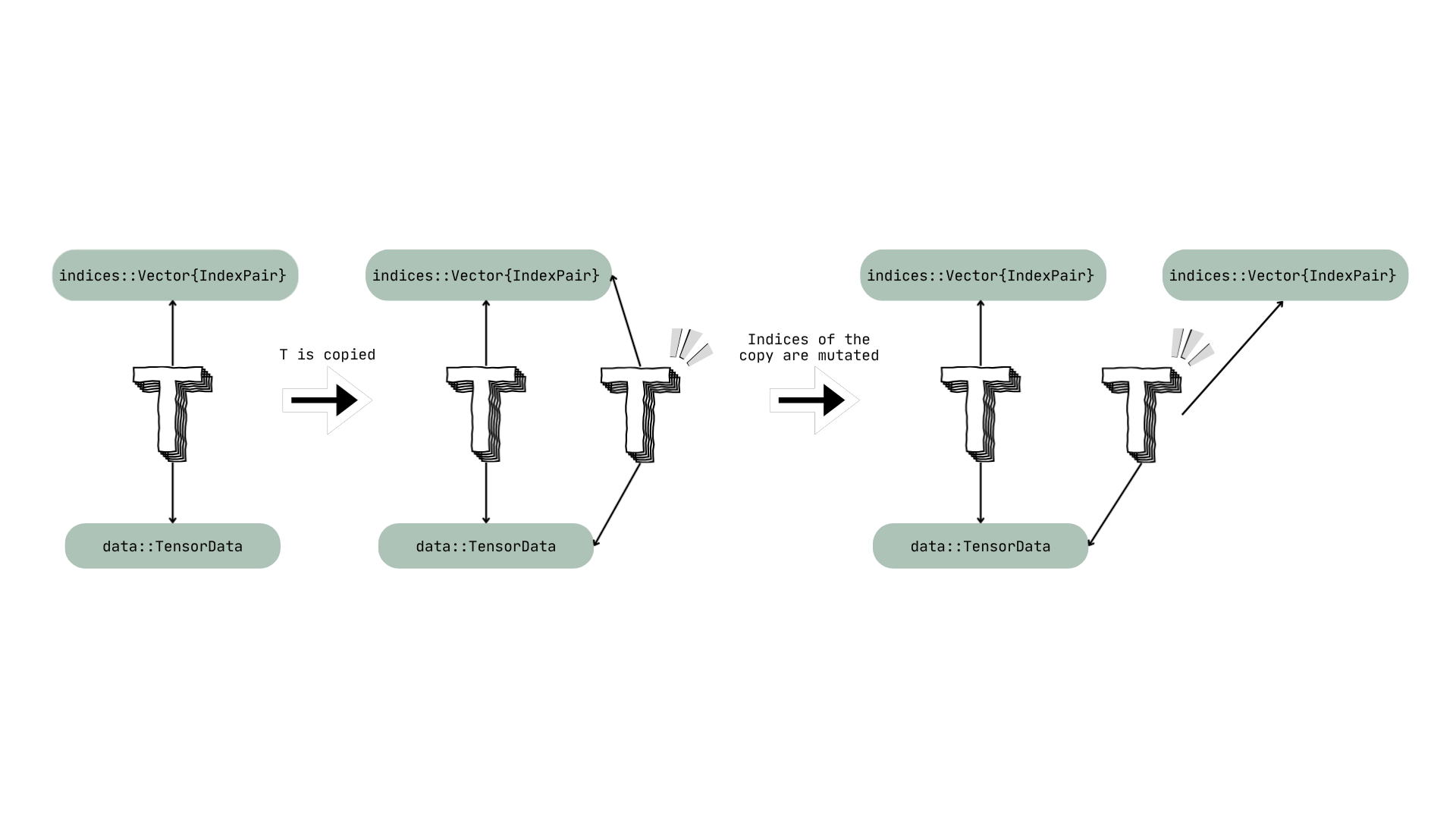}
    \caption{When copied, \jil{Tensor}s share their data and indices. When a change is needed, the appropriate tensor points to a new object in memory.}
    \label{fig:datasharing}
\end{figure}

Before optimising, instrumenting a representative superspace calculation showed that of approximately 28 million tensor copies, fewer than 25,000 ($<0.1\%$) ever mutated their index vector. The remaining copies existed solely to guarantee isolation that was never exercised. To minimise these allocations, Alakazam adopts a strict Copy-On-Write policy. The standard copy of a tensor shares references to both \jil{indices} and \jil{TensorData}. The copy is safe to read but must not be written to in place. When assigning new indices, the indices field of the instance can be pointed towards a new vector, leaving the original reference unchanged, and does not modify the reference to the \jil{TensorData}, see Figure~\ref{fig:datasharing}. This saves on memory hugely. However, it does introduce fragility as a drawback - if one edits the underlying vector of indices explicitly in an unsafe way, this change will apply to all copies, potentially invalidating other expressions. Internally, we avoid this, but one should be careful as a user - this is an unenforced contract, though when using the implemented functions in the expected way the user is protected. A second routine, \jil{copy_owned}, explicitly produces a copy with its own index vector, and is used at the small number of sites that mutate indices directly in place. Routines that replace the index vector directly like \jil{set_indices!} and \jil{update_indices!} construct a fresh vector and assign it, and so are safe under either type of copy.

\subsubsection{Unchecked Operations}
When adding together two tensors, Alakazam must check that the index structure is compatible, i.e. both terms have the same free indices. This is automatically implemented in the \jil{+} and \jil{-} arithmetic operations accessible to the user. However, in many instances internally, we know addition is safe by construction. Thus, in some places, we use an unchecked addition operation \jil{⨥} which bypasses safeties. This is used only selectively in places where the checks would be redundant, and gives minor speed improvements.

Similarly, upon \jil{TensorExpression} creation, we have to make analogous consistency checks as well as checking if terms or coefficients can be merged. In some cases, where safe to do so, we can bypass the checks by using the \jil{TensorExpression} constructor with the \jil{_RAW} keyword as the first argument. This is also employed selectively.

In the same vein, in many cases it is faster to construct \jil{TensorExpression}s and \jil{TensorTerm}s directly, rather than build them iteratively using arithmetic. We deploy this strategy in some targeted places in the code which are used repeatedly.

\subsubsection{Filtering}
The checks that guard index structure are intrinsically pairwise. Multiplying two factors requires
knowing whether any index of one collides with an index of the other, and if so whether a repeated index
sits at opposite levels as Einstein summation requires. We additionally have to check if the indices are lightcone indices, in which case repetition is permitted. Written naively this is a nested loop over both index lists, running on every
multiplication operation and every term construction, which is costly.

Alakazam aims to minimise this cost by firstly assigning a \jil{uid} to hash each index. Comparing these integer \jil{uid}s is far faster than comparing the string names of the indices, and allows for a fast path for rejection. 

For each factor, we compute three $64$-bit words using a single pass recording which indices are upper, lower, and not lightcone respectively. Each index is assigned a (not necessarily unique) fingerprint from the \jil{uid}, and we set the bit in fingerprint position for each word with an appropriate value. This value is determined by a bitwise-or of the word with conjunction of the fingerprint \jil{&} a mask, or its complement. This is constructed from the arithmetic negation of the cast of the index upper/lower position boolean to an integer for the upper/lower words, or the equivalent operation on the lightcone flag. This is done without if statement branching for performance. This is lossy in terms of encoding, but acts as a rejection filter.

This allows us to collapse the naive quadratic pairwise screening to a few bitwise operations. Whether an index is a candidate dummy can be determined by a simple \jil{&} operation, and can easily be flagged if the filter suggests that it appears too many times, or in the wrong positions. Since this is only a filter, when we detect a possible invalidity, we resort to an explicit check using the quadratic scan. In practice the overwhelming majority of multiplications are cleared by the words alone and never touch the exact path.

The same idea appears in \jil{update_indices!}, where a factor list must be checked for any
index used more than twice. Here a population count suffices, so we accumulate one mask over the
non-lightcone indices while counting them, and if every index is distinct then each contributes a
fresh bit so the counts match, again resorting to the exact check if needed.

A related filtering is also included in \jil{collect_terms} (see~\ref{sec:collect-terms}).

\subsubsection{Arithmetic}
Thanks to the above optimisations of copying and the like, we are able to achieve very fast speeds on primitive operations in Alakazam. This has a flow-on effect to all other functions which use arithmetic. We present below in Table~\ref{tab:primitives} some indicative speeds to demonstrate the rapidity of primitive operations.

\begin{table}[H]
    \centering
    \begin{tabular}{lr}
        \toprule
        Operation & Time (\si{\nano\second}) \\
        \midrule
       \jil{A_μ*B_ν} & 37.02 \\
       \jil{A_μ+A_μ} & 37.35 \\
        \jil{A_μ+C_μ} & 49.79 \\
       \jil{A_μ*C^μ} & 37.23 \\
       \jil{A_μ*B_ν+B_ν*A_μ} & 104.65\\
       \jil{D_μν^ρ*E^σ_ρ} & 39.30\\
        \bottomrule
    \end{tabular}
    \caption{Indicative timings for some primitive operations, median timings across 10000 samples.}
    \label{tab:primitives}
\end{table}

\subsubsection{Expression Storage}
Symbolic systems commonly represent expressions as trees, with each node an operation and its children the operands. Alakazam does not. Each expression is held in a fixed two-level form consisting of coefficients and \jil{TensorTerm}s, and the levels are lists rather than nodes.  A sum of products is therefore stored as exactly that, expanding out all bracketing.

The cost is as expected: common sub-expressions are not shared, and factorisation cannot be performed. An expression that would compress well as a tree does not
compress at all here. This could be a limiting factor for some computations where expanding explicitly a factorised form leads to an intractable problem - for example an expression raised to an exceptionally high power. Such expressions are commonly found in General Relativity calculations, presenting an obstacle for Alakazam.

Nesting occurs only for \jil{Operator}s, which themselves hold \jil{TensorExpression}s.

\subsubsection{Precompilation}
The package also implements a precompilation workload via \jil{PrecompileTools} during installation to minimise waiting times on first call. This runs a wide selection of functions and operations in Alakazam, that allows the compiler to explore the methods, meaning Alakazam is more responsive to users. It might be noted that a large portion of the compilation time comes from our usage of the \jil{Symbolics} package, which is seemingly unavoidable.

\subsection{Testing and Correctness}
Alakazam implements a test suite consisting of over 1000 tests across over 30 identified areas of the key features. These have been selected to cover as many scenarios as possible, to guarantee integrity of the operations. In addition, we have tested the ability to reproduce more lengthy, non-trivial results, such as those seen in~\ref{sec:extended-example-compton-scattering} and~\ref{sec:ttbarexample}. These will be continually extended as new features are added. The Grassmann decomposition is also verified to be correct a posteriori, as described in~\cite{woods2026decomposing}.

\subsection{\jil{hard_simplify} Implementation}\label{sec:hardsimpimp}
Two products of tensors may admit manipulations through permutations of indices, terms and dummy indices that can reduce them to an equivalent form. Recognising these equivalences is a highly non-trivial problem and is well studied in the literature. In general no polynomial-time algorithm is known when dummy indices are at play. One approach that is common amongst CAS is the index-canonicalisation approach~\cite{rodionov1987combinatorial}. A common algorithm for doing this is the Butler-Portugal algorithm~\cite{butler1991fundamental,MANSSUR_2002}. This is what is used by other Computer Algebra Systems, like Cadabra's \jil{canonicalise} and xAct's \jil{toCanonical} via the implementation in xPerm~\cite{MartinGarcia:2008xPerm}. As the symmetry of each tensor is expressible as a signed permutation of its index slots, the full symmetry group of a monomial can be generated. The Butler-Portugal algorithm canonicalises a monomial using this group, modulo the dummy symmetry group~\cite{MANSSUR_2002}, which is equivalent to the double coset membership problem. This is known to be graph-isomorphism hard. For a more comprehensive description of the Butler-Portugal algorithm than we mention here, as well as improvements, see~\cite{niehoff2018faster}.

We approach the same problem by a different route, encoding each \jil{TensorTerm} directly as a graph and searching for (graded) isomorphisms. While graph isomorphism has no known polynomial bound~\cite{babai2016graph}, there exist several algorithms which can solve such problems with efficiency, especially for ``small'' graphs (relatively speaking, compared to networking problems) which is the regime in which we work, such as VF2~\cite{cordella2004sub,juttner2018vf2++}. We find our implementation (which consists of a cheap Weisfeiler-Leman~\cite{leman1968reduction} colour refinement pre-filter computed once per graph followed by a modified VF2) to be well suited to the graphs under consideration, affording us the ability to encode information about the ordering of nested operators and grading directly. The prefilter performs a similar role (though cannot alone decide on isomorphism~\cite{cai1992optimal}) to the canonicalisation provided by something like Nauty or Traces~\cite{mckay2014practical} which are the natural alternatives whose advantage is fast amortised comparisons after the canonicalisation cost. We note also that we encode information in both edges and nodes in contrast to these algorithms which normally colour vertices. We find that generic graphs in our computations are sufficiently small and sparse with degree bounded by the symmetries on the index slots~\cite{luks1982isomorphism}, and tend to be in the regime in which VF2 backtracking can terminate quickly. For totally (anti-)symmetric tensors, the number of edges may grow, but as we iterate isomorphisms lazily (we need generally only the first isomorphism and rarely two or more before finding a valid mapping), rather than count the isomorphisms, the cost from increasing the rank is somewhat balanced by the ease of finding an isomorphism due to symmetry. The zero detection search however, which has to explore the automorphisms, has a budget so that when the group becomes too large the result is safely reported as non-zero, rather than hanging. 

We note that Redberry also uses graph isomorphism for simplification, though their encoding and implementation are different.

The algorithm proceeds as follows. Firstly, \jil{TensorTerm}s with coefficients equal to zero are filtered out and removed. The remaining terms are iterated over and added to a fresh expression. At each step, the \jil{TensorTerm} is encoded to a graph, and it is checked whether or not there is a 4-cycle that implies that this term is zero (the exact details of the encoding are to follow). Then, the current term is compared against terms in the new sum by checking if they are equal up to a sign. 

To do this, each term is encoded as a labelled weighted multigraph. The construction preserves all algebraically relevant information:
\begin{itemize}
    \item tensor nodes represent individual tensor factors, and each tensor vertex is labelled by the tensor name
    \item operator nodes represent nested differential or algebraic operators (which we will mention soon), where operator vertices are labelled by the operator name, while their position in the nesting hierarchy is encoded through their connectivity.
\item index nodes represent tensor indices, where free indices are labelled by their names, and dummy indices are assigned an empty label so that their identities become irrelevant,
\item tensor-index edges encode index attachment,
\item contraction edges encode dummy-index contractions,
\item weighted edges encode tableau symmetries,
\item self-loops distinguish upper and lower contracted indices.
\end{itemize}
Dummy-index names are deliberately omitted from the graph labels. Consequently, two graphs differing only by dummy-index relabelling are isomorphic.
Nested operators are represented as additional nodes connected according to the operator hierarchy, allowing operator permutations and graded commutation relations to be handled within the same framework as tensor permutations. Thus, graph isomorphism becomes a test for tensor equality modulo index renaming and tensor symmetries.
Each tensor vertex is connected to the vertices corresponding to its indices by edges of weight zero. These incidence edges specify exactly which indices belong to which tensor factor. Since indices occupy ordered positions within a tensor, the graph naturally preserves index ordering through the attachment pattern together with the symmetry edges described below. For operator vertices, an identical construction is used: every operator is connected to the vertices representing its own operator indices using weight-zero edges.
Intrinsic tensor symmetries are encoded using weighted edges between index vertices.

For every Young tableau associated with a tensor, the corresponding adjacency matrix is traversed. If two index positions belong to a symmetric relation, an edge of weight $+1$ is inserted, and for an antisymmetric relation, an edge of weight $-1$ is inserted. Consequently, the graph contains the complete symmetry information of every tensor without explicitly storing permutation groups.
This representation allows subsequent graph isomorphisms to recover admissible index permutations directly from the graph structure.
Dummy-index contractions are represented independently of tensor symmetry. Every contracted upper-lower index pair generates an edge of weight 2 joining the corresponding index vertices and
a self-loop of weight -2 on the upper index. The contraction edge records that two indices belong to the same dummy pair, while the self-loop distinguishes upper and lower indices. This prevents graph isomorphisms from exchanging covariant and contravariant indices while still allowing arbitrary renaming of dummy labels. Because dummy-index names are omitted from the vertex labels, any relabelling of contracted indices leaves the graph unchanged up to isomorphism.

Due to this labelling, identically vanishing contractions of symmetric indices and antisymmetric indices are encoded by a 4-cycle with edge weights $(1,2,-1,2)$. The simplification algorithm first searches for these cycles to identify if the term is zero.

Nested operators are incorporated by introducing additional operator vertices. Suppose an operator chain has the form
\begin{equation}
    \mathcal{O}_1(\mathcal{O}_2(...\mathcal{O}_n(T))...).
\end{equation}
Each operator becomes its own graph vertex. Every operator vertex is connected to the vertices representing its operator indices,
to every other operator in the same nesting chain, and, for the deepest operator, to the tensor vertex representing the operand. These operator to operand edges have weight $3$.

The resulting subgraph uniquely determines both which operators belong to the same chain, and their relative nesting order. During graph isomorphism these operator vertices therefore determine how nested operators must be permuted (if permitted), allowing the corresponding graded commutation signs to be reconstructed afterwards.

Internally, vertices are assigned deterministic integer identifiers chosen to distinguish different object classes
\begin{itemize}
    \item positive even integers denote tensor vertices,
\item positive odd integers denote index vertices,
\item negative integers denote operator vertices.

\end{itemize}
Operator identifiers are chosen as
\begin{equation}
    -2^t3^d
\end{equation}
where $t$ denotes the tensor position and $d$ the depth within the operator chain.
Since powers of two and three are uniquely recoverable from prime factorisation, every operator vertex can later be mapped back to both its tensor factor and its nesting depth. This allows graph isomorphisms to be translated directly into permutations of operator chains.

\begin{figure}[h]
    \centering
    \includegraphics[width=\linewidth]{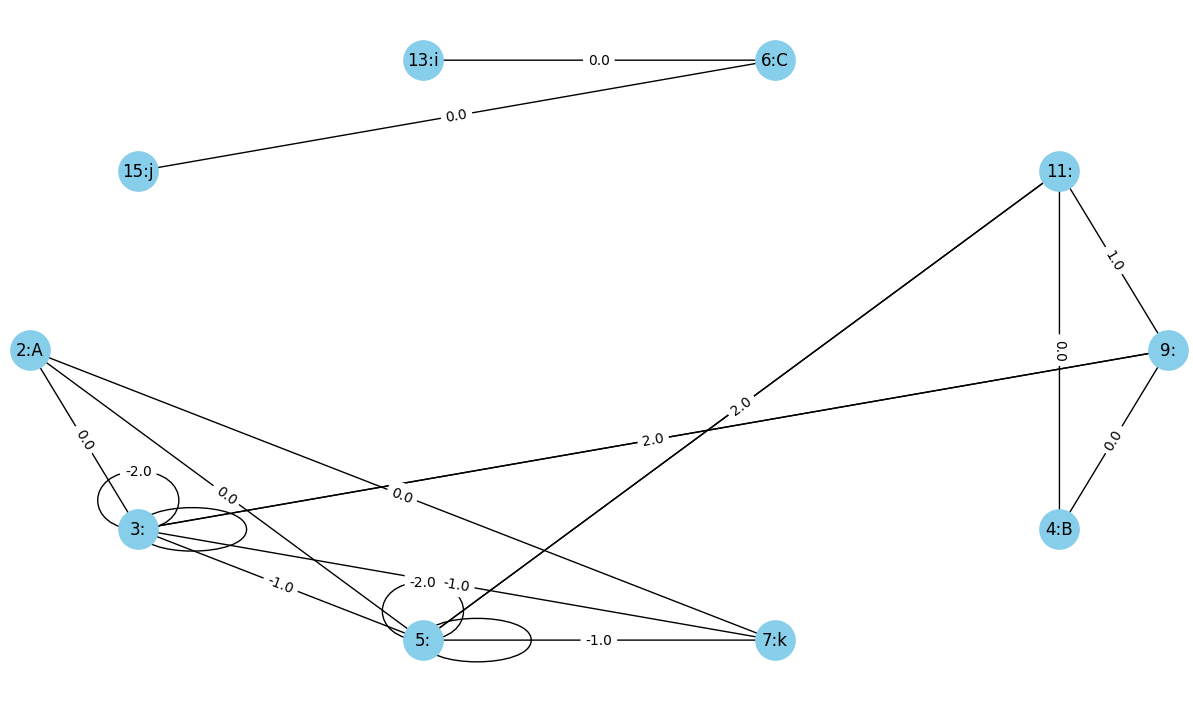}
    \caption{Encoding of a basic \jil{TensorTerm} for $A^{[lm}{}_{k]}B_{(lm)}C_{ij}$ as a graph. Index exchange symmetries are encoded by edges of weight $1$, antisymmetries of weight $-1$, and Einstein summation contractions by $2$.}
    \label{fig:graph1}
\end{figure}

For example, the \jil{TensorTerm}
\begin{Verbatim}[commandchars=\\\{\},xleftmargin=\parindent,numbers=left,bgcolor=bg]
\PYG{+w}{    }\PYG{p}{(}\PYG{n}{AntisymmetricTensor}\PYG{p}{(}\PYG{l+s}{\PYGZdq{}}\PYG{l+s}{A}\PYG{l+s}{\PYGZdq{}}\PYG{p}{,}\PYG{p}{[}\PYG{n}{l}\PYG{o}{=\PYGZgt{}}\PYG{n}{upper}\PYG{p}{,}\PYG{n}{m}\PYG{o}{=\PYGZgt{}}\PYG{n}{upper}\PYG{p}{,}\PYG{n}{k}\PYG{o}{=\PYGZgt{}}\PYG{n}{lower}\PYG{p}{]}\PYG{p}{)}
\PYG{+w}{    }\PYG{o}{*}\PYG{n}{SymmetricTensor}\PYG{p}{(}\PYG{l+s}{\PYGZdq{}}\PYG{l+s}{B}\PYG{l+s}{\PYGZdq{}}\PYG{p}{,}\PYG{p}{[}\PYG{n}{l}\PYG{o}{=\PYGZgt{}}\PYG{n}{lower}\PYG{p}{,}\PYG{n}{m}\PYG{o}{=\PYGZgt{}}\PYG{n}{lower}\PYG{p}{]}\PYG{p}{)}
\PYG{+w}{    }\PYG{o}{*}\PYG{n}{Tensor}\PYG{p}{(}\PYG{l+s}{\PYGZdq{}}\PYG{l+s}{C}\PYG{l+s}{\PYGZdq{}}\PYG{p}{,}\PYG{p}{[}\PYG{n}{i}\PYG{o}{=\PYGZgt{}}\PYG{n}{lower}\PYG{p}{,}\PYG{n}{j}\PYG{o}{=\PYGZgt{}}\PYG{n}{lower}\PYG{p}{]}\PYG{p}{)}\PYG{p}{)}
\end{Verbatim}
is encoded to the graph in Figure~\ref{fig:graph1}.
Consequently, two tensor terms are graph isomorphic if and only if they have identical contraction structure, tensor content, operator structure, and symmetry data up to dummy-index renaming. The graph isomorphism therefore provides precisely the permutation required to reconstruct the graded transformation relating the two tensor terms.

For every graph isomorphism returned by the graph matcher, the corresponding permutation of tensor factors, nested operators, and tensor indices is reconstructed. The dummy indices of one term are explicitly renamed to match those of the other, reducing the remaining comparison to one involving only free-index permutations and tensor symmetries. Indices are rearranged according to the graph isomorphism and tested against the tensor's symmetry group. This stage determines whether the induced permutation is allowed or forbidden, and the sign corresponding to the swaps. If no valid symmetry exists, the current graph isomorphism is discarded and the next candidate is examined.

Whenever two equivalent tensor terms are identified, their coefficients are combined according to the computed graded sign.
The representative tensor term is retained, while its coefficient is updated.

After all summands have been processed, any newly generated zero coefficients are removed. If every coefficient cancels, the expression is replaced by the zero tensor expression. Note that if a coefficient of a term in the new summand has become zero at some point in the calculations, further comparisons with it are skipped, which is safer than directly removing it in a way that may affect the iteration.

\subsection{Pattern Matching Implementation}\label{sec:explanation-of-the-algorithm}

In general, pattern-matching problems are computationally difficult, especially so in our case. Indeed, the fact the expressions are (anti)commutative means that the usual improvements in runtime gained in text matching algorithms from techniques like preprocessing cannot be performed, as the expression has no definite order. Preprocessing usually reduces the time complexity of pattern matching in text significantly - for algorithms like KMP the runtime reduces to $O(n+m)$ for an expression of $n$ characters and a pattern of $m$ characters. It should be noted that the naive brute-force approach to pattern matching for a(n) (anti)commutative expression and (anti)commutative pattern is computationally intensive, running in exponential time. In fact, the problem reduces to a three-partition problem, which is known to be NP-complete.

To see this equivalence, consider a pattern built from $n$ wildcard tensor names $?P_1,\dots,?P_n$, in which name $?P_i$ can occur with multiplicity $s_i$, matched against an expression built from $n$ distinct tensor names each
occurring $T$ times, where $\sum_i s_i = nT$. Because a wildcard binds consistently, a match assigns each $?P_i$ to a tensor name whose $T$ occurrences must accommodate its multiplicity, so a match
exists precisely when the multiset $\{s_i\}$ partitions into $n$ groups each summing to $T$. In the case
$T/4 < s_i < T/2$, then each group is forced to contain three elements - this is the 3-partition problem, which is strongly NP-complete
~\cite{garey1975complexity,garey2002computers}; a proposed match is verifiable in polynomial time, so the matching problem is
NP-complete. Note that this argument uses only the multiplicities of tensor names, and not any property of the indices, and thus is inherently a problem due to the commutativity.

No complete matcher can therefore run in polynomial time, and the useful question is not how to avoid exponential behaviour but where to confine it. Alakazam performs a backtracking search over assignments of pattern factors to expression factors, costing $O(n^m)$ in the worst case for an expression of $n$ factors and a pattern of $m$. This is polynomial in the length of the expression for any fixed pattern, and exponential only in the length of the pattern, which is precisely where the hardness established above resides. Its practical speed comes from pruning rather than from that bound: a necessary condition computed once per term discards almost every candidate assignment before any structural comparison is made. This is exhaustive over the assignments within each name class.

This algorithm is one of the most complex features of the package. Our algorithm runs in a few main steps, following a backtracking search with early rejection approach to solving the problem. For a tensor term of $n$ tensors, and a pattern of $m$ tensors,
\begin{enumerate}
    \item \textbf{Construct Histograms}: Firstly, the expression and the pattern are converted to histograms, by firstly counting the number of occurrences of each tensor with a given name, then inverting the histogram so that the result has keys denoting frequencies, and values that are a set of tensor names (as well as their associated positions in the original expression). This allows for quick comparison between the pattern and the expression to find possible candidates for matches based on frequency. This step runs in $O(n+m)$ (amortised) time.
    \item \textbf{Find a Match}: The inverted histograms give a cheap necessary condition for a match. A
    pattern tensor occurring $k$ times can only match an expression tensor occurring at least $k$
    times, so the great majority of candidates can be discarded before a single index is examined. The frequency classes most likely to fail are tested first.
    The algorithm then recursively checks for matches, by checking combinations of tensors with the correct frequency. If there is a single frequency in the pattern, then two cases are considered
    \begin{itemize}
        \item \textit{One type of tensor in the single pattern frequency}: If there is only one type of tensor (name) in the pattern, we loop through all frequencies in the expression histogram where the frequency of the pattern is less than the expression frequency (a requirement for a match). For each of these valid expression frequencies, the algorithm loops over the associated tensors. If the names match (or if the tensor in the pattern is a dummy),  we then assign the pattern's slots to the occurrences of that tensor in the expression. Each slot is tried against every occurrence not already taken, backtracking whenever the indices are incompatible (bearing in mind any wildcards from \jil{DummyPatternIndex}, and any labels). A match is therefore found whenever one exists within that class, whatever order the factors happen to sit in within the term. If one is found, it is registered along with information about which wildcards match to which index/tensor. If no assignment succeeds and the pattern name was concrete, the search terminates in the negative; if it was a wildcard, the next tensor name is tried instead.

        \item \textit{Many types of tensors in the single pattern frequency}: The algorithm runs similarly, by first calling itself on the expression, but on only the first type of tensor in the pattern. This runs the algorithm and returns information about wildcards. If a match was found, it is removed from the pattern, and the rest of the pattern is now checked against the expression, now with the appropriate wildcards fixed. If no match is found at any step, then we return up a level in the algorithm and consider the next possible match of the pattern tensors with the expression (as perhaps there is a match that exists with a different set of wildcard matches).  In this sense, the algorithm is doing a backtracking search for matches, at the level of tensor names as well as within each name.
    \end{itemize}
    Otherwise, if there are multiple frequencies in the pattern histogram, we loop over the frequencies and recursively call the algorithm on the single frequency case, and record information about the wildcards to use at each step. Note that if any frequency in the pattern has no matches with the expression, the search can terminate in the negative.
     In the worst case, a class of $n_c$ occurrences with $m_c$ pattern slots must examine every injective assignment between them, which recovers the $O(n^m)$ bound quoted above - but in reality it is far faster, as this assumes that every tensor in the pattern shares a single name and that no partial assignment is ever rejected by its index checks.
      Whenever the pattern names distinct concrete tensors, which is the usual case, each class holds a single slot and the assignment reduces to one scan.
    Note also that only the first match found is substituted. The alternatives recorded along the way are competing ways of binding the same occurrence rather than separate summands - adding them would multiply-count, and for a wildcard name matching several different tensors they need not even share the same free indices - so further occurrences are reached by repeating the substitution instead.

    \item \textbf{Make the replacement}: If a match is found, the replacement must then be made. Firstly, the concerned terms must be dragged together. When every factor lying between them commutes freely with them, this contributes only
    a graded sign, computed by counting the odd-graded unmatched factors each matched factor must
    cross. When some intervening factor carries a registered commutation relation, the factors are
    dragged together using that relation instead, and the commutator remainders are emitted as
    additional summands, to be matched against on a later pass. Then the replacement must be constructed by creating new tensors given information about the wildcards. These are instantiated from the recorded bindings, including inside nested
    operators, and spliced in at the anchor position, with the unmatched factors restored before
    and after it. In total, this runs in roughly $O(nm)$ time in the absence of dragging. Replacement tensors are recorded
    in an exclusion set so that a repeated substitution does not rewrite its own output.
\end{enumerate}
By default the search enumerates every match in the term and the substitution is repeated until no
match remains; both are controlled by the \jil{repeat} argument.
In summary, the histogram stage is linear and the replacement stage is $O(mn)$, while the search
stage is exponential in the worst case and, in ordinary use, closer to linear, as the frequency
filter and the consistency requirement on wildcard bindings between them eliminate almost all
candidate assignments before any structural comparison is performed. Thus, in usual operation, we expect roughly $O(n^2)$ timings, as $m$ is usually fixed and small relative to expression lengths. 

As a sketch to support this analysis, we time four classes of replacement, each chosen to isolate one stage of the
algorithm.
 The first matches a pattern a single time:
\begin{equation}\label{eq:pm-single}
    \text{given } X_{a_1}X_{a_2}\ldots X_{a_n}, \quad \text{find \jil{X_{?a}}}
\end{equation}
This essentially times histogram creation, which we expect to be linear, as the matching is
trivial and only takes one pass. The next class is similar, but with repeated replacements to substitute every occurrence:
\begin{equation}\label{eq:pm-repeat}
    \text{given } X_{a_1}X_{a_2}...X_{a_n}, \quad \text{find \emph{all} \jil{X_{?a}}}
\end{equation}
This measures the cost of traversing the expression for matches which we expect to be linear in $n$, giving overall quadratic in $n$ behaviour. For a more complicated example, we consider pair matching:
\begin{equation}\label{eq:pm-pair}
    \text{given } X_{a_1}X_{a_2}...X_{a_n}Y^{a_1}Y^{a_2}...Y^{a_n}, \quad \text{find \emph{all} \jil{X_{?a}Y^{?a}}}
\end{equation}
This lets us test wildcard matching across dummy indices. Finally we test against contractions with both slots on the same named tensor, so that the frequency filter cannot separate
the candidates by name
\begin{equation}\label{eq:pm-sameclass}
    \text{given } X_{a_1}X_{a_2}...X_{a_n}X^{a_1}X^{a_2}...X^{a_n}, \quad \text{find \emph{all} \jil{X_{?a}X^{?a}}}.
\end{equation}

\begin{figure}[H]
    \centering
\begin{tikzpicture}
\begin{loglogaxis}[
    width=13cm,
    height=9cm,
    xlabel={term length $n$},
    ylabel={time (\si{\micro\second})},
    grid=both, ymin=1, xmin=1, xmax=1000,
]
  \patternplot[yshift=14pt]{0.291982319
}{1.013576894
}{single}{blue}{300}
  \patternplot[yshift=-12pt]{0.193694237
}{2.068291017
}{repeat}{red}{350}
  \patternplot[yshift=44pt]{0.324897108
}{2.060131090
}{pair}{mygreen}{70}
  \patternplot[yshift=30pt]{0.321085615
}{2.060320503
}{sameclass}{orange}{400}
\end{loglogaxis}
\end{tikzpicture}
    \caption{Asymptotic growth estimations for pattern matching. Blue: Single pattern; Red: Repeat Matching; Green: Pair Matching; Orange: Same-class Matching, as defined in text. Exponents are fitted over the last $200$ points of each sweep.}
    \label{fig:patternmatching}
\end{figure}
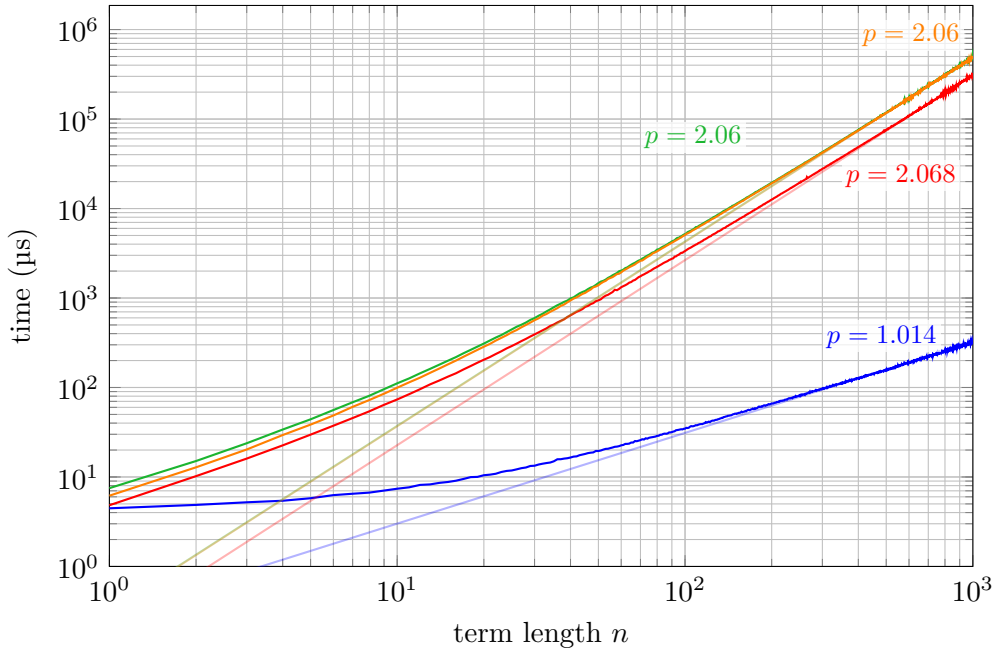

We include some representative timings in Figure~\ref{fig:patternmatching}.

\begin{figure}[H]
    \centering
\begin{tikzpicture}
\begin{axis}[name=benefit, width=7.6cm, height=6cm,
    xlabel={decoys $k$}, ylabel={time (\si{\micro\second})},
    grid=both, xmin=0, xmax=1000, ymin=500,
    title={\small benefit: $n=1000$ fixed}, legend pos=north west, legend cell align=left]
  \addplot[only marks, mark size=0.5pt, color=blue]
    table[col sep=comma, x=n, y=nameclass] {ims/nameclass.csv};
  \addlegendentry{measured}
  \addplot[thick, color=red, domain=1:998, samples=2] {576.3 + 0.2961*x};
  \addlegendentry{fit}
\end{axis}
\begin{axis}[at={(benefit.right of south east)}, anchor=left of south west,
    width=7.6cm, height=6cm,
    xlabel={term length $n$}, ylabel={time (\si{\micro\second})},
    grid=both, xmin=0, xmax=1000, ymin=0,
    title={\small cost: $k=2$ fixed}, legend pos=north west, legend cell align=left]
  \addplot[only marks, mark size=0.5pt, color=blue]
    table[col sep=comma, x=n, y=nameclassn] {ims/patterns3.csv};
  \addlegendentry{measured}
  \addplot[thick, color=red, domain=4:1000, samples=2] {-11.0 + 0.5738*x};
  \addlegendentry{fit}
\end{axis}
\end{tikzpicture}
    \caption{The two sides of the frequency filter. Left: cost of a single search against $k$
    candidates that are offered and then rejected, at fixed term length $n=1000$; linear in $k$. Right: the same search with the candidate set held at two while the term grows;
    linear in $n$. The filter costs $O(n)$ to build and returns $O(1)$ per candidate
    discarded.}
    \label{fig:patterndecoy}
\end{figure}
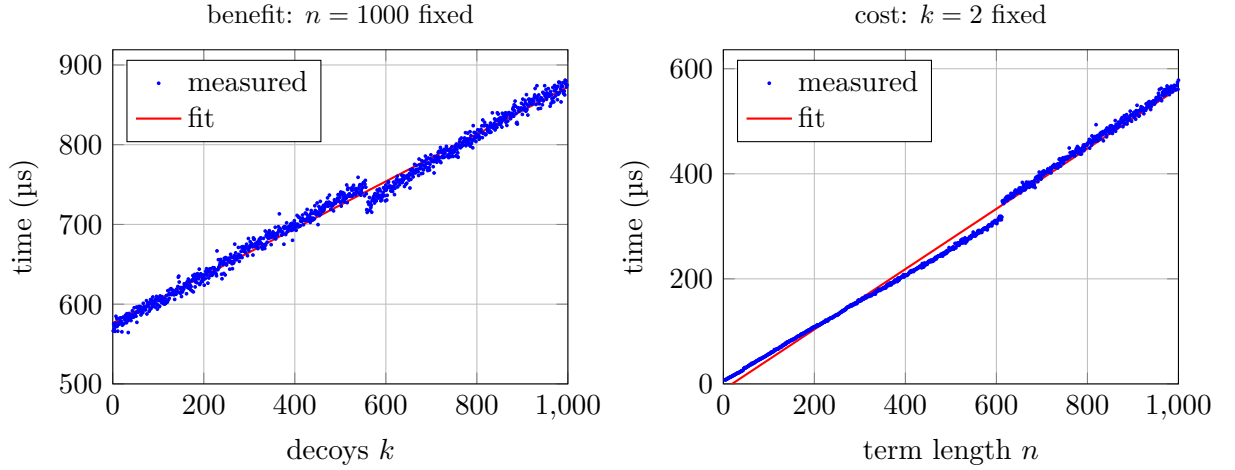

The fitted exponents are $1.01$, $2.07$, $2.06$ and $2.06$ respectively.
The first two behave exactly as the cost model predicts. The two wildcard classes land on the same exponent. This particular two-slot search therefore costs nothing measurable above the histogram construction that every pass performs anyway. In \eqref{eq:pm-pair} and \eqref{eq:pm-sameclass} every candidate for the first slot has a partner, so the first one attempted succeeds, and since only one match is needed per pass no
second candidate is ever examined. We can instead consider a more adversarial example, where
\begin{equation}\label{eq:pm-decoy}
    \text{given } X_{a_1}X_{a_2}\ldots X_{a_k}\,F_{1}\ldots F_{n-k-2}\,X_{a_0}Y^{a_0}, \quad
    \text{find \jil{X_{?a}Y^{?a}}}
\end{equation}
Here $a_1,\ldots,a_k$ are free, so each of the $k$ leading factors carries the pattern's name, is
offered by the frequency filter, and is then rejected once the binding of \jil{?a} of this initial find yields a failure by not matching the index of  $Y$. The remaining $F_j$ are named apart and are not considered as a match. The term length $n$ is held fixed at $1000$ while $k$ grows, with the actual match placed last in the term. We can see this plotted in Figure~\ref{fig:patterndecoy}.

The slope is the price of one rejection.
The cost side of the same filter is obtained by holding the number of decoys at two and growing the term
instead. Rejection is
therefore constant-time, and a candidate set of size $k$ is disposed of in linear time.

\begin{figure}
    \centering
\begin{tikzpicture}
\begin{loglogaxis}[
    width=13cm, height=8cm,
    xlabel={pattern length $m$ (contracted factors)},
    ylabel={time (\si{\micro\second})},
    grid=both, xmin=2, xmax=1000,
    legend pos=north west, legend cell align=left,
]
  \addplot[only marks, mark size=0.6pt, color=blue]
    table[col sep=comma, x=n, y=chaink] {ims/chaink.csv};
  \addlegendentry{term fixed at $n=1000$}
  \addplot[dashed, color=blue, domain=2:81, samples=2, forget plot] {1071.4};
  \addplot[dotted, gray, forget plot] coordinates {(81,20) (81,25000)};
  \addplot[only marks, mark size=0.6pt, color=orange]
    table[col sep=comma, x=n, y=chain] {ims/patterns3.csv};
  \addlegendentry{term length $n=2m$}
  \addplot[thick, color=orange, opacity=.4, domain=100:1000, samples=2, forget plot]
    {0.014432792*x^(2.067263048)};
\end{loglogaxis}
\end{tikzpicture}
    \caption{Cost of a single search for a sub-chain of $m$ contracted factors, the structure in
    which neither the frequency filter nor the slot-position check can separate any candidate from another. Blue: term held at $1000$ factors; the dashed line marks the mean cost over
    $m \leq 81$, from which the measurements depart by less than $10\%$. Orange: pattern and term
    grown together, fitted at $O(m^{2.07})$.}
    \label{fig:patternsize}
\end{figure}
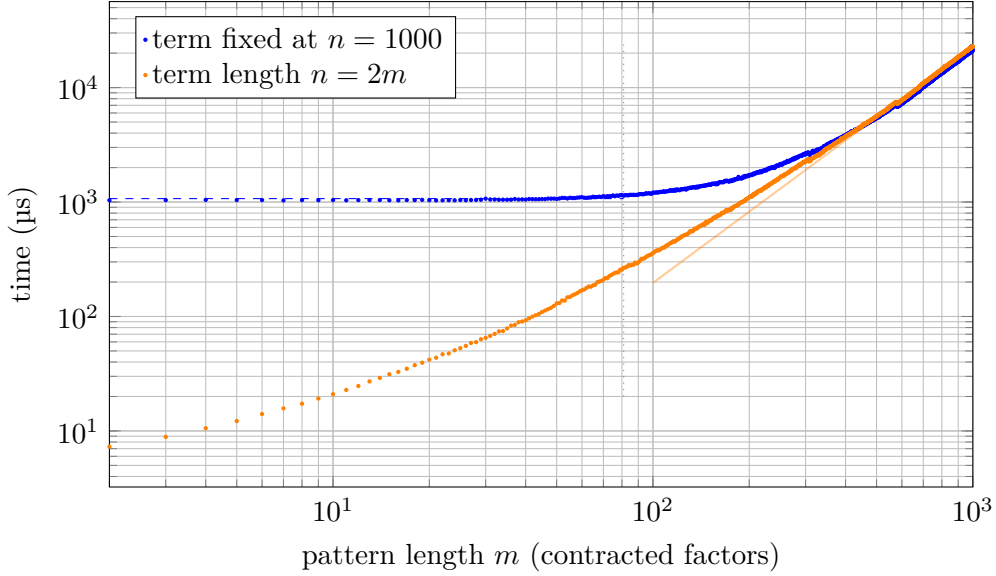

All of the above holds the pattern length $m$ fixed. Two further
measurements test what happens when it does not. Take a chain of $n$ contracted factors
$A^{a_1}{}_{a_2}A^{a_2}{}_{a_3}\cdots$ and search for a sub-chain of length $m$. This is the least
favourable structure available as every factor carries the same name and the same pair of index
positions, so neither the frequency filter nor the slot-position check separates any candidate from
any other, and the shared dummy indices alone distinguish the true match. We can see the behaviour of such a search in Figure~\ref{fig:patternsize}.

Holding the term at $1000$ factors, a single search is flat to within $10\%$ for all lengths $m$ up to $81$,
roughly $8\%$ of the full term length. A pattern of factors like this admits some $10^{243}$ ordered
assignments against a term of this length and is not measurably more expensive to match than a
pattern of two. Letting pattern and term grow together, at $m$ and $2m$ factors respectively, a
single search costs $O(m^{2.07})$: quadratic rather than exponential, with every cheap test disabled
by construction. The assumption that $m$ is small relative to $n$ is therefore comfortable.

To summarise these sweeps, the key observation is that the search stage does not appear in the timings. What is being measured, in every case, is the term being walked. Reaching the exponential worst case would require a term in which large numbers
of candidates agree in name, index position, and dummy structure deep enough to survive to structural comparison. The chain of contractions is the nearest such structure we are able to construct, and it remains approximately quadratic. For practical purposes, a single
substitution into a term of fifty factors costs some $20\,\si{\micro\second}$.

These tests are by no means extensive, however they should be indicative of the performance of our algorithm. It may be possible to improve the runtime of the algorithm slightly by using memoisation techniques; however, it is unlikely as the algorithm should by construction avoid repeated computations.

\newpage
\section{Benchmarks}\label{sec:benchmarks}
In this section we provide some indicative benchmarks for a small set of operations, comparing against Cadabra, xAct, and Redberry, to demonstrate Alakazam's speed and robustness. These operations have been chosen to give users a rough idea of the sort of performance they can expect between packages. Note that not all packages have exactly the same feature set and often take different approaches to solving problems. We have therefore tried to implement equivalent functionality across these benchmarks as much as possible. We comment throughout on limitations, particularly when no equivalent operation is available to make a fair comparison, in which case we try the next best thing. The implementations across packages were written by us, to the best of our knowledge of each system.

Note that all numbers for speeds and benchmarks were obtained by running the code in a fresh Windows Subsystem for Linux (WSL) instance on a system with an Intel Core i9-14900K processor. The timings measured for Alakazam are reported for the ``warm'' state after a single initial cold run of the relevant code to ensure Just-In-Time compilation has been completed, to mirror what a real user would experience in usual operation. We afford similar treatment to Redberry, which requires many runs before reaching a warm state due to how the Java Virtual Machine compiles.

 The version numbers used for testing can be found in Table~\ref{tab:software_versions}.
\begin{table}[ht]
\centering
\begin{tabular}{ll}
\hline
\textbf{Software} & \textbf{Version} \\
\hline
OS & Ubuntu 26.04 LTS\\
Julia & v1.12.6 \\
Alakazam & v0.2.15 \\
Cadabra  & v2.5.15 4037.0afcc65266 \\
WolframScript & v1.13.0 for Linux x86 (64-bit) \\
xAct-xPerm    &   v1.2.4, $\{2025, 12, 29\}$\\  xAct-xTensor & v1.3.0, $\{2025, 12, 29\}$ \\
Groovy & v2.4.21\\
Java VM & v$1.8.0\_492$ \\
Redberry & v1.1.9 \\
\hline
\end{tabular}
\caption{Software versions used in this work.}
\label{tab:software_versions}
\end{table}

\subsection{General Details}\label{sec:genbench}
We present a short set of benchmarks of Alakazam compared with a few popular CAS capable of manipulating tensors with similar capabilities. These include Cadabra2~\cite{Peeters:2007wn,Peeters2018}: \textit{A field-theory motivated approach to symbolic computer algebra}, with LaTeX-style scripting, programmable with Python, and underlying C++ core; xAct~\cite{xAct}: \textit{Efficient tensor computer algebra for the Wolfram Language}, a Mathematica package; and Redberry~\cite{Bolotin:2013qgr}: \textit{open source computer algebra system for algebraic manipulations with tensors},  with LaTeX-style scripting and written in Java. Given the different platforms for each package, we attempt to provide a fair benchmark by paying heed to the execution model of each system. In particular, this means for Redberry, we provide two numbers. This is because Redberry runs on Java, and the Java Virtual Machine initially interprets bytecode and compiles hot methods only after repeated execution, so timings depend strongly on how many iterations precede measurement. For this reason, we provide a ``warm'' number, after first running the code a large $N$ number of times before starting the timing measurements, and a ``cold'' timing after only $N=1$ prior execution. This value for the large $N$ is chosen depending on the test based on what we observe is needed for performance to saturate - for basic operations, this is $10^6$, while for more complicated operations like simplification the behaviour saturates at much lower $N$ as they often involve multiple invocations of functions per execution. We include both of these hot and cold metrics, as it is not always probable that the average user will run their script sufficiently many times to achieve a warm state. The cold numbers here additionally lend themselves to better comparison with Julia's numbers, where we also perform a single run before timing to ensure Just-In-Time compilation of methods is complete. In this vein for Julia, the timed Alakazam code is also placed inside a function rather than at the top level to make the most of Julia's compiler optimisations. This is because Julia's compiler specialises on argument types within a function body, whereas top-level code operates on untyped global bindings and is dispatched dynamically. Since code should be called from functions, this enclosed form is closer to normal operation, and is a Julia-specific distinction. For Mathematica, we clear the system cache between measurements, since it memoises aggressively and would otherwise return stored results. For Cadabra, whose declarations accumulate in a global kernel state, we declare all indices and tensor properties once outside the timed region (aside from when we want to explicitly test this) and rebuild only the expression within it. 

Process startup and package loading are excluded in all cases. Alakazam's tests were run within a .jl file, Cadabra's via .cdb, xAct from .wl Wolfram scripts, and Redberry from .groovy files as recommended. Scripts used for benchmarking can be found in the supplementary materials. We note also that the published expression files from the Redberry paper~\cite{Bolotin:2013qgr} are no longer available, so our tests are not direct reproductions of theirs, and are thus not directly comparable.  

We present all plots in logarithmic scale.

\subsection{Riemann Tensor Creation}\label{sec:riemann-tensor-creation}
As a rudimentary test we time the creation of a basic, but nontrivial tensor - the Riemann tensor. 

While an unusual benchmark, we provide it nonetheless as an indicator for how quick it is to get started working with a tensor. It also provides context for the later benchmarks, and future comparisons with other software - for example, if any particular package front-loads and performs aggressive canonicalisation and caches it under the hood at tensor creation, later simplification algorithms would become exceptionally fast.

This tensor carries symmetries, and so in Alakazam's case necessitates the creation of extra vectors to accommodate the Young tableau data. We include for this test alone the overhead time required to create the attached indices. In Julia, the tested block corresponds to the code
\begin{Verbatim}[commandchars=\\\{\},xleftmargin=\parindent,numbers=left,bgcolor=bg]
\PYG{+w}{    }\PYG{n}{st}\PYG{+w}{ }\PYG{o}{=}\PYG{+w}{ }\PYG{n}{IndexSet}\PYG{p}{(}\PYG{l+s}{\PYGZdq{}}\PYG{l+s}{spacetime}\PYG{l+s}{\PYGZdq{}}\PYG{p}{,}\PYG{+w}{ }\PYG{l+m+mi}{4}\PYG{p}{,}\PYG{+w}{ }\PYG{n}{lower}\PYG{p}{)}
\PYG{+w}{    }\PYG{n}{a}\PYG{+w}{ }\PYG{o}{=}\PYG{+w}{ }\PYG{n}{Index}\PYG{p}{(}\PYG{l+s}{\PYGZdq{}}\PYG{l+s}{a}\PYG{l+s}{\PYGZdq{}}\PYG{p}{,}\PYG{+w}{ }\PYG{n}{st}\PYG{p}{)}\PYG{p}{;}\PYG{+w}{ }\PYG{n}{b}\PYG{+w}{ }\PYG{o}{=}\PYG{+w}{ }\PYG{n}{Index}\PYG{p}{(}\PYG{l+s}{\PYGZdq{}}\PYG{l+s}{b}\PYG{l+s}{\PYGZdq{}}\PYG{p}{,}\PYG{+w}{ }\PYG{n}{st}\PYG{p}{)}
\PYG{+w}{    }\PYG{n}{c}\PYG{+w}{ }\PYG{o}{=}\PYG{+w}{ }\PYG{n}{Index}\PYG{p}{(}\PYG{l+s}{\PYGZdq{}}\PYG{l+s}{c}\PYG{l+s}{\PYGZdq{}}\PYG{p}{,}\PYG{+w}{ }\PYG{n}{st}\PYG{p}{)}\PYG{p}{;}\PYG{+w}{ }\PYG{n}{d}\PYG{+w}{ }\PYG{o}{=}\PYG{+w}{ }\PYG{n}{Index}\PYG{p}{(}\PYG{l+s}{\PYGZdq{}}\PYG{l+s}{d}\PYG{l+s}{\PYGZdq{}}\PYG{p}{,}\PYG{+w}{ }\PYG{n}{st}\PYG{p}{)}
\PYG{+w}{    }\PYG{n}{RiemannTensor}\PYG{p}{(}\PYG{l+s}{\PYGZdq{}}\PYG{l+s}{R}\PYG{l+s}{\PYGZdq{}}\PYG{p}{,}\PYG{+w}{ }\PYG{p}{[}\PYG{n}{a}\PYG{o}{=\PYGZgt{}}\PYG{n}{lower}\PYG{p}{,}\PYG{+w}{ }\PYG{n}{b}\PYG{o}{=\PYGZgt{}}\PYG{n}{lower}\PYG{p}{,}\PYG{+w}{ }\PYG{n}{c}\PYG{o}{=\PYGZgt{}}\PYG{n}{lower}\PYG{p}{,}\PYG{+w}{ }\PYG{n}{d}\PYG{o}{=\PYGZgt{}}\PYG{n}{lower}\PYG{p}{]}\PYG{p}{)}
\end{Verbatim}
By running this benchmark, we see that Alakazam performs much faster than the other packages for establishing a tensor from scratch, though we stress that this is more of a contextual benchmark.
\begin{table}[htbp]
  \centering
  \caption{Benchmark: \emph{Create a Riemann Tensor} ($n=100$, times in \si{\micro\second}).}
  \label{tab:create-riemann}
  \begin{tabular}{lrrr}
    \toprule
    Software & Min & Median & Mean $\pm$ 95\%\,CI \\
    \midrule
    Cadabra              & 182.83    & 199.09     & $204.32 \pm 3.50 $  \\
    xAct                 & 1417.00     & 1497.00    & $1540.76 \pm 33.24$ \\
    Alakazam             & 2.45    & 2.55    & $2.68 \pm 0.10$     \\
    Redberry ($N=10^{6}$) & 10.55   & 13.14   & $18.58 \pm 3.82$    \\
    Redberry ($N=1$)     & 62.67   & 110.66  & $151.44 \pm 25.70$  \\
    \bottomrule
  \end{tabular}
\end{table}

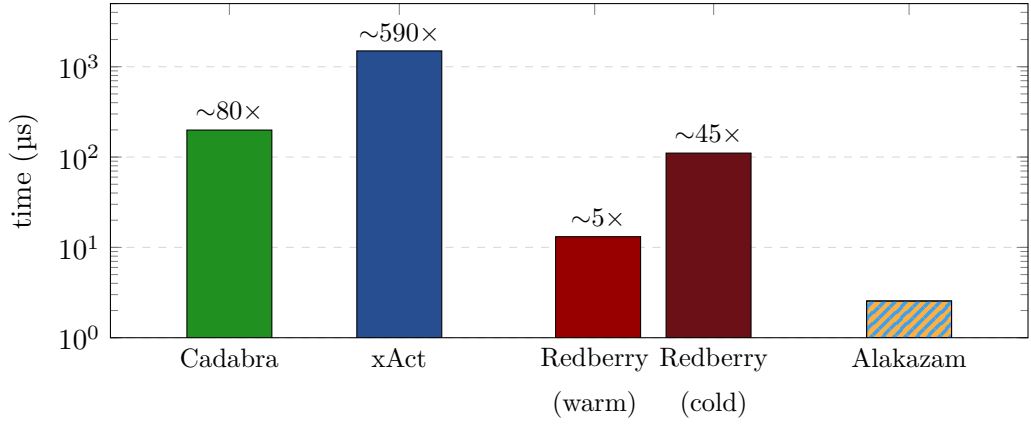
\begin{figure}[H]
    \centering
\begin{tikzpicture}
\begin{axis}[
    width=0.9\linewidth,
    height=6cm,
    ymode=log,
    log origin=infty,
    ymin=1, ymax=5000,
    ylabel={time (\si{\micro\second})},
    xmin=0.3, xmax=5.7,
    xtick={1, 2, 3.15, 3.85, 5},
    xticklabels={Cadabra, xAct, {Redberry\\(warm)}, {Redberry\\(cold)}, Alakazam},
    xticklabel style={align=center, font=\small},
    ymajorgrids=true,
    grid style={dashed, gray!30},
]
\addplot[ybar, bar width=0.5, fill=cadabcol, draw=black]
  coordinates {(1, 199.09)};
\addplot[ybar, bar width=0.5, fill=xactcol, draw=black]
  coordinates {(2, 1497.00)};
\addplot[ybar, bar width=0.5, fill=rbhotcol, draw=black]
  coordinates {(3.17, 13.143)};
\addplot[ybar, bar width=0.5, fill=rbcol, draw=black]
  coordinates {(3.82, 110.655)};
\addplot[ybar, bar width=0.5, fill=akzo, draw=black,
         postaction={pattern={Lines[angle=45, distance=3pt, line width=1.2pt]},
                     pattern color=akzb}]
  coordinates {(5, 2.55)};

\node[anchor=south, font=\small] at (axis cs:1, 201.25)     {$\sim$80$\times$};
\node[anchor=south, font=\small] at (axis cs:2, 1480.17)    {$\sim$590$\times$};
\node[anchor=south, font=\small] at (axis cs:3.17, 13.143)  {$\sim$5$\times$};
\node[anchor=south, font=\small] at (axis cs:3.82, 110.655) {$\sim$45$\times$};
\end{axis}
\end{tikzpicture}
    \caption{Riemann Tensor Creation, medians.}
    \label{fig:bench-riem}
\end{figure}

\subsection{Non-trivial Symmetry}\label{sec:bianchi-identity}
To test the reduction of an expression based on non-trivial symmetries, we can examine the algebraic Bianchi identity of the Riemann tensor
\begin{equation}
    R_{abcd}+R_{acdb}+R_{adbc}=0.
\end{equation}
Already at this level we encounter diverging functionality between the packages. Benchmarking this allows us to see the cumulative cost of not only performing basic addition, but also of a specific simplification. In this case, simplification of this sum occurs in each CAS implicitly via computing the Young projection,
\begin{equation}
    YP(R_{abcd})=\frac{2}{3}R_{abcd}+\frac{1}{3}R_{acbd}-\frac{1}{3}R_{adbc}
\end{equation}
which is defined by
\begin{equation}
    YP=\frac{1}{12}(1-(12))(1-(34))(1+(13))(1+(24)).
\end{equation}
In Alakazam, this Young projector is not stored internally and is computed directly from the Young tableau upon demand, which we include in the time cost.

To perform this benchmark, we pre-construct
\begin{Verbatim}[commandchars=\\\{\},xleftmargin=\parindent,numbers=left,bgcolor=bg]
\PYG{+w}{    }\PYG{n}{st}\PYG{+w}{ }\PYG{o}{=}\PYG{+w}{ }\PYG{n}{IndexSet}\PYG{p}{(}\PYG{l+s}{\PYGZdq{}}\PYG{l+s}{spacetime}\PYG{l+s}{\PYGZdq{}}\PYG{p}{,}\PYG{+w}{ }\PYG{l+m+mi}{4}\PYG{p}{,}\PYG{+w}{ }\PYG{n}{lower}\PYG{p}{)}
\PYG{+w}{    }\PYG{n}{a}\PYG{+w}{ }\PYG{o}{=}\PYG{+w}{ }\PYG{n}{Index}\PYG{p}{(}\PYG{l+s}{\PYGZdq{}}\PYG{l+s}{a}\PYG{l+s}{\PYGZdq{}}\PYG{p}{,}\PYG{+w}{ }\PYG{n}{st}\PYG{p}{)}\PYG{p}{;}\PYG{+w}{ }\PYG{n}{b}\PYG{+w}{ }\PYG{o}{=}\PYG{+w}{ }\PYG{n}{Index}\PYG{p}{(}\PYG{l+s}{\PYGZdq{}}\PYG{l+s}{b}\PYG{l+s}{\PYGZdq{}}\PYG{p}{,}\PYG{+w}{ }\PYG{n}{st}\PYG{p}{)}
\PYG{+w}{    }\PYG{n}{c}\PYG{+w}{ }\PYG{o}{=}\PYG{+w}{ }\PYG{n}{Index}\PYG{p}{(}\PYG{l+s}{\PYGZdq{}}\PYG{l+s}{c}\PYG{l+s}{\PYGZdq{}}\PYG{p}{,}\PYG{+w}{ }\PYG{n}{st}\PYG{p}{)}\PYG{p}{;}\PYG{+w}{ }\PYG{n}{d}\PYG{+w}{ }\PYG{o}{=}\PYG{+w}{ }\PYG{n}{Index}\PYG{p}{(}\PYG{l+s}{\PYGZdq{}}\PYG{l+s}{d}\PYG{l+s}{\PYGZdq{}}\PYG{p}{,}\PYG{+w}{ }\PYG{n}{st}\PYG{p}{)}
\PYG{+w}{    }\PYG{n}{R1}\PYG{o}{=}\PYG{n}{RiemannTensor}\PYG{p}{(}\PYG{l+s}{\PYGZdq{}}\PYG{l+s}{R}\PYG{l+s}{\PYGZdq{}}\PYG{p}{,}\PYG{+w}{ }\PYG{p}{[}\PYG{n}{a}\PYG{o}{=\PYGZgt{}}\PYG{n}{lower}\PYG{p}{,}\PYG{+w}{ }\PYG{n}{b}\PYG{o}{=\PYGZgt{}}\PYG{n}{lower}\PYG{p}{,}\PYG{+w}{ }\PYG{n}{c}\PYG{o}{=\PYGZgt{}}\PYG{n}{lower}\PYG{p}{,}\PYG{+w}{ }\PYG{n}{d}\PYG{o}{=\PYGZgt{}}\PYG{n}{lower}\PYG{p}{]}\PYG{p}{)}
\PYG{+w}{    }\PYG{n}{R2}\PYG{o}{=}\PYG{n}{RiemannTensor}\PYG{p}{(}\PYG{l+s}{\PYGZdq{}}\PYG{l+s}{R}\PYG{l+s}{\PYGZdq{}}\PYG{p}{,}\PYG{+w}{ }\PYG{p}{[}\PYG{n}{a}\PYG{o}{=\PYGZgt{}}\PYG{n}{lower}\PYG{p}{,}\PYG{+w}{ }\PYG{n}{c}\PYG{o}{=\PYGZgt{}}\PYG{n}{lower}\PYG{p}{,}\PYG{+w}{ }\PYG{n}{d}\PYG{o}{=\PYGZgt{}}\PYG{n}{lower}\PYG{p}{,}\PYG{+w}{ }\PYG{n}{b}\PYG{o}{=\PYGZgt{}}\PYG{n}{lower}\PYG{p}{]}\PYG{p}{)}
\PYG{+w}{    }\PYG{n}{R3}\PYG{o}{=}\PYG{n}{RiemannTensor}\PYG{p}{(}\PYG{l+s}{\PYGZdq{}}\PYG{l+s}{R}\PYG{l+s}{\PYGZdq{}}\PYG{p}{,}\PYG{+w}{ }\PYG{p}{[}\PYG{n}{a}\PYG{o}{=\PYGZgt{}}\PYG{n}{lower}\PYG{p}{,}\PYG{+w}{ }\PYG{n}{d}\PYG{o}{=\PYGZgt{}}\PYG{n}{lower}\PYG{p}{,}\PYG{+w}{ }\PYG{n}{b}\PYG{o}{=\PYGZgt{}}\PYG{n}{lower}\PYG{p}{,}\PYG{+w}{ }\PYG{n}{c}\PYG{o}{=\PYGZgt{}}\PYG{n}{lower}\PYG{p}{]}\PYG{p}{)}
\end{Verbatim}

and time the call
\begin{Verbatim}[commandchars=\\\{\},xleftmargin=\parindent,numbers=left,bgcolor=bg]
\PYG{+w}{    }\PYG{n}{young\PYGZus{}projection}\PYG{p}{(}\PYG{n}{R1}\PYG{o}{+}\PYG{n}{R2}\PYG{o}{+}\PYG{n}{R3}\PYG{p}{)}
\end{Verbatim}
explicitly including the addition step before projection. For Redberry, Young projection is not implemented, so instead we resorted to a substitution of each term in the sum with its projected form. This favours Redberry, as it firstly doesn't need to compute the Young projector and apply it from scratch, and secondly it doesn't need to simplify the resulting large sum of redundant terms from the naive young projection.

\begin{table}[htbp]
  \centering
  \caption{Benchmark: \emph{Bianchi Identity} ($n=100$, times in \si{\micro\second}).}
  \label{tab:canon-riemann}
  \begin{tabular}{lrrr}
    \toprule
    Software & Min & Median & Mean $\pm$ 95\%\,CI \\
    \midrule
    Cadabra              & 1163.60     & 1203.83 & $1217.15 \pm 11.80$  \\
    xAct                 & 6706.00 & 6915.50 & $6997.01 \pm 60.04$  \\
    Alakazam             & 99.98   & 102.96  & $139.51 \pm 63.77$   \\
    Redberry ($N=10^{6}$) & 70.40   & 84.60   & $137.67 \pm 74.65$   \\
    Redberry ($N=1$)     & 177.04  & 388.77  & $746.33 \pm 302.92$  \\
    \bottomrule
  \end{tabular}
\end{table}

By the numbers, this test looks quite favourable to Redberry - roughly matching Alakazam when in the warm state. Though, again, this is not really comparable, as Redberry is doing an easier computation in stead of computing the Young projection. Despite this, Alakazam is still on parity in performance.

\begin{figure}[H]
    \centering
\begin{tikzpicture}
\begin{axis}[
    width=0.9\linewidth,
    height=6cm,
    ymode=log,
    log origin=infty,
    ymin=10, ymax=100000,
    ylabel={time (\si{\micro\second})},
    xmin=0.3, xmax=5.7,
    xtick={1, 2, 3.15, 3.85, 5},
    xticklabels={Cadabra, xAct, {Redberry\\(warm)}, {Redberry\\(cold)}, Alakazam},
    xticklabel style={align=center, font=\small},
    ymajorgrids=true,
    grid style={dashed, gray!30},
]
\addplot[ybar, bar width=0.5, fill=cadabcol, draw=black]
  coordinates {(1, 1203.83)};
\addplot[ybar, bar width=0.5, fill=xactcol, draw=black]
  coordinates {(2, 6915.50)};
\addplot[ybar, bar width=0.5, fill=rbhotcol, draw=black]
  coordinates {(3.17, 84.60)};
\addplot[ybar, bar width=0.5, fill=rbcol, draw=black]
  coordinates {(3.82, 388.77)};
\addplot[ybar, bar width=0.5, fill=akzo, draw=black,
         postaction={pattern={Lines[angle=45, distance=3pt, line width=1.2pt]},
                     pattern color=akzb}]
  coordinates {(5, 102.96)};

\node[anchor=south, font=\small] at (axis cs:1, 1203.83)   {$\sim$12$\times$};
\node[anchor=south, font=\small] at (axis cs:2, 6915.50)    {$\sim$70$\times$};
\node[anchor=south, font=\small] at (axis cs:3.17, 84.60) {$\sim$same};
\node[anchor=south, font=\small] at (axis cs:3.82, 388.77) {$\sim$4$\times$};
\end{axis}
\end{tikzpicture}
    \caption{Bianchi Identity / Young Projection, medians. Note that Redberry has no Young projection function, so this number is not directly comparable.}
    \label{fig:bench-bianchi}
\end{figure}
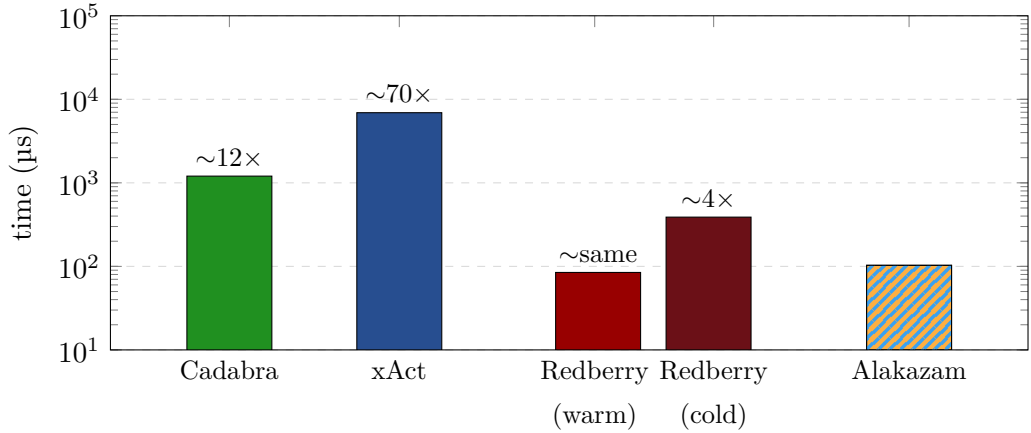

\subsection{Basic Multiterm Symmetry}\label{sec:multiterm-symmetry}
Any contraction of a pair of antisymmetric tensor indices with a pair of symmetric tensor indices vanishes identically.
\begin{equation}
    S^{(ab)}A_{[abcd]}=0.
\end{equation}
This is less easy to detect for CAS.
This is the test that should, in theory, be not so favourable to us, since we have to use our graph isomorphism algorithm to detect this reduction.

To benchmark, we create
\begin{Verbatim}[commandchars=\\\{\},xleftmargin=\parindent,numbers=left,bgcolor=bg]
\PYG{+w}{    }\PYG{n}{st}\PYG{+w}{ }\PYG{o}{=}\PYG{+w}{ }\PYG{n}{IndexSet}\PYG{p}{(}\PYG{l+s}{\PYGZdq{}}\PYG{l+s}{spacetime}\PYG{l+s}{\PYGZdq{}}\PYG{p}{,}\PYG{+w}{ }\PYG{l+m+mi}{4}\PYG{p}{,}\PYG{+w}{ }\PYG{n}{lower}\PYG{p}{)}
\PYG{+w}{    }\PYG{n+nd}{@indices}\PYG{+w}{ }\PYG{n}{a}\PYG{o}{:}\PYG{n}{e}\PYG{+w}{ }\PYG{o}{∈}\PYG{+w}{ }\PYG{n}{st}
\PYG{+w}{    }\PYG{n}{A1}\PYG{o}{=}\PYG{n}{AntisymmetricTensor}\PYG{p}{(}\PYG{l+s}{\PYGZdq{}}\PYG{l+s}{A}\PYG{l+s}{\PYGZdq{}}\PYG{p}{,}\PYG{+w}{ }\PYG{p}{[}\PYG{n}{a}\PYG{o}{=\PYGZgt{}}\PYG{n}{lower}\PYG{p}{,}\PYG{+w}{ }\PYG{n}{b}\PYG{o}{=\PYGZgt{}}\PYG{n}{lower}\PYG{p}{,}\PYG{+w}{ }\PYG{n}{c}\PYG{o}{=\PYGZgt{}}\PYG{n}{lower}\PYG{p}{,}\PYG{+w}{ }\PYG{n}{d}\PYG{o}{=\PYGZgt{}}\PYG{n}{lower}\PYG{p}{]}\PYG{p}{)}
\PYG{+w}{    }\PYG{n}{S1}\PYG{o}{=}\PYG{n}{SymmetricTensor}\PYG{p}{(}\PYG{l+s}{\PYGZdq{}}\PYG{l+s}{S}\PYG{l+s}{\PYGZdq{}}\PYG{p}{,}\PYG{+w}{ }\PYG{p}{[}\PYG{n}{a}\PYG{o}{=\PYGZgt{}}\PYG{n}{upper}\PYG{p}{,}\PYG{+w}{ }\PYG{n}{b}\PYG{o}{=\PYGZgt{}}\PYG{n}{upper}\PYG{p}{]}\PYG{p}{)}
\PYG{+w}{    }\PYG{n}{expr}\PYG{o}{=}\PYG{n}{A1}\PYG{o}{*}\PYG{n}{S1}
\end{Verbatim}

and time the code
\begin{Verbatim}[commandchars=\\\{\},xleftmargin=\parindent,numbers=left,bgcolor=bg]
\PYG{+w}{    }\PYG{n}{hard\PYGZus{}simplify}\PYG{p}{(}\PYG{n}{expr}\PYG{p}{)}
\end{Verbatim}

\begin{table}[htbp]
  \centering
  \caption{Benchmark \emph{Multiterm Symmetry} ($n=100$, times in \si{\micro\second}).}
  \label{tab:multiterm}
  \begin{tabular}{lrrr}
    \toprule
    Software & Min & Median & Mean $\pm$ 95\%\,CI \\
    \midrule
    Cadabra              & 245.56     & 256.36    & $259.42 \pm 2.62$    \\
    xAct                 & 1085.00 & 1131.50 & $1166.26 \pm 14.63$ \\
    Alakazam             & 6.64    & 6.95    & $7.14 \pm 0.25$   \\
    Redberry ($N=10^{6}$) & 51.33   & 56.79   & $69.01 \pm 6.16$    \\
    Redberry ($N=1$)     & 256.95  & 588.86  & $878.02 \pm 272.01$ \\
    \bottomrule
  \end{tabular}
\end{table}

Again this test was difficult to compare to Redberry. While claiming to recognise symmetries such as these (and indeed, similar examples from the documentation are reproducible) it was unable to reduce this particular expression. To find a remotely comparable benchmark, we assist it by manually antisymmetrising the indices of $A$ and multiplying it with $S$. Thus, this is again not a literal operational comparison, but still suggests how an equivalent operation might perform.

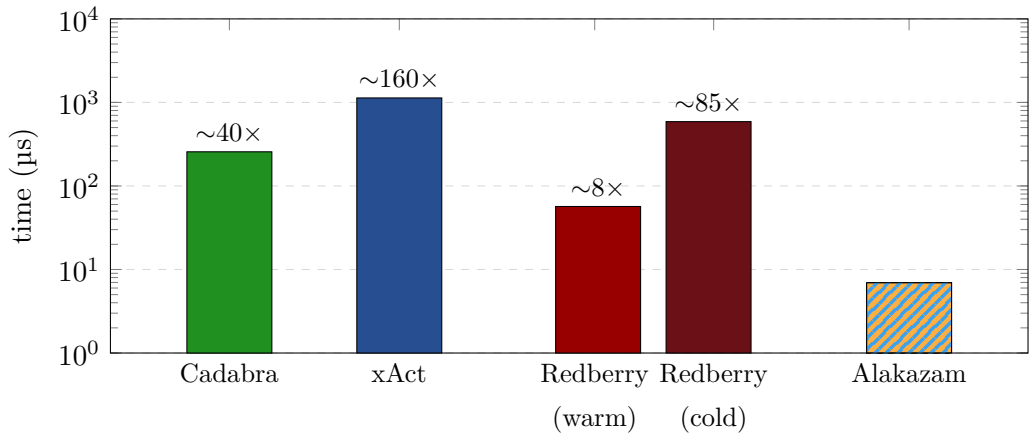
\begin{figure}[H]
    \centering
\begin{tikzpicture}
\begin{axis}[
    width=0.9\linewidth,
    height=6cm,
    ymode=log,
    log origin=infty,
    ymin=1, ymax=10000,
    ylabel={time (\si{\micro\second})},
    xmin=0.3, xmax=5.7,
    xtick={1, 2, 3.15, 3.85, 5},
    xticklabels={Cadabra, xAct, {Redberry\\(warm)}, {Redberry\\(cold)}, Alakazam},
    xticklabel style={align=center, font=\small},
    ymajorgrids=true,
    grid style={dashed, gray!30},
]
\addplot[ybar, bar width=0.5, fill=cadabcol, draw=black]
  coordinates {(1, 256.36)};
\addplot[ybar, bar width=0.5, fill=xactcol, draw=black]
  coordinates {(2,1131.5)};
\addplot[ybar, bar width=0.5, fill=rbhotcol, draw=black]
  coordinates {(3.17, 56.79)};
\addplot[ybar, bar width=0.5, fill=rbcol, draw=black]
  coordinates {(3.82, 588.86)};
\addplot[ybar, bar width=0.5, fill=akzo, draw=black,
         postaction={pattern={Lines[angle=45, distance=3pt, line width=1.2pt]},
                     pattern color=akzb}]
  coordinates {(5,  6.95)};

\node[anchor=south, font=\small] at (axis cs:1, 256.36)     {$\sim$40$\times$};
\node[anchor=south, font=\small] at (axis cs:2, 1131.5)    {$\sim$160$\times$};
\node[anchor=south, font=\small] at (axis cs:3.17, 56.79)  {$\sim$8$\times$};
\node[anchor=south, font=\small] at (axis cs:3.82, 588.86) {$\sim$85$\times$};
\end{axis}
\end{tikzpicture}
    \caption{Multiterm Symmetry, medians.}
    \label{fig:bench-symmetry}
\end{figure}

\subsection{Differentiation}\label{sec:differentiation}
A very important benchmark is the performance of basic calculus, as this operation grows in time complexity with term length. It also necessitates various operations such as replacements of tensors by their derivative, elimination of Kronecker deltas, merging of like-terms, in order to obtain a final canonical form. 

Here, we first test the performance of differentiation of a basic tensor polynomial
\begin{equation}
    A_{ab}X^a(x)X^b(x)+B_aX^a(x)+C
\end{equation}
and restoration to the canonical form. We first set up the problem with
\begin{Verbatim}[commandchars=\\\{\},xleftmargin=\parindent,numbers=left,bgcolor=bg]
\PYG{+w}{    }\PYG{n}{st}\PYG{+w}{ }\PYG{o}{=}\PYG{+w}{ }\PYG{n}{IndexSet}\PYG{p}{(}\PYG{l+s}{\PYGZdq{}}\PYG{l+s}{spacetime}\PYG{l+s}{\PYGZdq{}}\PYG{p}{,}\PYG{+w}{ }\PYG{l+m+mi}{4}\PYG{p}{,}\PYG{+w}{ }\PYG{n}{lower}\PYG{p}{)}
\PYG{+w}{    }\PYG{n}{a}\PYG{+w}{ }\PYG{o}{=}\PYG{+w}{ }\PYG{n}{Index}\PYG{p}{(}\PYG{l+s}{\PYGZdq{}}\PYG{l+s}{a}\PYG{l+s}{\PYGZdq{}}\PYG{p}{,}\PYG{+w}{ }\PYG{n}{st}\PYG{p}{)}\PYG{p}{;}\PYG{+w}{ }\PYG{n}{b}\PYG{+w}{ }\PYG{o}{=}\PYG{+w}{ }\PYG{n}{Index}\PYG{p}{(}\PYG{l+s}{\PYGZdq{}}\PYG{l+s}{b}\PYG{l+s}{\PYGZdq{}}\PYG{p}{,}\PYG{+w}{ }\PYG{n}{st}\PYG{p}{)}
\PYG{+w}{    }\PYG{n}{c}\PYG{+w}{ }\PYG{o}{=}\PYG{+w}{ }\PYG{n}{Index}\PYG{p}{(}\PYG{l+s}{\PYGZdq{}}\PYG{l+s}{c}\PYG{l+s}{\PYGZdq{}}\PYG{p}{,}\PYG{+w}{ }\PYG{n}{st}\PYG{p}{)}\PYG{p}{;}\PYG{+w}{ }\PYG{n}{d}\PYG{+w}{ }\PYG{o}{=}\PYG{+w}{ }\PYG{n}{Index}\PYG{p}{(}\PYG{l+s}{\PYGZdq{}}\PYG{l+s}{d}\PYG{l+s}{\PYGZdq{}}\PYG{p}{,}\PYG{+w}{ }\PYG{n}{st}\PYG{p}{)}
\PYG{+w}{    }\PYG{n}{x}\PYG{+w}{ }\PYG{o}{=}\PYG{+w}{ }\PYG{n}{Coordinate}\PYG{p}{(}\PYG{l+s}{\PYGZdq{}}\PYG{l+s}{x}\PYG{l+s}{\PYGZdq{}}\PYG{p}{,}\PYG{+w}{ }\PYG{p}{[}\PYG{n}{st}\PYG{p}{]}\PYG{p}{)}
\PYG{+w}{    }\PYG{n}{D}\PYG{o}{=}\PYG{n}{PartialDerivative}\PYG{p}{(}\PYG{l+s}{\PYGZdq{}}\PYG{l+s}{∂}\PYG{l+s}{\PYGZdq{}}\PYG{p}{,}\PYG{+w}{ }\PYG{n}{x}\PYG{p}{,}\PYG{n}{a}\PYG{o}{=\PYGZgt{}}\PYG{n}{lower}\PYG{p}{)}
\PYG{+w}{    }\PYG{n}{X}\PYG{o}{=}\PYG{n}{Tensor}\PYG{p}{(}\PYG{l+s}{\PYGZdq{}}\PYG{l+s}{X}\PYG{l+s}{\PYGZdq{}}\PYG{p}{,}\PYG{n}{x}\PYG{p}{,}\PYG{n}{b}\PYG{o}{=\PYGZgt{}}\PYG{n}{upper}\PYG{p}{)}
\PYG{+w}{    }\PYG{n+nd}{@action}\PYG{+w}{ }\PYG{n}{D}\PYG{p}{(}\PYG{n}{X}\PYG{p}{)}\PYG{o}{=}\PYG{n}{KroneckerDelta}\PYG{p}{(}\PYG{n}{a}\PYG{o}{=\PYGZgt{}}\PYG{n}{lower}\PYG{p}{,}\PYG{n}{b}\PYG{o}{=\PYGZgt{}}\PYG{n}{upper}\PYG{p}{)}
\PYG{+w}{    }\PYG{n}{eq}\PYG{o}{=}\PYG{p}{(}\PYG{n}{SymmetricTensor}\PYG{p}{(}\PYG{l+s}{\PYGZdq{}}\PYG{l+s}{A}\PYG{l+s}{\PYGZdq{}}\PYG{p}{,}\PYG{+w}{ }\PYG{p}{[}\PYG{n}{a}\PYG{+w}{ }\PYG{o}{=\PYGZgt{}}\PYG{+w}{ }\PYG{n}{lower}\PYG{p}{,}\PYG{+w}{ }\PYG{n}{b}\PYG{+w}{ }\PYG{o}{=\PYGZgt{}}\PYG{+w}{ }\PYG{n}{lower}\PYG{p}{]}\PYG{p}{)}\PYG{o}{*}
\PYG{+w}{    }\PYG{n}{Tensor}\PYG{p}{(}\PYG{l+s}{\PYGZdq{}}\PYG{l+s}{X}\PYG{l+s}{\PYGZdq{}}\PYG{p}{,}\PYG{+w}{ }\PYG{p}{[}\PYG{n}{a}\PYG{+w}{ }\PYG{o}{=\PYGZgt{}}\PYG{+w}{ }\PYG{n}{upper}\PYG{p}{]}\PYG{p}{,}\PYG{+w}{ }\PYG{n}{x}\PYG{p}{)}\PYG{o}{*}\PYG{n}{Tensor}\PYG{p}{(}\PYG{l+s}{\PYGZdq{}}\PYG{l+s}{X}\PYG{l+s}{\PYGZdq{}}\PYG{p}{,}\PYG{+w}{ }\PYG{p}{[}\PYG{n}{b}\PYG{+w}{ }\PYG{o}{=\PYGZgt{}}\PYG{+w}{ }\PYG{n}{upper}\PYG{p}{]}\PYG{p}{,}\PYG{+w}{ }\PYG{n}{x}\PYG{p}{)}
\PYG{+w}{    }\PYG{o}{+}\PYG{n}{Tensor}\PYG{p}{(}\PYG{l+s}{\PYGZdq{}}\PYG{l+s}{B}\PYG{l+s}{\PYGZdq{}}\PYG{p}{,}\PYG{+w}{ }\PYG{p}{[}\PYG{n}{a}\PYG{+w}{ }\PYG{o}{=\PYGZgt{}}\PYG{+w}{ }\PYG{n}{lower}\PYG{p}{]}\PYG{p}{)}\PYG{o}{*}\PYG{n}{Tensor}\PYG{p}{(}\PYG{l+s}{\PYGZdq{}}\PYG{l+s}{X}\PYG{l+s}{\PYGZdq{}}\PYG{p}{,}\PYG{+w}{ }\PYG{p}{[}\PYG{n}{a}\PYG{+w}{ }\PYG{o}{=\PYGZgt{}}\PYG{+w}{ }\PYG{n}{upper}\PYG{p}{]}\PYG{p}{,}\PYG{+w}{ }\PYG{n}{x}\PYG{p}{)}
\PYG{+w}{    }\PYG{o}{+}\PYG{n}{Tensor}\PYG{p}{(}\PYG{l+s}{\PYGZdq{}}\PYG{l+s}{C}\PYG{l+s}{\PYGZdq{}}\PYG{p}{)}\PYG{p}{)}
\end{Verbatim}
and benchmark the operation
\begin{Verbatim}[commandchars=\\\{\},xleftmargin=\parindent,numbers=left,bgcolor=bg]
\PYG{n}{rename\PYGZus{}dummies}\PYG{p}{(}\PYG{n}{eliminate\PYGZus{}kronecker}\PYG{p}{(}\PYG{n}{collect\PYGZus{}terms}\PYG{p}{(}\PYG{n}{apply\PYGZus{}derivative}\PYG{p}{(}\PYG{n}{D}\PYG{p}{(}\PYG{n}{eq}\PYG{p}{)}\PYG{p}{)}\PYG{p}{)}\PYG{p}{)}\PYG{p}{)}
\end{Verbatim}
and similarly the minimal equivalent operation in the other systems.

\begin{table}[htbp]
  \centering
  \caption{Benchmark: \emph{Differentiation} ($n=100$, times in \si{\micro\second}).}
  \label{tab:differentiation}
  \begin{tabular}{lrrr}
    \toprule
    Software & Min & Median & Mean $\pm$ 95\%\,CI \\
    \midrule
    Cadabra              & 1086.42 & 1131.30 & $1143.43 \pm 8.78$  \\
    xAct                 & 2069.00 & 2301.00 & $2325.61 \pm 34.44$ \\
    Alakazam             & 19.29  & 19.94   & $22.59 \pm 3.20$    \\
    Redberry ($N=10^{6}$) & 71.26   & 74.77   & $86.76 \pm 5.73$    \\
    Redberry ($N=1$)     & 232.51  & 481.28  & $673.45 \pm 99.18$  \\
    \bottomrule
  \end{tabular}
\end{table}

We see that Alakazam is able to perform differentiation and reduce the resulting expression to canonical form between one to two orders of magnitude faster than Cadabra and xAct, and almost four times as fast as Redberry. This is an important benchmark for performing calculations in superspace. It is useful to note that in the case of Redberry, any expression with derivatives parses directly to the differentiated form. This would make covariant derivatives employed as in Alakazam, tricky.

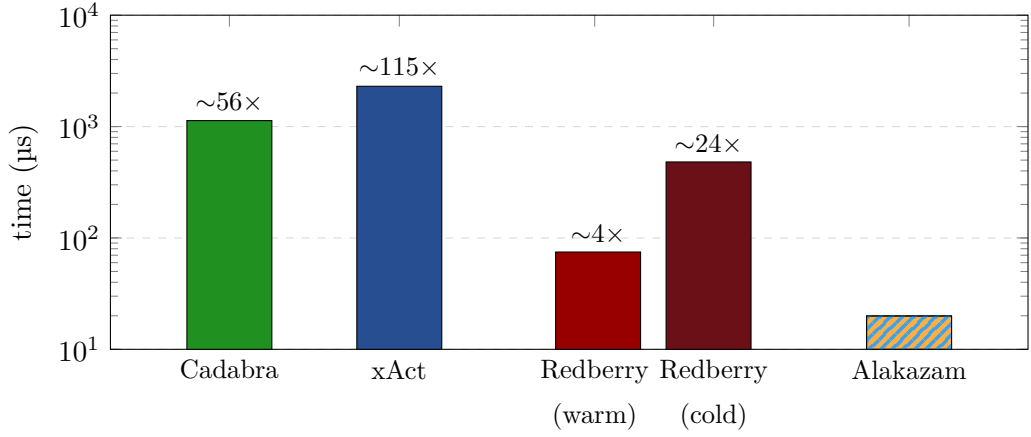
\begin{figure}[H]
    \centering
\begin{tikzpicture}
\begin{axis}[
    width=0.9\linewidth,
    height=6cm,
    ymode=log,
    log origin=infty,
    ymin=10, ymax=10000,
    ylabel={time (\si{\micro\second})},
    xmin=0.3, xmax=5.7,
    xtick={1, 2, 3.15, 3.85, 5},
    xticklabels={Cadabra, xAct, {Redberry\\(warm)}, {Redberry\\(cold)}, Alakazam},
    xticklabel style={align=center, font=\small},
    ymajorgrids=true,
    grid style={dashed, gray!30},
]
\addplot[ybar, bar width=0.5, fill=cadabcol, draw=black]
  coordinates {(1, 1131.3)};
\addplot[ybar, bar width=0.5, fill=xactcol, draw=black]
  coordinates {(2,2301)};
\addplot[ybar, bar width=0.5, fill=rbhotcol, draw=black]
  coordinates {(3.17, 74.77 )};
\addplot[ybar, bar width=0.5, fill=rbcol, draw=black]
  coordinates {(3.82, 481.28)};
\addplot[ybar, bar width=0.5, fill=akzo, draw=black,
         postaction={pattern={Lines[angle=45, distance=3pt, line width=1.2pt]},
                     pattern color=akzb}]
  coordinates {(5, 19.94)};

\node[anchor=south, font=\small] at (axis cs:1, 1131.3)    {$\sim$56$\times$};
\node[anchor=south, font=\small] at (axis cs:2, 2301)      {$\sim$115$\times$};
\node[anchor=south, font=\small] at (axis cs:3.17, 74.77) {$\sim$4$\times$};
\node[anchor=south, font=\small] at (axis cs:3.82, 481.28) {$\sim$24$\times$};
\end{axis}
\end{tikzpicture}
    \caption{Basic Differentiation, medians.}
    \label{fig:bench-diff}
\end{figure}

\begin{figure}[H]
\centering
\begin{tikzpicture}
\begin{axis}[
    width=0.9\linewidth,
    height=7cm,
    ymode=log,
    log origin=infty,
    xlabel={Monomial order, $X^n$},
    ylabel={time (\si{\micro\second})},
    xmin=1, xmax=100,
    grid=major,
    grid style={dashed, gray!30},
    legend style={
        at={(0.5,-0.25)},
        anchor=north,
        legend columns=-1
    },
]

\addplot[
    thick,
    color=cadabcol
] table[
    x=order,
    y=cadabra,
    col sep=comma
] {ims/plotdata.csv};
\addlegendentry{Cadabra}

\addplot[
    thick,
    color=xactcol
] table[
    x=order,
    y=xact,
    col sep=comma
] {ims/plotdata.csv};
\addlegendentry{xAct}

\addplot[
    thick,
    color=rbhotcol
] table[
    x=order,
    y=redberry_warm,
    col sep=comma
] {ims/plotdata.csv};
\addlegendentry{Redberry (warm)}

\addplot[
    thick,
    color=rbcol
] table[
    x=order,
    y=redberry_cold,
    col sep=comma
] {ims/plotdata.csv};
\addlegendentry{Redberry (cold)}

\addplot[
    thick,
    color=akzo
] table[
    x=order,
    y=alakazam,
    col sep=comma
] {ims/plotdata.csv};
\addlegendentry{Alakazam}

\end{axis}
\end{tikzpicture}
\caption{Median timings for monomial orders 1--100.}
\label{fig:bench-monomial}
\end{figure}
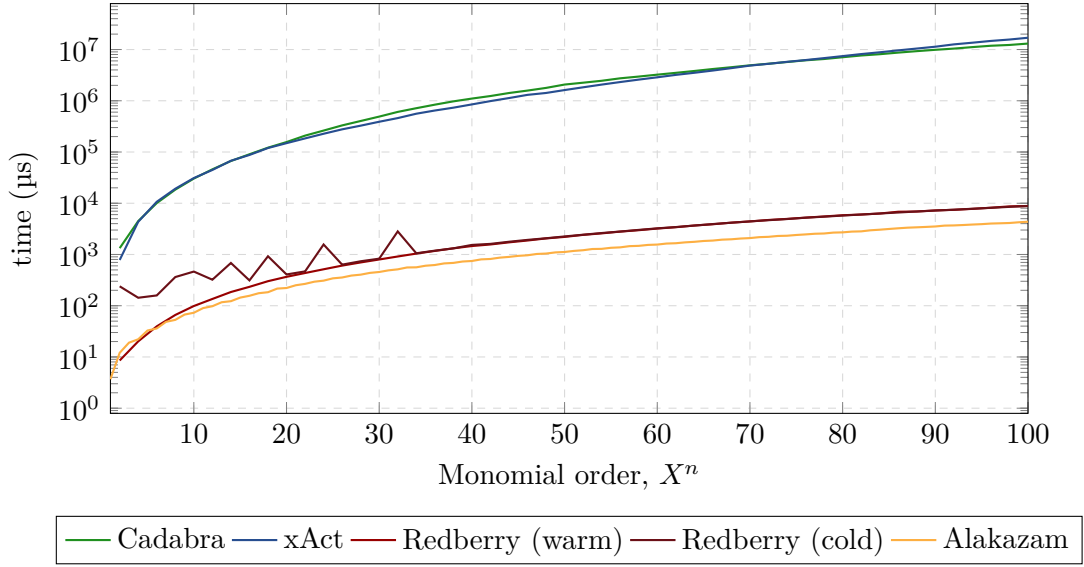

A useful metric with regard to differentiation is how the algorithm scales with term size. For this, we consider differentiating $n$-term monomials
\begin{equation}
    \del_\mu\left[\prod_{i=1}^{n/2}(X^{\nu_i}(x)X_{\nu_{i}}(x))\right]
\end{equation}
with 
\begin{equation}
    \del_\mu X^\nu(x)=\delta_\mu{}^\nu, \quad  \del_\mu X_\nu(x)=\eta_\mu{}_\nu,
\end{equation}
and simplifying to the canonical form. In Alakazam, this is done via
\begin{Verbatim}[commandchars=\\\{\},xleftmargin=\parindent,numbers=left,bgcolor=bg]
\PYG{+w}{    }\PYG{n}{eliminate\PYGZus{}kronecker}\PYG{p}{(}\PYG{n}{contract\PYGZus{}metrics}\PYG{p}{(}\PYG{n}{collect\PYGZus{}terms}\PYG{p}{(}\PYG{n}{apply\PYGZus{}derivative}\PYG{p}{(}\PYG{n}{D}\PYG{p}{(}\PYG{n}{eq}\PYG{p}{)}\PYG{p}{)}\PYG{p}{)}\PYG{p}{)}\PYG{p}{)}
\end{Verbatim}
We can collect the median times for each system, and plot in Figure~\ref{fig:bench-monomial}. Here, we use $N=10,000$ runs before taking data for the Redberry warm state. We take 100 samples for Alakazam and Redberry, and 5 samples for xAct and Cadabra for the interest of runtime of the benchmark.

In this plot, we see that both Redberry and Alakazam perform much better than Cadabra and xAct for any order monomial.
Note that for Cadabra, we must \jil{sort_product} on such an expression instead of \jil{canonicalise}, otherwise the naive Butler-Portugal becomes too expensive at around $n=16$ and never finished on our system, hanging for an unreasonable amount of time. A similar story holds for xAct, for which instead the pairs of tensors with dummy contractions must be reframed as independent scalars to ignore the dummy index repeats. Without doing this, for the case of xAct, in our benchmarks 16-term monomials cause WSL to run out of memory (16GB) and crash the operating system. This similar breakdown is likely because both Cadabra and xAct perform the core of this operation using the same xPerm implementation, and also explains the very similar performance curves.

Note that Redberry differentiates expressions on parse, as opposed to in Alakazam, where derivatives are genuine objects that can be manipulated. This allows derivatives to be further manipulated before evaluation - for example, commuting the order of derivatives to simplify an expression without evaluating on their operands. This advantage comes at little-to-no performance cost. We also see that the Redberry timings saturate and the warm and cold timings converge at around 30-40 terms.

At low order monomials, Alakazam and Redberry in its warm state roughly have parity. However, Redberry exhibits faster growth, meaning that for large monomials Alakazam runs in approximately half the time. Note that this statistic measures just the derivative and canonicalisation step, excluding the initial parsing of the expression, which when included would favour Alakazam especially at low monomial length, as these load times are comparable in magnitude to the timings for small numbers of terms.

\subsection{Simplification}\label{sec:simplification}
We finally would like to compare the ability of each package to recognise when expressions are equivalent, and merge them. To benchmark the performance, we consider two cases each with two branches:
\begin{itemize}
    \item Simplification time as a function of product length, with and without symmetries on the indices
    \item Simplification time as a function of number of indices, with and without symmetries on the indices
\end{itemize}
To test this, we wrote a script to randomly generate a product of tensors of a fixed length (or fixed number of indices). Our script then shuffles the terms in the product to obtain another monomial. For the symmetry case, we additionally allow for a permitted random permutation of the indices of the tensors. We then subtract this from the original expression. Each seed term is additionally tested by our script to certify that it is non-zero, so we truly test recognition of symmetries. We repeat until we have an expression of 200 summands consisting of these pairs, which is mathematically equal to zero. We shuffle the resulting summands before exporting them to each CAS. We then test the ability for each package to reduce the resulting expression to zero. For each product length/number of indices, we randomly generate 10 different expressions in this way for our benchmark. We impose a $\sim$60s limit for each simplification procedure, and if the system fails to reduce the expression in this time, it is counted as a fail.
For example, without symmetries, one summand of a tested expression was
\begin{equation}
   B_{g}{}^{f}\,B^{b}{}_{h}\,C_{f}{}^{ch}\,A^{d}\,C^{aeg}
\end{equation}
from which an equivalent summand is generated,
\begin{equation}
    A^{d}\,C_{g}{}^{cf}\,B_{h}{}^{g}\,C^{aeh}\,B^{b}{}_{f}
\end{equation}
which is a reordered term with renamed dummies, for which the difference must vanish.
With symmetries, we can generate a term like
\begin{equation}
       C_{(f}{}^{ab)}\,B^{[e}{}_{g]}\,A^{f}\,B_{[h}{}^{c]}\,C^{(hdg)}
\end{equation}
which is equivalent to
\begin{equation}
     A^{h}\,C_{(h}{}^{ba)}\,B_{[f}{}^{e]}\,C^{(fdg)}\,B^{[c}{}_{g]}
\end{equation}
and so the difference must vanish. Here, the brackets denote the declared symmetries. We also allow for antisymmetric tensors up to rank 3 in these tests.

For Alakazam and Redberry, each sample point corresponds to the median of $20$ trials. For xAct and Cadabra, each point is a single trial, for the interest of speed of the test suite. Here, we use $N=50$ runs before measuring the Redberry warm state.

\begin{figure}[H]
    \centering

    \begin{subfigure}{0.43\linewidth}
        \centering
        \includegraphics[width=\linewidth, trim=0cm 0cm 5cm 0cm,
    clip]{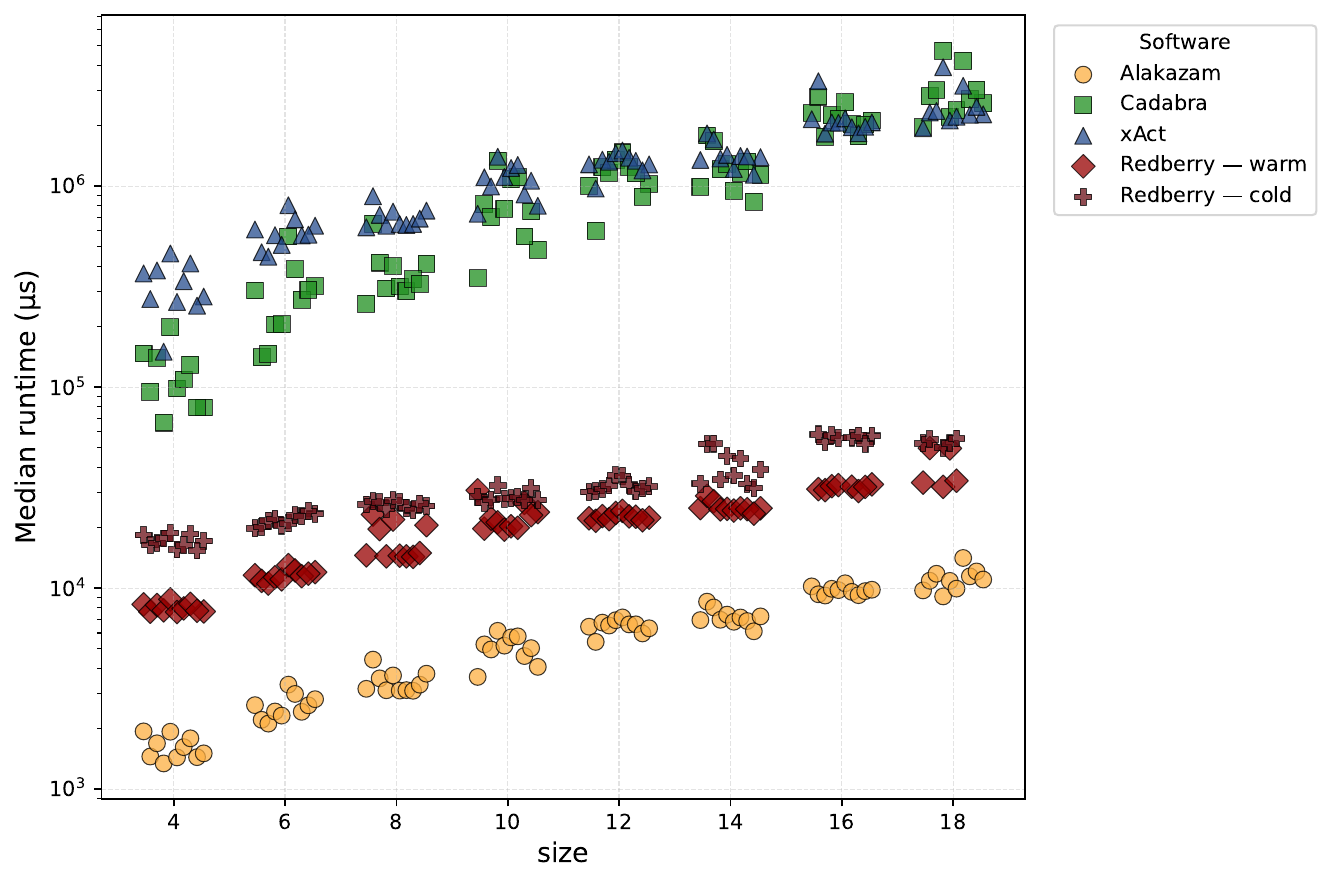}
        \caption{Runtime vs $\#$ tensors.}
        \label{fig:prodsbenchmarknosym}
    \end{subfigure}
    \hfill
    \begin{subfigure}{0.56\linewidth}
        \centering
        \includegraphics[width=\linewidth]{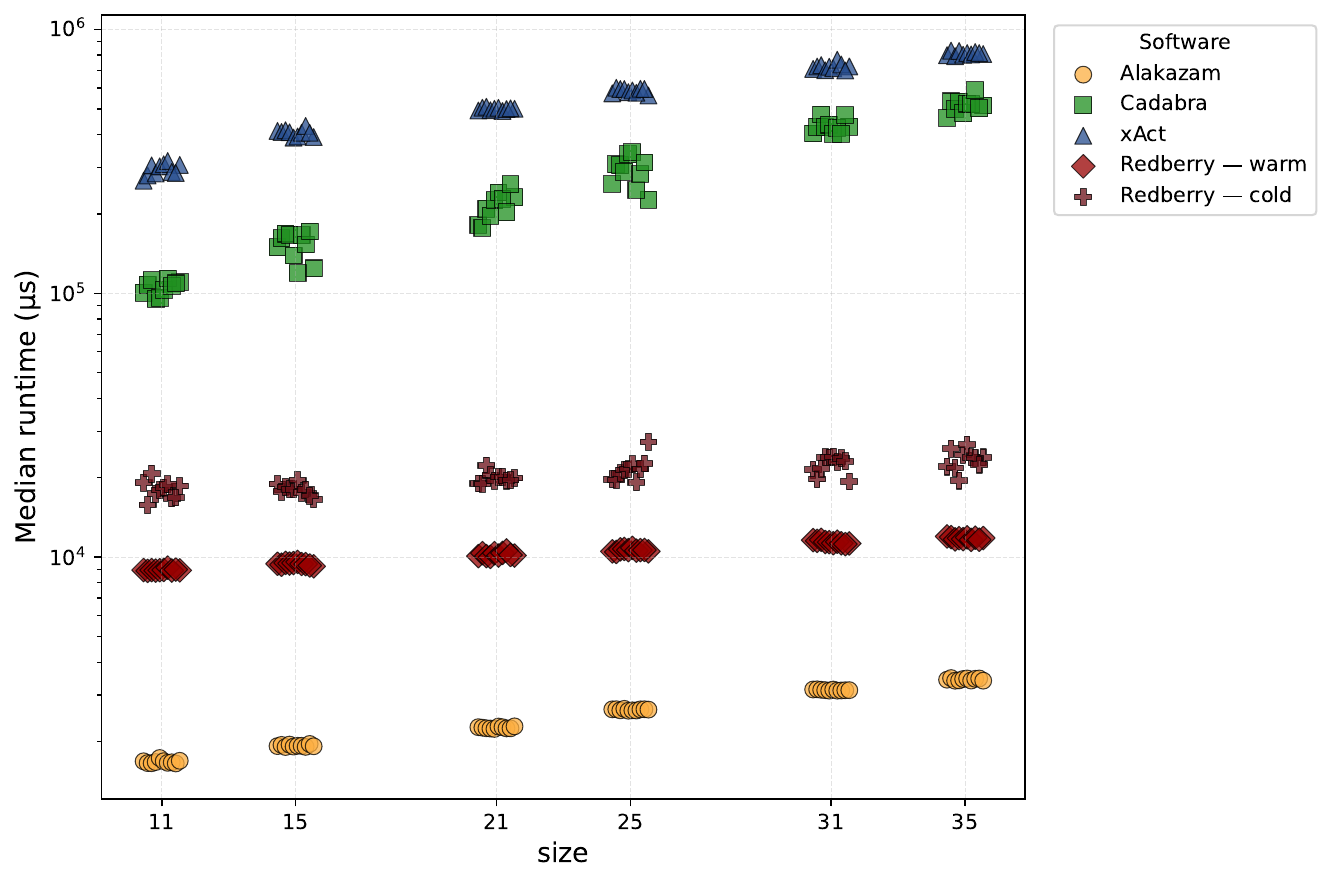}
        \caption{Runtime vs $\#$ indices.}
        \label{fig:indsbenchmarknosym}
    \end{subfigure}

    \caption{Comparison of simplification with no symmetries.}
    \label{fig:benchmarksnosym}
\end{figure}
For the case of no symmetries, all software packages passed all runs. We see that Alakazam outperforms the other platforms by roughly an order of magnitude at the minimum.
\begin{figure}[H]
    \centering
    \begin{subfigure}{0.43\linewidth}
        \centering
        \includegraphics[width=\linewidth,trim=0cm 0cm 5cm 0cm,
    clip]{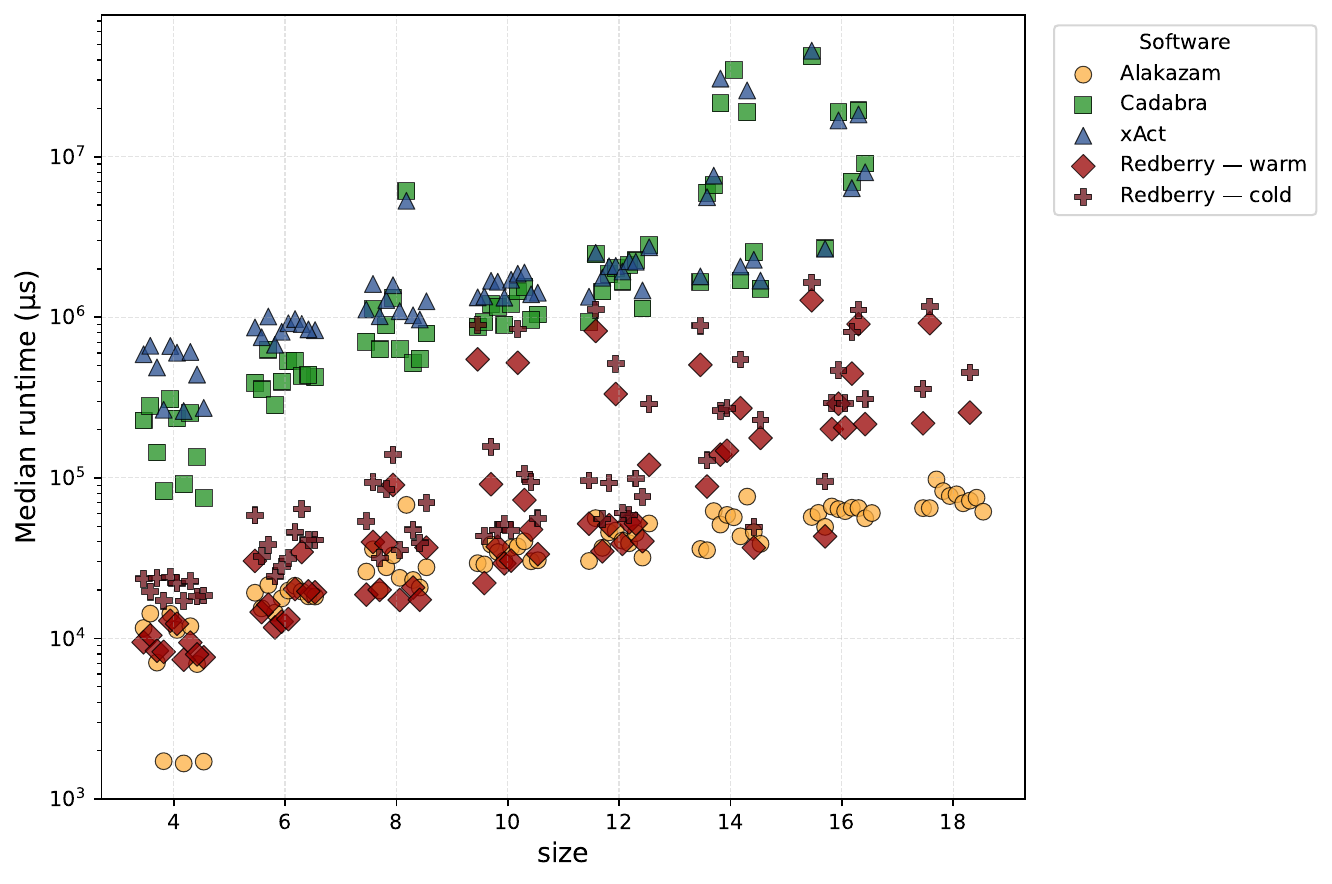}
        \caption{Runtime vs $\#$ tensors, w/ symmetries.}
        \label{fig:prodsbenchmarksyms}
    \end{subfigure}
        \hfill
   \begin{subfigure}{0.56\linewidth}
        \centering
        \includegraphics[width=\linewidth]{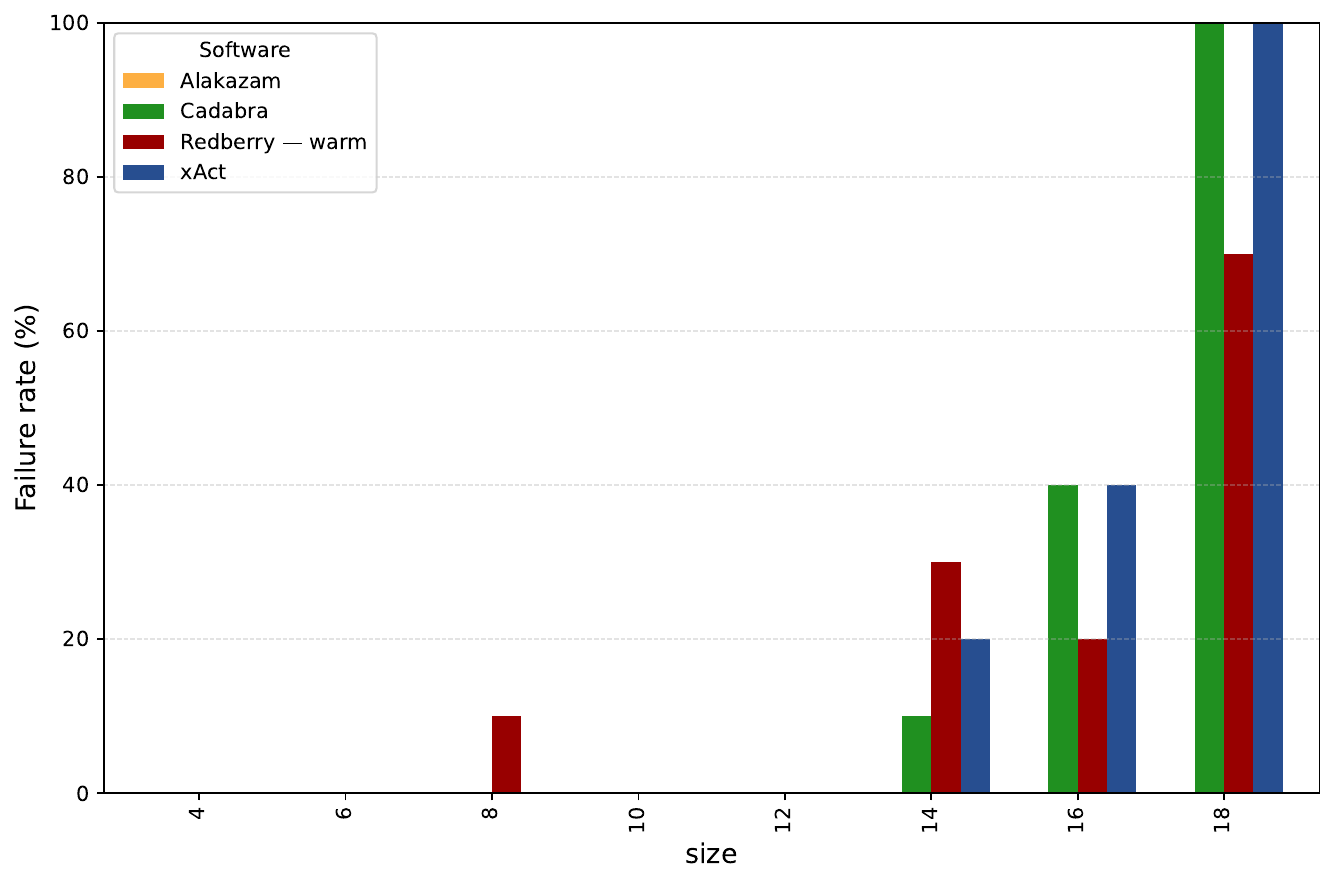}
        \caption{Product simplification failure rate.}
        \label{fig:prodsbenchmarkfails}
    \end{subfigure}
    \caption{Simplification vs $\#$ tensors with symmetries.}
    \label{fig:prodssyms}
\end{figure}
We see that, for the case where tensors have symmetries, Alakazam has parity with Redberry for small products, while being faster than xAct and Cadabra by roughly an order of magnitude. For larger products, the behaviour diverges, as the runtimes for Redberry become much higher, as well as increasingly failing to simplify the expressions. For large products, xAct and Cadabra both fail due to timeout, likely due to the increased runtime of the Butler Portugal algorithm needed for simplification. Note that Alakazam records no failures to simplify expressions.
\begin{figure}[H]
    \centering
    \begin{subfigure}{0.43\linewidth}
        \centering
        \includegraphics[width=\linewidth,trim=0cm 0cm 5cm 0cm,
    clip]{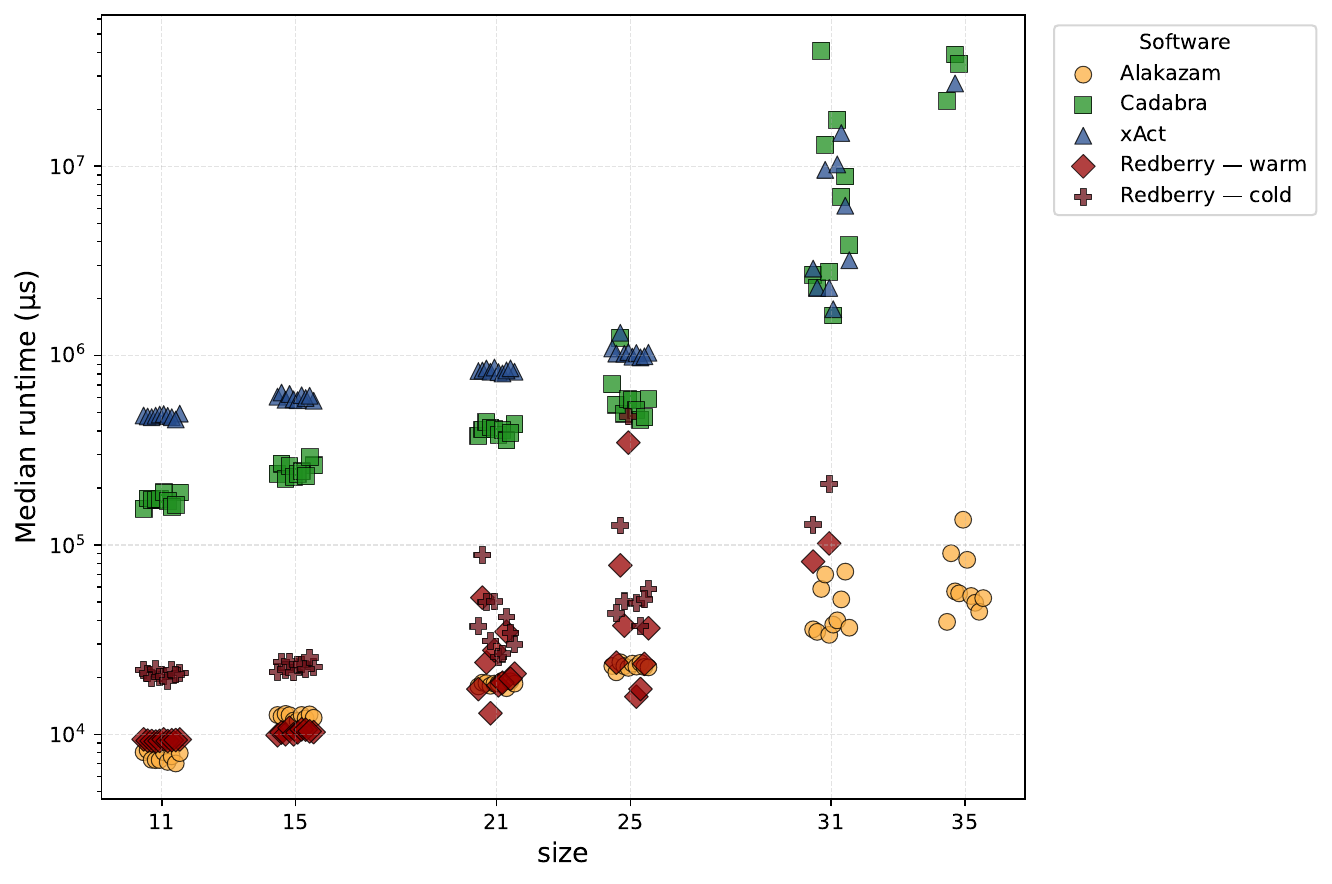}
        \caption{Runtime vs $\#$ indices, w/ symmetries.}
        \label{fig:indsbenchmarksyms}
    \end{subfigure}
        \hfill
   \begin{subfigure}{0.56\linewidth}
        \centering
        \includegraphics[width=\linewidth]{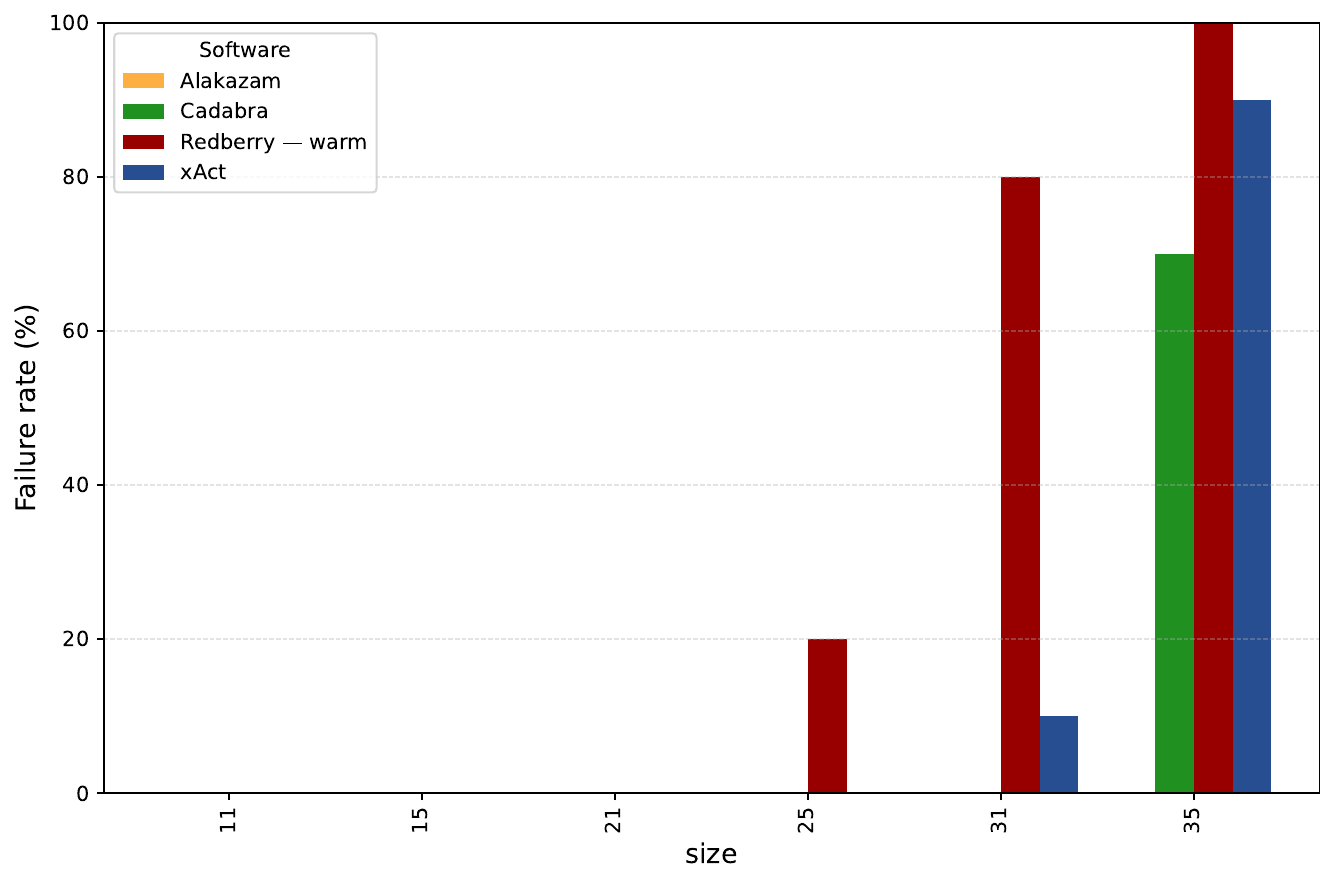}
        \caption{Indices simplification failure rate.}
        \label{fig:indsbenchmarkfails}
    \end{subfigure}
    \caption{Simplification vs $\#$ indices with symmetries.}
    \label{fig:indssyms}
\end{figure}
A similar story holds for simplification as a function of the number of indices. Alakazam and Redberry have parity for low numbers of indices, and both are roughly an order of magnitude faster than Cadabra and xAct. We see however an increased failure rate here for Redberry at a high number of indices, in agreement with the observations made by the benchmarks in the Redberry paper~\cite{Bolotin:2013qgr}. We again make the note that Alakazam passes all simplification tests with no failure to reduce or encountering a timeout.

\newpage
\section{Conclusion}\label{sec:conclusion}
In this paper, we introduced Alakazam, an open-source Julia package designed from the ground up to manipulate symbolic, potentially graded, tensorial expressions. We have implemented several key features that are important for dealing with abstract tensors in a computer algebra system, for instance custom algorithms for manipulating expressions with distinct free and dummy indices. We have support for basic tensors, fermionic tensors, non-commuting matrices, covariant differentiation, replacement pattern matching, symmetry recognition, simplification, traces, and superspace integration. Everything in Alakazam was designed with supersymmetry in mind, so all operations will produce consistent grading. We have explicitly given examples throughout the manual illustrating non-trivial use cases for the algorithms. Given the large feature set, and fast speeds, we hope that many users will find Alakazam useful to their work. We invite interested users to keep up with the development of Alakazam on Gitlab. Being open-source, it is open to community contributions.

In order to be useful as a computer algebra system for tensor computations, there are several features that are generally agreed upon that should ideally be present. A set of criteria for effectiveness has been proposed for example in~\cite{Korolkova:2013diz}. They note that almost no systems have satisfactory support of tensors. Our implementation focuses mainly on abstract index computations, which we believe is satisfactorily addressed. Additionally, we do have partial support of non-index computation via grading and non-commuting matrix types. In future, we would like to implement component computations for tensors, if there is demand. This is the one criterion of~\cite{Korolkova:2013diz} that we do not address at all - noting that the authors there declare that no currently available systems address all their requirements. However, given that our package lives in the Julia environment and is not an isolated system, such component calculations are most likely easily implementable in combination with other packages in the Julia ecosystem. In this direction, there are several features that we would additionally consider for implementation in future versions, including coordinate-free differential forms, differentiable functions of tensors, operations on Young tableau, and increased support for quantum mechanics via explicit bra and ket structures. For many of these ideas, the groundwork exists already. Many additional features for General Relativity and spinors are already in progress.

We note that Alakazam is new, and not yet mature. We see this as an opportunity for further optimisation and development of new features based on the community demand. Additionally, Julia receives frequent updates which often increase performance. Thus, we can optimistically hope that over time, our benchmarks will become even better.

\newpage
\section{Acknowledgements}\label{sec:acknowledgements}
The author would like to thank Gabriele Tartaglino-Mazzucchelli for several important comments on the manuscript. The author would like to thank Valentin Hirschi and Lucien Huber for discussions on the manuscript, algorithms, and symbolics. 
The author would like to thank Cian Luke Martin for the suggestion of macros during the early stages of development. The author would like to thank Alejandro Soto Franco for several useful merge requests. 
The author is supported by the Swiss National Science Foundation under grant number 200021\_219267. The author would like to thank the Ramsay Centre for Western Civilisation for its support. The author would also like to thank The University of Queensland for the Duncan McNaughton Scholarship, which supported part of this work. Credits to Kasper Peeters, the creator of Cadabra, whose project inspired the author to create Alakazam.

\newpage

\newpage

\appendix
\section{Index of Exports}\label{sec:export-index}
This section is a compilation of all the structures and functions exported to the user to use within Alakazam after running \jil{using Alakazam}. We include a short description and a back-link to the relevant location in the text. We include here also several aliases/shorthands for common functions.

\subsection*{Concrete Types}
\begin{longtable}{@{}p{0.36\linewidth}p{0.49\linewidth}p{0.1\linewidth}@{}}
\toprule
Name & Description & Section \\
\midrule
\endfirsthead
\toprule
Name & Description & Section \\
\midrule
\endhead
\jil{Boson} & Basic Boson & \ref{sec:bosonfermion} \\
\jil{Bra} & Bra used in Quantum Mechanics & \ref{sec:braket} \\
\jil{CoeffPair} & One coefficient paired with its term & \ref{sec:efficiency-considerations} \\
\jil{Commutator} & Stores a commutation relation & \ref{sec:lie-brackets} \\
\jil{Coordinate} & A coordinate that tensors may be functions of & \ref{sec:coordinate} \\
\jil{CovariantCommutator} & Commutation of derivative operators & \ref{sec:derivative} \\
\jil{Dagger} & Hermitian conjugation, held unevaluated until applied & \ref{sec:dagger-and-complex-conjugation} \\
\jil{Delta} & Alias of \jil{KroneckerDelta} & \ref{sec:kroneckerdelta} \\
\jil{Derivative} & Generic derivative object & \ref{sec:derivatives} \\
\jil{DummyPatternIndex} & A wildcard index, bound when a pattern is matched & \ref{sec:basic-replacement-objects} \\
\jil{EpsilonTensor} & The Levi-Civita tensor & \ref{sec:epsilontensor} \\
\jil{Fermion} & Basic Fermion & \ref{sec:bosonfermion} \\
\jil{FreeIndexMismatch} & Error: summands do not carry the same free indices & \ref{sec:tensorexpression} \\
\jil{GammaM} & A gamma matrix & \ref{sec:gamma-matrices} \\
\jil{GammaT} & A gamma matrix component, a commuting tensor & \ref{sec:gamma-matrices} \\
\jil{Gamma5} & The chirality matrix & \ref{sec:gamma-matrices} \\
\jil{GrassmannStructure} & An independent structure in a Grassmann expansion & \ref{sec:superspace-integration} \\
\jil{IdMatrix} & The identity matrix & \ref{sec:traces} \\
\jil{Index} & A single index, belonging to one IndexSet & \ref{sec:index} \\
\jil{IndexGrading} & Whether an index set is bosonic or fermionic & \ref{sec:indexset} \\
\jil{IndexPosition} & Structure storing an \jil{Int}. Position is 0 for upper index, 1 for lower index & \ref{sec:indexset} \\
\jil{IndexSet} & A class of indices: range, position, metric, grading and conjugate & \ref{sec:indexset} \\
\jil{InvalidCommutatorDefinition} & Error: a commutator was declared inconsistently & \ref{sec:lie-brackets} \\
\jil{InvalidDerivativeBasis} & Error: an unusable basis was given to a derivative & \ref{sec:partialderivative} \\
\jil{InvalidDifferentialForm} & Error: the object is not a well-formed differential form & \ref{sec:differential-geometry} \\
\jil{InvalidEpsilonIndices} & Error: an epsilon was given the wrong number or type of indices & \ref{sec:epsilontensor} \\
\jil{InvalidKroneckerIndices} & Error: a Kronecker delta was given unusable indices & \ref{sec:kroneckerdelta} \\
\jil{InvalidReplacementPattern} & Error: the replacement pattern cannot be matched & \ref{sec:pattern-matching-and-replacements} \\
\jil{Ket} & Ket used in Quantum Mechanics & \ref{sec:braket} \\
\jil{KroneckerDelta} & The Kronecker delta & \ref{sec:kroneckerdelta} \\
\jil{Lie} & Lie Bracket Operator & \ref{sec:lie-brackets} \\
\jil{LieGenerator} & A generator of a Lie algebra & \ref{sec:lie-algebra-generators} \\
\jil{Mat} & Basic Matrix object & \ref{sec:general-matrix-relations} \\
\jil{Metric} & Raises and lowers indices, default signature (1,d-1) & \ref{sec:indexset} \\
\jil{MixedWeightOperandTerms} & Error: an operand mixes terms of different weight & --- \\
\jil{Operator} & A named operator acting on an expression & \ref{sec:operators} \\
\jil{OperatorData} & Internal information an operator carries & \ref{sec:operators} \\
\jil{OperatorPrecedenceError} & Error: operators were combined in an unusable order & --- \\
\jil{OutOfIndices} & Error: the index set has no unused index left & \ref{sec:indexset} \\
\jil{OverusedIndices} & Error: an index appears more times than summation allows & \ref{sec:einstein-summation-and-dummy-indices} \\
\jil{Partial} & Alias of \jil{PartialDerivative} & \ref{sec:derivatives} \\
\jil{PartialDerivative} & Generic (anti)commuting partial derivatives & \ref{sec:derivatives} \\
\jil{PauliM} & A Pauli matrix & \ref{sec:pauli-matrices} \\
\jil{PauliT} & A Pauli matrix component, a commuting tensor & \ref{sec:pauli-matrices} \\
\jil{PD} & Alias of \jil{PartialDerivative} & \ref{sec:derivatives} \\
\jil{RegisteredExpansion} & A stored expansion of a derivative acting on a field & \ref{sec:derivative} \\
\jil{SameLevelDummyError} & Error: a contracted pair sits at the same level & \ref{sec:einstein-summation-and-dummy-indices} \\
\jil{SLAlgebra} & The special linear algebra & \ref{sec:indexset} \\
\jil{SOAlgebra} & The special orthogonal algebra & \ref{sec:indexset} \\
\jil{SpAlgebra} & The symplectic algebra & \ref{sec:indexset} \\
\jil{Spinor} & Explicitly anticommuting matrix & \ref{sec:spinors} \\
\jil{SUAlgebra} & The special unitary algebra & \ref{sec:indexset} \\
\jil{Superspace} & A superspace: its Grassmann basis and derivatives & \ref{sec:superspace-and-superfields} \\
\jil{SymorNum} & Union of a symbolic and a numeric coefficient & \ref{sec:tensorexpression} \\
\jil{SymplecticForm} & The symplectic invariant form & \ref{sec:indexset} \\
\jil{Tensor} & Basic Tensor object & \ref{sec:tensor} \\
\jil{TensorData} & Mutual tensor data that all tensors should have in common & \ref{sec:tensor} \\
\jil{TensorExpression} & A sum of terms with coefficients & \ref{sec:tensorexpression} \\
\jil{TensorTerm} & A product of tensors & \ref{sec:tensorterm} \\
\jil{Trace} & (Super) Trace operator & \ref{sec:traces} \\
\jil{VariationalDerivative} & Variational Derivative & \ref{sec:variational-calculus} \\
\jil{YoungTableau} & Young Tableau diagram that stores information about symmetries and irreducible representations & \ref{sec:symmetries-and-youngtableau} \\
\jil{ℂ} & The complexes & \ref{sec:tensor} \\
\jil{ℍ} & The quaternions & \ref{sec:tensor} \\
\jil{iℝ} & The imaginary numbers & \ref{sec:tensor} \\
\jil{ℚ} & The rationals & \ref{sec:tensor} \\
\jil{ℝ} & The reals & \ref{sec:tensor} \\
\jilg{$\mathbb{O}$} & The octonions & \ref{sec:tensor} \\
\bottomrule
\end{longtable}

\subsection*{Abstract Types}
\begin{longtable}{@{}p{0.36\linewidth}p{0.49\linewidth}p{0.1\linewidth}@{}}
\toprule
Name & Description & Section \\
\midrule
\endfirsthead
\toprule
Name & Description & Section \\
\midrule
\endhead
\jil{AntimorphismSuperType} & Operators with $O(AB)=O(B)O(A)$, such as transpose or Hermitian conjugation & \ref{sec:dagger-and-complex-conjugation} \\
\jil{AntisymMetricSuperType} & Metrics with an antisymmetric invariant form & \ref{sec:epsilontensor} \\
\jil{BraSuperType} & Dirac Bra vectors & \ref{sec:braket} \\
\jil{CovDerivSuperType} & Covariant derivatives & \ref{sec:derivative} \\
\jil{CyclicSuperType} & Operators with $O(ABC)=O(BCA)$ & \ref{sec:traces} \\
\jil{DaggerSuperType} & Hermitian conjugation specifically & \ref{sec:dagger-and-complex-conjugation} \\
\jil{DerivativeSuperType} & Derivative operators & \ref{sec:typing-hierarchy} \\
\jil{EpsilonTensorSuperType} & Levi-Civita tensors, an antisymmetric form & \ref{sec:epsilontensor} \\
\jil{Gamma5SuperType} & The chirality matrix & \ref{sec:gamma-matrices} \\
\jil{GammaSuperType} & Gamma matrices & \ref{sec:gamma-matrices} \\
\jil{GeneratorDerivSuperType} & Derivatives valued in a Lie algebra generator & \ref{sec:lie-algebra-generators} \\
\jil{HermMetricSuperType} & Hermitian forms, whose two slots carry different index types & --- \\
\jil{HomomorphismSuperType} & Operators with $O(AB)=O(A)O(B)$, such as a determinant & --- \\
\jil{IndexSuperType} & Supertype of the index types & \ref{sec:index} \\
\jil{InvolutionSuperType} & Antimorphisms with $O(O(A))=A$ & \ref{sec:dagger-and-complex-conjugation} \\
\jil{KetSuperType} & Dirac Ket vectors & \ref{sec:braket} \\
\jil{KroneckerSuperType} & Kronecker deltas & \ref{sec:kroneckerdelta} \\
\jil{LeibnizOperatorSuperType} & Operators with the graded Leibniz property & \ref{sec:derivatives} \\
\jil{LieAlgebraSuperType} & Supertype of the Lie algebras & \ref{sec:indexset} \\
\jil{LieBracketSuperType} & Lie brackets, which also obey a derivation property & \ref{sec:lie-brackets} \\
\jil{MatrixSuperType} & Non-commuting matrix-valued objects & \ref{sec:general-matrix-relations} \\
\jil{MetricSuperType} & Supertype of the invariant forms used to raise and lower & \ref{sec:metric} \\
\jil{NumberSet} & Supertype of the number fields a tensor may be defined over & \ref{sec:tensor} \\
\jil{OperatorSuperType} & Supertype of operators acting on an expression & \ref{sec:operators} \\
\jil{PartialSuperType} & Partial derivatives & \ref{sec:partialderivative} \\
\jil{PauliSuperType} & Pauli matrices & \ref{sec:pauli-matrices} \\
\jil{SymMetricSuperType} & Metrics with a symmetric invariant form & \ref{sec:metric} \\
\jil{TensorDataSuperType} & Supertype of the internal information a tensor carries & \ref{sec:tensor} \\
\jil{Tensorial} & Root of the hierarchy, everything tensorial descends from it & \ref{sec:typing-hierarchy} \\
\jil{TensorSuperType} & Supertype of the tensor-like objects & \ref{sec:tensor} \\
\jil{TraceSuperType} & Traces & \ref{sec:traces} \\
\jil{VariationalSuperType} & Variational (functional) derivatives & \ref{sec:variational-calculus} \\
\bottomrule
\end{longtable}

\subsection*{Functions}
\begin{longtable}{@{}p{0.36\linewidth}p{0.49\linewidth}p{0.1\linewidth}@{}}
\toprule
Name & Description & Section \\
\midrule
\endfirsthead
\toprule
Name & Description & Section \\
\midrule
\endhead
\jil{add_commutator!} & Registers the commutator of two matrices & \ref{sec:general-matrix-relations} \\
\jil{add_derivative_action} & Define action of a derivative operator on some basis of tensors & \ref{sec:derivatives} \\
\jil{add_global_property!} & Add something to the list of global properties associated with the environment & \ref{sec:serialisation-and-saving} \\
\jil{antisym_contraction} & Raising/lowering rule for an antisymmetric form & \ref{sec:epsilontensor} \\
\jil{AntisymmetricTensor} & Construct a tensor antisymmetric in its slots & \ref{sec:symmetrictensor-and-antisymmetrictensor} \\
\jil{append} & Alias of \jil{×} & \ref{sec:tensor-appension} \\
\jil{apply_dagger} & Carry out the conjugation a \jil{Dagger} represents, rather than leaving it as an unevaluated operator & \ref{sec:dagger-and-complex-conjugation} \\
\jil{apply_derivative} & Carry out the derivatives left standing in an expression & \ref{sec:derivatives} \\
\jil{canonicalise} & Bring an expression to canonical form & \ref{sec:canonicalise} \\
\jil{canonicalize} & Alias of \jil{canonicalise} & \ref{sec:canonicalise} \\
\jil{clear_commutators!} & Forget every declared commutator & \ref{sec:clearing} \\
\jil{clear_coords!} & Forget every declared coordinate & \ref{sec:clearing} \\
\jil{clear_covariant!} & Forget every declared covariant derivative & \ref{sec:clearing} \\
\jil{clear_global!} & Clears all global registrations, like commutators, sort order, index sets & \ref{sec:clearing} \\
\jil{clear_indexsets!} & Forget every declared index set & \ref{sec:clearing} \\
\jil{clear_sort_order!} & Reset the factor ordering & \ref{sec:clearing} \\
\jil{coef} & The coefficient of a term & \ref{sec:accessors-and-properties} \\
\jil{coef_string} & A coefficient rendered as a string & \ref{sec:display-and-to-latex} \\
\jil{coefficients} & The coefficients of every summand & \ref{sec:accessors-and-properties} \\
\jil{collect_terms} & Collect summands that are equal up to dummy index naming and the ordering of commuting factors & \ref{sec:collect-terms} \\
\jil{commute} & Commute two factors past one another & \ref{sec:lie-algebra-generators} \\
\jil{commute_right} & Move a factor rightwards through a term & \ref{sec:lie-algebra-generators} \\
\jil{cong} & Alias of \jil{≅} & \ref{sec:equality-and-congruence} \\
\jil{conj_convention} & The conjugation convention for indices in a given \jil{IndexSet} & \ref{sec:dagger-and-complex-conjugation} \\
\jil{construct_tensorterm_graph} & Explicitly construct a labeled graph from a tensor term & \ref{sec:hard-simplify} \\
\jil{contract_adjacent} & Moves matrices of the given type to be adjacent, then contracts them when possible using simplification rules & \ref{sec:gamma-matrices} \\
\jil{contract_metrics} & Eliminates explicit metrices from an expression to raise and lower indices & \ref{sec:metric} \\
\jil{contract_sigma} & Contract Pauli or sigma matrix indices & \ref{sec:pauli-matrices} \\
\jil{contraction} & Contract a pair of indices & \ref{sec:interior-product} \\
\jil{coord} & Returns which coordinates a derivative is with respect to & \ref{sec:accessors-and-properties} \\
\jil{custom_sort_order!} & Set a custom ordering for factors & \ref{sec:sort} \\
\jil{data} & Get the data associated to a tensor object & \ref{sec:accessors-and-properties} \\
\jil{derive} & Fully expand, apply derivatives and simplify & \ref{sec:derivative} \\
\jil{det} & Returns the det of a metric associated with this index type & \ref{sec:indexset} \\
\jil{dimension} & The range of an index set, or of an algebra & \ref{sec:indexset} \\
\jil{division_algebra} & The endomorphism algebra of the defining representation, \jilg{ℝ}, \jilg{ℂ} or \jilg{ℍ} & \ref{sec:dagger-and-complex-conjugation} \\
\jil{drop_nilpotent} & Discard every term carrying a higher power of some Grassmann in \jilg{basis} than it can hold & \ref{sec:supersymmetry-and-superspace} \\
\jil{drop_total_derivatives} & Reduce \jilg{exp} modulo total derivatives & \ref{sec:ibp} \\
\jil{drop_total_derivatives_general} & Reduce \jilg{exp} modulo total derivatives at any degree, by constructing the total-derivative subspace of each sector explicitly and reducing against it & \ref{sec:ibp} \\
\jil{drop_weight_above} & Get all summands in a TensorExpression of below a particular weight & \ref{sec:weights} \\
\jil{drop_weight_below} & Get all summands in a TensorExpression of above a particular weight & \ref{sec:weights} \\
\jil{dummies} & Gets the set of repeated summation indices in a Tensor & \ref{sec:accessors-and-properties} \\
\jil{dummies_positional} & Gets the set of repeated summation index pairs in a Tensor & \ref{sec:accessors-and-properties} \\
\jil{EL} & Computes the Euler-Lagrange equations for a given Lagrangian and tensor field & \ref{sec:variational-calculus} \\
\jil{eliminate_epsilon} & Eliminate the Levi-Civita epsilon tensor by contracting indices or taking traces & \ref{sec:epsilontensor} \\
\jil{eliminate_kronecker} & Eliminate Kronecker deltas by contracting indices or taking traces & \ref{sec:kroneckerdelta} \\
\jil{eliminate_matrices} & Eliminate matrices from an expression using matrix product rules & \ref{sec:general-matrix-relations} \\
\jil{eliminate_trace} & Eliminate traces by expanding applying matrix reduction rules & \ref{sec:traces} \\
\jil{epsilon} & The invariant epsilon of an algebra & \ref{sec:dagger-and-complex-conjugation} \\
\jil{expand_commutators} & Expand declared commutators within an expression & \ref{sec:lie-brackets} \\
\jil{expand_derivative} & Replaces a symbolic derivative with its full form, e.g. substituting a covariant D with its connections & \ref{sec:derivative} \\
\jil{explain_structure} & Describe how a Grassmann structure was built & \ref{sec:superspace-and-superfields} \\
\jil{exterior_derivative} & Exterior derivative of a p-form & \ref{sec:ext-deriv} \\
\jil{factor_trace} & Factors out scalars from a trace, leaving only <:MatrixSuperType inside & \ref{sec:traces} \\
\jil{find_weighted_square} & True iff the graph contains a 4-cycle whose consecutive edges carry weights \jilg{-1, 2, 1, 2} & \ref{sec:hard-simplify} \\
\jil{first} & Gets the first argument of a Lie bracket & \ref{sec:lie-brackets} \\
\jil{first_coef} & First numerical/symbolic coefficient in the expression & \ref{sec:accessors-and-properties} \\
\jil{first_summand} & First TensorTerm in the expression & \ref{sec:accessors-and-properties} \\
\jil{flip_all} & Return a copy of \jilg{T} with every free index position reversed, upper becoming lower and lower upper & \ref{sec:flip-all-raise-all-lower-all} \\
\jil{free_indices} & Gets the set of free index/↑↓position pairs attached to an expression & \ref{sec:accessors-and-properties} \\
\jil{free_indices_nopos} & The free indices, ignoring their position & \ref{sec:accessors-and-properties} \\
\jil{fresh_indices} & Rename so that no index of one expression clashes with another & \ref{sec:tensor-appension} \\
\jil{from_JSON} & Deserialise from JSON & \ref{sec:serialisation-and-saving} \\
\jil{function_of} & The coordinates a tensor depends on & \ref{sec:accessors-and-properties} \\
\jil{Gamma} & Creates a gamma matrix or tensor & \ref{sec:gamma-matrices} \\
\jil{generate_antisymmetric_tableau} & The tableau of a fully antisymmetric tensor & \ref{sec:symmetries-and-youngtableau} \\
\jil{generate_conformal_algebra} & Generates the algebra so(d,2) & \ref{sec:lie-algebra-generators} \\
\jil{generate_grassmann_product} & Generates a product of Grassmann variables & \ref{sec:superspace-and-superfields} \\
\jil{generate_independent_structure} & Decomposes Grassmann monomials & \ref{sec:superspace-and-superfields} \\
\jil{generate_riemann_tableau} & The tableau of the Riemann symmetries & \ref{sec:symmetries-and-youngtableau} \\
\jil{generate_superfield} & Expand a superfield, reduced to independent structures & \ref{sec:superspace-and-superfields} \\
\jil{generate_superfield_naive} & Generates a superfield expansion over the specified superspace, without performing any reductions based on symmetries & \ref{sec:superspace-and-superfields} \\
\jil{generate_symmetric_tableau} & The tableau of a fully symmetric tensor & \ref{sec:symmetries-and-youngtableau} \\
\jil{get_commutator} & Gets the result of (anti)commuting A with B & \ref{sec:lie-brackets} \\
\jil{get_global_values} & The registries and rules currently declared & --- \\
\jil{get_index} & Look an index up by name within a set & \ref{sec:indexset} \\
\jil{get_index_sets} & Gets the set of IndexSets associated to a tensorial object & --- \\
\jil{get_sort_order} & The ordering currently in force & \ref{sec:sort} \\
\jil{get_unused_indices} & Indices of a set not yet used in an expression & \ref{sec:accessors-and-properties} \\
\jil{get_weighted_terms} & Get all summands in a TensorExpression of a particular weight & \ref{sec:weights} \\
\jil{grading} & $\mathbb{Z}_2$ grading. Returns 0 if the object is bosonic, 1 if fermionic & \ref{sec:accessors-and-properties} \\
\jil{grassmann_coefficient} & The coefficient of Grassmann monomial in a superfield expansion & \ref{sec:superspace-projection} \\
\jil{grassmann_order} & The highest power of the Grassmann that can be non-zero, namely its number of independent components & \ref{sec:superspace-and-superfields} \\
\jil{hard_simplify} & Merge terms up to symmetry and dummy relabelling, more aggressively than \jilg{simplify} & \ref{sec:hard-simplify} \\
\jil{has_commutation_rel} & Whether a commutation relation has been registered for \jilg{a} and \jilg{b} & \ref{sec:lie-brackets} \\
\jil{hasIndices} & Whether the object carries any indices & \ref{sec:accessors-and-properties} \\
\jil{in_default_pos} & Whether an index sits at its set's default position & \ref{sec:flip-all-raise-all-lower-all} \\
\jil{independent_structures} & All independent Grassmann structures at degree n & \ref{sec:superspace-and-superfields} \\
\jil{indices} & Gets the vector of index/position pairs attached to an expression & \ref{sec:accessors-and-properties} \\
\jil{integrate} & Perform superspace integration & \ref{sec:superspace-integration} \\
\jil{is_const} & Whether the expression carries no indices or coordinates & \ref{sec:accessors-and-properties} \\
\jil{is_riemann_metric} & Returns true when the metric has two indices in the same level & \ref{sec:metric} \\
\jil{isAntisymmetric} & Checks a tensor's Young Tableau to see if it is a totally antisymmetric tensor & \ref{sec:symmetries-and-youngtableau} \\
\jil{isbarred} & Conjugation maps a barred name to its unbarred partner and back, using the same combining macron as the barred index alphabet & \ref{sec:dagger-and-complex-conjugation} \\
\jil{isComplex} & Whether the object is defined over the complexes & \ref{sec:accessors-and-properties} \\
\jil{isForm} & Whether the object is a differential form & \ref{sec:differential-geometry} \\
\jil{isImaginary} & Whether the object is imaginary & \ref{sec:accessors-and-properties} \\
\jil{isReal} & \jilg{ℚ} counts as real & \ref{sec:accessors-and-properties} \\
\jil{isrealfield} & Whether conjugation can stop at this leaf & \ref{sec:accessors-and-properties} \\
\jil{isSymmetric} & Checks a tensor's Young Tableau to see if it is a totally symmetric tensor & \ref{sec:symmetries-and-youngtableau} \\
\jil{isTensor} & Check if a TensorExpression is a single summand of one tensor with a 1 coefficient & \ref{sec:accessors-and-properties} \\
\jil{isTensorTerm} & Returns false, or the TensorTerm & \ref{sec:accessors-and-properties} \\
\jil{isZeroTens} & Whether the expression is identically zero & \ref{sec:accessors-and-properties} \\
\jil{labels} & The labels carried by a tensor & \ref{sec:accessors-and-properties} \\
\jil{load} & Loads Alakazam objects from ``filename''.json & \ref{sec:serialisation-and-saving} \\
\jil{load_defaults!} & Load default package commutators and other properties to restore to a clean session & \ref{sec:clearing} \\
\jil{load_gamma_defaults!} & Register the parts of the gamma algebra that hold in every dimension and for every index set & \ref{sec:clearing} \\
\jil{load_header} & Read back declared sets, algebras and rules & \ref{sec:serialisation-and-saving} \\
\jil{lower_all} & Return a copy of \jilg{T} with every free index moved to the lower position; the counterpart of \jil{raise_all} & \ref{sec:flip-all-raise-all-lower-all} \\
\jil{metric} & Returns the metric-like invariant tensor for a given algebra & \ref{sec:indexset} \\
\jil{multiterm_reduce} & Reduce using multi-term identities, within a budget & \ref{sec:multiterm} \\
\jil{name} & Return the name String of a tensor & \ref{sec:accessors-and-properties} \\
\jil{numberfield} & Which number field the components live in & \ref{sec:accessors-and-properties} \\
\jil{op_data} & The payload of an operator & \ref{sec:operators} \\
\jil{op_function_of} & The coordinates an operator depends on & \ref{sec:operators}  \\
\jil{op_grading} & Grading of the operator, not including the operand & \ref{sec:operators}  \\
\jil{op_indices} & The indices of an operator & \ref{sec:operators}  \\
\jil{op_labels} & The labels of an operator & \ref{sec:operators}  \\
\jil{op_pos} & Reverse the positions of the free indices of an expression; an alias for \jil{flip_all} & \ref{sec:flip-all-raise-all-lower-all} \\
\jil{op_tableaux} & The tableaux of an operator & \ref{sec:operators}  \\
\jil{op_weights} & The weights of an operator & \ref{sec:operators}  \\
\jil{operand} & The expression an operator acts on & \ref{sec:operators} \\
\jil{operand_indices} & The indices carried by that operand & \ref{sec:operators}  \\
\jil{parse_latex} & Read TeX markup held in a \jilg{String} at run time, as @LT\_str does at parse time & \ref{sec:macros} \\
\jil{Pauli} & Creates a Pauli matrix or tensor & \ref{sec:pauli-matrices} \\
\jil{pop_metric} & ``Pop'' metrics off all tensors, putting indices in default position and returning the popped metric as a prefactor & \ref{sec:metric} \\
\jil{product_rule} & Perform the graded Leibniz product rule for all derivatives in the expression & \ref{sec:derivatives} \\
\jil{project_to_component} & Partially integrates over superspace to project to the desired component & \ref{sec:superspace-projection} \\
\jil{raise_all} & Return a copy of \jilg{T} with every free index moved to the upper position & \ref{sec:flip-all-raise-all-lower-all} \\
\jil{rebuild_nested} & Rebuild a nested chain of operators over \jilg{root}, \jilg{ops} given outermost-first & \ref{sec:ibp} \\
\jil{reduce_bp_with} & Reduce \jilg{exp} modulo total derivatives while applying replacement pattern \jilg{rules}, iterating until neither step changes anything & \ref{sec:reducing-expressions-replacements-and} \\
\jil{reduce_matrix} & Generic reduce matrix function, can be overloaded to define new identities & \ref{sec:general-matrix-relations} \\
\jil{reduce_trace} & Generic trace relation, overloaded by \jilg{@mrule} when its left-hand side is a \jilg{Trace} & \ref{sec:traces} \\
\jil{remove_commutator!} & Remove one declared commutator & \ref{sec:clearing} \\
\jil{rename_dummies} & Rename dummy Einstein summation indices to be more lexicographically uniform across expressions & \ref{sec:rename-dummies} \\
\jil{rename_dummies!} & Rename dummy pairs in place, lexicographically & \ref{sec:rename-dummies} \\
\jil{render_compact} & Compact expansion at degree n & \ref{sec:superspace-and-superfields} \\
\jil{render_expression} & Fully explicit rendering of one structure & \ref{sec:superspace-and-superfields} \\
\jil{replace_pattern} & Find \jilg{pattern} in \jilg{exp} and substitute \jilg{replace} for it, repeating until no match remains unless \jilg{repeat} is false & \ref{sec:pattern-matching-basic-usage} \\
\jil{Riemann} & Alias for \jil{Riemann} & \ref{sec:riemann-tensor} \\
\jil{RiemannTensor} & Construct a tensor with the symmetries of the Riemann tensor & \ref{sec:riemann-tensor} \\
\jil{safe_pop!} & Gets an unused index from a set, or generate a new index if we've run out & \ref{sec:indexset} \\
\jil{same_tensorterm_mod_syms_grade} & Are tensors equal, up to the graded sign flip to turn T2 into T1 & \ref{sec:equality-and-congruence} \\
\jil{save} & Save an Alakazam object to "filename".json & \ref{sec:serialisation-and-saving} \\
\jil{save_header} & Write the declared sets, algebras and rules & \ref{sec:serialisation-and-saving} \\
\jil{second} & The second argument of a Lie bracket & \ref{sec:lie-brackets} \\
\jil{set_index!} & Sets the index on a vector in place & \ref{sec:accessors-and-properties} \\
\jil{set_indices!} & Only updates the external indices, not those of the underlying tensors & \ref{sec:accessors-and-properties} \\
\jil{set_indices_deep!} & Overwrite the indices in a tensorterm according to ref dict & \ref{sec:accessors-and-properties} \\
\jil{set_name!} & Rename a tensor & \ref{sec:accessors-and-properties} \\
\jil{set_op_indices!} & Replace an operator's indices & \ref{sec:accessors-and-properties} \\
\jil{set_operand!} & Replace an operator's operand & \ref{sec:accessors-and-properties} \\
\jil{set_operand_deep!} & Replace the operand throughout a nested operator & \ref{sec:accessors-and-properties} \\
\jil{show_function_of} & Whether coordinate dependence is printed & --- \\
\jil{simp_ops_n_sym} & Simplify operators together with symmetries & \ref{sec:simplify} \\
\jil{simplify} & Hand-selected series of operations to reduce an expression & \ref{sec:simplify} \\
\jil{singleterm} & Check if a TensorExpression is a single TensorTerm & \ref{sec:accessors-and-properties} \\
\jil{solve_linear} & Basic function to solve the linear equation \jilg{eqn = 0} for \jilg{target} & \ref{sec:solving-equations} \\
\jil{spacetime_dimension} & Spacetime dimension a spin algebra belongs to, inverting Spin for the \jilg{SL(2,$\mathbb{F}$)} family & \ref{sec:indexset} \\
\jil{Spin} & \jil{Spin(d)} $\to$ \jil{SpinAlgebra(1,d-1)}, or \jil{Spin(p,q)} the spin algebra of a space with \jilg{p} timelike and \jilg{q} spacelike directions, for any signature & \ref{sec:indexset} \\
\jil{spinor_dimension} & The number of components a Dirac spinor has over the spacetime \jilg{st}, namely \jilg{2\textasciicircum{}⌊d/2⌋} & \ref{sec:indexset} \\
\jil{spinor_indexset} & A helper indexset & \ref{sec:indexset} \\
\jil{star} & Take the Hodge star of a form & \ref{sec:hodge-star} \\
\jil{strip_derivatives} & Similar to \jil{root_operand}, but returns the operator tree and fails if not deriv of a single tensor & \ref{sec:ibp} \\
\jil{summands} & The summands of an expression & \ref{sec:accessors-and-properties} \\
\jil{swap} & Signed flip of the arguments of the Lie bracket & \ref{sec:lie-brackets} \\
\jil{SymmetricTensor} & Construct a tensor symmetric in its slots & \ref{sec:symmetrictensor-and-antisymmetrictensor} \\
\jil{symmetrise_partials} & Perform a symmetrisation of nested partial derivatives & \ref{sec:simplify} \\
\jil{tableaux} & The Young tableaux carried by a tensor & \ref{sec:accessors-and-properties} \\
\jil{term_weight} & Extract the overall weight value from a TensorTerm for a given weight label & \ref{sec:weights} \\
\jil{terms} & Vector of tensors inside TensorTerm & \ref{sec:accessors-and-properties} \\
\jil{to_dumless_same_tensor_mod_syms} & Whether two tensors agree up to dummies and symmetry & \ref{sec:hard-simplify} \\
\jil{to_JSON} & Serialise to JSON & \ref{sec:serialisation-and-saving} \\
\jil{to_latex} & Render an expression as TeX & \ref{sec:display-and-to-latex} \\
\jil{togglebar} & Add or remove the bar decorating a name & \ref{sec:dagger-and-complex-conjugation} \\
\jil{update_indices!} & Updates external indices of a tensor term, expression, or operator, by looking at the indices of all the terms it contains & \ref{sec:accessors-and-properties} \\
\jil{wedge} & Take the wedge exterior product between two forms & \ref{sec:wedge-product} \\
\jil{weights} & The weights carried by a tensor & \ref{sec:weights} \\
\jil{young_projection} & Apply the Young projector of a tableau & \ref{sec:symmetries-and-youngtableau} \\
\jil{young_projector_size} & Number of terms the projector of a tableau expands to & \ref{sec:multiterm} \\
\jil{×} & Appension, product of tensors with fresh indices & \ref{sec:tensor-appension} \\
\jil{ι} & Interior product (contraction) of a vector with a differential form & \ref{sec:interior-product} \\
\jil{∧} & Take the wedge exterior product between two forms & \ref{sec:wedge-product} \\
\jil{∫} & Alias of \jil{integrate} & \ref{sec:superspace-integration} \\
\jil{≅} & \jilg{≅} is the deep equality test, so it opts into the multi-term (Young projector) comparison that \jil{hard_simplify} leaves off by default & \ref{sec:equality-and-congruence} \\
\jil{⋆} & Take the Hodge star of a form & \ref{sec:hodge-star} \\
\jil{⨼} & Interior product (contraction) of a vector with a differential form & \ref{sec:interior-product} \\
\jilg{\sfd} & Alias of \jil{exterior_derivative} & \ref{sec:ext-deriv} \\
\bottomrule
\end{longtable}

\subsection*{Macros}
\begin{longtable}{@{}p{0.36\linewidth}p{0.49\linewidth}p{0.1\linewidth}@{}}
\toprule
Name & Description & Section \\
\midrule
\endfirsthead
\toprule
Name & Description & Section \\
\midrule
\endhead
\jil{@action} & Declare how a derivative acts on a given tensor & \ref{sec:partialderivative} \\
\jil{@anticommutator} & Adds an anticommutation relation \{A,B\}=C & \ref{sec:general-matrix-relations} \\
\jil{@commutator} & Adds a commutation relation [A,B]=C & \ref{sec:general-matrix-relations} \\
\jil{@conjugates} & Sets two \jil{IndexSet}s as conjugates, or one as self-conjugate & \ref{sec:dagger-and-complex-conjugation} \\
\jil{@covariant} & Declare a covariant derivative and its connection & \ref{sec:derivative} \\
\jil{@indices} & Declare a range of indices in \jilg{set} and bind each to a variable of the same name & \ref{sec:index} \\
\jil{@Lie} & Recursively rewrites nested [] / \{\} into Lie(A, B, grading) calls & \ref{sec:lie-brackets} \\
\jil{@load} & Convenience macro for importing from JSON & \ref{sec:serialisation-and-saving} \\
\jil{@LT_str} & Read genuine TeX markup as a tensor & \ref{sec:macros} \\
\jil{@matrix} & Macro to define a new \jil{MatrixType<:MatrixSuperType}, with basic constructors & \ref{sec:general-matrix-relations} \\
\jil{@mrule} & Macro to generate a function \jil{reduce_matrix} that reduces a product of matrices & \ref{sec:general-matrix-relations} \\
\jil{@syms} & Re-exported from SymbolicUtils, for symbolic coefficients & --- \\
\jil{@T_str} & Build a tensor from a string, the full parser & \ref{sec:macros} \\
\jil{@Tensor} & Build a simple tensor & \ref{sec:tensor} \\
\bottomrule
\end{longtable}

\subsection*{Constants}
\begin{longtable}{@{}p{0.36\linewidth}p{0.49\linewidth}p{0.1\linewidth}@{}}
\toprule
Name & Description & Section \\
\midrule
\endfirsthead
\toprule
Name & Description & Section \\
\midrule
\endhead
\jil{bosonic} & Grading of a commuting index & \ref{sec:indexset} \\
\jil{COORD_REGISTRY} & Every declared Coordinate, by name & \ref{sec:clearing} \\
\jil{derivative_actions} & Registered actions of derivatives on particular tensors & \ref{sec:derivatives} \\
\jil{derivative_algebra} & Registered commutators among the derivatives & \ref{sec:derivative} \\
\jil{fermionic} & Grading of an anticommuting index & \ref{sec:indexset} \\
\jil{function_of_on} & Should we print the \jil{function_of} coordinates? Includes both \jil{@show} and \jil{to_latex} printing & \ref{sec:clearing} \\
\jil{lower} & Index position: covariant & \ref{sec:setup} \\
\jil{SET_REGISTRY} & Every declared IndexSet, by name & \ref{sec:clearing} \\
\jil{sort_order} & The ordering used when sorting factors within a term & \ref{sec:sort} \\
\jil{upper} & Index position: contravariant & \ref{sec:setup} \\
\jil{↑} & Alias of \jil{upper} & \ref{sec:setup} \\
\jil{↓} & Alias of \jil{lower} & \ref{sec:setup} \\
\jil{✝} & Postfix conjugation, equivalent to postfix \jil{'} & \ref{sec:dagger-and-complex-conjugation} \\
\bottomrule
\end{longtable}

\printbibliography

\end{document}